\documentstyle[epsf,epsf]{aipproc2}
\input psfig.sty

\def\wtil{\widetilde}
\def\bit{\begin{itemize}}
\def\eit{\end{itemize}}
\def\chitil{\wt\chi}
\def\cnone{\wt\chi^0_1}
\def\cntwo{\wt\chi^0_2}
\def\cnthree{\wt\chi^0_3}
\def\cnfour{\wt\chi^0_4}
\def\snu{\wt\nu}
\def\snul{\wt\nu_L}
\def\msnul{m_{\snul}}

\def\snue{\wt\nu_e}
\def\snuel{\wt\nu_{e\,L}}
\def\msnuel{m_{\snul}}

\def\snubar{\ov{\snu}}
\def\msnu{m_{\snu}}
\def\mcnone{m_{\cnone}}
\def\mcntwo{m_{\cntwo}}
\def\mcnthree{m_{\cnthree}}
\def\mcnfour{m_{\cnfour}}
\def\h{h}
\def\mh{m_{\h}}
\def\wt{\widetilde}
\def\cpone{\wt \chi^+_1}
\def\cmone{\wt \chi^-_1}
\def\cpmone{\wt \chi^{\pm}_1}
\def\mcpone{m_{\cpone}}
\def\mcpmone{m_{\cpmone}}

\def\cptwo{\wt \chi^+_2}
\def\cmtwo{\wt \chi^-_2}
\def\cpmtwo{\wt \chi^{\pm}_2}
\def\mcptwo{m_{\cptwo}}
\def\mcpmtwo{m_{\cpmtwo}}

\def\stauone{\wt \tau_1}
\def\mstauone{m_{\stauone}}

\def\dmchi{\Delta m_{\wtil\chi}}
\def\stau{\wtil\tau}

\def\runfig#1#2#3{
\begin{figure}
\centerline{\protect\psfig{file=#2,width=10cm}}   
\caption{#3}
\label{#1}
\end{figure}}

\def\runfigd#1#2#3{
\begin{figure}
\centerline{\protect\psfig{file=#2,width=7cm}}   
\caption{#3}
\label{#1}
\end{figure}}

\def\rotaterunfig#1#2#3{
\begin{figure}
\centerline{
\psfig{figure=#2,height=12cm,width=17cm,angle=-90}
}   
\caption[]{\label{#1} #3}
\end{figure}}

\begin{document}

\title{Report of the Beyond the MSSM Subgroup 
for the Tevatron Run II SUSY/Higgs Workshop}

\author{
S.~Ambrosanio$^{1,2}$,
H.~Baer$^{15}$,
A.~Brignole$^3$,
A.~Castro$^{3}$,
M.~Chertok$^{**,7}$,
K.~Cheung$^4$,
L.~Clavelli$^6$,
D.~Cutts$^8$,
M.~Cveti\v c$^5$,
D.~Dooling$^8$,
H. Dreiner$^{**,17}$,
B.~Dutta$^9$,
U.~Ellwanger$^{11}$,
L.~Everett$^{12}$,
F.~Feruglio$^{3}$,
G.F.~Giudice$^1$,
J.~F. Gunion$^{**,***,4}$,
J.L.~Hewett$^{10}$,
C.~Hugonie$^{11}$,
K.~Kang$^8$,
S.K.~Kang$^8$,
G.~Landsberg$^{**,8}$,
P.~Langacker$^5$, 
M.~Mangano$^1$,
D.~McKay$^{14}$,
R.N.~Mohapatra$^{13}$,
S.~Mrenna$^{4}$,
D.J.~Muller$^9$,
R.~Rattazzi$^{16}$,
T.~Rizzo$^{10}$,
J.W.~Wang$^5$,
J.D. Wells$^{**,4}$,
F.~Zwirner$^3$}
\address{
$^{***}$ EDITOR \\
$^{**}$ BTMSSM CONVENORS \\
$^1$ CERN, Theory Division, Geneva, Switzerland \\
$^2$ Deutsches Elektronen-Synchrotron, DESY, Hamburg, Germany \\
$^3$ INFN, Padova and 
        Universit\` a di Padova, Padua, Italy \\
$^4$ Department of Physics, University of California, Davis, California \\
$^5$ University of Pennsylvania,   Philadelphia, Pennsylvania \\
$^6$ University of Alabama,    Tuscaloosa, Alabama \\
$^7$ Texas A\& M University, College Station, Texas \\
$^8$ Brown University, Providence, Rhode Island \\
$^9$ Oklahoma State University, Stillwater, Oklahoma  \\
$^{10}$ Stanford Linear Accelerator Center, Stanford, California \\
$^{11}$ Universit\' e de Paris XI, Orsay, France \\
$^{12}$ University of Michigan, Ann Arbor, Michigan \\
$^{13}$ University of Maryland, College Park, Maryland \\
$^{14}$ University of Kansas, Lawrence, Kansas \\
$^{15}$ Florida State University, Tallahassee, Florida \\
$^{16}$ INFN and Scuola Normale Superiore, Pisa, Italy \\
$^{17}$ Rutherford Appleton Laboratory, Chilton, Didcot, Great Britain} 

\maketitle

\def\gtino{\widetilde G}
\def\mgtino{m_{\gtino}}
\def\mz{m_Z}
\def\mgut{M_U}
\def\tanb{\tan\beta}
\def\vev#1{\langle #1 \rangle}

\def\beq{\begin{equation}}
\def\eeq{\end{equation}}
\def\eq{\end{equation}}
\def\to{\rightarrow}
\def\gsim{\lower.7ex\hbox{$\;\stackrel{\textstyle>}{\sim}\;$}}
\def\lsim{\lower.7ex\hbox{$\;\stackrel{\textstyle<}{\sim}\;$}}
\def\mev{\hbox{\rm\,MeV}}            
\def\gev{\hbox{\rm\,GeV}}            
\def\tev{\hbox{\rm\,TeV}}
\def\agt{\stackrel{>}{\sim}}
\def\alt{\stackrel{<}{\sim}}
\def\order#1{{\cal O}(#1)}
\def\EPC#1#2#3{Eur. Phys. J. C {\bf #1}, #3 (19#2)}
\def\NPB#1#2#3{Nucl. Phys. B {\bf #1}, #3 (19#2)}
\def\PLB#1#2#3{Phys. Lett. B {\bf #1}, #3 (19#2)}
\def\PLBold#1#2#3{Phys. Lett. B {\bf#1} (19#2) #3}
\def\PRD#1#2#3{Phys. Rev. D {\bf #1}, #3 (19#2)}
\def\PRL#1#2#3{Phys. Rev. Lett. {\bf#1}, #3 (19#2)}
\def\PRT#1#2#3{Phys. Rep. {\bf#1} (19#2) #3}
\def\ARAA#1#2#3{Ann. Rev. Astron. Astrophys. {\bf#1} (19#2) #3}
\def\ARNP#1#2#3{Ann. Rev. Nucl. Part. Sci. {\bf#1} (19#2) #3}
\def\MODA#1#2#3{Mod. Phys. Lett. A {\bf #1} (19#2) #3}
\def\ZPC#1#2#3{Zeit. f\"ur Physik C {\bf #1}, #3 (19#2)}
\def\APJ#1#2#3{Ap. J. {\bf#1} (19#2) #3}
\def\MPL#1#2#3{Mod. Phys. Lett. A {\bf #1} (19#2) #3}
%
%
\def\beq{\begin{equation}}
\def\eeq{\end{equation}}
\def\bea{\begin{eqnarray}}
\def\eea{\end{eqnarray}}
%
%
\def\slashchar#1{\setbox0=\hbox{$#1$}           
   \dimen0=\wd0                                 
   \setbox1=\hbox{/} \dimen1=\wd1               
   \ifdim\dimen0>\dimen1                        
      \rlap{\hbox to \dimen0{\hfil/\hfil}}      
      #1                                        
   \else                                        
      \rlap{\hbox to \dimen1{\hfil$#1$\hfil}}   
      /                                         
   \fi}                                         %
%
%
\catcode`@=11
\long\def\@caption#1[#2]#3{\par\addcontentsline{\csname
  ext@#1\endcsname}{#1}{\protect\numberline{\csname
  the#1\endcsname}{\ignorespaces #2}}\begingroup
    \small
    \@parboxrestore
    \@makecaption{\csname fnum@#1\endcsname}{\ignorespaces #3}\par
  \endgroup}
\catcode`@=12
\def\jfig#1#2#3{
 \begin{figure}
 \centering
 \epsfysize=3.0in
 \hspace*{0in}
 \epsffile{#2}
 \caption{#3}
 \label{#1}
 \end{figure}}
\def\sfig#1#2#3{
 \begin{figure}
 \centering
 \epsfysize=2.5in
 \hspace*{0in}
 \epsffile{#2}
 \caption{#3}
 \label{#1}
 \end{figure}}

\begin{abstract}

There are many low-energy models of supersymmetry breaking parameters
which are motivated by theoretical and experimental considerations.
Some of these approaches have gained more proponents than others over
time, and so have been studied in greater detail.  In this contribution
we discuss some of the lesser-known theories of low-energy supersymmetry,
and outline their phenomenological consequences.  In some cases, these
theories have more gauge symmetry or particle content than the 
Minimal Supersymmetric Standard Model.  In other cases, the parameters
of the Lagrangian are unusual compared to commonly accepted norms
(e.g., Wino LSP, heavy gluino LSP, light gluino, etc.). 
The phenomenology of supersymmetry varies greatly between
the different models. Correspondingly, particular aspects of the
detectors assume greater or lesser importance. Detection of supersymmetry
and the determination of all parameters
may well depend upon having the widest possible view of supersymmetry
phenomenology.

\end{abstract}

\def\lsim{\;\raise0.3ex\hbox{$<$\kern-0.75em\raise-1.1ex\hbox{$\sim$}}\;}
\def\gsim{\;\raise0.3ex\hbox{$>$\kern-0.75em\raise-1.1ex\hbox{$\sim$}}\;}
\def\ZZ{\hbox{\it Z\hskip -4.pt Z}} 
\def\ba{\begin{array}}
\def\ea{\end{array}} 

\section{Introduction}

\medskip

Most of the studies performed to assess the discovery reach for supersymmetry 
and most of the current limits on the masses of supersymmetric particles
have been obtained assuming R-parity conservation,
the minimal matter content of the
Minimal Supersymmetric Model (MSSM), and universal
boundary conditions at $\mgut$ for the soft-SUSY-breaking parameters:
$m_0$ for the scalar masses; $M_0$ for the SU(3),SU(2) and U(1)
gaugino masses $M_{3,2,1}$; and $A_0$ for the tri-linear scalar
field couplings. Additional parameters of the MSSM include: $\tanb$,
the ratio of Higgs field vacuum expectation values $\vev{H_u}/\vev{H_d}$; 
$\mu$, the coefficient of the bilinear $\widehat H_u \widehat H_d$ superpotential
term; and $B$, which specifies the strength of the corresponding
$H_uH_d$ scalar field mixing term.  By requiring correct electroweak
symmetry breaking after evolution down to the scale $\mz$, the magnitude
of $\mu$ is fixed and only its sign remains undetermined.
This boundary condition scenario is often referred to as the mSUGRA
or CMSSM model.
If unification of the $\tau$ and $b$ Yukawa couplings at $\mgut$
is required, then correct EWSB strongly constrains $\tanb$ as well.

\renewcommand{\arraystretch}{1.3}
\setlength{\tabcolsep}{0.01in}
\begin{table}[htb]
\centering
\caption{Parameter count in the SM and MSSM extensions: BV=baryon-number
violating; LV=lepton-number violating; BLV=both B and L violating.}
\vskip6pt
\begin{tabular}{|c|c|c|c|}
\hline
          & masses and       & CP-violating  & \\[-5pt]
model & mixing angles& phases & TOTAL \\
\hline
Standard Model & 17& 2 & 19 \\
\phantom{(}MSSM\phantom{)$_{\rm RPV}$} & 79 & 45 &124 \\(MSSM)$_{\rm BV\phantom{L}}$ & 97 & 62 & 159 \\
(MSSM)$_{\rm LV\phantom{B}}$ & 157 & 122 &279 \\
(MSSM)$_{\rm BLV}$ & 175 & 140 & 315 \\
\hline
\label{paramcount}
\end{tabular}
\end{table}

While the matter content and boundary conditions 
of the MSSM have the virtue of simplicity
and can be reasonably motivated in the context of 
several typ{es of gravity-mediated supersymmetry
breaking, other possibilities should certainly be considered.
For a model with MSSM matter content and R-parity conservation,
the most general form of soft-SUSY-breaking allows for
a total of 124 parameters (not counting certain additional parameters
expected to be suppressed by $1/\mgut$ factors). 
This is to be compared to the 19 parameters of the Standard Model.
The most general form of R-parity violation increases
the parameter count to 315. Some of these parameters are associated
with phases and CP violation. A summary appears in Table~\ref{paramcount}.
A final parameter is the mass of the gravitino, $\gtino$.  
If the scale of supersymmetry breaking is sufficiently
low, $\mgtino$ can be small enough that it is the LSP.  In particular,
this is the expectation in models where supersymmetry breaking
is mediated by gauge interactions.

Although almost all of the most general 
parameter space is excluded by various phenomenological
constraints (no FCNC, proton stability, small EDM's, etc.),
there are sub-spaces that differ drastically from mSUGRA while
maintaining consistency with all existing data. 
Examples include models with
R-parity violation and the standard GMSB phenomenology, both of which
will be covered in separate
reports.  We also do not consider non-zero phases. 
Aside from a discussion of the phenomenology of an extremely light $\gtino$,
our focus will be on models with a heavy $\gtino$
and soft-SUSY-breaking parameters that conserve R-parity and CP.
Even with these restrictions, there are many well-motivated 
theories with MSSM matter content that yield vastly different
phenomenology than the mSUGRA model.  In addition, we will consider
a number of models in which the matter/gauge content of the MSSM
is extended. We will focus in particular on implications for supersymmetry
discovery and study at the upgraded Tevatron.

We now give brief motivation and an introduction to the models considered.
\bit
\item
One of the least satisfactory features of the MSSM is the ad hoc
nature of the parameter $\mu$, which a priori is most naturally
of order $\mgut$, but which is expected to be $\lsim 1\tev$ for
natural EWSB. The addition of a singlet superfield $\widehat S$
provides a very compelling and natural origin
of the $\mu \widehat H_u\widehat H_d$ superpotential term. Such a term arises
if the scalar component of $\widehat S$ ($S$) 
acquires a vacuum expectation value. The result is 
an effective superpotential interaction of the form
$\vev{S}\widehat H_u\widehat H_d$. The $S$ quantum degrees of freedom
result in 1 CP-even and 1 CP-odd Higgs bosons beyond 
the 2 CP-even and 1 CP-odd Higgs bosons of the MSSM.
The spin-1/2 component of $\widehat S$ provides
an additional neutralino, $\widetilde S$,
that can mix with the usual four neutralinos of the MSSM. It is
very natural for the LSP of this model to be the $\widetilde S$.
All supersymmetric particles then cascade decay down to the $\widetilde S$.
The phenomenology of SUSY detection is then significantly altered
compared to mSUGRA. A review of this phenomenology is given in 
Sec.~\ref{section:Ellwanger}.

\item
In mSUGRA, the LSP is essentially always a light bino-like neutralino, 
$\cnone\sim \widetilde B$. However, there is substantial motivation
for the possibility that the LSP is a massive gluino. 
This occurs if $M_3\ll M_{1,2}$, as is possible in several well-motivated
SUSY-breaking scenarios. Current limits
on a heavy gluino LSP are summarized and discovery prospects
are discussed in Sec.~\ref{section:Baer}.

\item
Another alternative arrangement of the gaugino masses that arises
in string and brane models is $M_2<M_1<M_3$. In this case,
the LSP $\cnone$ is wino-like and is highly degenerate with the 
lightest chargino. (This assumes $|\mu|$ is large, as usually
implied by RGE EWSB.)
The resulting phenomenology differs greatly from mSUGRA phenomenology.
If $\dmchi\equiv \mcpmone-\mcnone$ is not too much larger than $m_\pi$,
striking background-free signals for $\cpone\cmone+\cpmone\cnone$
production will be present. For $\dmchi\gsim 300\mev$, detection
of these processes will be very difficult; one will have to
hope that other SUSY particles are light. Section~\ref{section:Mrenna}
gives a discussion of the phenomenology and of some very critical detector
issues and related discovery strategies.

\item
Although the scalar masses have the universal value $m_0$ at the high scale,
significant flavor violation can arise via RGE evolution
if this scale is not the same as $\mgut$. More generally, FCNC will be
a problem unless the SUSY breaking mechanism yields either universality
for the scalar masses or flavor alignment.  An interesting exception to
this statement is the possibility that the scalar masses for all the
sfermions of the first two generations are simply extremely heavy ($>10\tev$)
and, thus, have greatly suppressed FCNC effects. Very heavy scalars
are also helpful for unifying with greater precision the
strong coupling constant with the SU(2) and U(1) couplings
for the somewhat `low' value, $\alpha_s(\mz)\sim 0.12$, preferred
by existing data. Of course, to maintain naturalness for the Higgs sector
the 3rd generation squarks should be below $\sim 1\tev$. This scenario
is sometimes referred to as Superheavy Supersymmetry or
More Minimal Supersymmetry. Section~\ref{section:Wells}
reviews some theoretical issues and constraints on this scenario,
including the apparent necessity to have GMSB-like boundary conditions
in order to preserve anomaly cancellations.

\item
As noted earlier, the mass of the $\gtino$ is another crucial
parameter of supersymmetry.  If $\mgtino$ is very small, the couplings
of the $\gtino$ are sufficiently large that processes
in which the $\gtino$ is directly produced have observable rates.
Since the $\gtino$ is undetectable, the most basic signature is
jets plus missing energy. If these processes are detected,
they provide a measurement of the scale of supersymmetry breaking,
possibly the most important parameter of supersymmetry.
The phenomenology and discovery prospects for direct production
of a very light $\gtino$ are reviewed in Sec.~\ref{section:Brignole}. 

\item
An interesting question is whether superstring theory provides any
guidance as regards boundary conditions and matter content for low
energy supersymmetry.
The detailed predictions of one sample superstring model are outlined
in Sec.~\ref{section:Cvetic}. The model considered has a plethora
of additional matter, including exotics, extra Higgs bosons, and extra
gauge bosons. This provides further warning against being complacent
in our approach to SUSY phenomenology.

\item
The possibility that left-right symmetry is restored at a high energy scale
is very attractive.  In LR-symmetric models, proper symmetry breaking
requires introduction of triplet Higgs representations
that contain a doubly-charged Higgs field.   In the supersymmetric
context, these doubly-charged scalars have a doubly-charged higgsino
partner.  Careful investigation reveals that these are likely
to be one of the lightest states in the superparticle spectrum.
They will appear in cascade decays and can also be directly produced.
The phenomenology of the doubly-charged higgsino states is reviewed
in Sec.~\ref{section:Dutta}.

\item
Recently, the possibility that the compactified extra dimensions
of the string/brane world are large and that the Kaluza-Klein excited
states are within experimental reach has received much attention.
Some of the indirect signals for such extra dimension are reviewed
in Sec.~\ref{section:Hewett}. In addition, Sec.~\ref{section:Brignole}
considers external KK gravitons production, which provides a 
signature similar to that of the very light gravitino through jets plus missing
energy.

\item
It is well-known that supersymmetry predicts a rather low mass
for the lightest CP-even Higgs boson.  Thus, one should ask
whether supersymmetry can be rescued if a sufficiently light Higgs
boson is not discovered. One means for increasing the upper
bound on the light Higgs is to introduce a 4th family. The contraints upon
and implications for the Higgs sector of supersymmetry in a 4-family
model are discussed in Sec.~\ref{section:Dooling}.

\item
Is there room for a 4th family in supersymmetry?  If the Yukawa couplings
associated with the 4th family are to remain perturbative 
in evolving from $\sim \mz$ up to some high scale, one finds that the
leptons and quarks of the 4th family must be quite light.
Experimental constraints are becoming very restrictive. 
In Sec.~\ref{section:McKay}, the current situation is reviewed with
the conclusion that the 4th generation will almost certainly be
either discovered or eliminated as a possibility during Run II
at the Tevatron.

\item
Could the gluino be very light? Remarkably, this scenario cannot yet
be absolutely excluded. In addition, it might explain some detailed
features of Run I jet data at very high $p_T$. Section~\ref{section:Clavelli}
presents the case for a very light gluino.

\item
The D\O~detector has been upgraded dramatically for Run II. It is important
to understand the extent to which it will be able to probe some of
the more exotic supersymmetry scenarios that are discussed here
and elsewhere. Particularly interesting are signals associated with
long-lived charged particles, photons, vertices etc.  The capabilities
of the D\O~detector as regards such exotic phenomena are discussed
in Sec.~\ref{section:Cutts}.

\item
The other major detector at the Tevatron, CDF, has also been upgraded.
Section~\ref{section:Chertok} reviews its capabilities for 4th generation
searches via: looking for a long-lived ($b^\prime$) parent of the $Z$;
looking for prompt $b^\prime\to b Z$ production; and searching for
a long-lived heavy quark or similar object.

\item New gauges bosons are a common feature of supersymmetric
models motivated by string theory.  The ability to detect  
such gauge bosons and to determine their couplings 
during Run II at the Tevatron is considered
in Section~\ref{section:Rizzo}

\item
We end with some brief concluding remarks in
Section~\ref{section:Conclusions}.
\eit

\section{Cascade decays in the NMSSM}
\noindent
\centerline{\large\it U. Ellwanger and C Hugonie}
\medskip
\label{section:Ellwanger}


The NMSSM (Next-to-minimal SSM, or (M+1)SSM) is defined by the addition of
a gauge singlet superfield $S$ to the MSSM. The superpotential $W$ is scale
invariant, i.e. there is no $\mu$-term. Instead, two Yukawa couplings
$\lambda$ and $\kappa$ appear in $W$. Apart from the standard quark and
lepton Yukawa couplings, $W$ is given by
\bea W = \lambda H_1 H_2 S + \frac{1}{3} \kappa S^3 + \ldots \label{sp} \eea
and the corresponding trilinear couplings $A_{\lambda}$ and $A_{\kappa}$
are added to the soft susy breaking terms. The vev of $S$ generates an
effective $\mu$-term with $\mu = \lambda \langle S \rangle $.

The constraint NMSSM (CNMSSM)~\cite{ell1} 
is defined by universal soft susy breaking
gaugino masses $M_0$, scalar masses $m_0^2$ and trilinear couplings
$A_0$ at the GUT scale, and a number of phenomenological constraints:

\noindent
- Consistency of the low energy spectrum and couplings with negative
Higgs and sparticle searches.

\noindent
- In the Higgs sector, the minimum of the effective potential with 
$\langle H_1 \rangle $ and $ \langle H_2 \rangle \neq 0$ has to be 
deeper than any minimum with $ \langle H_1 \rangle $ and/or 
$ \langle H_2 \rangle = 0$. Charge and colour breaking minima induced by
trilinear couplings have to be absent. (However, deeper charge and 
colour breaking minima in "UFB" directions are allowed, since the
decay rate of the physical vacuum into these minima is usually large
compared to the age of the universe~\cite{ell2}.)

Cosmological constraints as the correct amount of dark matter are not
imposed at present. (A possible domain wall problem due to the discrete
$Z_3$ symmetry of the model is assumed to be solved by, e.g., embedding
the $Z_3$ symmetry into a $U(1)$ gauge symmetry at $M_{GUT}$, or by
adding non-renormalisable interactions which break the $Z_3$ symmetry
without spoiling the quantum stability~\cite{ell3}.)

The number of free parameters of the CNMSSM, ($M_{1/2}$, $m_0$, $A_0$,
$\lambda$, $\kappa$ + standard Yukawa couplings), is the same as in the
CMSSM ($M_{1/2}$, $m_0$, $A_0$, $\mu$, $B + \ idem$). The new physical
states in the CNMSSM are one additional neutral Higgs scalar and Higgs
pseudoscalar, respectively, and one additional neutralino. In general
these states mix with the corresponding ones of the MSSM with a mixing
angle proportional to the Yukawa coupling $\lambda$. However, in the
CNMSSM $\lambda$ turns out to be quite small, $\lambda \lsim 0.1$ (and
$\lambda \ll 1$ for most allowed points in the parameter space)~\cite{ell1}
Thus the new physical states are generally almost pure gauge singlets with
very small couplings to the standard sector.


The new states in the Higgs sector can be very light, a few GeV or
less, depending on $\lambda$~\cite{ell4}. 
Due to their small couplings to the $Z$
boson they will escape detection at LEP and elsewhere, i.e. the lightest
``visible'' Higgs boson is possibly the next-to-lightest Higgs of the
NMSSM. The upper limits on the mass of this visible Higgs boson (and its
couplings) are, on the other hand, very close to the ones of the MSSM,
i.e. $\lsim 140$ GeV depending on the stop masses~\cite{ell4}.

The phenomenology of sparticle production in the CNMSSM can differ
considerably from the MSSM, depending on the mass of the additional
state $\tilde S$ in the neutralino sector: If the $\tilde S$ is not the
LSP, it will hardly be produced, and all sparticle decays proceed as in
the MSSM with a LSP in the final state. If, on the other
hand, the $\tilde S$ is the LSP, the sparticle decays will proceed
differently: First, the sparticles will decay into the NLSP, 
because the couplings to the $\tilde S$ are too small. Only
then the NLSP will realize that it is not the true LSP, and decay
into the $\tilde S$ plus an additional cascade. 

The condition for a singlino LSP scenario can be expressed relatively
easily in terms of the bare parameters of the CNMSSM: Within the allowed 
parameter space of the CNMSSM, the lightest non-singlet neutralino is 
essentially a bino $\tilde B$. Since the masses of $\tilde S$ and $\tilde B$
are proportional to $A_0$ and $M_{1/2}$, respectively, one finds,
to a good approximation, that the $\tilde S$ is the true LSP if the bare
susy breaking parameters satisfy $|A_0| \lsim 0.4 M_{1/2}$. Since $A_0^2
\gsim 9 m_0^2$ is also a necessary condition within the CNMSSM, the
singlino LSP scenario corresponds essentially to the case where the
gaugino masses are the dominant soft susy breaking terms.

Note, however, that the $\tilde B$ is not necessarily the NLSP in this case:
Possibly the lightest stau $\tilde\tau_1$ is lighter than the $\tilde B$,
since the lightest stau can be considerably lighter than the sleptons of the 
first two generations. Nevertheless, most sparticle decays will proceed via
the $\tilde B \to \tilde S + \ldots$ transition, which will give rise to
additional cascades with respect to decays in the MSSM. The properties of 
this cascade have been analysed in~\cite{ell5}, and in the following we will
briefly discuss the branching ratios and the $\tilde B$ life times in
the different parameter regimes:

a) $\tilde B\to \tilde S \nu\bar\nu$: This invisible process is mediated
dominantly by sneutrino exchange. Since the sneutrino mass, as the mass of
$\tilde B$, is essentially fixed by $M_{1/2}$~\cite{ell5}, the
associated branching ratio varies in a predictable way with $M_{\tilde B}$: It
can become up to 90\% for $M_{\tilde B} \sim 30$~GeV, but decreases with
$M_{\tilde B}$ and is maximally 10\% for $M_{\tilde B} \gsim 65$~GeV. 

b) $\tilde B \to \tilde S l^+l^-$: This process is mediated dominantly by the
exchange of a charged slepton in the s-channel. If the lightest stau
$\tilde\tau_1$ is considerably lighter than the sleptons of the first two
generations, the percentage of taus among the charged leptons can well exceed
$\frac{1}{3}$. If $\tilde\tau_1$ is lighter than $\tilde B$, it is produced
on-shell, and the process becomes $\tilde B \to \tilde\tau_1 \tau \to \tilde S
\tau^+ \tau^-$. Hence we can have up to 100\% taus among 
the charged leptons and
the branching ratio of this channel can become up to 100\%. 

c) $\tilde B\to \tilde S S$: This two-body decay is kinematically allowed if
both $\tilde S$ and $S$ are sufficiently light. (A light $S$ is not excluded by
Higgs searches at LEP1, if its coupling to the $Z$ is too
small~\cite{ell4}.) However, the coupling $\tilde B \tilde S S$ is
proportional to $\lambda^2$, whereas the couplings appearing in the decays a)
and b) are only of $O(\lambda)$. Thus this decay can only be important for
$\lambda$ not too small. In~\cite{ell5}, we found that its branching ratio can
become up to 100\% in a window $10^{-3} \lsim \lambda \lsim 10^{-2}$. Of
course, $S$ will decay immediately into $b\bar b$ or $\tau^+ \tau^-$, depending
on its mass. (If the branching ratio $Br(\tilde B\to \tilde S S)$ is
substantial, $S$ is never lighter than $\sim 5$~GeV.) If the singlet is heavy 
enough, its $b\bar b$ 
decay gives rise to 2 jets with $B$ mesons, which are easily detected with
$b$-tagging. In any case, the invariant mass of the $b\bar b$  or the 
$\tau^+ \tau^-$ system would be peaked at $M_S$, making
this signature easy to search for.

d) $\tilde B\to \tilde S \gamma$: This branching ratio can be important if the
mass difference $\Delta M = M_{\tilde B} - M_{\tilde S}$ is small ($\lsim
5$~GeV). 

Further possible final states like $\tilde B \to \tilde S q\bar q$ via $Z$
exchange have always branching ratios below 10\%. (The two-body decay $\tilde B
\to \tilde S Z$ is never important, even if $\Delta M$ is larger than $M_Z$: In
this region of the parameter space $\tilde\tau_1$ is always the NLSP, and thus
the channel $\tilde B \to \tilde\tau_1 \tau$ is always preferred.)

The $\tilde B$ life time depends strongly on the Yukawa coupling $\lambda$,
since the mixing of the singlino $\tilde S$ with gauginos and higgsinos is
proportional to $\lambda$. Hence, for small $\lambda$ (or a small mass 
difference $\Delta M$) the $\tilde B$ can 
be so long lived that it decays only after a macroscopic length of flight 
$l_{\tilde B}$. An approximate formula for $l_{\tilde B}$ (in meters) is given 
by 
\bea l_{\tilde B}[m] \simeq 2\cdot 10^{-10} \frac{1}{\lambda^2 \cdot
M_{\tilde B}[GeV]}\ ,\eea
and $l_{\tilde B}$ becomes $>\ 1$ mm for $\lambda \lsim 6\cdot 10^{-5}$.

\vspace{.4cm}
To summarize, the following unconventional signatures are possible
within the CNMSSM, compared to the MSSM:

\noindent
a) additional cascades attached to the original vertex (but still
missing energy and momentum): one or two additional $l^+ l^-$, $\tau^+ \tau^-$
or $b \bar b$ pairs or photons, with the corresponding branching ratios
depending on the parameters of the model.

\noindent
b) one or two additional $l^+ l^-$ or $\tau^+ \tau^-$ pairs or photons with
macroscopically displaced vertices, with distances varying from
millimeters to several meters. These displaced vertices do not point
towards the interaction point, since an additional invisible particle is
produced.

More details on the allowed branching ratios and life times can be found
in~\cite{ell5}, applications to sparticle production processes et LEP 2 are
published in~\cite{ell6}, and differential (spin averaged) cross
sections of the $\tilde B \to \tilde S$ decay are available upon request.

\def\glsp{$\wtil g$-LSP}
\def\lsim{\mathrel{\raise.3ex\hbox{$<$\kern-.75em\lower1ex\hbox{$\sim$}}}}
\def\gsim{\mathrel{\raise.3ex\hbox{$>$\kern-.75em\lower1ex\hbox{$\sim$}}}}
\def\ifmath#1{\relax\ifmmode #1\else $#1$\fi}
\def\half{\ifmath{{\textstyle{1 \over 2}}}}
\def\threehalf{\ifmath{{\textstyle{3 \over 2}}}}
\def\quarter{\ifmath{{\textstyle{1 \over 4}}}}
\def\sixth{\ifmath{{\textstyle{1 \over 6}}}}
\def\third{\ifmath{{\textstyle{1 \over 3}}}}
\def\twothirds{{\textstyle{2 \over 3}}}
\def\fivethirds{{\textstyle{5 \over 3}}}
\def\fourth{\ifmath{{\textstyle{1\over 4}}}}

\def\ejet{E_{\rm jet}}
\def\thetamuid{\theta(\mu\mbox{id})}
\def\mrecoil{M_{\rm recoil}}
\def\sigp{\sigma_{\rm P}^{\rm ann}}
\def\signp{\sigma_{\rm NP}^{\rm ann}}
\def\alsp{\alpha_s^{\rm P}}
\def\alsnp{\alpha_s^{\rm NP}}
\def\mpi{m_{\pi}}
\def\sigann{\sigma^{\rm ann}}
\def\water{{\rm H}_2{\rm 0}}
\def\nmess{N_m}
\def\vev#1{\langle #1 \rangle}
\def\lam{\lambda}
\def\lamu{\lam_u}
\def\lamd{\lam_d}
\def\lamud{\lam_u^\dagger}
\def\lamdd{\lam_d^\dagger}
\def\lampr{\lam^\prime}
\def\lampp{\lam^{\prime\prime}}

\def\Eq#1{Eq.~(\ref{#1})}
\def\Ref#1{Ref.~\cite{#1}}

\def\bml{\hbox{$B\!\!-\!\!L$}}
\def\gtino{\wt G}
\def\mgtino{m_{\gtino}}
\def\mplanck{M_{\rm P}}
\def\mpl{\mplanck}
\def\sur{{\wt u_R}}
\def\msur{{m_{\sur}}}
\def\stl{{\wt t_L}}
\def\str{{\wt t_R}}
\def\mstl{m_{\stl}}
\def\mstr{m_{\str}}
\def\sbl{{\wt b_L}}
\def\sbr{{\wt b_R}}
\def\msbl{m_{\sbl}}
\def\msbr{m_{\sbr}}
\def\sq{\wt q}
\def\sqbar{\ov{\sq}}
\def\msq{m_{\sq}}
\def\slep{\wt \ell}
\def\slepbar{\ov{\slep}}
\def\mslep{m_{\slep}}
\def\slepl{\wt \ell_L}
\def\mslepl{m_{\slepl}}
\def\slepr{\wt \ell_R}
\def\mslepr{m_{\slepr}}

\def\sel{\wt e}
\def\selbar{\ov{\sel}}
\def\msel{m_{\sel}}
\def\sell{\wt e_L}
\def\msell{m_{\sell}}
\def\selr{\wt e_R}
\def\mselr{m_{\selr}}

\def\cptwo{\wt \chi^+_2}
\def\cmtwo{\wt \chi^-_2}
\def\cpmtwo{\wt \chi^{\pm}_2}
\def\mcptwo{m_{\cptwo}}
\def\mcpmtwo{m_{\cpmtwo}}
\def\stautwo{\wt \tau_2}
\def\mstautwo{m_{\stauone}}

\def\mth{m_{3/2}}
\def\delgs{\delta_{GS}}
\def\kpr{K^\prime} 

\def\caln{{\cal N}}
\def\cald{{\cal D}}
\def\DM{D$^-$}
\def\DP{D$^+$}
\def\NSM{NS$^-$}
\def\NSP{NS$^+$}
\def\HSM{HS$^-$}
\def\HSP{HS$^+$}

\def\twoloop{two-loop/RGE-improved}
\def\Twoloop{Two-loop/RGE-improved}

\def\mhi{m_{h_1^0}}
\def\etmiss{/ \hskip-7pt E_T}
\def\etmin{/ \hskip-7pt E_T^{\rm min}}
\def\etjet{E_T^{\rm jet}}
\def\ptmiss{/ \hskip-7pt p_T}
\def\mslash{/ \hskip-7pt M}
\def\rslash{/ \hskip-7pt R}
\def\susyslash{\susy\hskip-24pt/\hskip19pt}
\def\mmissl{M_{miss-\ell}}
\def\mhalf{m_{1/2}}
\def\gl{\wt g}
\def\mgl{m_{\gl}}

\def\lefft{L_{\rm eff}(t\anti t\h)}
\def\leffzh{L_{\rm eff}(Z\h)}
\def\sigmat{\sigma_T(t\anti t\h)}
\def\sigmazh{\sigma_T(Z\h)}
\def\thdm{2HDM}
\def\caln{{\cal N}}
\def\cald{{\cal D}}
\def\DM{D$^-$}
\def\DP{D$^+$}
\def\NSM{NS$^-$}
\def\NSP{NS$^+$}
\def\HSM{HS$^-$}
\def\HSP{HS$^+$}
\def\etc{{\it etc.}}
\def\leff{L_{\rm eff}}
\def\sign{{\rm sign}}

\def\h{h}
\def\a{a}
\def\mh{m_{\h}}
\def\ma{m_{\a}}
\def\gamh{\Gamma_{\h}^{\rm tot}}

\def\etc{{\em etc.}}
\def\chisq{\chi^2}
\def\cale{{\cal E}}
\def\calo{{\cal O}}
\def\eg{{\it e.g.}}
\def\etal{{\it et al.}}
\def\mhalf{m_{1/2}}
\def\dmm{\Delta^{--}}
\def\dm{\Delta^{-}}
\def\mdmm{m_{\dmm}}
\def\hdmm{h^{\dmm}}
\def\dpp{\Delta^{++}}
\def\delp{\Delta^{+}}
\def\delm{\Delta^{-}}
\def\mdelm{m_{\delm}}
\def\hzero{\Delta^0}
\def\sigdmmbar{\overline\sigma_{\dmm}}
\def\gamdmm{\Gamma_{\dmm}}

\def\stop{\wt t}
\def\stopone{\wt t_1}
\def\stoptwo{\wt t_2}
\def\mstop{m_{\stop}}
\def\msquark{m_{\wt q}}
\def\mstopone{m_{\stopone}}
\def\mstoptwo{m_{\stoptwo}}

\def\sbot{\wt b}
\def\sbotone{\wt b_1}
\def\sbottwo{\wt b_2}
\def\msbot{m_{\sbot}}
\def\msbotone{m_{\sbotone}}
\def\msbottwo{m_{\sbottwo}}

\def\stl{{\wt t_L}}
\def\str{{\wt t_R}}
\def\mstl{m_{\stl}}
\def\mstr{m_{\str}}
\def\sbl{{\wt b_L}}
\def\sbr{{\wt b_R}}
\def\msbl{m_{\sbl}}
\def\msbr{m_{\sbr}}
\def\sqbar{\ov{\sq}}
\def\slep{\wt \ell}
\def\slepbar{\ov{\slep}}
\def\mslep{m_{\slep}}
\def\slepl{\wt \ell_L}
\def\mslepl{m_{\slepl}}
\def\slepr{\wt \ell_R}
\def\mslepr{m_{\slepr}}

\def\To{\Rightarrow}
\def\lra{\leftrightarrow}
\def\msusy{m_{\rm SUSY}}
\def\msusyslash{m_{\susyslash}}
\def\susy{{\rm SUSY}}
\def\caln{{\cal N}}
\def\cald{{\cal D}}
\def\DM{D$^-$}
\def\DP{D$^+$}
\def\NSM{NS$^-$}
\def\NSP{NS$^+$}
\def\HSM{HS$^-$}
\def\HSP{HS$^+$}
\def\etc{{\it etc.}}
\def\leff{L_{\rm eff}}
\def\sign{{\rm sign}}

\def\rpm{R^{\pm}}
\def\mrpm{m_{\rpm}}
\def\rzero{R^0}
\def\mrzero{m_{\rzero}}
\def\etc{{\em etc.}}
\def\chisq{\chi^2}
\def\cale{{\cal E}}
\def\calo{{\cal O}}
\def\eg{{\it e.g.}}
\def\etal{{\it et al.}}
\def\mhalf{m_{1/2}}
\def\gl{\wt g}
\def\mgl{m_{\gl}}
\def\stop{\wt t}
\def\stopone{\wt t_1}
\def\mstop{m_{\stop}}
\def\msquark{m_{\wt q}}
\def\mstopone{m_{\stopone}}
\def\sqbar{\ov{\sq}}
\def\slep{\wt \ell}
\def\slepbar{\ov{\slep}}
\def\mslep{m_{\slep}}
\def\slepl{\wt \ell_L}
\def\mslepl{m_{\slepl}}
\def\slepr{\wt \ell_R}
\def\mslepr{m_{\slepr}}
\def\sbot{\wt b}
\def\msbot{m_{\sbot}}
\def\hsm{h_{\rm SM}}
\def\mhsm{m_{\hsm}}
\def\hl{h^0}
\def\hh{H^0}
\def\ha{A^0}
\def\hp{H^+}
\def\hm{H^-}
\def\hpm{H^{\pm}}
\def\mhl{m_{\hl}}
\def\mhh{m_{\hh}}
\def\mha{m_{\ha}}
\def\mhp{m_{\hp}}
\def\mhpm{m_{\hpm}}
\def\tanb{\tan\beta}
\def\cotb{\cot\beta}
\def\mt{m_t}
\def\mb{m_b}
\def\mz{m_Z}
\def\mw{m_W}
\def\mgut{M_U}
\def\mx{M_X}
\def\mstring{M_S}
\def\wp{W^+}
\def\wm{W^-}
\def\wpm{W^{\pm}}
\def\wmp{W^{\mp}}
\def\chitil{\wt\chi}
\def\cnone{\wt\chi^0_1}
\def\cnonestar{\wt\chi_1^{0\star}}
\def\cntwo{\wt\chi^0_2}
\def\cnthree{\wt\chi^0_3}
\def\cnfour{\wt\chi^0_4}
\def\snu{\wt\nu}
\def\snul{\wt\nu_L}
\def\msnul{m_{\snul}}

\def\snue{\wt\nu_e}
\def\snuel{\wt\nu_{e\,L}}
\def\msnuel{m_{\snul}}

\def\snubar{\ov{\snu}}
\def\msnu{m_{\snu}}
\def\mcnone{m_{\cnone}}
\def\mcntwo{m_{\cntwo}}
\def\mcnthree{m_{\cnthree}}
\def\mcnfour{m_{\cnfour}}
\def\h{h}
\def\mh{m_{\h}}
\def\wt{\widetilde}
\def\wh{\widehat}
\def\cpone{\wt \chi^+_1}
\def\cmone{\wt \chi^-_1}
\def\cpmone{\wt \chi^{\pm}_1}
\def\mcpone{m_{\cpone}}
\def\mcpmone{m_{\cpmone}}

\def\cptwo{\wt \chi^+_2}
\def\cmtwo{\wt \chi^-_2}
\def\cpmtwo{\wt \chi^{\pm}_2}
\def\mcptwo{m_{\cptwo}}
\def\mcpmtwo{m_{\cpmtwo}}

\def\staur{\wt \tau_R}
\def\staul{\wt \tau_L}
\def\stau{\wt \tau}
\def\mstaur{m_{\staur}}
\def\stauone{\wt \tau_1}
\def\mstauone{m_{\stauone}}
\def\emem{e^-e^-}
\def\dmm{\Delta^{--}}
\def\mdmm{m_{\dmm}}
\def\hhmm{h^{\dmm}}
\def\dpp{\Delta^{++}}
\def\delp{\Delta^{+}}
\def\delm{\Delta^{-}}
\def\mdelm{m_{\delm}}
\def\hzero{\Delta^0}
\def\sigdmmbar{\overline\sigma_{\dmm}}
\def\gamdmm{\Gamma_{\dmm}}

\def\MPL #1 #2 #3 {{\sl Mod.~Phys.~Lett.}~{\bf#1} (#3) #2}
\def\NPB #1 #2 #3 {{\sl Nucl.~Phys.}~{\bf #1} (#3) #2}
\def\PLB #1 #2 #3 {{\sl Phys.~Lett.}~{\bf #1} (#3) #2}
\def\PR #1 #2 #3 {{\sl Phys.~Rep.}~{\bf#1} (#3) #2}
\def\PRD #1 #2 #3 {{\sl Phys.~Rev.}~{\bf #1} (#3) #2}
\def\PRL #1 #2 #3 {{\sl Phys.~Rev.~Lett.}~{\bf#1} (#3) #2}
\def\RMP #1 #2 #3 {{\sl Rev.~Mod.~Phys.}~{\bf#1} (#3) #2}
\def\ZPC #1 #2 #3 {{\sl Z.~Phys.}~{\bf #1} (#3) #2}
\def\IJMP #1 #2 #3 {{\sl Int.~J.~Mod.~Phys.}~{\bf#1} (#3) #2}
\def\NIM #1 #2 #3 {{\sl Nucl.~Inst.~and~Meth.}~{\bf#1} {#3} #2}
\def\mVV{M_{VV}}
\def\ep{e^+}
\def\em{e^-}
\def\mup{\mu^+}
\def\mum{\mu^-}
\def\taup{\tau^+}
\def\taum{\tau^-}
\def\wpm{W^{\pm}}
\def\hpm{H^{\pm}}
\def\mhm{m_{\hm}}
\def\call{{\cal L}}
\def\calm{{\cal M}}
\def\wtil{\widetilde}
\def\what{\widehat}
\def\tauptaum{\tau^+\tau^-}
\def\mbb{m_{b\anti b}}

\def\ltot{L_{\rm tot}}
\def\taup{\tau^+}
\def\taum{\tau^-}
\def\lam{\lambda}
\def\br{BR}
\def\tauptaum{\tau^+\tau^-}
\def\mbb{m_{b\anti b}}
\def\sprime{{s^\prime}}
\def\rtsprime{\sqrt{\sprime}}
\def\shat{{\widehat s}}
\def\rtshat{\sqrt{\shat}}
\def\gam{\gamma}
\def\sigrts{\sigma_{\tiny\rts}^{}}
\def\sigrtssq{\sigma_{\tiny\rts}^2}
\def\sigrtsprime{\sigma_{E}}
\def\nsigrts{n_{\sigrts}}
\def\betao{{\beta_0}}
\def\rhoo{{\rho_0}}
\def\etal{{\it et al.}}
\def\etc{{\it etc.}}
\def\sighbar{\overline \sigma_{\h}}
\def\sighlbar{\overline \sigma_{\hl}}
\def\sighhbar{\overline \sigma_{\hh}}
\def\sighabar{\overline \sigma_{\ha}}
\def\anti{\overline}
\def\epem{e^+e^-}
\def\zstar{Z^\star}
\def\wstar{W^\star}
\def\zstarp{Z^{(\star)}}
\def\wstarp{W^{(\star)}}
\def\mupmum{\mu^+\mu^-}
\def\lplm{\ell^+\ell^-}
\def\brwweff{\br_{WW}^{\rm eff}}
\def\brzzeff{\br_{ZZ}^{\rm eff}}
\def\mstar{M^{\star}}
\def\mstarmin{M^{\star\,{\rm min}}}
\def\drts{\Delta\sqrt s}
\def\rts{\sqrt s}
\def\ie{{\it i.e.}}
\def\eg{{\it e.g.}}
\def\eps{\epsilon}
\def\anti{\overline}
\def\wp{W^+}
\def\wm{W^-}
\def\mw{m_W}
\def\mz{m_Z}
\def\h{h}
\def\mh{m_{\h}}
\def\gamh{\Gamma_{\h}^{\rm tot}}
\def\gamsnu{\Gamma_{\snu}^{\rm tot}}
\def\a{a}
\def\ma{m_{\a}}
\def\hsm{h_{SM}}
\def\mhsm{m_{\hsm}}
\def\gamhsm{\Gamma_{\hsm}^{\rm tot}}
\def\tanb{\tan\beta}
\def\hl{h^0}
\def\mhl{m_{\hl}}
\def\gamhl{\Gamma_{\hl}^{\rm tot}}
\def\ha{A^0}
\def\mha{m_{\ha}}
\def\gamha{\Gamma_{\ha}^{\rm tot}}
\def\hh{H^0}
\def\mhh{m_{\hh}}
\def\gamhh{\Gamma_{\hh}^{\rm tot}}
\def\fbi{~{\rm fb}^{-1}}
\def\fb{~{\rm fb}}
\def\pbi{~{\rm pb}^{-1}}
\def\pb{~{\rm pb}}
\def\mev{~{\rm MeV}}
\def\gev{~{\rm GeV}}
\def\tev{~{\rm TeV}}
\def\stop{\widetilde t}
\def\mstop{m_{\stop}}
\def\mt{m_t}
\def\mb{m_b}

\def\hi{\h_1}
\def\hii{\h_2}
\def\hiii{\h_3}
\def\mhi{m_{\hi}}
\def\mhii{m_{\hii}}
\def\mhiii{m_{\hiii}}
\def\mpp{m_{PP}}

\def\dmm{\Delta^{--}}
\def\mdmm{m_{\dmm}}
\def\hdmm{h^{\dmm}}
\def\dpp{\Delta^{++}}
\def\mdm{m_{\dm}}
\def\hzero{\Delta^0}
\def\sigdmmbar{\overline\sigma_{\dmm}}
\def\gamdmm{\Gamma_{\dmm}}

\newcommand{\nc}{\newcommand}
\nc{\baa}{\begin{array}}      \nc{\eaa}{\end{array}}
\def\bit{\begin{itemize}}    
\def\eit{\end{itemize}}
\nc{\ben}{\begin{enumerate}}  \nc{\een}{\end{enumerate}}
\nc{\bce}{\begin{center}}     \nc{\ece}{\end{center}}
\def\beqa{\begin{eqnarray}}
\def\eeqa{\end{eqnarray}}

\def\Twoloop{Two-loop/RGE-improved}

\def\mm{\mu^+\mu^-}
\def\ee{e^+e^-}
\def\rta{\rightarrow}
\def\tanb{\tan\beta}
\def\mVV{M_{VV}}
%
%
\def\hf{\hfill}
\def\ie{{\it i.e.}}
\def\etal{{\it et al.}}
\def\9{\phantom 0}      
\renewcommand\linebreak{\unskip\break} 
\newlength{\captsize} \let\captsize=\small 
\newlength{\captwidth}                     

\section{Report on the gluino-LSP scenario}
\noindent
\centerline{\large\it H Baer, K. Cheung, J.F. Gunion}
\medskip
\label{section:Baer}

\subsection{Introduction}
This contribution will present a brief overview of the results
of Ref.~\cite{bcg}.
Most GUT scale boundary conditions (\eg\ the 
mSUGRA universal boundary conditions in which the gaugino
masses are assumed to have a common value at $\mgut$) lead to the
gluino being much heavier than the lightest neutralino.  However,
there are several models in which the gluino is heavy but is yet
the lightest supersymmetric particle, denoted \glsp.\\
$\phantom{aaa}$(a) A \glsp\ can arise in
models in which the gaugino
masses are given by one-loop corrections plus a contribution
from Green-Schwarz mixing (parameterized by $\delgs$)
\cite{nonuniv,guniondrees2}. An example is
the O-II string model in the limit where all SUSY breaking
arises from the size-modulus field and none from the dilaton.
At $\mgut$ one has:
\beq
M_3:M_2:M_1\stackrel{\scriptstyle O-II}{\sim}-(3+\delgs):(1-\delgs):
({33\over 5}-\delgs)\,,
\label{oiibc}
\eeq
and after evolution down to 1 TeV or below
a heavy gluino is the LSP when $\delgs\sim -3$ (a preferred
range for the model). \\
$\phantom{aaa}$(b)
In the GMSB context, the possibility of a 
heavy \glsp\ has been stressed in Ref.~\cite{raby}. In the model
constructed, the $\gl$ is the LSP as a result of mixing between the 
Higgs fields and the messenger fields.

In fact, there are three significant indications that a light gluino
is to be preferred over the universal gaugino
mass result of $\mgl\sim 7\mcnone$.

$\bullet$~The first such hint relates to the magnitude of $\alpha_s$ predicted
by requiring precise gauge coupling unification at $\mgut\sim 10^{16}\gev$.
For universal masses at $\mgut$, gauge coupling unification typically
requires $\alpha_s(\mz)\sim 0.125-0.13$ when sparticle
masses are $\lsim 1\tev$, a value that is uncomfortably
high relative to the best fit value of $\alpha_s(\mz)\lsim 0.12$.
Although unified gauge couplings at $\mgut$ can be made consistent
with $\alpha_s(\mz)\lsim 0.12$ if the sparticle masses
important in the gauge coupling running are all $\gsim 10\tev$
(for which fine tuning is regarded as a problem),
a much more interesting possibility is that discussed in Ref.~\cite{lrshif}.
There it is shown that if the gluino mass is substantially below, or at
least comparable to the $\cpmone$ mass (these determine the two most
critical thresholds in gauge coupling running), then $\alpha_s(\mz)\lsim 0.12$
is much easier to achieve.

$\bullet$ The second hint is from fine tuning.  As discussed
most recently in Ref.~\cite{kaneking}, the most severe problem in fine
tuning arises from the fact that the magnitude of the standard
measure of fine tuning contains a term proportional to $M_3^2(\mgut)$
with a very large numerical coefficient, much larger
than the (possibly canceling) terms proportional to $M_1^2(\mgut)$
and $M_2^2(\mgut)$. (Very roughly, the relative size
of these coefficients are 
determined by the relative size of the corresponding gauge couplings squared,
although $\tanb$ is also an important ingredient.)
The fine tuning problem is greatly relaxed if $M_3(\mgut)$ is substantially
smaller than $M_{1,2}(\mgut)$. For example, using the $\tanb=2.5$
numerical coefficients given in \cite{kaneking}, one measure
of fine tuning is (all parameters at $\mgut$
with $\what M_i$ and $\what m_i$ 
denoting $M_i(\mgut)/\mz$ and $m_i(\mgut)/\mz$)
\beq
\Delta_\mu\sim 23.8 \what M_3^2-1.3 \what M_2^2 +0.01 \what
M_1^2
+1.66\what M_2\what M_3+0.206\what M_1\what M_3+0.045\what M_1\what M_2
+\ldots\,,
\label{delmu}
\eeq
Since the $\what M_2^2$
coefficient is not very large, taking $\what M_3=0$
(leading to a gluino LSP) avoids significant fine-tuning
problems. One can even arrange for the $\what M_3$ and $\what M_2$ terms 
to cancel ($\what M_3 \sim \what M_2/5$). 
Thus, the $\what M_3$ values predicted for O-II model boundary
conditions in the favored $\delgs\sim -3,-4,-5$ range would lead
to a considerable relaxation of the the fine tuning problem and
a gluino that is much lighter than normally anticipated.

$\bullet$
The third hint relates to vacuum stability.
This has been recently reviewed in Ref.~\cite{cim}, where
references to earlier work can be found. The most serious problem
associated with many of the soft-supersymmetry-breaking boundary conditions
motivated by string theory is the presence of directions in
field space for which the effective potential is unbounded from below
(UFB). Dangerous charge and color breaking (CCB) minima
are also possible. The strongest constraint typically arises from
the UFB-3 direction, which involves the fields
$H_2,\nu_{L_i},e_{L_j},e_{R_j}$ with $i\neq j$. After minimizing
the effective potential, the latter three fields can be expressed
in terms of $H_2$. The value of the potential in the UFB-3 direction
is then given by:
\beq
V_{UFB-3}=(m^2_{H_2}+m_{L_i}^2)|H_2|^2+{|\mu|\over \lam_{e_j}}
(m_{L_j}^2+m_{e_j}^2+m_{L_i}^2)|H_2|-{2m_{L_i}^4\over g^{\prime\,2}+g^2}\,,
\label{ubf3}
\eeq
where $\lam_{e_j}$ is the leptonic Yukawa coupling of the $j$-th generation.
We must have 
$V_{UFB-3}(Q=\what Q)>V_{\rm real~min}=-{1\over
8}(g^{\prime\,2}+g^2)(v_2^2-v_1^2)^2$,
where $\what Q\sim {\rm Max}\left[
\lam_{\rm top}|H_2|,m_{\wtil t}\right]$. The problem arises
when $m_{H_2}^2$ is negative (as happens under RGE electroweak
symmetry breaking) and is most restrictive for $j=3$
for which $\lam_{e_j}$ is largest. One finds
that any significant amount of supersymmetry breaking from the dilaton 
leads to violation of the condition. The modulus-dominated
(or one-loop) limit boundary conditions of Eq.~\ref{oiibc}
(or extremely close thereto) open up some allowed parameter space.
This happens as follows. The RGE for $m_{H_2}^2$ is (using $t=\ln(Q)$)
$dm_{H_2}^2/dt=12(\lam_{\rm top}/4\pi)^2m_{\wtil t}^2+\ldots$.
The smaller $m_{\wtil t}$, the less negative $m_{H_2}^2$ is driven
in evolving down from $\mgut$, and, since 
\beq
{dm_{\wtil q}^2\over dt} = \left({1\over 2\pi}\right)
\left(-{16\over 3} M_3^2\alpha_3+\ldots\right)\,,
\eeq
smaller $M_3$ implies smaller $m_{\wtil t}$.
Meanwhile, the positive terms in Eq.~(\ref{ubf3}) have evolution
\beq
{dm_e^2\over dt}=\left({1\over 2\pi}\right)\left(
-{12\over 5}M_1^2\alpha_1+\ldots\right)\,,\quad
{dm_L^2\over dt}=\left({1\over 2\pi}\right)\left(-3M_2^2\alpha_2
-{3\over 5}M_1^2\alpha_1+\ldots\right)\,,
\eeq
implying larger $m_e^2$ and $m_L^2$ in $V_{UFB-3}$ for larger $M_{1,2}$.
Thus, non-universal boundary conditions with small $M_3$ relative
to $M_{1,2}$ are crucial in the string model context and, more generally,
are quite useful in satisfying the UFB-3 constraint.

It is often stated that a stable \glsp\ is ruled out by virtue of relic
density constraints, especially the non-observation of anomalous
isotopes.  However, such constraints are inevitably model dependent.
In Ref.~\cite{bcg} it is shown that if the annihilation cross section
for gluinos is nonperturbatively enhanced 
near threshold (many models of this
type exist) then the relic density of gluinos could be very small.
If, in addition, they did not cluster with nucleons (\eg\ if
they are concentrated at galaxy cores), then
relic constraints would not rule out this scenario. Alternatively,
the reheating required to avoid the Polonyi problem would also
effectively eliminate the relic gluinos. 
It is also possible that the gluino could decay but with a lifetime
so long that it is effectively stable in the detector.
This is possible if there is a very weak violation of R-parity
or in gauge-mediated-SUSY-breaking (GMSB) models where $\gl\to g\gtino$
(where the $\gtino$ is the gravitino) can be very suppressed by
a large supersymmetric breaking scale.  Thus, it
is important to consider how to place constraints on a detector-stable
gluino (for which we use the generic
\glsp\ notation) using accelerator experiments. We will focus on constraints
that arise by looking directly for the $\gl$'s themselves. That
is, we do not include processes where other
supersymmetric particles are produced and then decay into $\gl$'s.

\subsection{Behavior of a \glsp\ in a Detector}

As soon as a \glsp\ is produced in a detector, it picks
up a gluon or quark-antiquark combination to form
an `R-hadron'; $R^0=\gl g$ is likely to be the lightest state,
but color-singlet $\gl q^\prime\anti q$ states could have very
similar mass, and if the difference in mass between such states
and the $R^0$ were $<m_\pi$ they would
be pseudo-stable in the detector.
The behavior of a \glsp\ in a typical detector depends very much
upon whether the dominant R-hadron fragment is charged (probability $P$)
or neutral (probability $1-P$). Simple quark counting models suggest
that $P<1/2$.
The important ingredients in determining the energy deposited
by the \glsp\ in the detector are:
\bit
\item the hadronic interaction length, $\lam_T$, as determined by $\sigma_T$;
\item the average energy deposited per hadronic collision, $\vev{\Delta E}$, 
as a function of the $\gl$'s velocity;
\item the amount of ionization energy deposited between
hadronic interactions and how the calibrated
detector measures this energy;\footnote{For example, in iron a given
amount of $E_{\rm ionization}$ is translated into measured
energy of $E_{\rm measured}=rE_{\rm ionization}$ with $r\sim 1.6$.}
\item the thickness (measured most conveniently by the number
of hadronic interaction lengths) of various components of the detector.
\eit
One should picture the \glsp\ as emerging from the
hard production process as a neutral (charged) R-hadron with probability
$1-P$ ($P$). At each hadronic interaction the light quarks and gluons
are presumed to be stripped away and the \glsp\ again fragments
into a neutral or charged R-hadron
with probability $1-P$ or $P$, respectively. Thus, for any $P\neq0,1$,
the charge of the R-hadron between hadronic collisions fluctuates.
In Ref.~\cite{bcg}, several models for $\lam_T$ (\ie\
for the \glsp\ total cross section) and for $\vev{\Delta E}$
are considered.  For the most likely case of
$P<1/2$, one finds that the energy deposited by the \glsp\ is dominated by
hadronic energy deposits rather than by ionization energy deposits.
Further, for $P<1/2$ only a small fraction of the gluino's energy
is actually deposited. The \glsp\ behaves like a bowling ball moving
through a sea of ping-pong balls.  The result is that the \glsp\
generally exits the detector, thereby leading to missing energy aligned with
a soft jet. For processes of interest, the $\gl$'s are always
produced in pairs and are seldom back to back.  As a result,
the net missing energy is usually large and not aligned with any
one of the jets observed in the detector. Thus, the crucial
signal for $P<1/2$ is jets + missing energy.  This signal is also
generally quite useful even for $P>1/2$ since the momentum
of a $\gl$-jet is never properly determined even if the energy deposited
via ionization is large. (In fact, the ionization energy deposit
is generally overestimated, and for $P\sim 1$ the `measured' gluino
momentum can even exceed its true momentum in the OPAL
analysis procedure used later.)

\begin{figure}[thb]
\vspace*{0.1in}
\centerline{\protect\psfig{figure=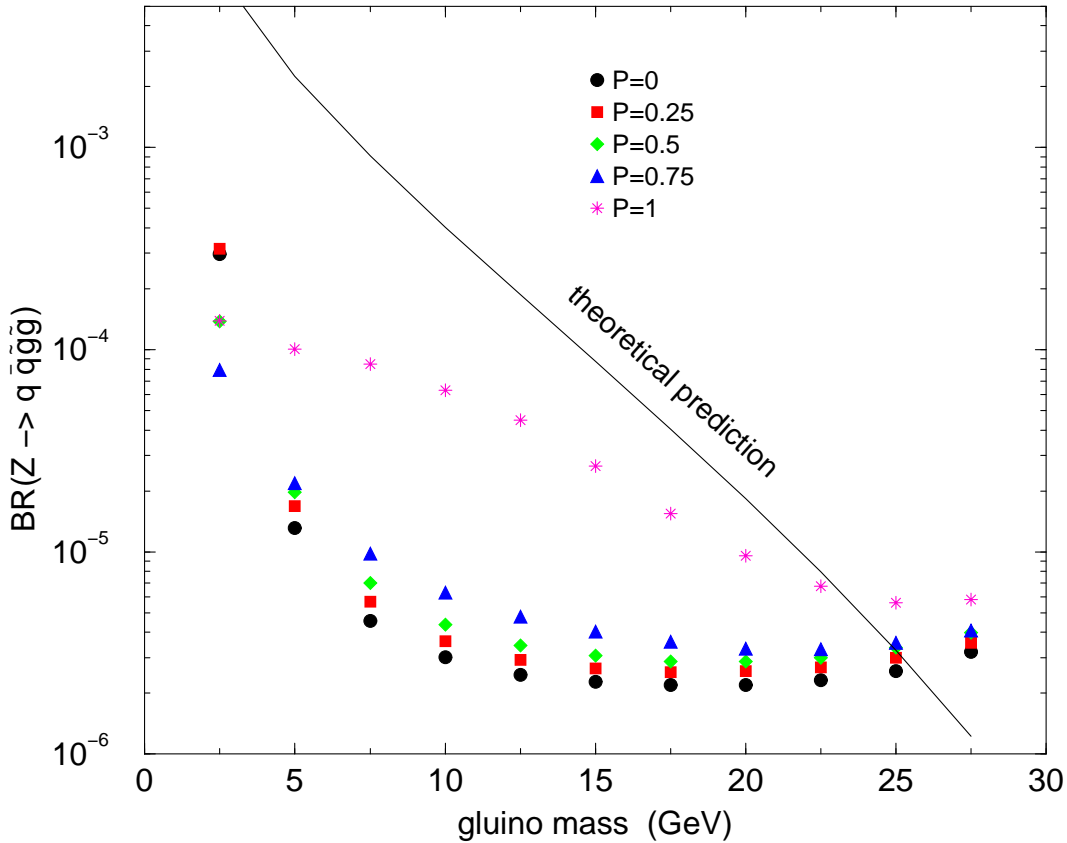,width=0.8\textwidth}}
\caption{
We consider $q\anti q \gl\gl$ 
for $P=0,1/4,1/2,3/4,1$ using event-by-event determination of 
the measured $\gl$ jet energies.
For $P\neq 0,1$, changes of the $R$-hadron
charge as it passes through the detector are randomly implemented.
Both smearing and fragmentation effects are included.
The figure gives, as a function of $\mgl$,
the OPAL 95\% CL upper limits compared to the
theoretical prediction for $\br(Z\to q\anti q \gl\gl)$. Results are
for $\lam_T=19$~cm and a middle-of-the-road model for $\vev{\Delta E}$.}
\label{pscannormal}
\end{figure}

\subsection{Constraints from LEP}

Gluinos can be directly produced via two processes:
$\epem\to q\anti q\gl\gl$ \cite{cer,css,mts}, 
which can take place at tree-level,
and $\epem \to \gl\gl$ \cite{nper,krol,css}, which takes place via loop
diagrams (involving squarks and quarks).
The latter process
is very model dependent and can be highly suppressed.  Thus, we
focus on the $q\anti q\gl\gl$ final state. An explicit calculation
of the cross section for this final state reveals that LEP2 running
will yield rather few $q\anti q\gl\gl$ events unless $\mgl$ is quite
small. However, the number of $Z$'s accumulated during LEP1 running
is sufficiently large that a significant number of $Z\to q\anti q\gl\gl$
events would be expected for $\mgl\lsim 25\gev$.

The only relevant LEP1 experimental analysis is the OPAL \cite{opal}
search for pair production of neutralinos, $Z\to \cnone\cntwo$,
with $\cntwo\to q\anti q\cnone$, in the
${\rm jets}+\ptmiss$ channel that is
potentially relevant for the $q\anti q\gl\gl$ final state.
Typically, $q\anti q\gl\gl$
events give $n({\rm jets})=2$, 3, or 4, depending upon
the amount of energy deposition by the $\gl$-jets. After implementing
the OPAL procedures in a detailed Monte Carlo simulation
of $q\anti q\gl\gl$ production (including a parameterization of experimental
resolutions and a Peterson form \cite{peterson} 
for $\gl\to$ R-hadron fragmentation),
we find that a \glsp\ is excluded for $3\lsim\mgl\lsim 24\gev$
if $P<1/2$. (For $P>1/2$, the excluded range only extends to $23\gev$.)
This is illustrated in Fig.~\ref{pscannormal} for our favored
choices of $\lam_T$ and $\vev{\Delta E}$ model. (The excluded mass
range is quite insensitive to these choices.) We note that for $P>1/2$,
a \glsp\ can also be excluded by OPAL over much the same mass range
by virtue of no excess of heavily ionizing tracks having been seen.

\subsection{Constraints from Tevatron Run I and Prospects for Run II}

At a hadron collider, gluinos are produced via $gg,q\anti q\to \gl\gl$.
Initial state radiation in association with the hard process yields
additional jets in the final state. In Ref.~\cite{bcg}, we explored
the limits that can be placed on a \glsp\ using the jets$+~\ptmiss$
analysis by CDF \cite{cdfcuts,cdffinal} of a portion of their Run I data.
We performed a Monte Carlo simulation of $\gl\gl$ events 
for the CDF jets$+~\ptmiss$ analysis cuts using
ISAJET, supplemented by a routine that models 
the behavior of the \glsp's in the CDF detector
for a given choice of $P$, $\lam_T$ and $\vev{\Delta E}$ model.
For $P>0$, one must discard events that
contain a `muonic' jet (\ie\ a jet that has substantial
ionization energy and minimal hadronic energy, as defined in the CDF analysis).

\begin{figure}[p]
\vspace*{0.1in}
\centerline{\protect\psfig{figure=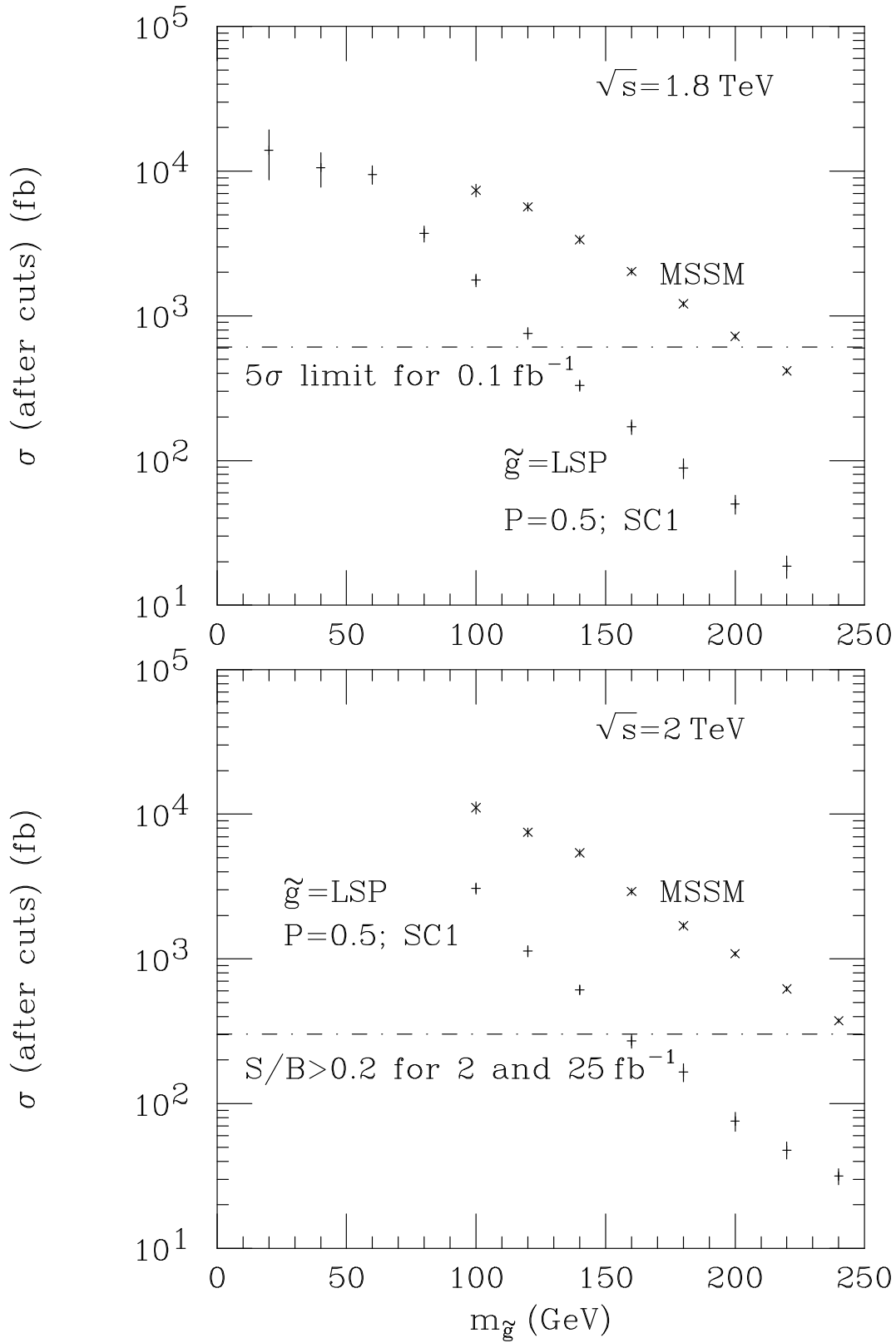,width=0.8\textwidth}}
\caption{
The cross section (after cuts) in the ${\rm jets}+\ptmiss$ channel
is compared to (a) the $5\sigma$ level for $L=0.1\fbi$
(also roughly the 95\% CL upper limit for $L=19\pbi$) at $\protect\rts=1.8\tev$
and (b) the $S/B=0.2$ level at Run II ($L\geq 2\fbi$, $\protect\rts=2\tev$)
as a function of $\mgl$ for $P=1/2$, using event-by-event
determination of the momentum (=energy) of each $\gl$-jet 
(including the probabilistic treatment of
charge-exchanges at each hadronic collision)
in events such that neither $\gl$-jet is ``muonic''. We use
$\lam_T=19$~cm and a middle-of-the-road $\vev{\Delta E}$ model.}
\label{phalfcdf}
\end{figure}

In Fig.~\ref{phalfcdf} (upper window) the predicted jets$+~\ptmiss$
cross section from $\gl\gl$ production
(assuming $P=1/2$ and after all cuts and efficiencies) is 
compared to the 95\% CL upper
limit obtained by analyzing $L=19\pbi$ of Run I data.  The range
$20\lsim\mgl\lsim 130\gev$ is clearly excluded. For $P\to 0$,
the upper limit of the excluded mass range increases
slowly to $150\gev$. These results are quite independent of $\lam_T$
and the $\vev{\Delta E}$ model. 
Note that the $130-150\gev$ lower limit on $\mgl$ obtained is
substantially below the lower limit that Run I data
places on $\mgl$ in a typical MSSM model.
For easy comparison, the figure shows the cross
section (after cuts) resulting from gluino pair production
in the MSSM model considered in Ref.~\cite{cdfcuts} with $\msq=1000\gev$,
$\mu=-400\gev$ and $\tanb=4$; one sees that Run I data
yields a 95\% CL limit of roughly $\mgl\gsim 210\gev$. This is
because the \glsp\ scenario yields fewer jets (in particular, none
from $\gl$ decay) as compared to a canonical MSSM scenario.
For $P\geq 3/4$, the ionization energy deposited 
by a $\gl$ jet increases significantly,
and for some $\lam_T$ and $\vev{\Delta E}$ model choices the hadronic
energy deposit is sufficiently small that one or both of the $\gl$ jets
are often declared to be `muonic' and the event discarded.
For such choices, the current jets$+~\ptmiss$ analysis does not 
constrain the \glsp\ scenario. (Note that a modified
analysis in which muonic jets are not discarded would,
and is highly recommended.)  However, the complementary CDF
search for events with heavily ionizing tracks does exclude $50\gev\leq \mgl$
(up to at least $200\gev$ for any $P\geq 3/4$).

The expected extension of the excluded range of $\mgl$ that will
result for $P=1/2$ by analyzing Run II 
jets$+~\ptmiss$ data in exactly the same way is also shown
in Fig.~\ref{phalfcdf}. For a requirement of $S/B>0.2$ (as possibly needed
for a reliable signal in the presence of systematic uncertainties),
the limits obtained (for $L>0.5\fbi$) will require
$\mgl\geq 160-180\gev$ for $P\leq 1/2$ (and also for $P\geq 1/2$
for many $\lam_T$ and $\vev{\Delta E}$ model choices).
If systematics could be controlled so
that a signal with $S/B\lsim 10\%$ becomes reliable, the
lower limits would be increased by about $30\gev$.
This, of course, is still substantially
lower than the $\mgl$ lower bound 
that can be achieved in the reference MSSM model
for the same $S/B$ criterion (\eg\ $250\gev$ for $S/B>0.2$). 
It is worth emphasizing that these Run II limits do not disappear
for large $P$ even for those $\lam_T$ and $\vev{\Delta E}$ model choices
that yield the largest probability for $\gl$ jets to be declared `muonic'.
For example, for $P=3/4$, the worst choice would still result
in excluding $\mgl\lsim 130\gev$ using Run II jets$+~\ptmiss$ data
and Run I analysis procedures. We anticipate that a more optimized
analysis procedure (in particular not throwing away muonic jets) will
do even better.

\subsection{Discussion and Conclusions}

To summarize, accelerator data places quite significant
constraints on a gluino LSP. Currently, for any reasonable
value of the probability $P$ for $\gl\to $ charged R-hadron fragmentation
($P\leq 1/2$), $3\lsim\mgl\lsim 130-150\gev$ is excluded at 95\% CL
by a combination of OPAL LEP1 and CDF Run I jets$+~\ptmiss$
analyses. For 
the theoretically much less likely $P\geq 3/4$ range, there is
a window $23\lsim \mgl\lsim 50\gev$ that (depending
upon the hadronic path length of the \glsp\ in the detector
and the average energy deposited in each hadronic collision)
might not be excluded
by the jets$+~\ptmiss$ analyses and would also not
be excluded by the OPAL and CDF
searches for heavily ionizing tracks. However, it is apparent
that more optimized CDF procedures are capable of easily excluding
this window. The increase in the lower bound
on $\mgl$ that will result from Run II Tevatron jets$+~\ptmiss$
data will be limited by the level of systematic uncertainty in the absolute
normalization of the background level.

For completeness, we also considered the scenario in which the gluino
is not the LSP, but rather the NLSP (next-to-lightest supersymmetric
particle), with the gravitino ($\gtino$) being the (now invisible) LSP.  
Such a situation
can arise in GMSB models, including that of Ref.~\cite{raby}.
In this scenario, $\gl\to g\gtino$.
Early universe/rare isotope limits are then irrelevant. 
Further, the decay will be prompt from the detector point of view
if $\mgtino$ is in the $\leq$ few eV region.
(If the scale of supersymmetry breaking is so large that the $\gl\to g\gtino$
decay lifetime is long enough that most $\gl$'s
exit the detector before decaying, then we revert to
the earlier \glsp\ results.) For a $\gl$-NLSP,
we find that the OPAL jets$+~\ptmiss$ analysis excludes
$\mgl\leq 26\gev$. The CDF Run I analysis excludes $\mgl\leq 240\gev$
(down to very low values), while Run II data can be expected to
exclude at the very least $\mgl\leq 280\gev$ (assuming $S/B>0.2$ is
required --- better if smaller $S/B$ can be excluded).

Given that there are arguments in favor of a light
gluino, it is unwise to simply assume that the gluino cannot be the lightest 
or next-to-lightest supersymmetric particle.  
Fortunately, present and future experiments can exclude or find
such a gluino for a significant range of $\mgl$.


\def\lsim{\mathrel{\raise.3ex\hbox{$<$\kern-.75em\lower1ex\hbox{$\sim$}}}}
\def\gsim{\mathrel{\raise.3ex\hbox{$>$\kern-.75em\lower1ex\hbox{$\sim$}}}}
\def\half{\ifmath{{\textstyle{1 \over 2}}}}
\def\threehalf{\ifmath{{\textstyle{3 \over 2}}}}
\def\quarter{\ifmath{{\textstyle{1 \over 4}}}}
\def\sixth{\ifmath{{\textstyle{1 \over 6}}}}
\def\third{\ifmath{{\textstyle{1 \over 3}}}}
\def\twothirds{{\textstyle{2 \over 3}}}
\def\fivethirds{{\textstyle{5 \over 3}}}
\def\fourth{\ifmath{{\textstyle{1\over 4}}}}
\def\square{\boxxit{0.4pt}{\fillboxx{7pt}{7pt}}\hskip-0.4pt}
    \def\boxxit#1#2{\vbox{\hrule height #1 \hbox {\vrule width #1
             \vbox{#2}\vrule width #1 }\hrule height #1 } }
    \def\fillboxx#1#2{\hbox to #1{\vbox to #2{\vfil}\hfil}    }

\def\ibid{{\it ibid.}}
\def\hf{\hfill}
\def\ie{{\it i.e.}}
\def\etal{{\it et al.}}

\def\dedx{dE/dx}
\def\ejet{E_{\rm jet}}
\def\thetamuid{\theta(\mu\mbox{id})}
\def\mrecoil{M_{\rm recoil}}
\def\sigp{\sigma_{\rm P}^{\rm ann}}
\def\signp{\sigma_{\rm NP}^{\rm ann}}
\def\alsp{\alpha_s^{\rm P}}
\def\alsnp{\alpha_s^{\rm NP}}
\def\mpi{m_{\pi}}
\def\sigann{\sigma^{\rm ann}}
\def\vev#1{\langle #1 \rangle}

\def\Eq#1{Eq.~(\ref{#1})}
\def\Ref#1{Ref.~\cite{#1}}

\def\sur{{\wt u_R}}
\def\msur{{m_{\sur}}}
\def\stl{{\wt t_L}}
\def\str{{\wt t_R}}
\def\mstl{m_{\stl}}
\def\mstr{m_{\str}}
\def\sbl{{\wt b_L}}
\def\sbr{{\wt b_R}}
\def\msbl{m_{\sbl}}
\def\msbr{m_{\sbr}}
\def\sq{\wt q}
\def\sqbar{\ov{\sq}}
\def\msq{m_{\sq}}

\def\sel{\wt e}
\def\selbar{\ov{\sel}}
\def\msel{m_{\sel}}
\def\sell{\wt e_L}
\def\msell{m_{\sell}}
\def\selr{\wt e_R}
\def\mselr{m_{\selr}}

\def\cptwo{\wt \chi^+_2}
\def\cmtwo{\wt \chi^-_2}
\def\cpmtwo{\wt \chi^{\pm}_2}
\def\mcptwo{m_{\cptwo}}
\def\mcpmtwo{m_{\cpmtwo}}
\def\stautwo{\wt \tau_2}
\def\mstautwo{m_{\stauone}}

\def\dmchi{\Delta m_{\tilde\chi_1}}

\def\mth{m_{3/2}}
\def\delgs{\delta_{GS}}
\def\kpr{K^\prime} 

\def\caln{{\cal N}}
\def\cald{{\cal D}}
\def\DM{D$^-$}
\def\DP{D$^+$}

\def\twoloop{two-loop/RGE-improved}
\def\Twoloop{Two-loop/RGE-improved}

\def\mhi{m_{h_1^0}}
\def\etmiss{/ \hskip-7pt E_T}
\def\emiss{/ \hskip-7pt E}
\def\etmin{/ \hskip-7pt E_T^{\rm min}}
\def\etjet{E_T^{\rm jet}}
\def\ptmiss{/ \hskip-7pt p_T}
\def\mslash{/ \hskip-7pt M}
\def\rslash{/ \hskip-7pt R}
\def\susyslash{\susy\hskip-24pt/\hskip19pt}
\def\mmissl{M_{miss-\ell}}
\def\mhalf{m_{1/2}}
\def\aeta{|\eta|}

\def\etc{{\it etc.}}
\def\leff{L_{\rm eff}}
\def\sign{{\rm sign}}

\def\chisq{\chi^2}
\def\cale{{\cal E}}
\def\calo{{\cal O}}
\def\eg{{\it e.g.}}
\def\mhalf{m_{1/2}}

\def\stop{\wt t}
\def\stopone{\wt t_1}
\def\stoptwo{\wt t_2}
\def\mstop{m_{\stop}}
\def\msquark{m_{\wt q}}
\def\mstopone{m_{\stopone}}
\def\mstoptwo{m_{\stoptwo}}

\def\sbot{\wt b}
\def\sbotone{\wt b_1}
\def\sbottwo{\wt b_2}
\def\msbot{m_{\sbot}}
\def\msbotone{m_{\sbotone}}
\def\msbottwo{m_{\sbottwo}}

\def\slep{\wt \ell}
\def\slepbar{\ov{\slep}}
\def\mslep{m_{\slep}}
\def\slepl{\wt \ell_L}
\def\mslepl{m_{\slepl}}
\def\slepr{\wt \ell_R}
\def\mslepr{m_{\slepr}}

\def\To{\Rightarrow}
\def\msusy{m_{\rm SUSY}}
\def\msusyslash{m_{\susyslash}}
\def\susy{{\rm SUSY}}

\def\gl{\wt g}
\def\mgl{m_{\gl}}

\def\tanb{\tan\beta}
\def\cotb{\cot\beta}
\def\mt{m_t}
\def\mb{m_b}
\def\mz{m_Z}
\def\mw{m_W}
\def\mgut{M_U}
\def\mx{M_X}
\def\mstring{M_S}
\def\wp{W^+}
\def\wm{W^-}
\def\wpm{W^{\pm}}
\def\wmp{W^{\mp}}
\def\chitil{\wt\chi}
\def\cnone{\wt\chi^0_1}
\def\cnonestar{\wt\chi_1^{0\star}}
\def\cntwo{\wt\chi^0_2}
\def\cnthree{\wt\chi^0_3}
\def\cnfour{\wt\chi^0_4}
\def\snu{\wt\nu}
\def\snul{\wt\nu_L}
\def\msnul{m_{\snul}}

\def\snue{\wt\nu_e}
\def\snuel{\wt\nu_{e\,L}}
\def\msnuel{m_{\snul}}

\def\snubar{\ov{\snu}}
\def\msnu{m_{\snu}}
\def\mcnone{m_{\cnone}}
\def\mcntwo{m_{\cntwo}}
\def\mcnthree{m_{\cnthree}}
\def\mcnfour{m_{\cnfour}}
\def\h{h}
\def\mh{m_{\h}}
\def\wt{\widetilde}
\def\wh{\widehat}
\def\cpone{\wt \chi^+_1}
\def\cmone{\wt \chi^-_1}
\def\cpmone{\wt \chi^{\pm}_1}
\def\cmpone{\wt \chi^{\mp}_1}
\def\mcpone{m_{\cpone}}
\def\mcpmone{m_{\cpmone}}

\def\cptwo{\wt \chi^+_2}
\def\cmtwo{\wt \chi^-_2}
\def\cpmtwo{\wt \chi^{\pm}_2}
\def\mcptwo{m_{\cptwo}}
\def\mcpmtwo{m_{\cpmtwo}}

\def\staur{\wt \tau_R}
\def\staul{\wt \tau_L}
\def\stau{\wt \tau}
\def\mstau{m_{\stau}}
\def\mstaur{m_{\staur}}
\def\stauone{\wt \tau_1}
\def\mstauone{m_{\stauone}}
\def\sigdmmbar{\overline\sigma_{\dmm}}
\def\gamdmm{\Gamma_{\dmm}}
\def\ep{e^+}
\def\em{e^-}
\def\mup{\mu^+}
\def\mum{\mu^-}
\def\taup{\tau^+}
\def\taum{\tau^-}
\def\wpm{W^{\pm}}
\def\hpm{H^{\pm}}
\def\mhm{m_{\hm}}
\def\call{{\cal L}}
\def\calm{{\cal M}}
\def\wtil{\widetilde}
\def\what{\widehat}

\def\ltot{L_{\rm tot}}
\def\taup{\tau^+}
\def\taum{\tau^-}
\def\lam{\lambda}
\def\br{BR}
\def\tauptaum{\tau^+\tau^-}
\def\mbb{m_{b\anti b}}
\def\sprime{{s^\prime}}
\def\rtsprime{\sqrt{\sprime}}
\def\shat{{\widehat s}}
\def\rtshat{\sqrt{\shat}}
\def\gam{\gamma}
\def\sigrts{\sigma_{\tiny\rts}^{}}
\def\sigrtssq{\sigma_{\tiny\rts}^2}
\def\sigrtsprime{\sigma_{E}}
\def\nsigrts{n_{\sigrts}}
\def\betao{{\beta_0}}
\def\rhoo{{\rho_0}}
\def\sighbar{\overline \sigma_{\h}}
\def\sighlbar{\overline \sigma_{\hl}}
\def\sighhbar{\overline \sigma_{\hh}}
\def\sighabar{\overline \sigma_{\ha}}
\def\anti{\overline}
\def\epem{e^+e^-}
\def\zstar{Z^\star}
\def\wstar{W^\star}
\def\zstarp{Z^{(\star)}}
\def\wstarp{W^{(\star)}}
\def\mupmum{\mu^+\mu^-}
\def\lplm{\ell^+\ell^-}
\def\brwweff{\br_{WW}^{\rm eff}}
\def\brzzeff{\br_{ZZ}^{\rm eff}}
\def\mstar{M^{\star}}
\def\mstarmin{M^{\star\,{\rm min}}}
\def\drts{\Delta\sqrt s}
\def\rts{\sqrt s}
\def\ie{{\it i.e.}}
\def\eg{{\it e.g.}}
\def\eps{\epsilon}
\def\anti{\overline}
\def\wp{W^+}
\def\wm{W^-}
\def\mw{m_W}
\def\mz{m_Z}
\def\h{h}
\def\mh{m_{\h}}
\def\gamh{\Gamma_{\h}^{\rm tot}}
\def\gamsnu{\Gamma_{\snu}^{\rm tot}}
\def\a{a}
\def\ma{m_{\a}}
\def\hsm{h_{SM}}
\def\mhsm{m_{\hsm}}
\def\gamhsm{\Gamma_{\hsm}^{\rm tot}}
\def\hl{h^0}
\def\mhl{m_{\hl}}
\def\gamhl{\Gamma_{\hl}^{\rm tot}}
\def\ha{A^0}
\def\mha{m_{\ha}}
\def\gamha{\Gamma_{\ha}^{\rm tot}}
\def\hh{H^0}
\def\mhh{m_{\hh}}

\def\cm{~\mbox{cm}}
\def\fbi{~{\rm fb}^{-1}}
\def\fb{~{\rm fb}}
\def\pbi{~{\rm pb}^{-1}}
\def\pb{~{\rm pb}}
\def\mev{~{\rm MeV}}
\def\gev{~{\rm GeV}}
\def\tev{~{\rm TeV}}
\def\stop{\widetilde t}
\def\mstop{m_{\stop}}
\def\mt{m_t}
\def\mb{m_b}
\def\mm{\mu^+\mu^-}
\def\ee{e^+e^-}

\def\MPL #1 #2 #3 {{\sl Mod.~Phys.~Lett.}~{\bf#1} (#3) #2}
\def\NPB #1 #2 #3 {{\sl Nucl.~Phys.}~{\bf #1} (#3) #2}
\def\PLB #1 #2 #3 {{\sl Phys.~Lett.}~{\bf #1} (#3) #2}
\def\PR #1 #2 #3 {{\sl Phys.~Rep.}~{\bf#1} (#3) #2}
\def\PRD #1 #2 #3 {{\sl Phys.~Rev.}~{\bf #1} (#3) #2}
\def\PRL #1 #2 #3 {{\sl Phys.~Rev.~Lett.}~{\bf#1} (#3) #2}
\def\RMP #1 #2 #3 {{\sl Rev.~Mod.~Phys.}~{\bf#1} (#3) #2}
\def\ZPC #1 #2 #3 {{\sl Z.~Phys.}~{\bf #1} (#3) #2}
\def\IJMP #1 #2 #3 {{\sl Int.~J.~Mod.~Phys.}~{\bf#1} (#3) #2}
\def\NIM #1 #2 #3 {{\sl Nucl.~Inst.~and~Meth.}~{\bf#1} {#3} #2}
\def\JHEP #1 #2 #3 {{\sl JHEP}~{\bf#1} (#3) #2}

\section{Detecting a highly degenerate lightest neutralino and
lightest chargino at the Tevatron}
\noindent
\centerline{\large\it J.F. Gunion, S.~Mrenna}
\label{section:Mrenna}
\medskip

For some choices of soft SUSY--breaking parameters, the LSP is a stable
neutralino $\cnone$, the NLSP is a chargino $\cpmone$ almost degenerate in
mass with the LSP ($\dmchi\equiv\mcpmone-\mcnone\sim \mpi-$few GeV),
and all other sparticles are relatively heavy.  In this case, detection 
of sparticles using the usual, mSUGRA--motivated signals will be difficult, 
since the visible decay products in $\cpmone\to\cnone+\ldots$
will be very soft,
and alternative signals must be considered. In this note, we summarize
the viability 
of signatures at the Tevatron based on highly--ionizing charged tracks, 
disappearing charged tracks, large impact parameters, missing transverse 
energy and a jet or a photon, and determine the
mass reach of such signatures assuming that only the $\cpmone$ and $\cnone$ 
are light. 
If $\dmchi$ is sufficiently big that $c\tau(\cpmone)\lsim$few cm and 
there are no other light superparticles, there is a significant possibility
that the limits on $\mcpmone$ based on LEP2 data cannot be extended at 
the Tevatron. If $c\tau(\cpmone)>$few cm, relatively background--free 
signals exist that will give a clear signal of $\cpmone$ production (for some 
range of $\mcpmone$).

\subsection{Introduction}
In mSUGRA models, gaugino masses are assumed to be universal at $\mgut$,
leading to $M_3:M_2:M_1\sim 6:2:1$ at the TeV energy scale, implying
relatively large mass splitting 
between the lightest chargino and 
the lightest neutralino (most often the LSP).
However, various attractive models exist for which $M_2<M_1$
at the TeV scale, which results in $\mcpmone\simeq\mcnone$
if $|\mu|\gg M_{1,2}$ (as is normally the case for correct EWSB).
In particular, $M_2<M_1$  when the gaugino masses are dominated by
or entirely generated by loop corrections. The first model of this
type to receive detailed attention was the O-II superstring model
proposed in Ref.~\cite{ibanez} and studied in 
Refs.~\cite{cdg1,cdg2,cdg3}.
For further review, see  Ref.~\cite{bcgrunii2}.
More recently, the one-loop boundary conditions have arisen
in the context of the conformal anomaly \cite{murayama,randall}.

In the O-II model, $M_{1,2,3}$ are determined both by the one-loop
beta functions and by the Green-Schwarz mixing parameter, $\delgs$ (required
in general to cancel anomalies). 
At the scale $\mgut$ ($\mz$), the O--II
model with $\delgs=-4$ yields $M_3:M_2:M_1=1:5:10.6$ ($6:10:10.6$);
 $\delgs=0$ (equivalent to the 
simplest version of the conformal anomaly approach) gives  
$M_3:M_2:M_1=3:1:33/5$ ($3:0.3:1$). After radiative corrections,
$\dmchi$ is  near in value to $m_\pi$ 
for $\delgs=0$, increasing to $\dmchi\sim1-2$ GeV
(depending on $|\mu|$) for $\delgs=-4$.
Since the typical values 
of $|\mu|$ required by RGE electroweak symmetry breaking are large,
the higgsino $\cpmtwo$, $\cnthree$ and $\cnfour$ states are very heavy.

When $M_2<M_1\ll|\mu|$, the $Z\to \cnone\cnone$, $Z\to\cnone\cntwo$,
$Z\to\cntwo\cntwo$, and $\wpm\to \cpmone\cntwo$ couplings 
are all small, while the $Z,\gam\to\cpone\cmone$ and 
$\wpm \to\cpmone\cnone$ couplings are full strength. Only cross sections
induced by the latter can have large rates.


The most critical ingredients in the phenomenology of such models
are the lifetime and the decay modes of the $\cpmone$, 
which in turn depend almost entirely
on $\dmchi$ when the latter is small. The $c\tau$ and branching ratios
of the $\cpmone$ as a function of $\dmchi$ have been computed
in Ref.~\cite{cdg3}. A tabulation of $c\tau$ values for the range
of $\dmchi$ of interest in this report is given in
Table~\ref{ctaus}. For $\dmchi<\mpi$, $c\tau$ is at least several meters;
once $\dmchi>\mpi$, $c\tau$ drops quickly. 
For $\dmchi<\mpi$, $\cpmone\to e\nu_e\cnone$ dominates, while for
$\dmchi>\mpi$ the $\cpmone\to \pi^\pm\cnone$ mode
turns on and is dominant for $\dmchi\lsim 800\mev$. For still larger $\dmchi$,
multi-pion modes become important merging eventually into
$\cpmone\to\ell\nu_\ell\cnone,q^\prime\anti q\cnone$.

\begin{table}[ht]
  \begin{center}
    \begin{tabular}[c]{|c|c|c|c|c|c|c|c|} \hline
$\dmchi(\mev)$ & 125 & 130 & 135 & 138 & 140 & 142.5 & 150 \\
$c\tau(\mbox{cm})$ & 1155 & 918.4 & 754.1 & 671.5 & 317.2 & 23.97 & 10.89 \\
\hline
$\dmchi(\mev)$ & 160 & 180 & 185 & 200 & 250 & 300 & 500 \\
$c\tau(\mbox{cm})$ & 6.865 & 3.719 & 3.291 & 2.381 & 1.042 & 0.5561 & 0.1023 \\
\hline
    \end{tabular}
    \caption{Summary of $c\tau$ values as a function of $\dmchi$
as employed in Monte Carlo simulations.}
    \label{ctaus}
  \end{center}
\end{table}

We now give a brief review of the results of Ref.~\cite{hithip}
for the case where  we assume  that only the
$\cpmone$ and $\cnone$ are light. We discuss the
 types of signals that will be important
for different ranges of $\dmchi$. 
(The modifications that arise if the $\gl$ is also light,
with $\mgl\gsim \mcpmone\simeq\mcnone$, are discussed in Ref.~\cite{hithip}.)
For this discussion, we ask
the reader to imagine a canonical detector (e.g. CDF or D\O~at Run II) with
the following components ordered according to increasing radial distance
from the beam.

\noindent (I) An inner silicon vertex (SVX) detector extending radially
from the beam axis. The CDF Run II vertex detector has layers at 
$r\sim 1.6$, 3, 4.5, 7, 8.5 and $11\cm$
(the first and second layers are denoted L00 and L0, respectively) 
extending out to $z=\pm 45$ cm
\cite{cdfextra}. The D\O~SVX has 4 layers (but 2 are double--sided),
with the first at 2.5 cm and the last at 11 cm.

\noindent (II) A central tracker (CT) extending from $15\cm$ to $73\cm$ (D\O)
or from roughly $20\cm$ to $130\cm$ (CDF).

\noindent (III) A thin pre--shower layer (PS).

\noindent (IV) An electromagnetic calorimeter (EC) and hadronic 
calorimeter (HC).

\noindent (V) The inner--most muon chambers (MC), 
starting just beyond the HC.  

\noindent (VI) Both CDF and D\O~will have a precise time--of--flight 
measurement (TOF) for any charged particle that makes it to the muon chambers.

It is important to note that the SVX, CT and PS can all give (independent)
measurements of the $dE/dx$ from ionization of a track passing through them.
This will be important to distinguish a heavily--ionizing chargino
(which would be $\geq$ twice minimal ionizing [2MIP] for $\beta\gamma\leq
0.85$) from an isolated minimally ionizing particle [1MIP]. For example,
at D\O~the rejection against isolated 1MIP tracks will be
${\rm few}\times 10^{-3}$, ${\rm few}\times 10^{-3}$, and $\sim 10^{-1}$
for tracks that pass through the SVX, CT and PS, respectively, with an
efficiency of 90\% for tracks with $\beta\gamma<0.85$
\cite{glandsberg}.
At CDF, the discrimination factors are  roughly similar \cite{dstuart}.
Because of correlations, one cannot simply multiply these numbers
together to get the combined discrimination power of the SVX, CT
and PS for an isolated track that passes through all three; for such
a track with $\beta\gamma<0.85$, 
the net discrimination factor would probably be
of order ${\rm few}\times 10^{-5}$. 
At LEP/LEP2, the detector structure is somewhat different and important
features will be noted where relevant.
We now list the possible signals.

\bigskip
\noindent {\bf (a) LHIT and TOF signals:} 
\medskip

For $\dmchi<\mpi$, 
a heavy chargino produced in a collision travels a distance of order
a meter or more and will often penetrate to the muon chambers. 
If it does, the chargino may be distinguished from a muon by
heavy ionization in the SVX, CT and PS.
There should be no hadronic energy deposits associated
with this track, implying that the energy deposited in the hadronic
calorimeter should be consistent with ionization energy losses
for the measured $\beta$. This type of
long, heavily--ionizing track signal will be denoted as an LHIT signal.

If the chargino penetrates to the muon chambers, its large mass
will also be evident from the time delay of its TOF signal.
This delay can substitute for the heavy ionization requirement.
The passage of the chargino through the muon
chamber provides an adequate trigger
for the events. In addition, the chargino will be clearly visible
as an isolated track in the CT, and this track could
also be used to trigger the event. In later analysis (off--line even),
substantial momentum can be required for the track without
loss of efficiency. (The typical transverse
momentum of a chargino when pair--produced 
in hadronic collisions is of order 1/2 the mass.)

After a reasonable cut on the $p_T$ of the chargino track, 
the LHIT and TOF signals will be background free.

\medskip
\noindent {\bf (b) DIT signals:} 
\medskip

For $\dmchi$ above but near $\mpi$, 
the chargino will often appear as an
isolated track in the central tracker but it will decay before
the muon chamber. (The appropriate mass range for which this
has significant probability is
$\mpi<\dmchi< 145\mev$, for which $c\tau\gsim 17\cm$.)
As such a chargino passes part way through the calorimeters
beyond the CT, it will deposit little energy. In particular,
any energy deposit in the hadronic calorimeter should be
no larger than that consistent with ionization energy deposits 
for the $\beta$ of the track as measured using ionization
information from the SVX+CT+PS. (In general,
the chargino will only deposit ionization energy up to the point of its decay.
Afterwards, the weakly--interacting neutralino will carry away most
of the remaining energy, leaving only a very soft pion or lepton remnant.)
Thus, we require that the track effectively disappear once it exits the CT.
(The point at which the ionization energy deposits end would typically
be observable in a calorimeter with sufficient radial segmentation,
but we do not include this in our analysis.)
Such a disappearing, isolated track will be called a DIT. The DIT
will have substantial $p_T$, which can be used to trigger the event.
A track with large $p_T$ from a background process
will either be a hadron, an electron or a muon. The first two will
leave large deposits in the calorimeters (EC and/or HC) and the latter
will penetrate to the muon chamber. Thus, the signal described
is very possibly a background--free signal. If not, a requirement
of heavy ionization in the SVX, CT and PS will certainly eliminate 
backgrounds, but with some sacrifice of signal events. Thus,
we will also consider the possibility of requiring that the DIT
track be heavily ionizing. In the most extreme case, we require
that the average ionization measured in the SVX, CT and PS correspond
to $\beta<0.6$ ($\beta\gamma<0.75$), which signal is denoted by DIT6.
For a DIT signal, this is a very
strong cut once $\dmchi$ is large enough that the average $c\tau$
is smaller than the radius of the CT.
This is because rather few events will have both large enough $\beta\gamma
c\tau$ to pass all the way through the CT and small enough $\beta$
to satisfy the heavy ionization requirement.

\medskip
\noindent {\bf (c) STUB and KINK signals, including STUB$+\etmiss$, 
or SMET signal:} 
\medskip

For $145\mev<\dmchi<160\mev$, $17\cm>c\tau>7\cm$.
For such $c\tau$, the probability for the chargino to pass all the
way through the central tracker will be small. The chargino will
be most likely to pass all the way through the SVX and decay somewhere in the
CT. Such a short SVX track we term a STUB. 
It will not be associated with any calorimeter energy deposits.
At a hadron collider,
the primary difficulty associated with a STUB signal is that
it will not provide its own Level--1 trigger. 
We have found that it is most efficient ($\eps\sim 10\%$) to trigger the event
by requiring substantial missing transverse energy ($\etmiss$).
Once an interesting event is triggered,
off--line analysis will provide a measurement of the ionization deposited
by the STUB in the SVX.\footnote{Note that for $\dmchi\gsim 160\mev$
(for which $c\tau\lsim 7$ cm), requiring heavy ionization, 
i.e. small $\beta$, begins to significantly conflict with the requirement 
that the chargino pass all the way through the SVX. 
For smaller $\dmchi$, the $c\tau$ of the chargino is larger and this conflict
is not very severe.}
Although we believe that the STUB signals will be background--free
without a $\beta$ cuts, we have also considered discovery reach 
after imposing a $\beta<0.6$ cut. Altogether, we will define 4 STUB--based
signals: (a) SNT -- a STUB track only, with no other trigger; 
(b) SNT6 -- a STUB track with $\beta<0.6$ and no other trigger.
(c) SMET -- a STUB track in an event with $\etmiss>35\gev$;
and, (d) SMET6 -- a STUB track with $\beta<0.6$ in an event with
$\etmiss>35\gev$. Only (c) and (d) are possible using the triggering
design planned by CDF and D\O~for Run II.

In addition to the STUB, most of the soft charged
pions from charginos that decay
after passing through the vertex detector will be seen in the tracker.
Typically, the soft pion track that intersects the STUB track will
do so at a large angle, a signature we call a KINK. We have not
explored this in detail, but believe that a KINK requirement 
in association with the STUB signals defined above would lead to
little loss of signal and yet make the signals background--free with high
certainty.

\medskip
\noindent {\bf (d) HIP signals:}
\medskip

For $160\mev<\dmchi<190\mev$, $7\cm>c\tau>3\cm$.
Some of the produced charginos will decay late compared to $c\tau$
and yield a STUB signature of the type discussed just above. More typically,
however, the $\cpmone$ will pass through two to three layers of the SVX.
The $\cpmone$ track will then end and turn into a single charged pion 
with substantially different momentum. Both the sudden disappearance of
and the lack of any calorimeter energy deposits associated
with the $\cpmone$ track will help to distinguish it from 
other light--particle tracks that would normally
register in all layers of the SVX and in the calorimeters.

For $160\mev<\dmchi<190\mev$, $p_\pi^*\sim 77-130\mev$.
The corresponding transverse impact parameter resolution of the SVX,
$\sigma_b$, is approximately  $300-170~\mu$m 
(taking $p_\pi^T\sim p_\pi^*$ and applying the $1\sigma$ values from 
Fig.~2.2 of \cite{cdfextra} when L00 is included),
and is much smaller than the typical impact
parameter (which is a sizeable fraction of $c\tau>3\cm$).
In this range, the KINK formed by the $\cpmone$ track and the soft
$\pi$ should be visible. In addition, the layers
that the $\cpmone$ passes through will provide an ionization
estimate for $\beta$ that could be used to help eliminate
backgrounds. However, we have not pursued either of these possibilities
since in the end the STUB signals are still viable in this $\dmchi$ range
and are probably superior. 

For $\dmchi>230\mev$, $c\tau<1.6\cm$
and the typical $\cpmone$ will not even pass through the innermost SVX layer
unless $\beta$ is very large . However, $p_\pi^*>180\mev$ and
the impact parameter resolution for the single emitted pion
moves into the $<150~\mu$m range.
For example, if $\dmchi=240,300,500,1000\mev$, $c\tau\sim 1.2,
0.37,0.09,0.007\cm$ while $p_\pi^T\sim 195,265,480,990\mev$ yields
$1\sigma$ impact parameter resolutions of $\sigma_b\sim 120,90,50,25~\mu$m. 
We thus considered the signal based on events defined by the $\gam+\etmiss$
trigger and the presence of one or more large--$b$ charged pions. 

Unlike the previous signals, the HIP signal has a large background even after
requiring the $\pi$'s impact parameter $b$ to satisfy $b>5\sigma_b$,
where $\sigma_b$ is the resolution. After imposing isolation criteria
and low-$p_T$  for the $\pi$, the main background arises from
production of long--lived baryons (e.g. $\Sigma^\pm,\ldots$) that decay
to a $\pi$ and a nucleon.

Once $\dmchi>1\gev$,
a HIP signature will not be useful and we must consider the chargino
decay to be prompt. This is because the largest possible impact parameter
is only a few times the $1\sigma$ value for the resolution and we will be
dominated by fakes. This leads us to one of two
completely different types of signal.

\medskip
\noindent {\bf (e) \boldmath $\gam+\etmiss$ and jets+$\etmiss$ signals:}
\medskip

For some interval of $\dmchi$ (e.g. $200\mev\lsim \dmchi\lsim
300\mev$ at the DELPHI LEP/LEP2 detector --- see later --- or, perhaps,
$1\gev\lsim\dmchi\lsim 10-20\gev$ at the Tevatron)
the decay products (hadron(s) or $\ell\nu$) produced along with the $\cnone$
will be too soft to be distinctively visible in the main part
of the detector and at the same time high--impact--parameter tracks associated
with chargino decay will
not be apparent. One will then have to detect chargino
production as an excess of events with an isolated photon or missing energy
above a large $\gam+\etmiss$ or jet(s)+$\etmiss$ background.
We find that, despite its lower rate, after appropriate cuts
the $\gam+\etmiss$ channel is superior to the jet(s)+$\etmiss$ signals. 
For some values of the chargino mass and $\dmchi$, an
excess in these channels could confirm the SVX signals discussed earlier.

\medskip
\noindent {\bf (f) standard mSUGRA signals:}
\medskip

For large enough $\dmchi$, the extra lepton or hadron 
tracks from $\cpmone$ decay will be sufficiently energetic to be detected
and will allow identification of chargino production events
when associated with a photon or missing energy trigger.
A detailed simulation
is required to determine exactly how large $\dmchi$ needs to be
for this signal to be visible above backgrounds. At LEP/LEP2,
backgrounds are sufficiently small that the extra
tracks are visible for $\dmchi\gsim 300\mev$ in association
with a photon trigger while standard mSUGRA searches
based on missing energy and jets/leptons require $\dmchi\gsim 3\gev$.
At a hadron
collider we estimate that $\dmchi\gsim 10-20$ GeV will be necessary to 
produce leptons or jets sufficiently energetic to produce a distinctive event
assuming a missing energy trigger.

\subsection{Collider Phenomenology of degenerate models}

Although our main focus will be on Tevatron Run II, it is useful
to summarize which of the above signals have been 
employed at LEP2 and the resulting constraints on the degenerate
scenarios we are considering.

\subsubsection{Lepton Colliders}

As discussed above and in Refs.~\cite{cdg1,cdg2,cdg3},
collider phenomenology depends crucially on $\dmchi$.
Most importantly, SUSY detection depends on which aspects (if any)
of the $\cpone\cmone$ final state are visible. 
 If the $\cpmone$ decay products are soft,  $\cpone\cmone$
production may be indistinguishable from the large
$\epem\to\epem\gam\gam\to\epem+{\rm soft}$ background.
Tagging $\cpone\cmone$ production
using a photon from initial or final state radiation (ISR) 
is necessary \cite{cdg1}. Even with an ISR tag, 
the $\cpone$ and $\cmone$ might be
invisible because of the softness of their decay products and
the lack of a vertex detector signal. 
In this case, $\gam\cpone\cmone$ production is observable only
as a $\gamma\mslash$ signature, where 
$\mslash=\sqrt{(p_{e^-}+p_{e^+}-p_{\gam})^2}$. Even after requiring
$\mslash>\mz$, the $\gam\nu\anti\nu$ background is large.
For $M_2<M_1\ll|\mu|$, 
Ref.~\cite{cdg1}  found that at LEP2 with $L=125\pbi$ per experiment,
no improvement over the $\mcpmone<45\gev$
limit coming from LEP1 $Z$--pole data was possible.
The experimental situation is greatly improved if LHIT and/or KINK
signals can be employed, or if the soft pions from the $\cpmone$
decays in $\gam\cpone\cmone$ events can be detected.
All of this is most clearly illustrated
by summarizing the analysis from DELPHI at LEP2 \cite{delphideg}.
\bit
\item
When $\dmchi\lsim 200\mev$, the 
charginos are sufficiently long--lived to produce 
either an LHIT or a KINK  signal for $\cpone\cmone$ production.
No additional trigger is required for either signal. As
a result, DELPHI is able to exclude
$\mcpmone$ out to nearly the kinematic limit (currently 90 GeV).
\item
When $\dmchi\gsim$ 3 GeV, the decay products of the $\cpmone$
become easily visible, and the standard mSUGRA search results apply; the
$\cpmone$ is excluded out to the kinematic limit (90 GeV for the data
sets analyzed), except for the case of a relatively
light sneutrino, for which the $\cpone\cmone$ cross
section is smaller and the limit does not extend past 75 GeV.
\item
For $200\mev\lsim\dmchi\lsim3\gev$, the chargino tracks are not long
enough to use the KINK signature, and the chargino decay
products are too soft to provide a clear signature on their own.
As proposed in Ref.~\cite{cdg1}, DELPHI employs an ISR photon tag. 
In order to essentially eliminate the $\gam\nu\anti\nu$
background, the event is required to contain soft charged tracks
consistent with the isolated pions expected from the chargino decays.
DELPHI observed no events after all cuts. 
For $M_2<M_1\ll|\mu|$, and a heavy (light) sneutrino, 
this excludes $\mcpmone\lsim 62\gev$ ($49\gev$)
for $0.3\lsim\dmchi\lsim 3\gev$ ($0.5\lsim\dmchi\lsim 3\gev$).
The gap from $0.2-0.3\gev$ ($0.2-0.5\gev$) arises because of the low
efficiency for detecting very soft pions.\footnote{With the ISR tag,
the $\gam\gam$ background is completely negligible.}
\eit
Thus, there is a gap from just above $\dmchi\sim 200\mev$ to at least 
$300\mev$ for which the chargino is effectively invisible. DELPHI finds
that the $\gam\mslash$ signature, discussed earlier, is
indeed insufficient to improve over the $\mcpmone\gsim 45\gev$
limit from $Z$ decays. We are uncertain whether DELPHI explored
the use of high--impact--parameter tracks in their vertex detector
(in association with the ISR trigger) to improve their sensitivity
(by sharply reducing the $\gam\nu\anti\nu$ background) in these gap regions.

\subsubsection{Hadron Colliders}

At hadron colliders, typical signatures of mSUGRA are tri--lepton events
from neutralino--chargino production, like--sign di--leptons from
gluino pair production, and multi-jets$+\etmiss$ from squark and
gluino production. The tri--lepton signal from $\cpmone\cntwo$
production and the like--sign di--lepton signal from $\gl\gl$ production
are both suppressed when $\dmchi$ is small by the softness of the leptons
coming from the $\cpmone$ decay(s). In $M_2<M_1\ll|\mu|$ scenarios,
the tri--lepton signal is further diminished
by the suppression of the $\cpmone\cntwo$ cross section.
Provided that $\mgl$ is light enough, the most obvious signal
for SUSY in degenerate models is jet(s) plus missing energy,
as studied in Refs.~\cite{cdg2,hithip}. 
 However, it is entirely possible that the gluino is
much heavier than the light $\cpmone,\cnone$ states
and that the $\gl\gl$ production rate (at the Tevatron at least)
will be quite suppressed.  In this case, the ability to detect events
in which the only directly produced SUSY particles
are light neutralino and chargino states could prove critical.  
In what follows, we assume that the
sfermion, gluino and heavier chargino and neutralino states are 
sufficiently heavy that their production rates at the Tevatron
are not useful, and investigate methods to probe $\cpone\cmone$
and $\cpmone\cnone$ production at the Tevatron.  
The possible signals were summarized earlier.
Details regarding cuts and triggering appear in Ref.~\cite{hithip}.

We performed  particle level studies using either
the processes contained in the PYTHIA 6.125 
event generator or by adding
external processes (several of the $\gamma+X$ processes considered here)
into PYTHIA.  
A calorimeter is defined out to $\eta=4.4$ with a Gaussian $E_T$ resolution
of $\sigma_{E_T}=80\%/\sqrt{E_T}$.  Jets with $E_T>5$ GeV and $R=0.5$ are
reconstructed to define $\etmiss$.
Non--Gaussian contributions will be estimated as described later.
Charged track momenta and impact parameters $b$ are unsmeared, but the
effects of detector resolution on $b$ are included.

We find that there is a natural boundary near a mass splitting 
of $\dmchi\sim 300\mev$, below which one or more
of the background--free signals are viable
but above which one must contend with large backgrounds.

\begin{figure}[ht]
\leavevmode
\begin{center}
\epsfxsize=4.7in
\hspace{0in}\epsffile{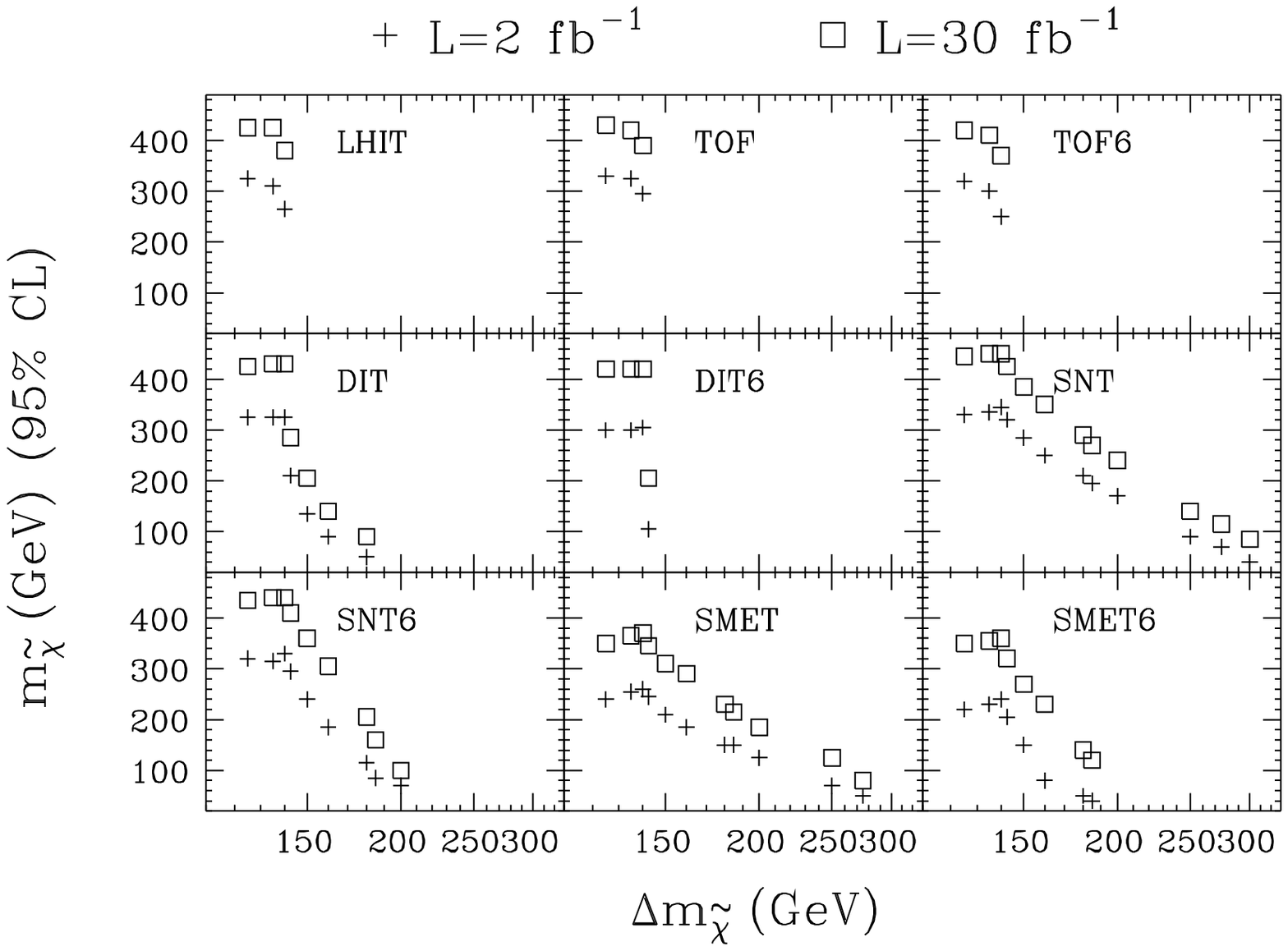}
\end{center}
\caption[]{95\% CL lower limits on $\mcpmone$ as a function of $\dmchi$
for ``background--free'' signatures at Run II with $L=2\fbi$
and $L=30\fbi$.}
\label{limits:summary} 
\end{figure}

\bigskip
\noindent {\bf Region (A)}
For $\dmchi$ values $\lsim 200-300\mev$, one considers
the background--free signals summarized above, which will have the most
substantial mass reach in $\mcpmone$.  
The $L=2\fbi$ and $L=30\fbi$ 95\% CL (3 events, no background) limits
on $\mcpmone$ deriving from these signals 
are summarized in Fig.~\ref{limits:summary}. We give a brief
verbal summary.

{\boldmath $\dmchi<\mpi$:}
For such $\dmchi$, the average $c\tau$ of the chargino
is of order a meter or more. The LHIT and TOF signals are prominent,
but the DIT and STUB signals appear if $\dmchi$ is not
extremely small. The relative weight between these signals
is determined by the exponential
form of the $c\tau$ distribution in the chargino rest frame
and the event--by--event variation of the
boosts imparted to the chargino(s) during production.
\bit
\item
The LHIT signature can probe masses in the range 
$260-325$ ($380-425)$ GeV for $L=2\fbi$ ($30\fbi$), the lower
reach applying for $\dmchi\sim\mpi$ and the highest reach applying
for any $\dmchi\lsim 125\mev$.  
The reach of the TOF signature is nearly identical to that of the LHIT
signature.
\item
The DIT signature has a reach of 320 (425) GeV for 
$ 120\mev\leq\dmchi\leq\mpi$,\footnote{We did not study lower $\dmchi$
values since they are highly improbable after including
radiative correction contributions to $\dmchi$.} and, 
in particular, is more efficient than the LHIT and
TOF signals for $\dmchi\sim\mpi$.
The DIT signature reach drops by about 20 GeV with a $\beta<0.6$ cut
(DIT6) designed to require that the chargino track be heavily--ionizing.  
\item
The STUB signature with no additional 
trigger (SNT) can reach to $\simeq 340$ (450) GeV
for $120\mev\leq\dmchi\leq\mpi$, which mass reach drops
by $10-20$ GeV if $\beta<0.6$ is required.
However, neither D\O~nor CDF can use
STUB information at Level--1 in their current design.  
\item
With the addition of a standard $\etmiss$
trigger, the resulting STUB signature (SMET) will be viable with 
the present detectors, reaching to
$240-260$ ($350-375$) GeV for $120\mev\leq\dmchi\leq\mpi$, 
which numbers drop by about
10 GeV if $\beta<0.6$ is required (SMET6).
\eit

{\boldmath $\mpi\lsim\dmchi\lsim 200-300\mev$:}
\bit
\item
The LHIT and TOF signatures disappear,
since almost all produced charginos decay before reaching the MC or TOF.  
\item
The DIT signature
remains as long as the $\beta<0.6$ (heavily--ionizing)
requirement is not necessary to eliminate backgrounds.  
If we require $\beta<0.6$, there is a mismatch with the requirement
that the chargino pass through the CT -- once $\dmchi$
is above $145\mev$, the entire signal is
generated by large boosts in the production process, which 
is in conflict with requiring small $\beta$.
\item
The SNT signature probes $\mcpmone\lsim 300\gev$ ($\lsim 400\gev$)
for $\dmchi\sim \mpi$ and $L=2\fbi$ ($L=30\fbi$). For $\dmchi$
as large as $300\mev$, it alone among the
background--free channels remains viable, probing $\mcpmone\lsim 70\gev$
($\lsim 95\gev$). Certainly, it would extend the $\sim 90\gev$ 
limit obtained by DELPHI at LEP2 that applies for $\dmchi<200\mev$
and the $\sim 45\gev$ limit from LEP data that is the only available
limit for $200\mev\leq\dmchi\leq 300\mev$.
But, as stated above, the SNT
signature will not be possible without a Level--1 SVX trigger.
\item The STUB+$\etmiss$, SMET and SMET6 signatures
are fully implementable at Run II and have a reach that
is only about 20 GeV lower than their SNT and SNT6 counterparts.
\eit

\bigskip
\noindent {\bf Region (B)}
For $300\mev\lsim \dmchi\lsim 600\mev$, the high--impact--parameter (HIP) 
signal (a $\gam+\etmiss$ tag for events yields
the smallest backgrounds) is very useful despite the large background from 
production of $\Sigma^\pm,\ldots$ hadrons. The luminosity
required to achieve 95\% CL exclusion or $5\sigma$ discovery
was evaluated in Ref.~\cite{hithip}, requiring also that   
$S/B>0.2$. We find that one can achieve
a 95\% CL lower bound of $95\gev$ ($75\gev$) on $\mcpmone$ for 
$\dmchi=300\mev$ ($\dmchi=600\mev$) for $L=30\fbi$. This would represent
some improvement over the $\sim 60\gev$ lower bound
obtained in the current DELPHI analysis of their LEP2 data 
for this same range of $\dmchi$ if the sneutrino is heavy.
(If the $\snu$ is light, then there is no useful LEP2 limit
if $300\mev\leq\dmchi\leq 500\mev$, but LEP data requires $\mcpmone>45\gev$.)
With only $L=2\fbi$ of data, the HIP analysis would only exclude
$\mcpmone<68\gev$ ($<53\gev$) for $\dmchi=300\mev$ ($\dmchi=600\mev$).

\bigskip
\noindent {\bf Region (C)}
For $\dmchi\gsim 600\mev$, up to some fairly large value (we estimate
at least 10 to 20 GeV), the chargino decay products are effectively
invisible at a hadron collider and 
the most useful signal is $\gamma+\etmiss$. However, 
this signal at best probes $\mcpmone\lsim 60\gev$ (for any $L>2\fbi$), 
whereas the DELPHI analysis of their LEP2 data
already excludes $\mcpmone\leq 60\gev$ for $500\mev\leq\dmchi\leq 3\gev$
(if the sneutrino is heavy --- only $\leq 48\gev$
if the sneutrino is light) and $\mcpmone\leq 90\gev$ for $\dmchi>3\gev$.

\begin{table}[h]
  \begin{center}
\sloppy
    \begin{tabular}[c]{|c|c|l|c|l|c|} 
\hline
~~$\dmchi$~~ & ~~$c\tau$~~ & \multicolumn{1}{c|}{Best Run II} & Trigger & \multicolumn{1}{c|}{Crucial measurements and} & ~~Reach~~ \\
(MeV) & (cm) & \multicolumn{1}{c|}{signature(s)} &   & \multicolumn{1}{c|}{associated detector components} & (GeV) \\
\hline\hline\hline
0     & $\infty$ & TOF   & MC  & TOF, $p_T$ (SVX+CT)  & 460 \\
      &          & LHIT  & MC  & $p_T$ (SVX+CT), $\dedx$ (SVX+CT+PS) & 450 \\
\hline\hline
125   &  1155    & TOF   & MC  & TOF, $p_T$ (SVX+CT)  & 430 \\
      &          & LHIT  & MC  & $p_T$ (SVX+CT),  $\dedx$ (SVX+CT+PS) & 425 \\
      &          & DIT   & CT  & $p_T$ (SVX+CT), HC veto &  425 \\
      &          & DIT6  & CT  & same + $\dedx$ (SVX+CT+PS), & 420 \\
\hline\hline
135   &   754    & LHIT  & MC  & $p_T$ (SVX+CT), $\dedx$ (SVX+CT+PS) & 425 \\
      &          & TOF   & MC  & TOF, $p_T$ (SVX+CT)  & 420 \\
      &          & DIT   & CT  & $p_T$ (SVX+CT), HC veto &  430 \\
      &          & DIT6  & CT  & same + $\dedx$ (SVX+CT+PS) & 420 \\
\hline\hline
140   &  317     & DIT   & CT  & $p_T$ (SVX+CT), HC veto &  430 \\
      &          & DIT6  & CT  & same + $\dedx$ (SVX+CT+PS) & 420 \\
\hline\hline
142.5 &   24     & SMET   & $\etmiss$ & $p_T$ (SVX), PS+EC+HC veto &  345 \\
      &          & SMET6  & $\etmiss$ & same + $\dedx$ (SVX) & 320 \\
\hline\hline
150   &   11     & SMET   & $\etmiss$ & $p_T$ (SVX), PS+EC+HC veto &  310 \\
      &          & SMET6  & $\etmiss$ & same + $\dedx$ (SVX) & 270 \\
\hline\hline
185   &   3.3    & SMET   & $\etmiss$ & $p_T$ (SVX), PS+EC+HC veto &  215 \\
      &          & SMET6  & $\etmiss$ & same + $\dedx$ (SVX) & 120 \\
\hline\hline
200 &   2.4      & SMET   & $\etmiss$ & $p_T$ (SVX), PS+EC+HC veto &  185 \\
\hline\hline
250 &   1.0     & SMET   & $\etmiss$ &  $p_T$ (SVX), PS+EC+HC veto  &  125 \\
\hline\hline
300 &   0.56    & HIP    & $\gamma,\etmiss$ & $b^\pi$ (SVX,L0), $p_T^\gam$, $\etmiss$, 
$p_T^\pi$ (CT),  EC+HC veto & 95 \\
\hline\hline
600 &   0.055   & HIP    & $\gamma,\etmiss$ & $b^\pi$ (SVX,L00), $p_T^\gam$, $\etmiss$, 
$p_T^\pi$ (CT), EC+HC veto & 75 \\
\hline\hline
$750-?$ &   $\sim 0$    & $\gam+\etmiss$ & $\gam,\etmiss$ & $p_T^\gam$, $\etmiss$ & $<60$ \\
\hline
    \end{tabular}
    \caption{Best signals at Run II for $\cpone\cmone$ 
and $\cpmone\cnone$ production
and important detector components and measurements as a function of $\dmchi$.
Mass reaches quoted are 95\% CL for $L=30\fbi$. 
The PS, EC, or HC veto requires
no preshower, small EC, or small HC
energy deposits in a $\Delta R<0.4$ cone around the $\cpmone$ track candidate. 
$p_T$ ($p_T^\pi$) is the $p_T$
of the $\cpmone$ ($\pi^\pm$ from $\cpmone\to\pi^\pm\cnone$). $b^\pi$
is the $\pi^\pm$ impact parameter.}
    \label{tab:summary}
\fussy
  \end{center}
\end{table}
An overall summary of the signals and their mass reach at the Tevatron
for detecting $\cpone\cmone$ and $\cpmone\cnone$ production
in the $M_2<M_1\ll|\mu|$ scenario appears in Table~\ref{tab:summary}.
Clearly, the very real possibility that $\dmchi\geq 300\mev$ would
present us with a considerable challenge.

For purposes of comparison, we note that in an mSUGRA scenario the tri--lepton
signature from $\cpmone\cntwo$ production allows one to probe
chargino masses up to about $160\gev$ for $L=30\fbi$ when the 
scalar soft--SUSY--breaking mass is large \cite{trilepref}.

Finally, we wish to note that the precise values of $\mcpmone$ and 
$\dmchi$ will be of significant theoretical interest. $\mcpmone$
will be determined on an event--by--event
basis if the chargino's momentum and velocity can both be measured.
This will be possible for the LHIT, TOF, DIT, and STUB
signals by combining tracking information with ionization information.
(Note that, in all these cases,
accepting only events roughly consistent with a given value of
$\mcpmone$ will provide further discrimination against backgrounds.)
However, for the HIP and $\gam+\etmiss$ signals $\mcpmone$ can
only be estimated from the absolute event rate. As regards $\dmchi$,
it will be strongly constrained by knowing which signals
are present and their relative rates. In addition,
if the the soft charged pion can be detected,
its momentum distribution, in particular the end--point thereof,
would provide an almost direct determination of $\dmchi$.

For large $\dmchi$ ($\dmchi>20-30\gev?$), one should
explore the potential of the tri--lepton signal coming from $\cpmone\cntwo$
production. However, this is a suppressed cross section when
both the lightest neutralino and lightest chargino are wino--like.
Standard mSUGRA studies do not apply without modification; the cross 
section must be rescaled and the lepton acceptance
recalculated as a function of $\dmchi$.
A detailed study is required to determine
the exact mass reach as a function of $\dmchi$.

Of course, additional SUSY signals
will emerge if some of the squarks, sleptons and/or 
sneutrinos are light enough 
(but still heavier than the $\cpmone$) that 
their production rates are substantial.
In particular, leptonic signals
from the decays [\eg\ $\wtil \ell_L^{\pm}\to \ell^{\pm}\cnone$ or 
$\wtil\nu_{\ell} \to \ell^{\pm} \cmpone$] would be present.

Given the possibly limited reach of the Tevatron when the lightest
neutralino and chargino are nearly degenerate, it will be very important
to extend these studies to the LHC.  A particularly important issue
is the extent to which the large $c\tau$ tails of the $\cpmone$
decay distributions can yield a significant rate in 
the background--free channels studied here.
Hopefully, as a result of the very high
event rates and boosted kinematics expected at the LHC, 
the background--free
channels will remain viable for significantly larger $\dmchi$ 
and $\mcpmone$ values than those to which one has sensitivity at the Tevatron.
In this regard, a particularly important issue 
for maximizing the mass reach of these channels will be the
extent to which tracks in the silicon vertex detector and/or
in the central tracker can be used for triggering in
a high--luminosity enviroment. 

\bigskip

While finalizing the details of this study, other papers 
\cite{randall2,wells}
appeared on the same topic.  Some of the signatures discussed here are
also considered in those papers. Our studies
are performed at the particle level and contain the most important
experimental details.

\bigskip

\def\beq{\begin{equation}}
\def\eeq{\end{equation}}
\def\eq{\end{equation}}
\def\to{\rightarrow}
\def\gsim{\lower.7ex\hbox{$\;\stackrel{\textstyle>}{\sim}\;$}}
\def\lsim{\lower.7ex\hbox{$\;\stackrel{\textstyle<}{\sim}\;$}}
\def\mev{\hbox{\rm\,MeV}}            
\def\gev{\hbox{\rm\,GeV}}            
\def\tev{\hbox{\rm\,TeV}}
\def\agt{\stackrel{>}{\sim}}
\def\alt{\stackrel{<}{\sim}}
\def\order#1{{\cal O}(#1)}
\def\EPC#1#2#3{Eur. Phys. J. C {\bf #1}, #3 (19#2)}
\def\NPB#1#2#3{Nucl. Phys. B {\bf #1}, #3 (19#2)}
\def\PLB#1#2#3{Phys. Lett. B {\bf #1}, #3 (19#2)}
\def\PLBold#1#2#3{Phys. Lett. B {\bf#1} (19#2) #3}
\def\PRD#1#2#3{Phys. Rev. D {\bf #1}, #3 (19#2)}
\def\PRL#1#2#3{Phys. Rev. Lett. {\bf#1}, #3 (19#2)}
\def\PRT#1#2#3{Phys. Rep. {\bf#1} (19#2) #3}
\def\ARAA#1#2#3{Ann. Rev. Astron. Astrophys. {\bf#1} (19#2) #3}
\def\ARNP#1#2#3{Ann. Rev. Nucl. Part. Sci. {\bf#1} (19#2) #3}
\def\MODA#1#2#3{Mod. Phys. Lett. A {\bf #1} (19#2) #3}
\def\ZPC#1#2#3{Zeit. f\"ur Physik C {\bf #1}, #3 (19#2)}
\def\APJ#1#2#3{Ap. J. {\bf#1} (19#2) #3}
\def\MPL#1#2#3{Mod. Phys. Lett. A {\bf #1} (19#2) #3}
%
%
\def\beq{\begin{equation}}
\def\eeq{\end{equation}}
\def\bea{\begin{eqnarray}}
\def\eea{\end{eqnarray}}
%
%
\def\slashchar#1{\setbox0=\hbox{$#1$}           
   \dimen0=\wd0                                 
   \setbox1=\hbox{/} \dimen1=\wd1               
   \ifdim\dimen0>\dimen1                        
      \rlap{\hbox to \dimen0{\hfil/\hfil}}      
      #1                                        
   \else                                        
      \rlap{\hbox to \dimen1{\hfil$#1$\hfil}}   
      /                                         
   \fi}                                         %
%
%
\catcode`@=11
\long\def\@caption#1[#2]#3{\par\addcontentsline{\csname
  ext@#1\endcsname}{#1}{\protect\numberline{\csname
  the#1\endcsname}{\ignorespaces #2}}\begingroup
    \small
    \@parboxrestore
    \@makecaption{\csname fnum@#1\endcsname}{\ignorespaces #3}\par
  \endgroup}
\catcode`@=12
\def\jfig#1#2#3{
 \begin{figure}
 \centering
 \epsfysize=3.0in
 \hspace*{0in}
 \epsffile{#2}
 \caption{#3}
 \label{#1}
 \end{figure}}
\def\sfig#1#2#3{
 \begin{figure}
 \centering
 \epsfysize=2.5in
 \hspace*{0in}
 \epsffile{#2}
 \caption{#3}
 \label{#1}
 \end{figure}}


\section{Superheavy Supersymmetry}
\noindent
\centerline{\large\it S.~Ambrosanio, J.~D. Wells}
\medskip
\label{section:Wells}

In the vast space of all viable physics theories, supersymmetry (SUSY) 
is not a point.  Any theory can be ``supersymmetrized'' almost trivially,
and the infinite array of choices for spontaneous SUSY
breaking just increases the scope of possibilities in the real world.
One thing that appears necessary, if SUSY has anything to
do with nature, is superpartners for the standard model particles
that we already know about: leptons, neutrinos, quarks, and gauge
bosons.  These superpartners must feel SUSY breaking and 
{\it a priori} can have arbitrary masses as a result.

Phenomenologically, the masses cannot be arbitrary.  There are several 
measurements that have been performed that effectively limit what
the SUSY masses can be. First, there are direct limits on
$Z\to {\rm SUSY}$, for example, that essentially require all superpartners 
to be above $m_Z/2$.  Beyond this, collider physics limits become model 
dependent, and it is not easy to state results simply in terms of the mass 
of each particle. Second, comparing softly broken SUSY model 
calculations with flavor changing neutral current (FCNC) 
measurements implies that superpartner masses cannot be light and arbitrary.
And finally, requiring that the $Z$ boson mass not result from a fine-tuned
cancellation of big numbers requires some of the particles masses 
be near $m_Z$ (less than about $1\tev$, say).

Numerous explanations for how the above criteria can be satisfied have
been considered.  Universality of masses, alignment of flavor matrices, 
flavor symmetries, superheavy supersymmetry, etc., have all been incorporated 
to define a more or less phenomenologically viable explanation of a softly 
broken SUSY description of nature.  

In this contribution, we would like to summarize some of the basic collider
physics implications of superheavy supersymmetry (SHS) at the Tevatron.  
Our understanding is that analyses of all the specific processes
that are mentioned here in principle are being pursued within other
subgroups. Therefore, our goal in this submission is to succinctly explain 
what SHS is and how some of the observables being studied within other 
contexts could be crucial to SHS.  We also hope that by enumerating
some of the variations of this approach that this contribution could help 
us anticipate and interpret results after discovery of SUSY, and help 
distinguish between theories.
The idea we are discussing goes under several names including ``decoupling 
supersymmetry'', ``more minimal supersymmetry'', ``effective supersymmetry'',
``superheavy supersymmetry'', etc.  
The core principle~\cite{cohen} is that very heavy superpartners do not 
contribute to low-energy FCNC or CP violating processes and therefore cannot 
cause problems. Furthermore, no fancy symmetries need be postulated to keep 
experimental predictions for them under control.

On the surface, it appears that decoupling superpartners is completely
irrelevant for the Tevatron.  After all, Tevatron phenomenology is 
limited to what the Tevatron can produce.  Superheavy superpartners,
which we define to be above at least $20\tev$, are of course not within
reach of a $2\tev$ collider.  However, not all sparticles need be superheavy
to satisfy constraints.  In fact, the third generation squarks
and sleptons need not be superheavy to stay within the boundaries of
experimental results on FCNC and CP violating phenomena.  
As an all important bonus, the third family squarks and sleptons
are the only ones that contribute significantly at one loop to the Higgs 
potential mass parameters.  By keeping the third generation sfermion light, 
we simultaneously can maintain a ``natural'' and viable lagrangian even after 
quantum corrections are taken into account.

In short, the first-pass description of SHS is to say that, in absence
of any alignment, special symmetry or other mechanism yielding 
flavor-horizontal degeneracy, all particles which are significantly coupled 
to the Higgs states should be light, and the rest heavy. 
The gluino does not by itself contribute to FCNC, nor does it couple directly 
to the Higgs bosons and so it could be heavy or light.  
However, the gauginos usually have a common origin, either in grand unified 
theories (GUTs), theories with gauge-mediated supersymmetry breaking (GMSB), 
or superstring theories, and so it is perhaps more likely that the gluino is 
relatively light with its other gaugino friends, the bino and the wino.
Furthermore, the $H_d$ could be superheavy as well, but that is not as 
relevant for Tevatron phenomenology.
Therefore, we can summarize the ``Basic Superheavy Supersymmetry'' (BSHS)
spectrum:
\begin{description}
\item[Superheavy] ($\gsim 20\tev$): $\tilde Q_{1,2}$, $\tilde u^c_{1,2}$, 
 $\tilde d^c_{1,2}$, $\tilde L_{1,2}$, $\tilde e^c_{1,2}$;

\item[Light] ($\lsim 1\tev$): $\tilde Q_{3}$, $\tilde t^c$, 
$\tilde B$, $\tilde W$, $H_u$, $\mu$ (higgsinos);

\item[Unconstrained] (either light or heavy):  
 $\tilde b^c$, $\tilde L_3$, $\tilde \tau^c$, $\tilde g$, $H_d$. 
 
\end{description}
Specific models of SUSY breaking will put the ``unconstrained'' 
fields in either the ``superheavy'' or ``light'' categories. 

Any question about relative masses within each category above can not be 
answered within this framework.
In fact, that is one of the theoretically pleasing aspect of this approach:
no technical details about the spectrum need be assumed to have a
viable theory. Another nice feature is that the mass pattern for 
the scalar partners across generations is somewhat opposite to that 
of the SM fermions. This might well inspire a profound connection 
between the physics of flavor and SUSY breaking. 
A possible theoretical explanation of such a large mass hierarchy in the 
scalar sector is that it could be a result of new gauge interactions
carried by the first two generations only, and which could be, e.g., 
involved in a dynamical breaking of SUSY. For Tevatron enthusiasts, it 
is a frustrating model, since we do not even know what phenomenology should 
be studied because things will change drastically depending on the relative 
ordering of states in the ``light'' category. 

However, there are several features about the BSHS spectrum which are 
interesting not because of the phenomena that it predicts at the Tevatron,
but rather for what it does not predict.
For example, $\tilde q_{1,2} \tilde g$ and $\tilde q_{1,2}\tilde q'_{1,2}$ 
production is not expected at the Tevatron.  This is a potentially large 
source of events in other scenarios, such as minimal supergravity (mSUGRA), 
but is not present here.  A more predictive feature is the expectation of
many bottom quarks and $\tau$ leptons in the final state of SUSY
production.  For example, $p\bar p \to \tilde \chi_2^0 \tilde \chi^\pm_1$ 
will not be allowed to cascade decay through $\tilde e_L$
for example, but may have hundred percent branching fractions to
$\tau$ final states. Therefore, while the ``golden tri-lepton'' signals
are generally suppressed in these models, efforts to look for specific 
$3\tau$ final states are relatively more important to study in the context 
of SHS compared to other models.
Furthermore, light $\tilde t$ and $\tilde b$ production either directly
or from gluino (chargino, stop) decays is of added interest in the BSHS 
spectrum, and may lead to high multiplicity $b$-jet final states.  
In short, drawing production and decay diagrams for all possible permutations 
of the BSHS spectrum always yields high multiplicity $\tau$ or $b$-jet final 
states.  From the BSHS perspective, preparation and analysis for $\tau$ and 
$b$-jet identification is of primary importance. 
For instance, while detection of selectrons and smuons would exclude BSHS, 
detection of many staus and no $\tilde e$ or $\tilde \mu$ would be a good 
hint for it (although one could think of other SUSY scenarios where the 
$m_{\tilde e}-m_{\tilde{\tau}}$ splitting is rather large, due e.g. to 
large values of $\tan\beta$). An interesting place to look for violations
of $e-\tau$ universality is $\chi_1^{\pm}$ or $\chi^0_2$ branching fractions,
after gaugino-pair 
($\tilde \chi_1^+ \tilde \chi_1^-$ or $\tilde\chi_1^{\pm}\tilde\chi_2^0$) 
production.

There are two main problems with the BSHS spectrum. The heavy
particles can generate a disastrously large hypercharge Fayet-Iliopoulos
term proportional to $g^2_1$Tr($Ym^2$).  In universal scalar mass scenarios
these terms are proportional to Tr($Y$) which is zero because of the
gravity--gravity--U(1)$_Y$ anomaly cancellation.  In minimal GMSB scenarios
$m^2\propto Y^2+\cdots$, and so Tr($Ym^2$) = Tr($Y^3$) + $\cdots$
vanishes because of the U(1)$_Y^3$ and SU(N)--SU(N)--U(1)$_Y$ anomaly
cancellation. No such principle exists in the BSHS ansatz given above,
and so the Tr($Ym^2$) is generically a problem.  Barring the possibility
of miraculous cancellations, we can cure the ``Tr($Ym^2$) problem'' by
postulating that the superheavy masses follow a GMSB hierarchy, or that the 
superheavy states come in complete multiplets of SU(5), and the masses of 
all states within an SU(5) representation are degenerate or nearly 
degenerate.  
We will consider both possibilities in the following. 
These requirements may lower the stock of ``superheavy supersymmetry'' ideas 
for some, or it may change how one perceives model building based on 
decoupling superpartners, but it has no direct effect on Tevatron 
phenomenology. 

The superheavy states are inaccessible anyway, so how they arrange their 
masses in detail is of little consequence to us here. On the other hand, 
the generic pattern and theoretical principles beyond this arrangement 
may affect the light sector of the model as well, both directly and indirectly
through higher-order mass corrections. Indeed, another more serious problem, 
which has direct consequence to Tevatron phenomenology is related to 
new two-loop logarithmic contributions to the light scalar masses in
SHS~\cite{arkani}.
For example, the relevant renormalization group equation has a term
\beq
\label{two-loop}
\frac{d \tilde m^2_{\rm light,f}}{d\ln Q} \propto \sum_i \alpha^2_i C^f_i 
\tilde m^2_{\rm heavy} + \cdots
\eeq
where $C^f_i$ are Casimirs for $f$, $i$ labels the indices of the 
SM gauge groups, and $m^2_{\rm heavy}$ is the characteristic superheavy 
mass scale. 
This renormalization group equation begins its running at the scale
where SUSY breaking is communicated to the superpartners.
In supergravity, this is the Planck scale, and so the shift in light
superpartner masses is proportional to the right side of eq.~\ref{two-loop}
multiplied by a large logarithm, of order $\ln M_{\rm Planck}/m_{Z}$. 
This term is so large that in order to keep, e.g., the top squark mass squared
from going negative, it must have a mass greater than several TeV at the high 
scale~\cite{arkani}. 
(Similar problems occur for the other ``light scalars'' which could potentially
put us in a charge or color breaking vacuum.)
Even though the top squark mass can be tuned to be light at the $Z$ scale, the 
renormalization group effects of the heavy top-squark at the high scale
feed into the Higgs sector and results in a fine-tuned Higgs potential.
Since fine-tuning is a somewhat subjective criteria, this problem 
may not be fundamental.  

A healing influence on the above two-loop malady is to make the SUSY 
breaking transmission scale much lower than the Planck scale.
This reduces the logarithm and allows for a more natural Higgs potential
without large cancellations. The most successful low-energy SUSY
breaking idea is GMSB~\cite{giudice}.  
There, the relevant scale is not tied to gravity ($M_{\rm Planck}$),
but rather to the scale of dynamical 
SUSY breaking. Transmission of this breaking to superpartner masses can take
place at scales as low as $\sim m_{\rm heavy}$ in this scheme.  

With some thought about the BSHS spectrum and the troubles that could
arise theoretically from it, we seem to be converging on something
that looks more or less like GMSB. In fact, we can think of the input 
parameters for our converging model to be the input parameters of minimal 
GMSB~\cite{giudice}, which are
\beq
\label{gmsb}
\Lambda, M, N_{\rm mess}, {\rm sign}(\mu), \tan\beta,~{\rm and}~\sqrt{F_0}
\eeq
where $\Lambda$ sets the overall mass scale of the superpartners, 
$M$ is the messenger scale, $N_{\rm mess}$ characterizes the number
of equivalent $5+\bar 5$ messenger representations, and $\sqrt{F_0}$ 
determines the interactions of the goldstino with matter.  
Then we add to these parameters, 
\beq
a_{1,2} = \frac{\tilde m^2_{f_{1,2}}(M)}{\tilde m^2_{f_3}(M)}
\eeq
where we define $\tilde m^2_{f_3}(M)$ to be the minimal GMSB 
values of the sfermion masses at the messenger scale excluding D-terms 
($f=\tilde Q$, $\tilde d^c$, $\tilde L$, $\tilde e^c$). 
The two $a_{1,2}$ parameters with the parameters of eq.~\ref{gmsb}
completely specify a gauge-mediated inspired superheavy SUSY  
(GMSS) model. (Another similar parameter might be introduced for 
the Higgs $H_d$ if this is heavy, but this is less relevant to 
Tevatron phenomenology).  
We suggest that analyses can use these input parameters to make
experimental searches and studies of SHS.
Adding some family dependent discrete symmetries on the superpartners and 
messengers would allow such a model to arise in a similar way as ordinary 
gauge-mediated models.
Recall also that in gauge mediation the Tr($Ym^2$) problem can be solved by 
the triple gauge anomaly rather than by the gravity-gauge anomaly
requirement as would be the case if we had heavy sparticles come in
degenerate remnants of $\bar 5$ and/or $10$ representation,
as a result of the presence of an approximate global SU(5) symmetry. 

The psychological disadvantage of this GMSS model is that it is overkill on 
the FCNC problem.  Gauge mediation cures this problem by itself, and there 
might not be strong motivation to further consider mechanisms that suppress 
it. However, gauge mediation does not automatically solve the CP problem, and 
so the heavy first two generations may help ameliorate it to some degree.  
As an aside, the above discussion can be reinterpreted as a powerful 
motivation for GMSB.
We started with no theory principles but rather only experimental constraints
and with some basic reasoning were drawn naturally to gauge mediation.  
However, we know of no compelling theoretical reason why 
$a_{1,2}\neq 1$.  We only know that if the heavy spectrum follows
a minimal gauge-mediated hierarchy, then the ``Tr($Ym^2$)'' problem can be 
solved. (However, it is possible to construct a more complex gauge-mediated
model that does not satisfy Tr$(Ym^2)=0$.) 
Gauge-mediation, of course, is not necessarily the only way to
transmit low-energy SUSY breaking.  From a phenomenological
point of view, one should be open to a more general low-energy
SUSY breaking framework.  

It must be said that in some cases, even when SUSY breaking is 
transmitted at low scales as in GMSS, one 
still could have a hard time avoiding 
color- and charge-breaking vacua.
Indeed, the contribution from the superheavy 
states in eq. \ref{two-loop} can still be large when loops from all the 
scalars of the first two generations add up. 

As anticipated, another possibility to cure the ``Tr($Ym^2$) problem'' and 
the ``two-loop problem'' is with the hybrid multi-scale SUSY models 
(HMSSM)~\cite{hybrid}, using the ``approximate global SU(5)'' pattern:

\begin{description}

\item[HMSSM-I:] The first two generations of the $10$ representation
of SU(5) ($\tilde Q_{1,2}$, $\tilde u^c_{1,2}$, 
$\tilde e^c_{1,2}$) are superheavy ($\tilde m_{10_{1,2}}$), while the
rest of the sparticles are light and approximately degenerate. 

\item[HMSSM-II:] In HMSSM-IIa
all three generations of the $\bar 5$ representation of SU(5) 
($\tilde d^c_{1,2,3}$,  $\tilde L_{1,2,3}$) are superheavy 
($\tilde m_{\bar 5_{1,2,3}}$), while the rest of the sparticles are light.  
In HMSSM-IIb just the first two generations of the $\bar 5$ are superheavy 
($\tilde m_{\bar 5_{1,2}}$).

\end{description}

In these models, one attempts a solution of the FCNC problem by using 
a combination of some decoupling (superheavy scalars) and some degeneracy.  
A theoretical motivation for this could be that due to an approximate 
SU(5) global symmetry of the SUSY breaking dynamics, only some of the 
quark/leptons superfields with the same SU(3)$\otimes$SU(2)$\otimes$U(1)
quantum numbers are involved in the SUSY breaking sector, carry an additional
quantum number under a new ``strong'' horizontal gauge group and are 
superheavy. 
The other superfields instead couple only weakly (but in a flavor-blind way) 
to SUSY breaking and are light and about degenerate.  
 
Actually, these ``hybrid'' models present many advantages compared  
to other SHS realizations. The reduced content of the superheavy 
sector considerably weakens the ``two-loop'' problem, since the negative 
contribution to the light scalar masses squared is less important. This 
is especially true for the HMSSM-II, and in particular the IIb version. 
Actually, it is in this case possible to raise the $m_{\rm heavy}$ scale 
up to $\sim 40\tev$, in a natural way. 
Most problems with FCNC phenomena come from $L-R$ operators, and since
these operators remain suppressed, the hybrid models
are phenomenologically viable and attractive versions of superheavy
supersymmetry.

The resulting spectrum is different than GMSS and BSHS in that some of 
first two generation states are now allowed to be light. 
For example, in the HMSSM-I model, the $\tilde L$ sleptons can be light, 
and on-shell decays of winos into $L+\tilde L$ can allow the trilepton 
signal $\tilde \chi^\pm_1 \tilde\chi^0_2\to 3l$ to have
near $100\%$ branching fraction.  This is not possible in the BSHS spectrum.
Also, it may be useful to study the total rate of jets plus missing energy 
and the kinematics of the events to discern that only $\tilde d^c$, 
$\tilde s^c$ and 3rd-generation squarks are light, and the remaining 
squarks are heavy. More detailed phenomenological studies might start
from observing that in the ``hybrid'' case too, one still needs 
low-energy SUSY breaking to deal with the ``two-loop problem''. 
Again, a GMSB-inspired spectrum for the light sector corrected by 
the (here reduced) presence of the heavy scalars seems relevant as a 
starting point. In this case, a parameterization along the lines described
above for the GMSS would involve new additional parameters such as 
$\tilde m_{10_{1,2}}$ for the HMSSM-I or 
$\tilde m_{\bar{5}_{1,2(,3)}}$ for the 
HMSSM-IIa(,b), plus possibly an analogous parameter for $H_d$.

Whether the spectrum is more minimal GMSB-like or is better 
described by the ``hybrid models'', there is one feature in common.  
Due to the ``two-loop problem'', SHS appears more natural with low-energy 
supersymmetry breaking, independent of how the SUSY breaking 
and transmission are accomplished 
(minimal gauge-mediation ideas or otherwise).  This implies that the lightest 
superpartner is the gravitino rather than the neutralino, as e.g. in mSUGRA. 

Depending on the details of SUSY breaking and the transmission of that 
breaking to superpartners (e.g., whether $\sqrt{F_0}$ in eq. \ref{gmsb} 
is much larger or smaller than about $100\tev$), the next-to-lightest 
superpartner (NLSP) will either decay promptly in the detector, or decay 
with a long lifetime outside the detector.  This may very well dominate the 
phenomenological implications of the model. Another important feature is the 
identity of the NLSP. It is well known that, e.g. in a GMSB-like spectrum, 
the best candidates are the $\tilde \chi^0_1$ and the lightest stau 
$\tilde \tau_1 \simeq \tilde \tau_R$. 
In SHS, a scenario with a neutralino NLSP, with associated decays such as 
$\tilde \chi^0_1\to \gamma \tilde G$ possibly inside the detector, is still 
an important possibility. In this case, multiple high-$p_T$ photons are the 
tags to spectacular events.
On the other hand, in the GMSS model and in the HMSSM models, the 
$\tilde \tau_R$ is always part of the light scalar sector. 
In addition, here the negative contributions from the heavy scalars
to its mass will tend to lower it compared to the neutralino mass. 

Further, in many realizations of SHS the big mass hierarchy between the 
Higgses $H_u$ and $H_d$ can trigger very large ${\cal O}(m_{\rm heavy}/M_Z)$ 
values of $\tan\beta$ (which might provide a reasonable explanation of
the large $m_t/m_b$ ratio without fine-tuning). 
As a side result, after $L-R$ mixing, the $\tilde \tau_1$ mass might 
turn out to be even lighter relative to the other scalars and the 
neutralino than in GMSB models. Hence, we believe that the possibility 
of a $\tilde \tau_1$ NLSP in SHS deserves very serious consideration. 
If this is the case, the NLSP is charged and might live beyond the detector
if $\sqrt{F_0}$ is relatively large.  
Then stable charged particle tracks in the calorimeter will 
be tags to even more spectacular events~\cite{dimo}.  
Many of the results in the
gauge-mediation literature will directly apply for {\it discovery}.
After discovery, the particles that come along with the spectacular
stable charged tracks (SCTs) or the high-$p_T$ photons can then be studied to 
find out with great confidence the light particle content of the theory, 
that could distinguish between the superheavy and the ``traditional''
models. 

As an example of distinguishing phenomenology, we can define 
$R_{\ell' \ell}$ to be
\beq
R_{\ell' \ell} \equiv \sum_{\stackrel{\scriptscriptstyle\ell=e,\mu }
{\scriptscriptstyle\ell'=e,\mu}}
\frac{\sigma[2 \; {\rm SCTs} + \ell'^+ \ell^-]}
{\sigma[2 \; {\rm SCTs} + X]}.
\eeq
From total SUSY production in HMSSM-IIb one expects 
$R_{\ell'\ell} < 1/10$ since $\tilde e$ and $\tilde \mu$ cannot participate
in the decays.  Most events will then have 
$X=\tau^+_{\rm hard}\tau^-_{\rm hard}$ accompanying the 2~SCTs.
However, in minimal GMSB the $\tilde e$ and $\tilde \mu$ are present
in the low-energy spectrum, and so
$\tilde \chi^+\to e^+ \nu_e \tau^\pm_{\rm soft}\tilde \tau^\mp$
may proceed with large branching fraction.  Although 
a precise number depends mainly on the number of
messenger representations and $\tan\beta$,
$R_{\ell'\ell}$ could be greater than $1/2$ in GMSB.
More generally, the unusually large 
$L-R$ mass hierarchies that are typical of ``hybrid'' models may allow
identification of observables suitable for discerning superheavy supersymmetry
from other more conventional forms of supersymmetry at the Tevatron.

\def\ebar{\hbox{E\kern-0.5em\lower-0.1ex\hbox{/}}} 


\section{Study of missing-$E_T$ plus jet signals at the Tevatron Run II}
\noindent
\centerline{\large\it A.~Brignole, A.~Castro, F.~Feruglio, G.F. Giudice, M. Mangano,}
\centerline{\large\it R. Rattazzi, J.D.~Wells, F.~Zwirner}
\medskip
\label{section:Brignole}

\subsection{Introduction}
Final states with large amounts of missing transverse energy ($\ebar_T$) are of
particular interest for searches of physics beyond the Standard Model in
hadronic collisions. The canonical signatures include, in addition to the
$\ebar_T$, the presence of 3 or more jets. In this case, the signal may come, 
for example, from the associated production of gluinos or scalar quarks,
which undergo cascade decays to stable, invisible neutralinos and to several
hadronic jets. It has been recently pointed out that signatures with just one
jet can be of interest for several classes of processes beyond the Standard
Model.  This
is the case, for example, of the associated production of a jet with a 
pair of very light gravitinos~\cite{brig}, or  of a
jet with one of the light Kaluza-Klein modes of the graviton from large 
extra dimensions~\cite{savas,grw,KK}.
                                             
The signature of one jet plus $\ebar_T$ has been neglected experimentally for a
long time. The parton-level studies presented in refs.~\cite{brig,grw,KK} 
indicate
however that this signal can be usefully exploited for searches of new physics,
in spite of the potentially large backgrounds. The purpose of this contribution
is to provide a realistic estimate of the discovery potential for this channel,
once a realistic simulation of the detector performance is included. We shall
estimate the expected backgrounds remaining after proper signal selection
criteria are applied, using current CDF measurements to extract the absolute
normalization of the backgrounds. We then extrapolate these estimates to the
2~TeV, 2~fb$^{-1}$ scenario expected at the Tevatron in the coming years, and
extract the number of events which can be excluded at the 95\%CL from an
observation consistent with the anticipated backgrounds. This result can then
be used to study the potential for discovery of arbitrary processes leading to 
jet+$\ebar_T$ final states. We present explicit results for the case 
of a very light-gravitino, studied in ref.~\cite{brig}, and of Kaluza-Klein
graviton excitations, studied in ref.~\cite{grw}.

\subsection{The strategy}
The topology we are interested in is the one with a high-$E_T$ jet plus a large
missing transverse energy ($\ebar_T$). In this study we  derive a set of
simple requirements which select effectively such a topology. This selection is
based on simple calorimetric and tracking 
informations. For this reason the full simulation of
the CDF detector \cite{cdf} operating at the Tevatron of Fermilab during Run I
is appropriate enough. For the time
being  we use no specific  lepton identification, so
we do not correct the $\ebar_T$
measurement for the presence of high-$P_T$ muons but only for jet energy
mismeasurement.
\par
We rely on simple requests on the minimum energy of the leading jet (clustering
cone 0.4, $E_T\ge 80$ GeV, $\vert\eta\vert\le 0.7$), 
on the missing transverse energy ($\ebar_T\ge 100$ GeV) and on the
azimuthal angle between the missing transverse energy direction and the closest
among all jets reconstructed with $E_T\ge 15$ GeV and $\vert\eta\vert\le 2.4$,
$\Delta\phi\ge 90^\circ$. Finally we reject events having one or more isolated
tracks with $P_T\ge 15$ GeV/c. Tracks are defined as isolated if the sum of the
$P_T$ of all other tracks within $\Delta R=0.4$ from the candidate is smaller
than 10 GeV/c.
The first requisite favors the selection of events where a large $\ebar_T$ is
associated to a high-$E_T$ jet, without being affected much by the presence of
additional radiation. The cut $\ebar_T\ge 100$ GeV is a
pre-requisite which reduces the number of 
background processes which need to be considered. The third cut
suppresses strongly the instrumental backgrounds, due to 
fluctuations in the jet energy
measurement, which concentrate at low $\Delta\phi$. The cut on the isolated
tracks helps in rejecting $W/Z+jet$ events where leptons from the bosons appear
as well-isolated tracks. While these cuts are appropriate to estimate the
potential for this measurement in Run~II, it goes without saying that it will
be possible to improve on them and to 
identify optimal selection criteria once the data are available.

\subsection{Standard Model backgrounds}
The main Standard Model sources of large $\ebar_T$ are $W+jet$, $Z+jet$,
$t\bar t$, $WW$, $WZ$ and $ZZ$ production. As a first step we evaluate their
production cross sections at $\sqrt{s}=2$ TeV by extrapolating the values
measured/calculated  at 1.8 TeV                  
 with the help
of the PYTHIA~\cite{PYTHIA} 
generator: $\sigma(\sqrt{s}=2~ {\rm TeV})=\sigma(\sqrt{s}=1.8~
 {\rm TeV})
\times\frac{\sigma_{PYTHIA}(2~ {\rm TeV})}{\sigma_{PYTHIA}(1.8~ {\rm TeV})}$.
We  generate with PYTHIA Monte Carlo samples of all these processes with a
full simulation of the CDF detector used for Run I. 
In the case of the leading source of backgrounds, namely the $W/Z$+jet
processes, we compare our results with the study of $W/Z$+jet production
documented in ref.~\cite{cdf_jet}, and rescale the PYTHIA result
by factors of the order of 1.2. It should be noted that this $K$ factor is
extracted from the study of jets with transverse energies in the 20~GeV, while
we shall be interested here in jets and $\ebar_T$ in the range of 100~GeV and
above. We shall therefore assume for this study that the relative factor
between the data and PYTHIA does not depend on the $E_T$ thresholds.
Ultimately, with the Run~II data available, it will be possible to determine
the absolute normalization of the $W/Z$+jet backgrounds by using the sample of 
$(Z\rightarrow \ell^+\ell^-)$+jet, and by accurately measuring the efficiency
of the isolated-track cuts on the $W$+jet events. The accuracy of these
determinations will be limited by the available statistics. Our preliminary
studies indicate that it should be possible to determine the absolute rate of
the $W/Z$+jet backgrounds to within 10\%. 
\par                            
For the Tevatron Run II, with $\sqrt{s}=2$ TeV and an expected integrated
luminosity of 2 fb$^{-1}$, we then expect
$5200\pm 520 $                     
$W+jet$ and $5600\pm 560$ $Z+jet$ events passing our selection. 
We notice that $W$+jet and $Z$+jet events have a similar rate.     
 $Z$+jet events come mainly from the                     
$Z\to\nu\bar\nu$ decay ($ 98\%$), while all $W$+jet events are  
from leptonic decays:                               
$W\to e\nu$ ($27\%$), $W\to \mu\nu$ ($31\%$) and $W\to \tau\nu$ ($42\%$).
In addition we notice that the irreducible background ($Z\to\nu\bar\nu+jet$)
represents $\approx 50\%$ of the total background, while
 the remaining backgrounds
(mainly $W\to \ell\nu+jet$) might be reduced by a more refined selection which
tries to identify the high-$P_T$ leptons from the $W$. 

The other background 
processes have much smaller cross sections. For $t\bar t$ we use the
CDF value of $7.6^{+1.8}_{-1.5}$ pb \cite{cdf_top}. For
the diboson production at 1.8 TeV we use the following values \cite{ohn}:
$\sigma_{WW}=9.5$ pb, $\sigma_{WZ}=2.5$ pb and $\sigma_{ZZ}=1.1$ pb, to which
we associate a systematic uncertainty of the order of $30\%$ from 
different PDF choices. While this uncertainty is probably over conservative, its
impact on the final result is marginal, since the overall contribution of these
processes is negligible, as will be shown.
\par                         
 Such a calculation provides the following values for $\sqrt{s}=2$ TeV:
$\sigma (t\bar t)=10.1^{+2.4}_{-2.0}$ pb, $\sigma (WW)=11.1\pm 3.3$ pb,
$\sigma (WZ)=3.1\pm 0.9$ pb and $\sigma (ZZ)=1.3\pm 0.4$ pb. 
For the Tevatron Run II, we then expect
$60\pm 25$ $t\bar t$ and $110\pm 45$ diboson
events passing our selection, accounting for an additional 25\% 
uncertainty on the selection efficiency.
 This represents of the order of 2\% of the total
background.
\par                                                            

Figs.~\ref{wlnu_rej} and~\ref{znunu_rej} show how our selection suppresses
the production rates of  $W+jet$ ($W\to \ell \nu$) and $Z+jet$ 
($Z\to \nu\bar\nu$) events. 
Fig. \ref{sm_2}  shows, for all  the sources separately, the production
rates  expected as a function of the  $\ebar_T$ threshold. 
{}From these rates we extract the number of non-SM events which can be excluded
at Run~II under the assumption that the observed rates will agree with the SM
expectations. We count the number of
background events expected above a given $\ebar_T$ cut and evaluate the excess
which can be excluded at the 95 \% confidence level. When       
the number of background events is much smaller than 1 this corresponds to a
limit of 3 signal events.  The result is presented in Fig.~\ref{ggj_n_2}, which
shows how the result depends upon the residual uncertainty on the absolute
background rate.

\subsection{The case of ${\tilde G}{\tilde G}+jet$ production}
All realistic models with global supersymmetry must contain the goldstino, a
massless and neutral spin-$\frac{1}{2}$ particle associated with the
spontaneous breaking of supersymmetry. When gravitation is introduced, and 
supersymmetry is realized locally, the spin-$2$ graviton is accompanied by the 
gauge particle of supersymmetry, the spin-$\frac{3}{2}$ gravitino. After
spontaneous supersymmetry breaking, the gravitino acquires a mass by absorbing
the would-be goldstino. The signals of pair-production of squarks, gluinos,
charginos and neutralinos have been extensively discussed at this workshop, in
the case of a heavy as well as of a light gravitino. In the latter case, the
gravitino is the Lightest Supersymmetric Particle, (LSP), and goes undetected
giving rise to $\ebar_T$.
\par
If the gravitino, $\tilde G$, is very light (much lighter than $10^{-4}$
eV/c$^2$), and all the other supersymmetric
particles are above the production threshold, supersymmetry may still be
 seen at
the Tevatron by looking at final states including gravitinos and ordinary
particles only. 
In particular,  we can have the following processes: 
$q\bar q\to \tilde G\tilde Gg$, $qg\to \tilde G\tilde Gq$, $\bar qg\to \tilde
G\tilde G\bar q$ and $gg\to \tilde G\tilde Gg$,  which all lead to a topology
with a single jet + $\ebar_T$. 
In this scenario the main parameters upon which these processes depend are the
supersymmetry-breaking scale $\sqrt{F}$ and the CMS energy $\sqrt{s}$.
We recall that the gravitino mass $m_{\tilde G}$ is related
to $F$ via $m_{\tilde G} M_{Pl}=F/{\sqrt{3}}$, where 
$M_{Pl} = 2.4 \times 10^{18}$ GeV is the reduced Planck mass.

For the production of very light gravitinos ($m_{\tilde G}\ll 10^{-4}$
eV/c$^2$) we 
consider the processes 
$p\bar p\to {\tilde G}{\tilde G}g, ~{\tilde G}{\tilde G}q$ 
leading to a {single jet + large $\ebar_T$} 
topology.
These processes are simulated  at $\sqrt{s}=2$  TeV 
in a private version of HERWIG V5.6~\cite{HERWIG}, 
which reproduces the calculations of
\cite{brig}, in the limit in which the supersymmetric particles of the Minimal
Supersymmetric Standard Model and other exotic particles are heavy. 
The production cross section depends on the supersymmetry-breaking 
parameter $F$: 
$\sigma\sim \frac{1}{F^4}$. For the generation we use  $\sqrt{F}=290$  GeV
as justified in the next section.
We choose a
renormalization/factorization scale $\mu=E_T$ and  the 
MRSD-$^\prime$~\cite{MRSD} set of  
structure functions. In order to reduce the computing time we require the hard
scatter momentum $P_{Thard}\ge 70$ GeV/c: this choice is justified by the fact
that events generated with $P_{Thard} < 70$ GeV/c have $\ebar_T$ below
100 GeV and do not pass our cuts.  With such choice of parameters, 
the production cross section amounts to
$2.62\pm 0.02$ pb for $\ebar_T\ge 100$ GeV. 
To populate the $\ebar_T$ region above 300 GeV we 
generate also
events with $P_{Thard} > 200$ and 300 GeV/c.
All samples are fed to the CDF detector simulation.
\par
Fig. \ref{ggj_sig_lev}  shows the ${\tilde G}{\tilde G}+jet$ 
production cross section  as a function of the $\ebar_T$ threshold, at
parton level, after the event reconstruction and after
our selection.
Fig. \ref{ggj_sig_eff} shows the effect of applying reconstruction and
selection cuts to the signal, relative to the naive results which would be
obtained at the parton level. The loss in efficiency is induced partly by the
shower evolution of the initial and final state, which slightly reduces the
amount of $\ebar_T$ relative to the parton-level configuration, and partly by
other purely detector-driven effects, such as the cuts on isolated tracks. 
It is reasonable to assume that the efficiencies presented in this figure apply
to the case of the signals studied in ref.~\cite{KK}, since the nature of both
the hadronic initial and final states are very similar to those encountered in
the light-gravitino case. Fig.~\ref{ggj_sig_eff} can therefore be used for
realistic estimates of the signal deterioration in the case of 
ref.~\cite{grw,KK},
and, when used in conjunction with the results of Fig.~\ref{ggj_n_2}, can
provide a realistic estimate of the exclusion potential of Run~II for signals
of millimeter-scale extra-dimensions. 
            
\subsection{Comparison ${\tilde G}{\tilde G}+jet$ production $vs$ 
Standard Model backgrounds}
In this Section 
we use the estimated background rate (see Fig. \ref{sm_ggj_2}) to 
 derive the sensitivity to the scale $\sqrt{F}$ in the case of a very
light gravitino.
The best lower 
limit on $\sqrt{F}$ which  can be obtained with our simple selection 
is $370$ GeV if we require $\ebar_T\ge 100$ GeV (see Fig. \ref{ggj_f_2}).
The corresponding lower limit on the gravitino mass is 
$3.3\times
10^{-5}$  eV/c$^2$. 
 We do not use lower thresholds
because the background increases strongly below 100 GeV due to the presence of
instrumental backgrounds.                       
This calculation is
repeated to account for the $\approx 10\%$ global uncertainty on the background
estimate. In this case the best lower limit on $\sqrt{F}$ 
is $\approx 290 $  GeV, reached for $\ebar_T\ge 200$ GeV, and the signal and
background rates are similar in size (see Fig. \ref{sm_ggj_2}). 
This limit corresponds
 to $m_{\tilde G}\ge 2.0\times 10^{-5}$  eV/c$^2$.   The limit has become much
worse because of the large systematic uncertainty on the background which
dominates over the statistical fluctuations up to $\ebar_T\approx 200$ GeV.
Finally, we remark that the existence of a very low supersymmetry-
breaking scale may also show up indirectly at the Tevatron Run II,
in the form of anomalous four-fermion interactions involving standard
quarks and leptons~\cite{brig2}.

\subsection{Production of jet plus Gravitons}

In ordinary gravity, the cross-section for $p\bar p\to {\rm jet}+$
graviton at the Tevatron is about $10^{-26}\, {\rm fb}$.
Needless to say, ordinary quantum gravity effects have no hope of
being seen at the Tevatron. However, 
the scale of quantum gravity may be as low as the weak scale, rather
than the Planck scale, if
gravity propagates in a higher dimensional space  with
$D$ total space-time dimensions~\cite{savas}.  
The
$\delta =D-4$ extra spatial dimensions must
be compactified with a large radius $R$ given by
\begin{equation}
M_{\rm Pl}^2=R^\delta M_D^{2+\delta},
\end{equation}
where $M_D$ is the characteristic quantum gravity scale. The probability
of producing a single light Kaluza-Klein excitation of the graviton
at a collision with typical energy $E$ is of the order of $E^2/M_{\rm Pl}^2$.
This small probability is compensated by the large number of available
graviton excitations. Indeed, 
the number of Kaluza-Klein excitations
of the graviton lighter than $E$ is equal to
\begin{equation}
N_G \simeq (ER)^\delta = E^\delta \frac{M_{\rm Pl}^2}{M_D^{2+\delta}}.
\end{equation}
With $\delta =2$ extra spatial dimensions and $M_D\simeq 1\, {\rm TeV}$,
the multiplicity of graviton states in typical Tevatron processes is 
about $10^{25-29}$. Therefore, the final cross-section may be roughly
comparable to SM cross-sections, and signs of quantum-gravity effects
may appear~\cite{savas}.

Single jet plus gravitons production at the Tevatron has
been calculated and compared to SM backgrounds, using
a parton-level Monte Carlo~\cite{grw,KK}.  
With the efficiencies presented in sections II and III, we can now improve
the parton-level results, relating them to a more realistic event
simulation that includes initial state radiation, jet clustering, and
jet energy reconstruction effects.  

Our parton-level calculation is very similar to that described
in ref.~\cite{grw}, except that here we follow the kinematic acceptance
criteria put forth in section II above.  We retain the 
$\ebar_T>150\,{\rm GeV}$ requirement in ref.~\cite{grw} in order to facilitate
a comparison between that publication and the present
analysis.  After the parton-level simulation, we multiply by a
factor of $0.55$ for the relative efficiency between full reconstruction
and the parton-level results. 
We expect the relative efficiency results for the gravitino signal to be
very similar to that of the gravitons signal considered here, and so this
number is read directly from
Fig.~\ref{ggj_sig_eff}b with $\ebar_T^{\rm min}=150\, {\rm GeV}$.

In Fig.~\ref{MDET} we plot the total number of signal events expected in 
$2\, {\rm fb}^{-1}$ of integrated luminosity, after all efficiencies
are taken into account, for both $\delta=2$ and $\delta=4$.
The {\bf b} curve indicates that we integrate 
the perturbative jet plus gravitons
amplitudes for all values of $\widehat s$ (partonic center of mass energy), and
the {\bf a} curve indicates
that we integrate the perturbative jet plus gravitons amplitudes only for
$\widehat s<M^2_D$ and set the amplitude to zero for $\widehat s>M^2_D$.
This is a way to parameterize our ignorance about the non-perturbative
quantum-gravity regime $\widehat s > M^2_D$.
We expect the full, unitarized cross-section in a complete quantum-gravity
theory
to be somewhere between the {\bf a} curve and the {\bf b} curve.
The limits on the quantum-gravity scale $M_D$ which can be obtained
from $2\, {\rm fb}^{-1}$ of data 
are presented in 
Table~\ref{graviton table}. The limits are derived under the more conservative
assumption of curve {\bf a}, {\it i.e.} assuming $\sigma (\widehat s > M^2_D)=0$.
In Table~\ref{graviton table}
we also show the original estimate of the limits based on the parton results
of ref.~\cite{grw}. As expected, the limits based on the full
jet reconstruction simulation with 10\% background uncertainty
are somewhat lower than the limits based on parton-level Monte Carlo
simulation. Nevertheless, probes of gravity above the TeV mass scale 
can be accomplished with $2\, {\rm fb}^{-1}$ integrated luminosity.
Furthermore, as the background becomes better understood, the range 
of $M_D$ that can be discovered quickly rises, as illustrated in
Table~\ref{graviton table}.  Finally we remark that
the existence of a low quantum-gravity scale can also be tested 
at the Tevatron Run II by studying the effects of virtual graviton exchange
in SM processes~\cite{grw,hew}, as reviewed in these proceedings~\cite{hewp}.

\subsection{Conclusions}
We have studied the production of events with a high-$E_T$ jet plus large
$\ebar_T$ at 2 TeV with the help of Monte Carlo simulations 
for the production of  very-light  gravitinos  and of 
Standard Model backgrounds. We have defined a set of simple selection 
criteria which are quite efficient on the signal ($\approx 50\%$ for
$\ebar_T\ge 100$ GeV) 
and reduce strongly the backgrounds (by a factor $\approx 6000$). 
Comparing  the estimated background to the expected signal as a
function of $\sqrt{F}$, and assuming a conservative 10\% uncertainty on the
absolute background rate, we derive a 95\% C.L. lower
limit on the 
supersymmetry-breaking scale $\sqrt{F}\ge 290$ GeV 
for the Tevatron  Run  II with 2 fb$^{-1}$.               
This limit corresponds to a lower limit on the gravitino mass of  $2.0\times
10^{-5}$  eV/c$^2$ and is expected to  improve once we                
 reduce the systematic uncertainty on the background estimate. 
We have also derived limits on the quantum-gravity scale $M_D$ in theories
with $\delta$ large extra dimensions. The Tevatron Run  II with 2 fb$^{-1}$
can reach the 95\% C.L. lower limits on $M_D$ presented in 
Table~\ref{graviton table}.

\runfig{wlnu_rej}{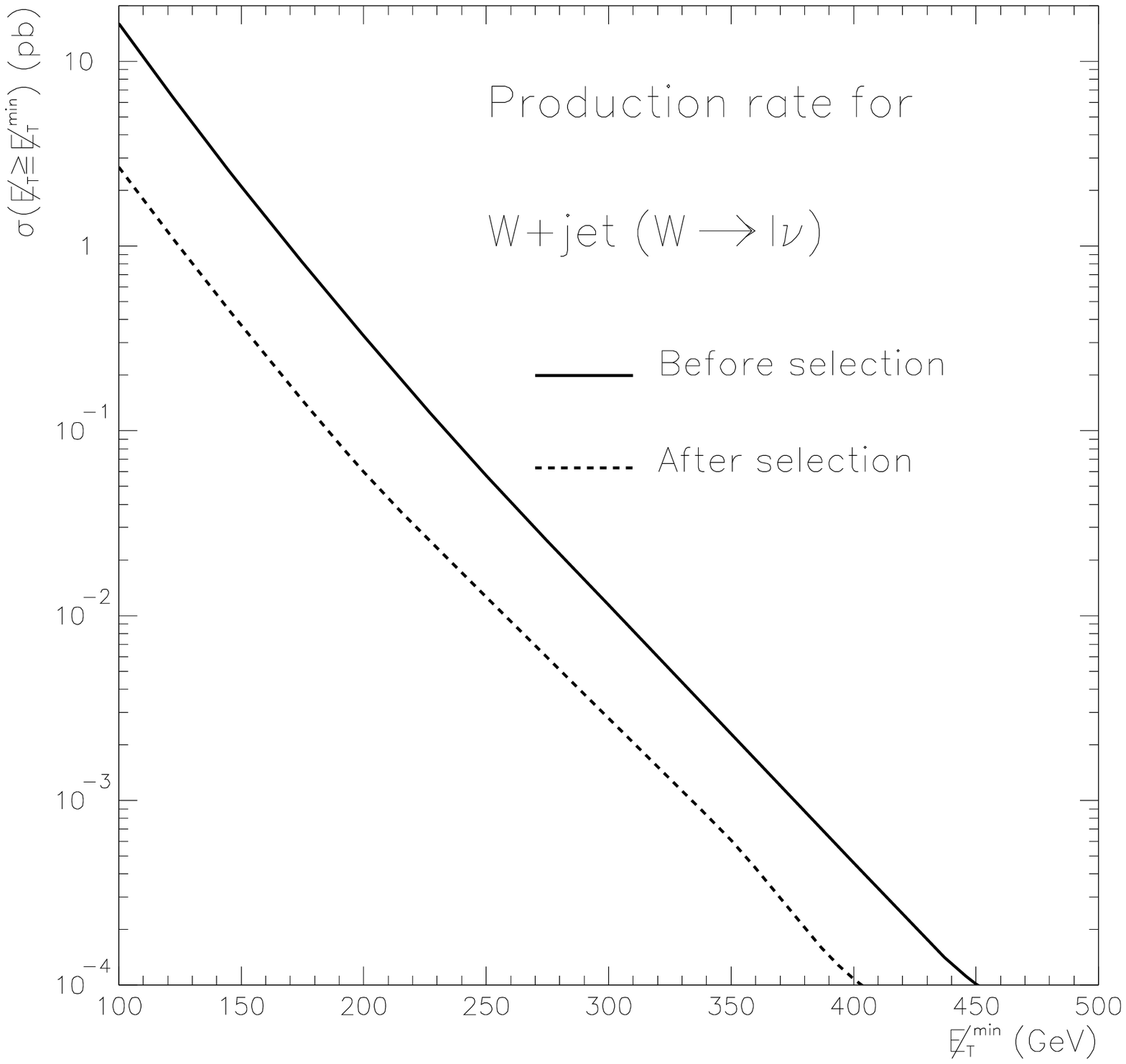}{
Production rate for $W+jet$ ($W\to \ell\nu$) events before
 (solid line) and  after (dashed line) the selection, as a function of
$\ebar_T^{min}$.}

\runfig{znunu_rej}{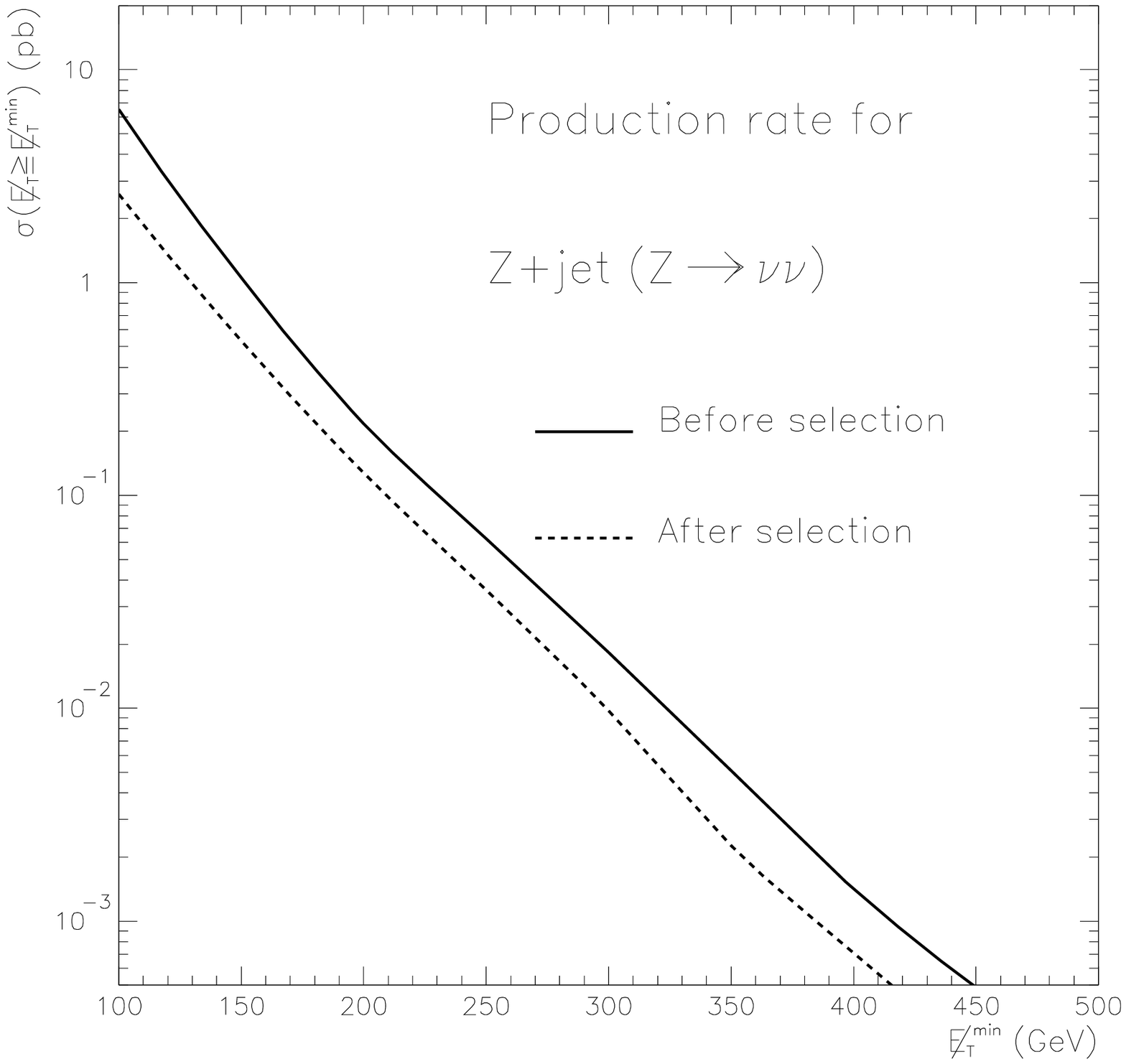}
{ Production rate for $Z+jet$ ($Z\to \nu\bar \nu$) events before
 (solid line) and  after (dashed line) the selection, as a function of
$\ebar_T^{min}$.}

\runfig{sm_2}{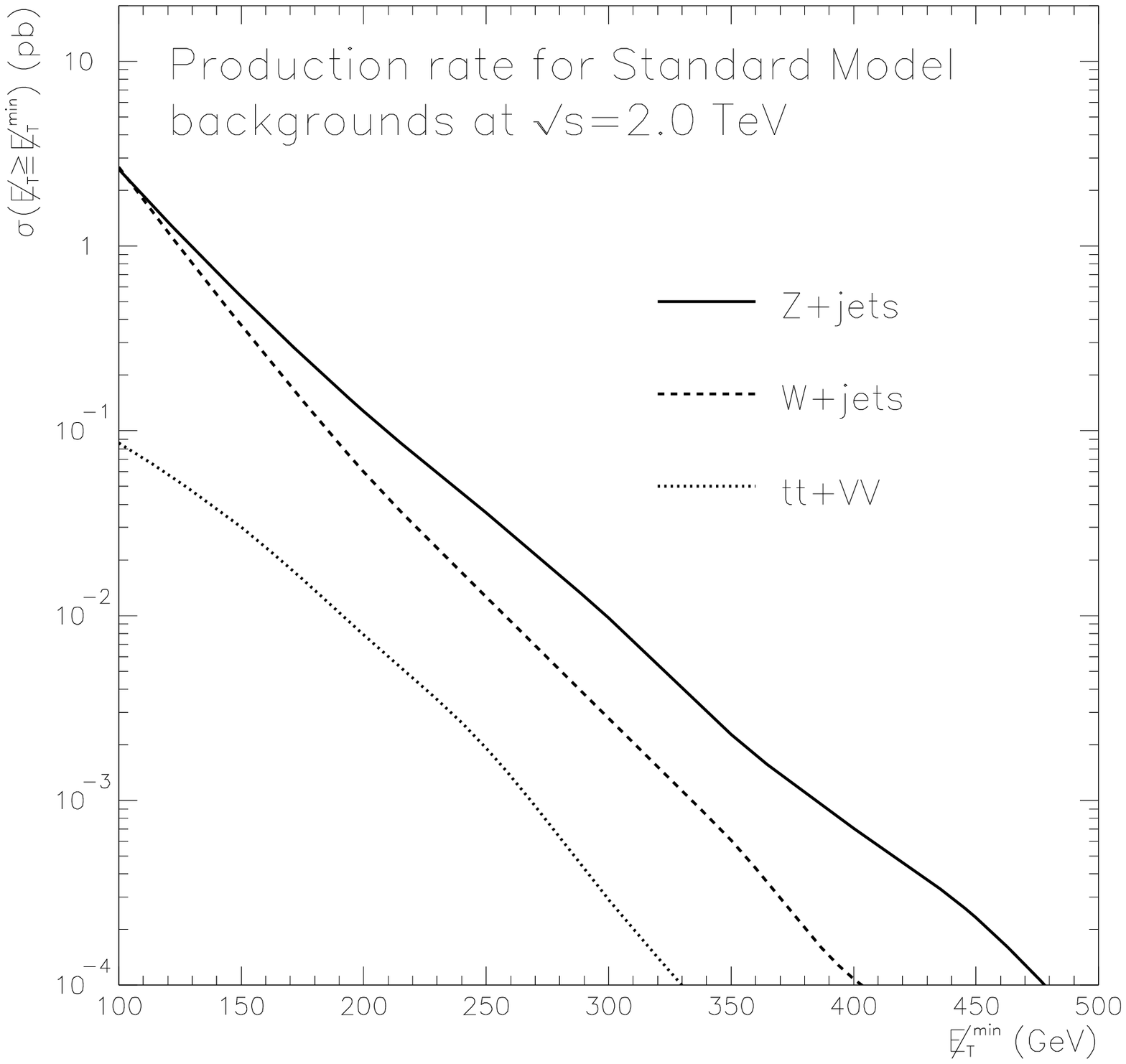}{
Production rate for $\ebar_T\ge \ebar_T^{min}$
for Standard Model backgrounds passing the selection. 
The uncertainties on the production rates are
dominated by  systematics of the order of $10\%$.}

\runfig{ggj_n_2}{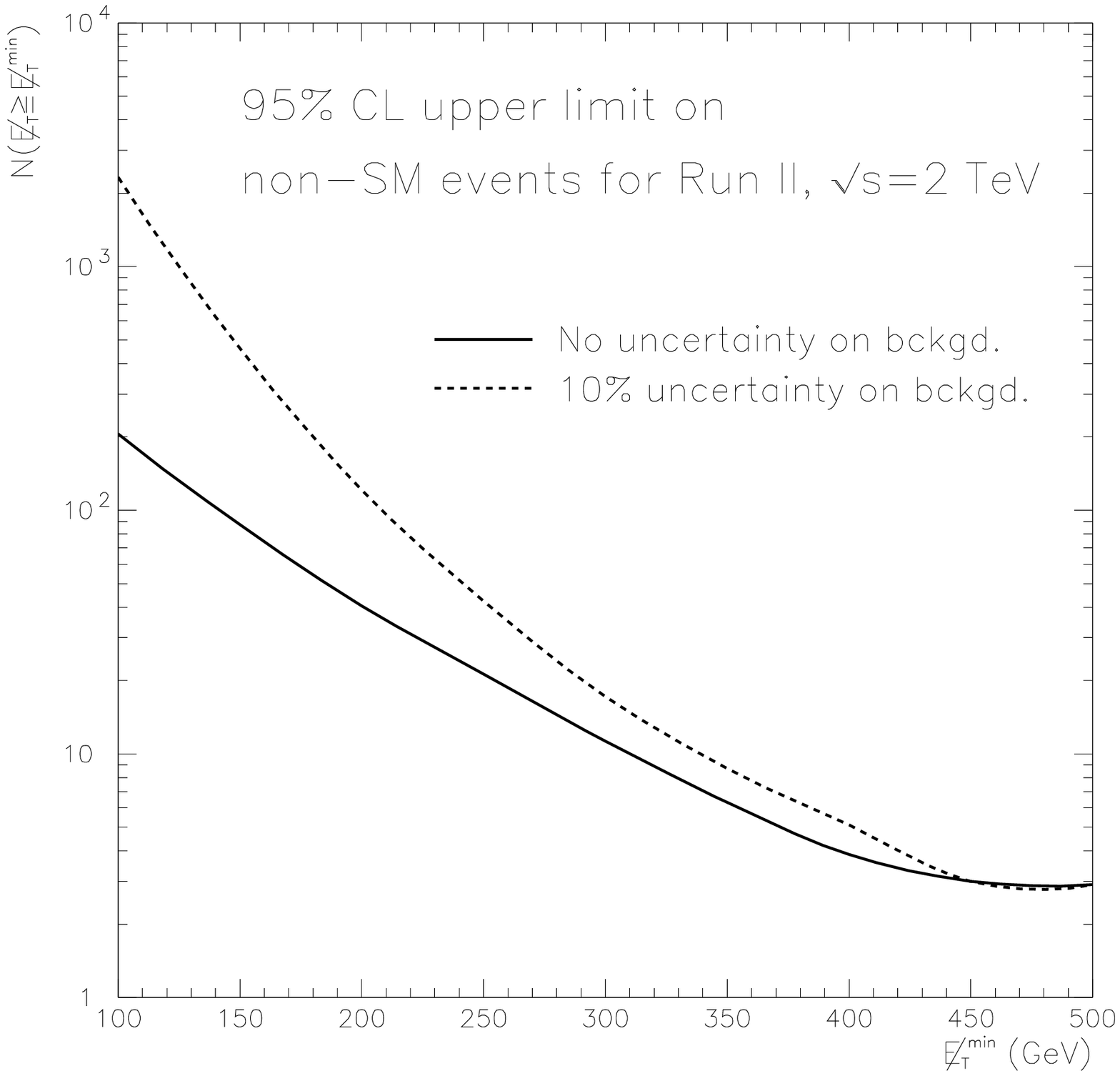}
{Number of non-SM events which can be excluded at
 95\% C.L. as a function of the $\ebar_T$ cut. 
The dashed line refers to the inclusion of a 10\% uncertainty
on the background rate.}

\runfig{ggj_sig_lev}{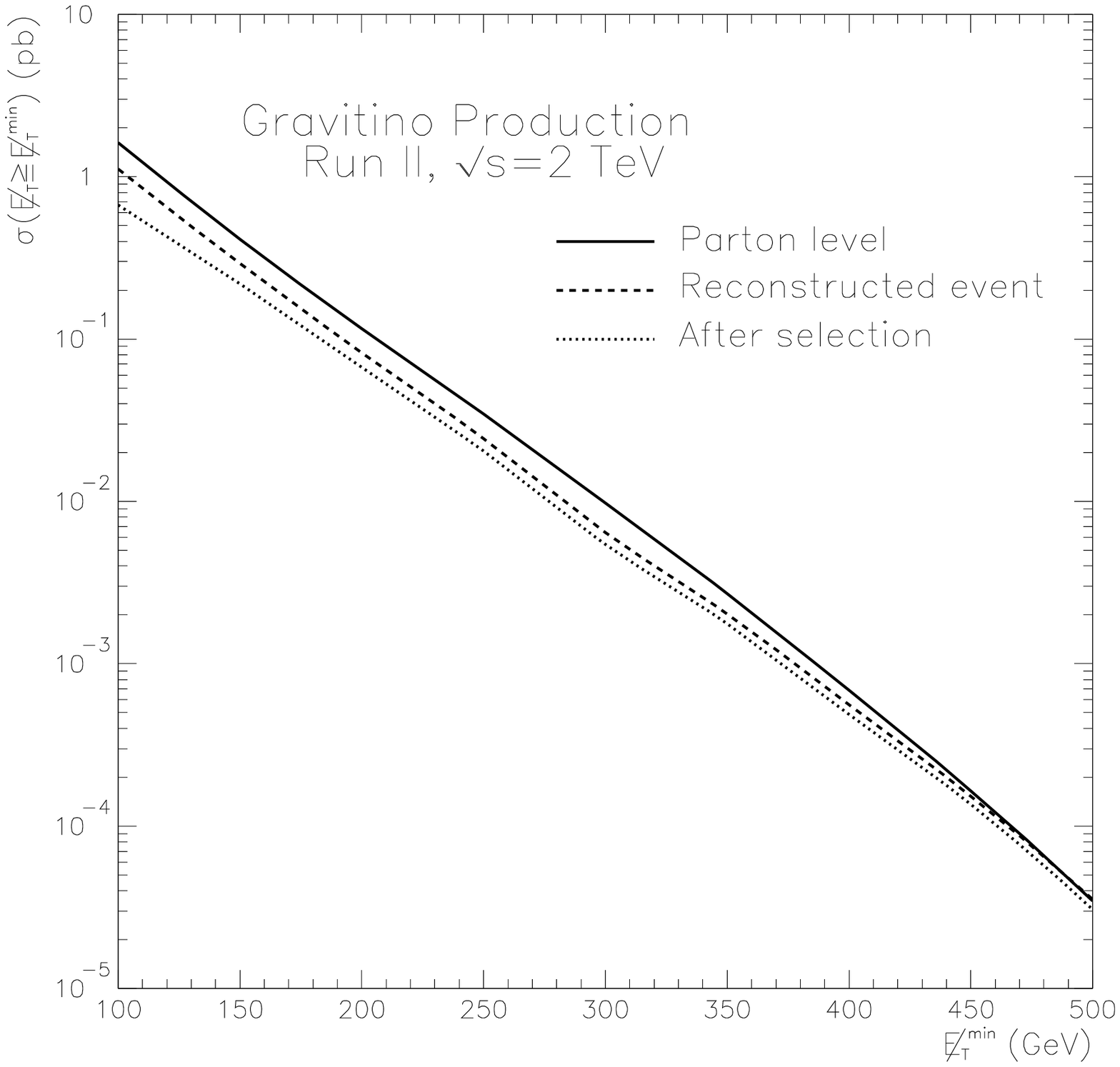}{
Cross sections for 
${\tilde G}{\tilde G}+jet$ ($\protect\sqrt{F}=290$ GeV) events as a function of 
$\ebar_T^{min}$. The solid line represents the parton level
($\ebar_T^{min}=E_{Thard}^{min}$), the dashed line refers to the measured
$\ebar_T$ after the event reconstruction, 
while the dotted line is for the production rate after our selection.}

\runfig{ggj_sig_eff}{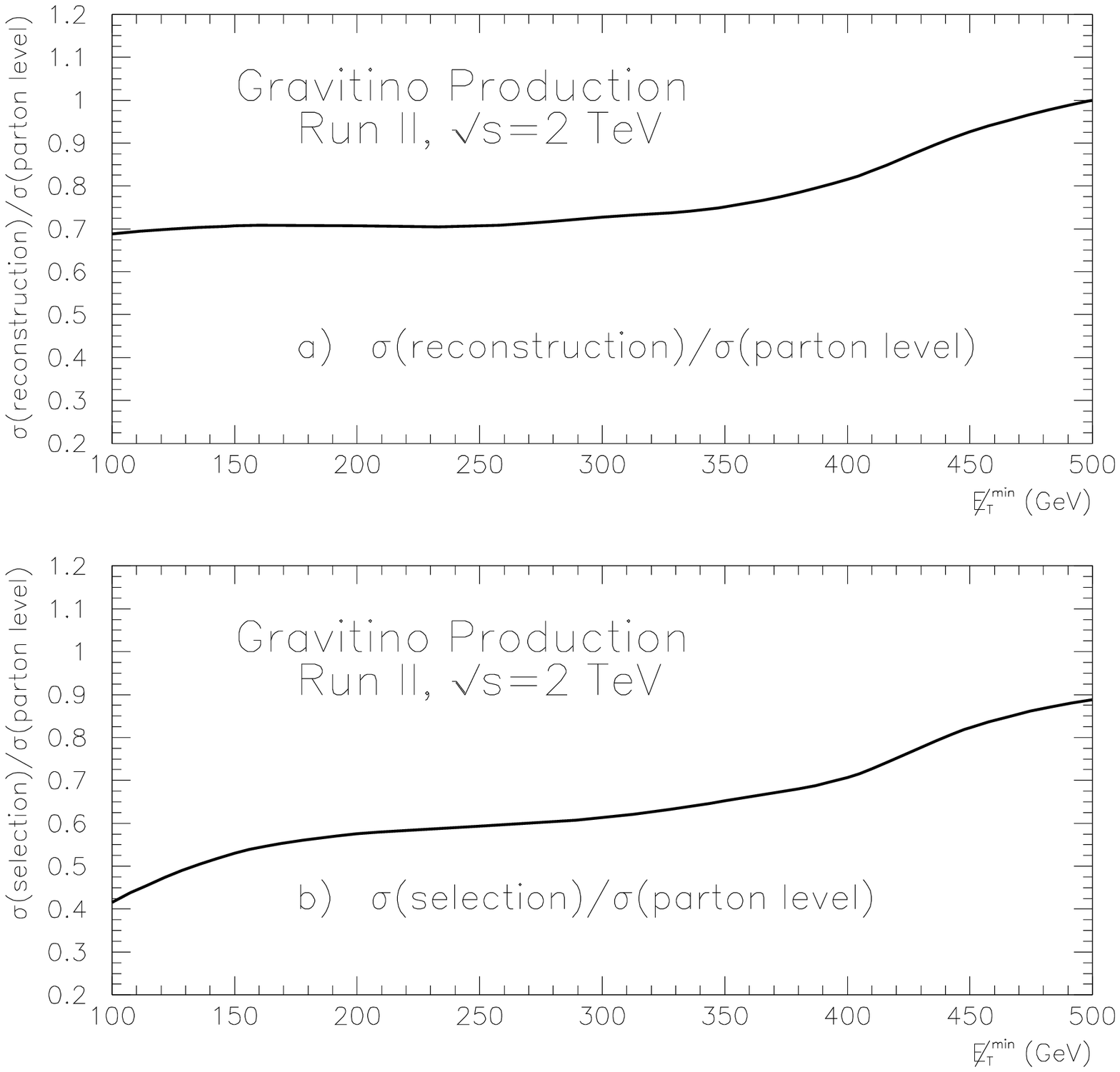}{
Efficiency as a function of $\ebar_T^{min}$ for (a) the event
 reconstruction and (b) the event selection.}

\runfig{sm_ggj_2}{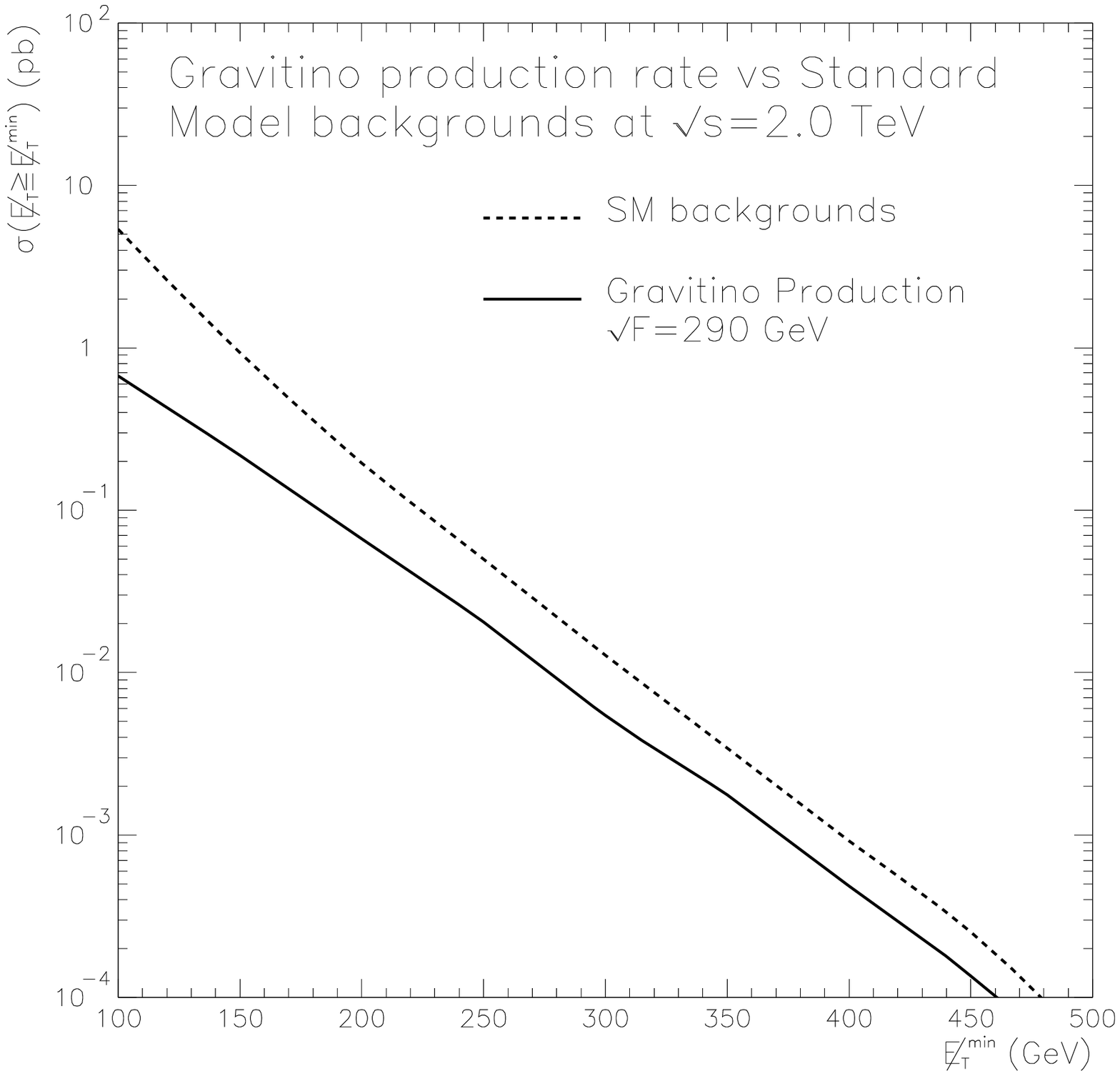}{
Production rate after the selection 
for $\ebar_T\ge \ebar_T^{min}$:
${\tilde G}{\tilde G}+jet$ events (solid line, 
$\protect\sqrt{F}=290$ GeV) and 
Standard Model backgrounds (dashed line). The uncertainties on the 
production rates are
dominated by systematics of the order of $15\%$ for the signal and
$10\%$ for the background.}

\runfig{ggj_f_2}{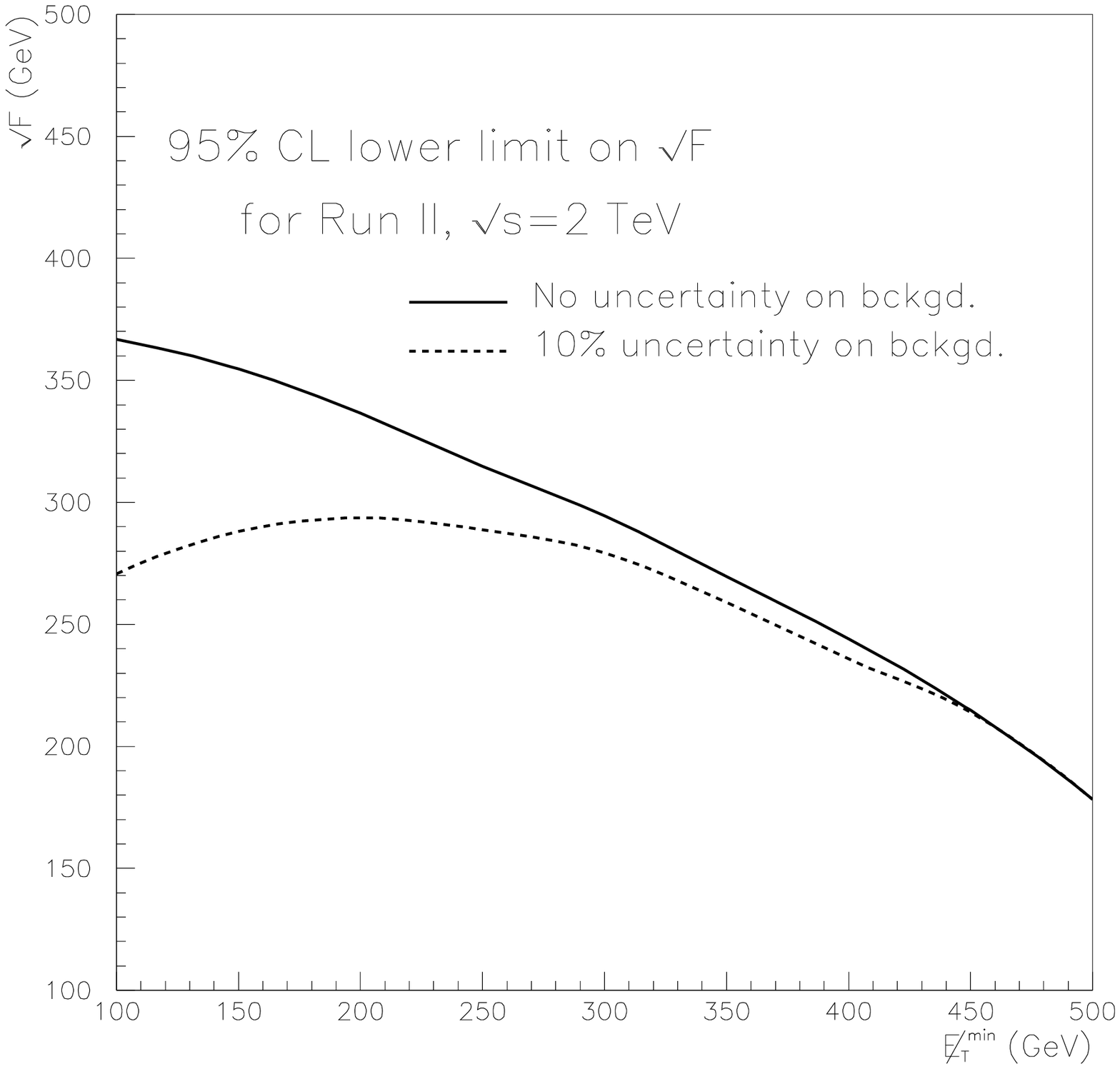}{
Estimated 95\% C.L. lower limit on $\protect\sqrt{F}$, assuming
no deviations from the background production rate, as a function of the
$\ebar_T$ cut. 
The dashed line refers to the inclusion of a 10\% uncertainty
on the background rate.}

\runfig{MDET}{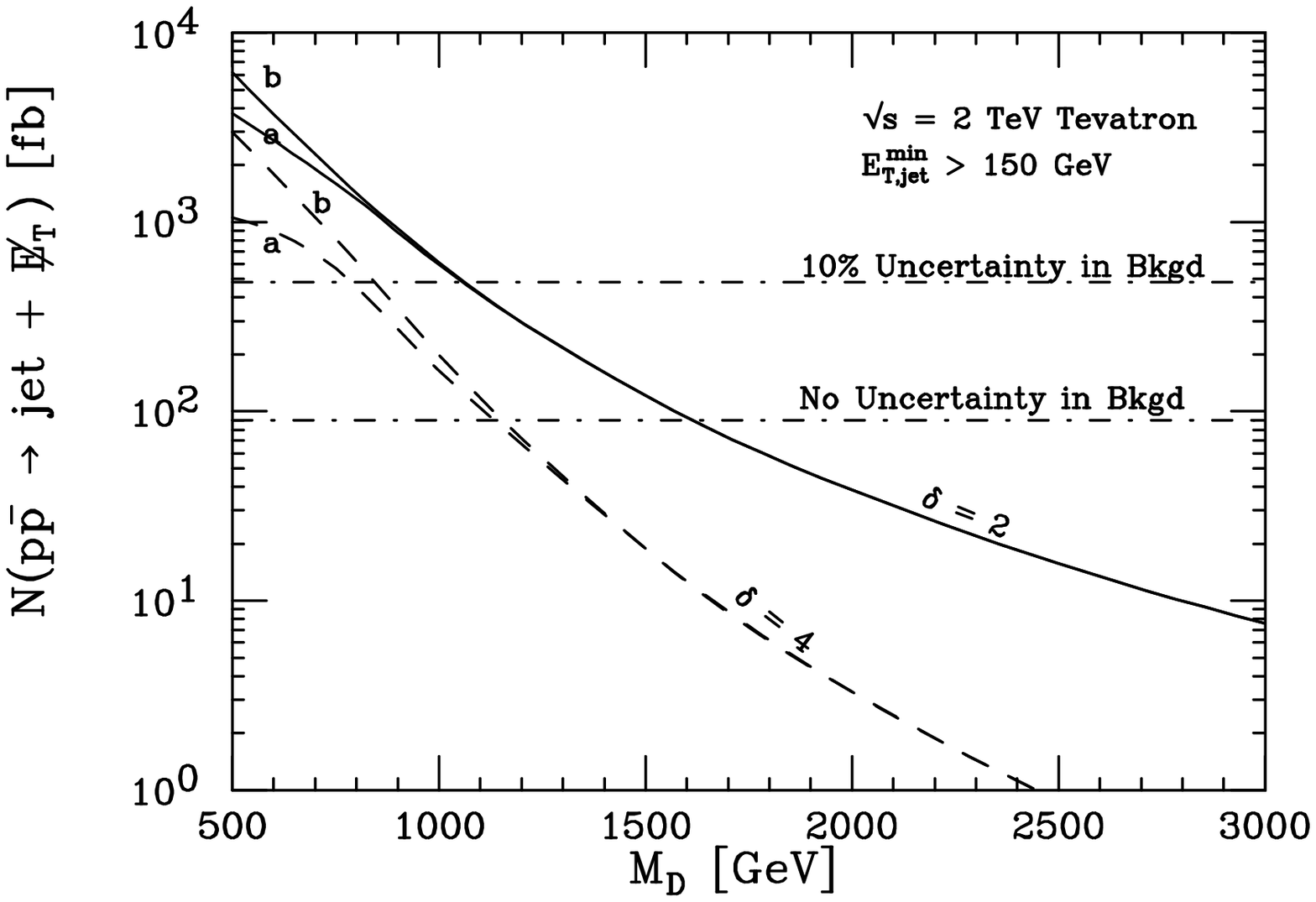}{
Total number of ${\rm jet}+\ebar_T$ selected signal events with
$2\, {\rm fb}^{-1}$ of data as a function of the $D$-dimensional
quantum gravity scale, $M_D$.  
The {\bf a} ({\bf b}) curves are constructed by integrating the cross
section over $\widehat s < M_D$ (all $\widehat s$).
The upper (lower) horizontal line indicates the
95\% C.L. upper limit on non-SM contributions, assuming
10\% (0\%) uncertainty of the SM rate.}

\begin{table}
\protect\protect
\begin{tabular}{cccc}
\hline
$\delta$  &  No Bkgd Uncertainty & 10\% Bkgd Uncertainty 
            & Results from~\cite{grw} \\
  &  $M_D$ Limit [GeV] & $M_D$ Limit [GeV] & $M_D$ Limit [GeV] \\
\hline
2 & 1600 & 1050 & 1400 \\
3 & 1300 & 900 & 1150 \\
4 & 1150 & 800 & 1000 \\
5 & 1000 & 650 & 900 \\
\hline
\end{tabular}
\caption[]{\label{graviton table}
\protect\protect Upper limits on the values of
$M_D$ that can be probed by
$2\, {\rm fb}^{-1}$ at the Fermilab Tevatron with $\sqrt{s}=2\,{\rm TeV}$
using jets plus missing energy.
The first column, $\delta$, indicates the number of extra dimensions,
the second (third) column indicates the limit on $M_D$ attainable 
with 0\% (10\%) uncertainty in the SM background rate.
The fourth column lists the $M_D$ sensitivity found in ref.~\cite{grw},
with a parton-level calculation and assuming a 10\% uncertainty in
the SM background rate.}
\end{table}

\def\simlt{\stackrel{<}{{}_\sim}}
\def\simgt{\stackrel{>}{{}_\sim}}
\def\be{\begin{equation}}
\def\ee{\end{equation}}
\def\bear{\begin{eqnarray}}
\def\eear{\end{eqnarray}}
\def\beqn{\begin{eqnarray}}
\def\eeqn{\end{eqnarray}}

\def\a{\alpha}
\def\b{\beta}
\def\g{\gamma}
\def\e{{\rm e}}
\def\MK{M}
\def\MP{M_{Pl}}
\def\MS{M_{\rm string}}
\def\IS{I^{singlets}}
\def\AS{A^{singlets}}
\def\beq{\begin{equation} }
\def\eeq{\end{equation} }
\def\nolabel{\nonumber }
\def\overbar{\overline}
\def\ol{\overline}
\def\vev#1{\langle #1\rangle}
\def\mvev#1{|\langle #1\rangle|^2}
\def\ben{\begin{eqnarray} }
\def\een{\end{eqnarray} }
\def\bS{{\bf S} }
\def\mod#1{{\rm (mod~2)} }
\def\et{ et.~al }
\def\Tr{{\rm Tr}}
\def\VEV#1{{\langle #1 \rangle} }
\def\ord#1{{\cal{O}}(#1) }
\def\simlt{\stackrel{<}{{}_\sim} }
\def\simgt{\stackrel{>}{{}_\sim} }
\def\Min{ $M_{\rm I}$}
\def\MI{ M_{\rm I}}
\def\Mrn{ $M_{\rm R}$}
\def\MR{ M_{\rm R}}
\def\MU{M_{\rm unif} }
\def\Mun{ $M_{\rm unif}$ }
\def\Mxst{ $M_{\rm X}$}
\def\MX{ M_{\rm X} }
\def\MS{M_{\rm string} }
\def\Mst{ $M_{\rm string}$ }

\def\uop{$U(1)$ }
\def\uops{$U(1)$'s }
\def\uopi{$U(1)_i$ }
\def\uopphi{$U(1)'_{\psi}$ }
\def\uopp{$U(1)'$}
\def\uopap{$U(1)'_A$}
\def\uopa{$U(1)'_A$ }
\def\uopab{$U(1)'_A$-breaking }

\def\ie{{\it i.e., }}
\def\third{\frac{1}{3}}
\def\half{\frac{1}{2}}
\def\o#1{\frac{1}{#1}}

\def\Tr{{\rm Tr}\,}
\def\tr{{\rm tr}\,}
\def\ln{{\rm ln}\,}

\def\KM{Ka\v c--Moody }

\def\phs{$\phantom{\surd}$}
\def\psu{$\phantom{\surd}$}
\def\su{$\surd$}

\def\fract#1#2{{#1\over #2}}
\def\mhalf{-{1\over 2}}
\def\mthird{-{1\over 3}}
\def\mfourth{-{1\over 4}}
\def\twothirds{{2\over 3}}
\def\mtwothirds{-{2\over 3}}
\def\isqrt{{1\over \sqrt{2}}}

\def\half{{1\over 2}}
\def\onehalf{{1\over 2}}
\def\threehalf{{3\over 2}}

\def\third{{1\over 3}}
\def\onethird{{1\over 3}}
\def\twothird{{2\over 3}}

\def\fourth{{1\over 4}}
\def\onefourth{{1\over 4}}
\def\twofourth{{2\over 4}}
\def\threefourth{{3\over 4}}

\def\fifth{{1\over 5}}
\def\onefifth{{1\over 5}}
\def\twofifth{{2\over 5}}
\def\threefifth{{3\over 5}}
\def\fourfifth{{4\over 5}}

\def\sixth{{1\over 6}}
\def\onesixth{{1\over 6}}
\def\twosixth{{2\over 6}}
\def\threesixth{{3\over 6}}
\def\foursixth{{4\over 6}}
\def\fivesixth{{5\over 6}}

\def\tenth{{1\over 10}}
\def\onetenth{{1\over 10}}
\def\twotenth{{2\over 10}}
\def\threetenth{{3\over 10}}
\def\fourtenth{{4\over 10}}
\def\fivetenth{{5\over 10}}
\def\sixtenth{{6\over 10}}
\def\sevententh{{7\over 10}}
\def\eighttenth{{8\over 10}}
\def\ninetenth{{9\over 10}}

\def\twelf{{1\over 12}}
\def\onetwelf{{1\over 12}}
\def\twotwelf{{2\over 12}}
\def\threetwelf{{3\over 12}}
\def\fourtwelf{{4\over 12}}
\def\fivetwelf{{5\over 12}}
\def\sixtwelf{{6\over 12}}
\def\seventwelf{{7\over 12}}
\def\eighttwelf{{8\over 12}}
\def\ninetwelf{{9\over 12}}
\def\tentwelf{{10\over 12}}
\def\eleventwelf{{11\over 12}}

\hyphenation{non-re-norm-al-iz-able re-norm-al-iz-able}
\def\B#1#2#3{\/ {\bf B#1} (19#2) #3}
\def\NPB#1#2#3{{\it Nucl.\ Phys.}\/ {\bf B#1} (19#2) #3}
\def\PLB#1#2#3{{\it Phys.\ Lett.}\/ {\bf B#1} (19#2) #3}   
\def\PRD#1#2#3{{\it Phys.\ Rev.}\/ {\bf D#1} (19#2) #3}
\def\PRL#1#2#3{{\it Phys.\ Rev.\ Lett.}\/ {\bf #1} (19#2) #3}
\def\PRT#1#2#3{{\it Phys.\ Rep.}\/ {\bf#1} (19#2) #3}
\def\MODA#1#2#3{{\it Mod.\ Phys.\ Lett.}\/ {\bf A#1} (19#2) #3}
\def\IJMP#1#2#3{{\it Int.\ J.\ Mod.\ Phys.}\/ {\bf A#1} (19#2) #3}
\def\nuvc#1#2#3{{\it Nuovo Cimento}\/ {\bf #1A} (#2) #3}
\def\RPP#1#2#3{{\it Rept.\ Prog.\ Phys.}\/ {\bf #1} (19#2) #3}
\def\etal{{\it et al\/}}


\section{Physics Implications of a Perturbative Superstring 
Construction}
\noindent
\centerline{\large\it M. Cveti\v c, L. Everett, P. Langacker, and J. Wang}
\medskip
\label{section:Cvetic}

\subsection{Introduction}

Predictions from superstring theory provide natural possible extensions of
the MSSM.
However, there are several problems to be resolved in attempting to
connect string theory to the observable, low energy world. 
First, many models can be derived from string theory, and there is no
dynamical principle as yet to select among them. Furthermore, no fully
realistic model, i.e., a model which contains just the particle content
and couplings of the MSSM, has been constructed.  In addition, there is no
compelling scenario for how to break supersymmetry in string theory, and
so soft supersymmetry breaking parameters must be introduced into the
model by hand, just as in the MSSM.  

We adopt a more modest strategy and consider a class of quasi-realistic
models constructed within weakly coupled heterotic string theory.  In
addition to the necessary ingredients of the MSSM, such models generically
contain an extended gauge structure that includes a number of $U(1)$ gauge
groups and ``hidden'' sector non-Abelian groups, and 
additional matter fields (including a number of SM exotics and SM
singlets). They predict gauge couplings unification (sometimes with
non-standard Ka\v c-Moody level) at the string scale
$M_{String} \sim 5 \times 10^{17}$ GeV. 
The most desirable feature of models in this class is that the
superpotential is explicitly calculable; in particular, the non-zero Yukawa
couplings are ${\cal O}(1)$, and can naturally accommodate the radiative
electroweak symmetry breaking scenario.  In addition, string selection
rules can forbid gauge-allowed terms, in contrast to the case in general
field-theoretic models.  

In addressing the phenomenology of these models, there are two
complementary approaches. The first is the "bottom-up" approach, in which
models with particle content and couplings motivated from
quasi-realistic string models are studied to provide insight into the
new physics that can emerge from string theory (such as additional $Z'$
gauge bosons)\cite{cl,SY,cdeel,cceel1,lw}.  In this work we adopt the
second (``top-down") approach, and analyze a prototype string
model (Model 5 of \cite{chl}) in detail.  The analysis of this class of
string models (done in collaboration with G. Cleaver and J. R.
Espinosa) proceeds in several stages, which will be briefly summarized
below and is documented in \cite{cceel2,cceel3,cceelw1,cceelw2}.  We then
focus on the main results: the determination of the mass spectrum and 
trilinear couplings at the string scale, the renormalization group
analysis, the low energy gauge symmetry breaking patterns and the 
mass spectrum of the model at the electroweak scale.

Our analysis shows that the prototype model is not fully realistic.  
In particular, many of the SM exotics remain massless in the low
energy theory.  However, we find there are other general features of the
model which have interesting phenomenological implications, including an
additional low-energy $U(1)'$ gauge group, $R$ parity violating couplings,
``mixed" effective $\mu$ terms, and extended chargino, neutralino, and
Higgs sectors (with patterns of mass spectra that differ substantially
from the case of the MSSM).  

In section II, we discuss the generation of the effective mass terms and
the trilinear couplings associated with the flat direction. In section
III, we present the effective couplings and the implications of the
effective theory along a particular flat direction as an illustrative
example. We conclude in section IV.   

\subsection{Flat Directions and Effective Couplings}

The model we have chosen as a prototype model to analyze is Model 5 of
\cite{chl}. Prior to vacuum restabilization, the model has the gauge
group
\begin{equation}
\{SU(3)_C\times SU(2)_L\}_{\rm obs}\times\{SU(4)_2\times SU(2)_2\}_{\rm hid}
\times U(1)_A\times U(1)^6,
\end{equation}
and a particle content that includes the following
chiral superfields in addition to the MSSM fields:
\begin{eqnarray}
&&6 (1,2,1,1) + (3,1,1,1) + (\bar{3},1,1,1) + \nonumber\\
&&4 (1,2,1,2) + 2 (1,1,4,1) + 10 (1,1,\bar{4},1) +\nonumber\\
&&8 (1,1,1,2) + 5 (1,1,4,2) + (1,1,\bar{4},2) +\nonumber\\
&& 8 (1,1,6,1) + 3 (1,1,1,3)+ 42 (1,1,1,1)\;\;,
\end{eqnarray}
where the representation under $(SU(3)_C,SU(2)_L,
SU(4)_2,SU(2)_2)$ is indicated. The SM hypercharge is determined as a
linear combination of the six non-anomalous $U(1)$'s.   

As the first step of the analysis, we address the presence of the 
anomalous $U(1)_A$ generic to this class of models. 
The standard anomaly cancellation mechanism generates a nonzero
Fayet-Iliopoulos  (FI) term of ${\cal O}(M_{String})$ to the $D-$ term of
$U(1)_A$. The FI term would appear to break supersymmetry at the string
scale, but certain scalar fields are triggered to acquire large VEV's
along $D-$ and $F-$ flat directions, such that the new ``restabilized''
vacuum is supersymmetric. The complete set of $D-$ and $F-$ flat
directions involving the non-Abelian singlet fields for Model 5 was
classified in
\cite{cceel2}.   

In a given flat direction, the rank of the gauge group is
reduced, and effective mass terms and trilinear couplings
may be generated from higher order terms in the superpotential:
\begin{eqnarray}
W_M&=&\frac{\alpha_{K+2}}{M_{Pl}^{K-1}}\Psi_i \Psi_j \langle \Phi^K
\rangle \\
W_3&=&\frac{\alpha_{K+3}}{M_{Pl}^K}\Psi_i \Psi_j \Psi_k \langle \Phi^K
\rangle,
\end{eqnarray}
in which the fields which are in the flat direction are denoted by
$\Phi$, and those which are not by $\{\Psi_i\}$.
Hence, some fields acquire superheavy masses and decouple.  The
effective Yukawa couplings of the remaining light fields are typically
suppressed~\footnote{However, in the prototype model considered the
effective trilinear couplings arising from fourth order terms are
comparable in strength to the original Yukawas.} compared with 
Yukawa couplings of the original superpotential (the $\alpha_K$
coefficients are in principle calculable; for details, see \cite{cew}).

This procedure has been carried out for the prototype model in
\cite{cceel2,cceelw1}. We carry out the analysis of the implications of
the model for the flat directions that break the maximal number of
$U(1)$'s, leaving $U(1)_Y$ and $U(1)^{'}$ unbroken. The list of matter
superfields and their $U(1)_Y$ and $U(1)'$ charges are presented in Table
1. 



\subsection{Example: Low Energy Implications of a Representative Flat
Direction}

We choose to present the analysis of the model along a particular flat
direction~\footnote{Other flat directions
involve other interesting features, such as fermionic textures,
baryon number violation, and the possibility of intermediate scale
$U(1)'$ breaking.}. 
The flat direction we consider is
the $P_1'P_2'P_3'$ direction (in the notation of
\cite{cceel2,cceelw1,cceelw2}),
which involves the set of fields
$\{\varphi_{2},\varphi_{5},\varphi_{10},\varphi_{13},
\varphi_{27},\varphi_{29},\varphi_{30}\}$. 

Along this flat direction, the effective mass terms which involve the
observable sector fields and the non-Abelian singlets~\footnote{We refer
the reader to \cite{cceelw1,cceelw2} for further details of the model,
such as the couplings involving the hidden sector fields.} are given by 
\begin{eqnarray}
\label{p123massw}
W_{M}&=&gh_f\bar{h}_b \langle \varphi_{27}\rangle+gh_g\bar{h}_d \langle
\varphi_{29}\rangle+
\frac{\alpha^{(1)}_{4}}{M_{Pl}}h_b\bar{h}_b \langle
\varphi_{5}\varphi_{10}\rangle +
\frac{\alpha^{(2)}_{4}}{M_{Pl}}h_b\bar{h}_b \langle
\varphi_{2}\varphi_{13}\rangle
\nonumber\\
&+&{g\over {\sqrt{2}}}(e^c_de_b+e^c_ge_a)\langle
\varphi_{30}\rangle +
{g\over {\sqrt{2}}}(\varphi_1 \varphi_{15}+\varphi_{4}\varphi_{9})\langle
\varphi_{10}\rangle +
{g\over {\sqrt{2}}}(\varphi_7 \varphi_{16}+\varphi_{9}\varphi_{12})\langle
\varphi_{2}\rangle\nonumber\\ &+&
{g\over {\sqrt{2}}}(\varphi_6
\varphi_{26}+\varphi_{8}\varphi_{23}+\varphi_{14}\varphi_{17})\langle
\varphi_{29}\rangle+\frac{\alpha^{(3)}_{4}}{M_{Pl}}\varphi_{21}\varphi_{25}
\langle \varphi_{27}\varphi_{29}\rangle
\, .
\end{eqnarray}
The  effective trilinear couplings involving
all fields which couple directly to the observable sector fields are
given by:
\begin{eqnarray}
\label{efftril}
W_{3}&=&gQ_cu^c_c\bar{h}_c+gQ_cd^c_bh_c
+\frac{\alpha^{(4)}_{4}}{M_{Pl}}Q_cd_d^ch_a\langle \varphi_{29} \rangle
+{g\over {\sqrt{2}}}e^c_ah_ah_c+
{g\over {\sqrt{2}}}e^c_fh_dh_c  \nonumber\\&+&
\frac{\alpha^{(1)}_{5}}{M^2_{Pl}}e^c_hh_eh_a\langle
\varphi_{5}\varphi_{27}\rangle+
\frac{\alpha^{(2)}_{5}}{M^2_{Pl}}e^c_eh_eh_a\langle
\varphi_{13}\varphi_{27}\rangle+
gh_{b'}\bar{h}_c\varphi_{20} \; .
\end{eqnarray}

In the observable sector, the fields which remain light include both the
usual  MSSM states and exotic states such as a fourth ($SU(2)_L$ singlet)
down-type quark, extra fields with the same quantum numbers as the lepton
singlets, and extra Higgs doublets. There are other
massless states with exotic quantum numbers (including fractional electric
charge) that also remain in the low energy theory.
The $U(1)'$ charges of the light fields are family
nonuniversal (and hence is problematic with respect to FCNC).

There are some generic features of the superpotential which
are independent of the details of the soft supersymmetry breaking
parameters.  In addition to a large top-quark Yuakwa coupling ($\sim {\cal
O} (1)$) which is necessary for radiative electroweak symmetry breaking,
the couplings indicate $t-b$ and (unphysical) $\tau -\mu$ Yukawa
unification, with the
identification of the fields  $\bar{h}_c$, $h_c$ as the standard
electroweak Higgs doublets. 
There is no  elementary or effective canonical $\mu$-term involving
$\bar{h}_c$ and $h_c$, but rather non-canonical effective $\mu$ terms
involving additional Higgs doublets.
Finally, there is also a possibility of lepton- number
violating couplings; thus this model violates $R$- parity, and has no
stable LSP. 

With the knowledge of the massless spectrum at the string scale, the gauge
coupling beta-functions can be determined, and the gauge couplings can
then be run from the string scale (where they are predicted to unify) to
the electroweak scale.  We determine the
gauge coupling constant $g=0.80$ at the string scale by assuming
$\alpha_{s}=0.12$ (the experimental value) at the electroweak scale, and
evolving $g_{3}$ to the string scale.
We then use this value as an input to determine the electroweak scale
values of the other gauge couplings by their (1-loop) renormalization
group equations (RGE's).
The low energy values of the gauge couplings are not
correct due to the exotic matter and non-standard $k_{Y}=11/3$ for this
model; however, it is surprising that $\sin^{2}{\theta_{W}} \sim 0.16$ and
$g_{2}=0.48$ are not too different from the experimental  values $0.23$ 
and $0.65$, respectively.

The string-scale values of the Yukawa couplings of
(\ref{efftril}) are calculable (with the knowledge of the VEV's
of the singlet fields in the flat direction). Utilizing the RGE's, we can
also determine the low energy values of the Yukawa coupling constants. The
running of the gauge couplings and the Yukawa couplings are shown in
Fig.~\ref{fig:cvetik1}.

\begin{figure}[htb]
\begin{minipage}[t]{.50\linewidth}
\centerline{\protect\psfig{figure=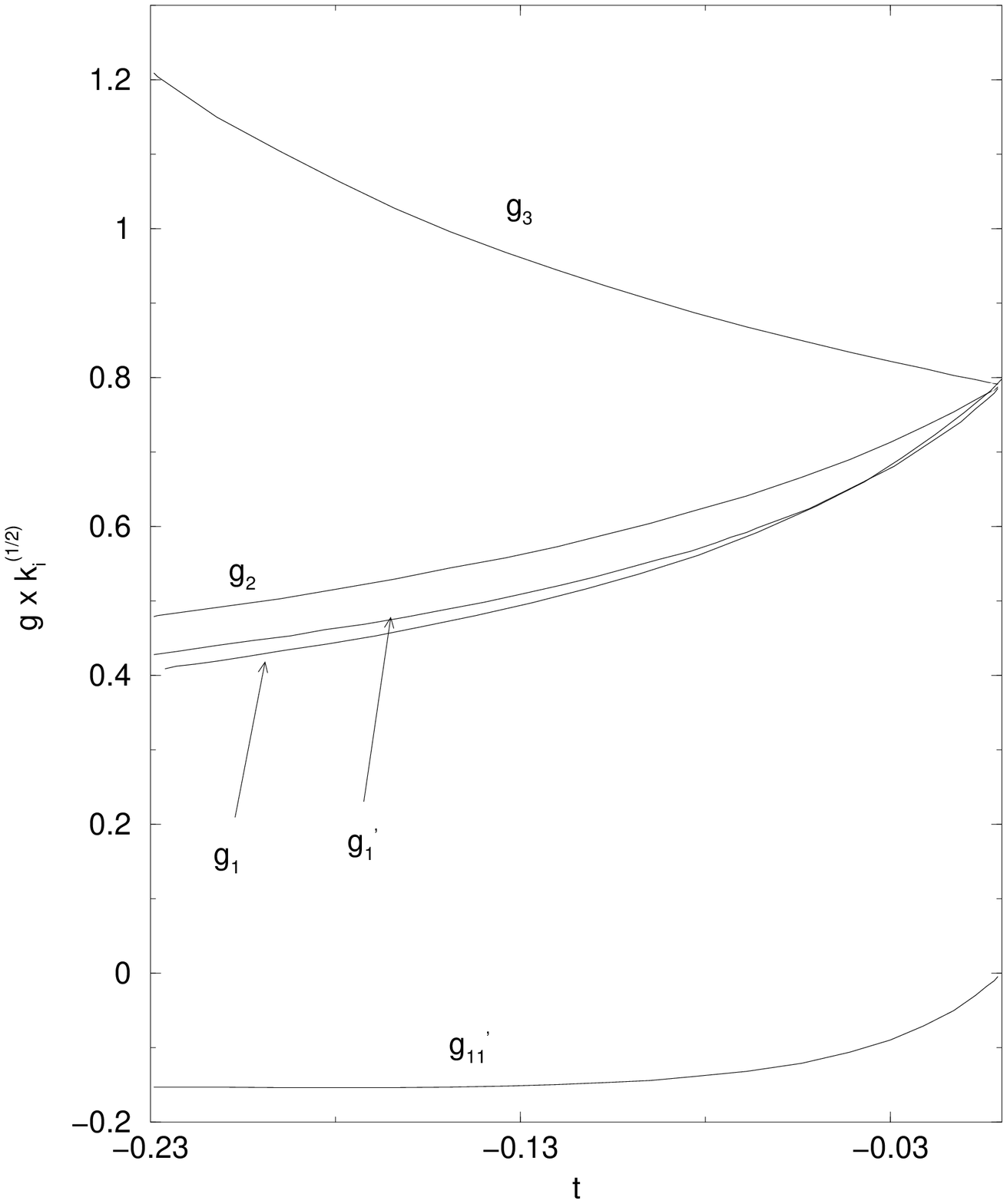,width=\linewidth}}
\end{minipage}\hfill
\begin{minipage}[t]{.50\linewidth}
\centerline{\protect\psfig{figure=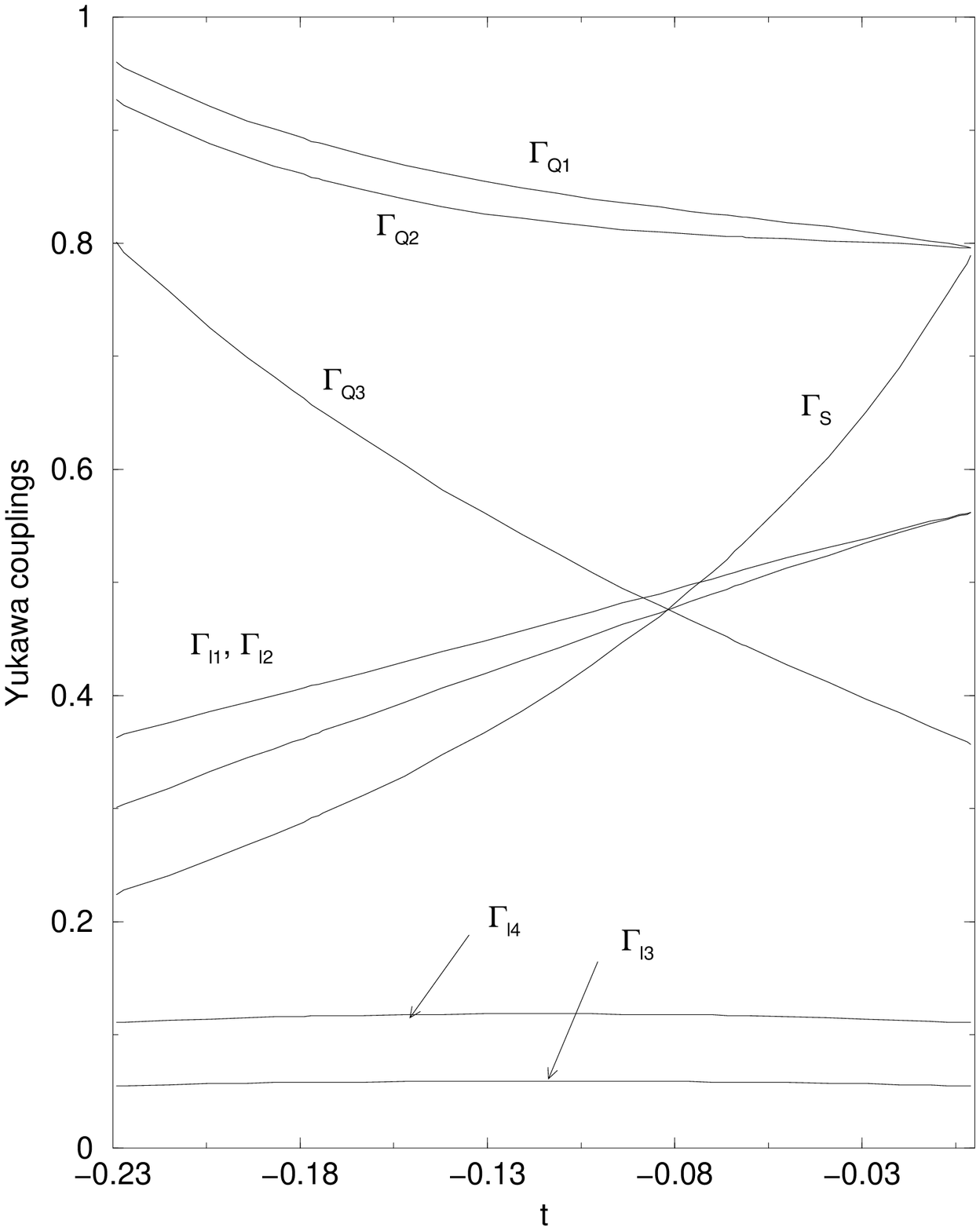,width=\linewidth}}
\end{minipage}
\caption{(a): Scale variation of the gauge couplings $\times
\protect\sqrt{k}$, with $t=(1/16\pi^2)\ln(\mu/M_{String})$, 
$M_{String}=5 \times
10^{17}$ GeV, and $g(M_{String})=0.80$. (b): The running of the Yukawa
couplings, in which the terms in  (\ref{efftril}) are denoted by
$\Gamma_{Q_{1,2,3}}$, $\Gamma_{l_{1,2,3,4}}$, and $\Gamma_S$,
respectively.}
\label{fig:cvetik1}
\end{figure}

To address the gauge symmetry breaking scenarios for this model, we
introduce soft supersymmetry breaking mass parameters
and run the RGE's from the string scale to
the electroweak scale. While the qualitative features of the analysis are
independent of the details of the soft breaking, we choose to illustrate
the analysis with a specific example with a realistic $Z-Z'$ hierarchy.    
General considerations \cite{cl,cdeel,cceel1,lw} and an inspection of
the $U(1)'$ charges of the light fields indicate that, in this example, 
the $U(1)'$ breaking is at the electroweak (TeV) scale.  Due to the lack
of a canonical effective $\mu$ term between $\bar{h}_c$, $h_c$ and a
singlet, an extended Higgs sector is required, with an
additional Higgs doublet and singlet ($\bar{h}_c$, $h_c$, $h_{b'}$, and $s
\equiv \varphi_{20}$). The symmetry breaking is characterized by a large 
(${\cal O}({\rm TeV})$) value of the SM singlet VEV, with the
electroweak symmetry breaking at a lower scale due to accidental
cancellations.  

We now present the mass spectrum for a concrete numerical example of this
scenario, which requires mild tuning of the soft supersymmetry breaking
mass parameters at the string scale. The initial and final values of the
parameters for this example are listed in Table 2.

\begin{itemize}

\item {\bf Fermion Masses:}
The masses for the $t$, $b$, $\tau$, and $\mu$ are due to Yukawa
couplings of the original superpotential, as shown in (\ref{efftril}). 
With the identification of $Q_c$ as the quark doublet of the third
family and $h_d$, $h_a$ as the lepton doublets of the third and second
families, respectively, 
$m_t=156$ GeV, $m_{b}=83$ GeV, $m_{\tau}=32$ GeV, and $m_{\mu}=27$ GeV. 
 The ratio $m_b/m_\tau$ is larger than in the usual $b-\tau$ unification
 because of the ratio $1:1/\sqrt{2}$ of the Yukawa couplings at the string
scale, and is probably inconsistent with experiment \cite{mbmtau} (of
course, the high values for $m_b$, $m_{\tau}$, and $m_\mu$ are
unphysical). Finally,  $u$, $d$, $c$, $s$, and $e^-$ remain massless.

\item {\bf Squarks/Sleptons:} 
To ensure a large $M_{Z'}$ in this model, the
squark and slepton masses have values in the
several TeV range, with $m_{\tilde{t}\,L}=2540$ GeV,
$m_{\tilde{t}\,R}=2900$
GeV; $m_{\tilde{b}\,L}=2600$ GeV, $m_{\tilde{b}\,R}=2780$ GeV;
$m_{\tilde{\tau}\,L}=2760$ GeV, $m_{\tilde{\tau}\,R}=3650$ GeV;
$m_{\tilde{\mu}\,L}=2790$ GeV, and $m_{\tilde{\mu}\,R}=3670$ GeV.

\item {\bf Charginos/Neutralinos:} 
The positively charged gauginos and higgsinos are
$\tilde{W}^{+}$, $\tilde{\bar{h}}_c$, $\tilde{\bar{h}}_a $,
and the negatively charged  gauginos and higgsinos are $\tilde{W}^{-}$,
$\tilde{h}_c$, $\tilde{h}_b'$.   There is one massless chargino, and the
other two have masses
$m_{\tilde{\chi}_{1}^{\pm}}= 591$ GeV, and
$m_{\tilde{\chi}_{2}^{\pm}}= 826$ GeV.

The neutralino sector consists of $\tilde{B}'$, $\tilde{B}$,
 $\tilde{W_3}$, $\tilde{\bar{h}}_{c}^0$, $\tilde{h}_{c}^0$,
$\tilde{h}_{b}^{'0}$,
 $\tilde{\varphi}^{'}_{20}$, and $\tilde{\bar{h}}_{a}^0$.
The mass eigenvalues are 
$m_{\tilde{\chi}_{1}}^0= 963$ GeV,
$m_{\tilde{\chi}_{2}}^0= 825$ GeV,
$m_{\tilde{\chi}_{3}}^0= 801$ GeV,
$m_{\tilde{\chi}_{4}}^0= 592$ GeV,
$m_{\tilde{\chi}_{5}}^0= 562$ GeV,
$m_{\tilde{\chi}_{6}}^0= 440$ GeV,
$m_{\tilde{\chi}_{7}}^0= 2$ GeV,
 and $m_{\tilde{\chi}_{8}}^0= 0$.

In both cases, the absence of couplings of the Higgs field $\bar{h}_a$
in the superpotential leads to a massless chargino and neutralino state.
The absence of an effective $\mu$ term involving $h_c$ leads to an
additional global $U(1)$ symmetry in the scalar potential, and an
ultralight neutralino pair in the mass spectrum.

\item {\bf Exotics:}
There are a number of exotic states, including
the $SU(2)_L$ singlet down-type quark, four $SU(2)_L$ singlets with
unit charge (the $e$ and extra $e^c$ states), and a number of
SM singlet ($\varphi$) states, as well as exotics
associated with the hidden sector. The scalar components of these
exotics are expected to acquire TeV-scale masses by soft supersymmetry
breaking. However, there is no mechanism within our assumptions to give
the fermions significant masses.

\item {\bf Higgs Sector:}
The non-minimal Higgs sector of three complex doublets and one
complex singlet leads to additional Higgs
bosons compared to the MSSM.  Four of the
fourteen degrees of freedom are  eaten to become the longitudinal
components of the $W^{\pm}$, $Z$, and $Z'$; and the global 
 $U(1)$ symmetry is broken, leading to a massless Goldstone boson (which,
however, acquires a small mass at the loop level) in the spectrum. 

The spectrum of the physical Higgs bosons after symmetry breaking
consists of two pairs of charged Higgs bosons $H_{1,2}^{\pm}$, four
neutral CP even Higgs scalars $(h^0_i\, ,i=1,2,3,4)$, and one CP odd Higgs
$A^0$, with masses  
$m_{H^{\pm}_1}=10$ GeV, 
$m_{H^{\pm}_2}=1650$ GeV, 
$m_{h^0_{1}}=33$
GeV, $m_{h^0_{2}}=47$ GeV, $m_{h^0_{3}}=736$ GeV, $m_{h^0_{4}}=1650$ GeV,
and $m_{A^0}=1650$ GeV.

In this model, the bound on the lightest Higgs scalar is
different that the traditional bound in the MSSM.  It is
associated with the breaking scale of the additional global $U(1)$
symmetry; since this scale is comparable to the electroweak
scale, not only one but two Higgs scalars will be
light in the decoupling limit.  In particular, the lightest Higgs
mass satisfies the (tree-level) bound~\cite{comelli}
\begin{eqnarray}
\label{bound1}
m^2_{h^0_{1}}\leq \frac{G^2}{4}v_1^2+g^2_{1'}Q_{1}^2v_1^2=(35 \, {\rm
GeV})^2.
\end{eqnarray}

\end{itemize}

\subsection{Conclusions}

The purpose of this work has been to explore the general features
of this class of quasi-realistic superstring models through a systematic,
``top-down" analysis of a prototype model.  The results of the
investigation of the low energy implications of the mass spectrum and
couplings predicted in a subset of the restabilized vacua of this model
demonstrate that in general, the TeV scale physics is more complicated
than that of the MSSM.  

In particular, we have found that non-canonical couplings, such as mixed
effective $\mu$ terms and $R$- parity violating operators are typically
present in the superpotential. In some other cases, there are
possibilities
for potentially interesting fermion textures.  The model is also
characterized by the presence of extra matter in the low energy theory
such as SM exotics, extra $Z'$ gauge bosons with TeV scale masses, and
additional charginos, neutralinos, and Higgs bosons with patterns of
masses that differ substantially from the MSSM.

The particular model we studied is not realistic, in part due to the
presence of (ultra-light or massless) extra matter.
However,  due to the large number of possible models that can be derived
from string theory, this result does not invalidate the potential
viability of string models, or the motivation for investigating their
phenomenological implications~\footnote{Progress
has been made in exploring models in which the exotic matter decouples
above the electroweak scale; e.g. see \cite{faraggi98}.}. We stress that
the features of this model are likely to be generic to this class of
quasi-realistic models based on weakly coupled heterotic string theory,
and thus warrant further consideration.

\begin{center}
\begin{tabular}{|c|c|c|c|}
\hline\hline
$(SU(3)_C,SU(2)_L,$& &
    $6Q_{Y}$&$100Q_{Y'}$\\
$SU(4)_2,SU(2)_2)$ &&      &\\
\hline\hline
(3,2,1,1):&$Q_a$& 1&68\\
&$Q_b$ &  1&68\\
&$Q_c$ & 1&$-$71\\
\hline
($\bar{3}$,1,1,1):&$u^c_a$&  $-$4&6\\
&$u^c_b$ & $-$4&6\\   
&$u^c_c$ & $-$4&$-$133\\  
&$d^c_a$ & 2&$-$3\\
&$d^c_b$ & 2&136\\
&$d^c_c$ & 2&$-$3\\
&$d^c_d$ & 2&$-$3\\
\hline
(1,2,1,1):&$\bar{h}_a$ &  3&$-$74\\
&$\bar{h}_b$ & 3&65\\
&$\bar{h}_c$ & 3&204\\
&$\bar{h}_d$ &  3&65\\
&$h_a$ &  $-$3&74\\
&$h_b$ & $-$3&$-$65\\
&$h_c$ &  $-$3&$-$65\\
&$h_d$ & $-$3&$-$65\\
&$h_e$ & $-$3&$-$204\\
&$h_f$ & $-$3&$-$65\\
&$h_g$ & $-$3&$-$65\\
\hline
(3,1,1,1):&${\cal D}_a$& $-$2&$-$136\\
\hline
\hline
\end{tabular}
\end{center}
\noindent Table Ia: List of non-Abelian non-singlet observable sector
fields in the model with their charges under hypercharge and $U(1)'$.

\begin{center}
\begin{tabular}{|c|c|c||c|c|c|}
\hline\hline 
 &  $6Q_{Y}$&$100Q_{Y'}$ & & $6Q_{Y}$ &$100Q_{Y'}$ \\
\hline\hline
$e^c_{a,c}$ &6&$-$9 & $e^c_b$  &6&$-$9\\
$e^c_{d,g}$ &6&130 & $e^c_e$ &6&130\\
$e^c_f$ &6&130 & $e^c_h$ &6&130\\
$e^c_i$ &6&$-$9 & $e_{a,b}$ &6&$-$130\\
$e_c$ &6&$-$130 & $e_{d,e}$ &$-$6&9\\
$e_f$ &$-$6&$-$269 & & & \\
$\varphi_{1}$ &0&0 & $\varphi_{2,3}$ &0&0 \\
$\varphi_{4,5}$ &0&0 & $\varphi_{6,7}$ &0&0 \\
$\varphi_{8,9}$ &0&0 & $\varphi_{10,11}$&0&0 \\
$\varphi_{12,13}$&0&0 & $\varphi_{14,15}$&0&0\\
$\varphi_{16}$ &0&0 & $\varphi_{17}$ &0&0 \\
$\varphi_{18,19}$&0&$-$139 & $\varphi_{20,21}$&0&$-$139 \\
$\varphi_{22}$ &0&$-$139 & $\varphi_{23}$ &0&0 \\
$\varphi_{24}$ &0&0 & $\varphi_{25}$ &0&139 \\
$\varphi_{26}$ &0&0 & $\varphi_{27}$ &0&0 \\
$\varphi_{28,29}$&0&0 & $\varphi_{30}$ &0&0 \\
\hline
\hline
\end{tabular}
\end{center}
\noindent Table Ib: List of non-Abelian singlet fields in
the model with their charges under hypercharge and $U(1)'$.


\begin{center}
\begin{tabular}{||c||c|c||c||c|c||}
\hline
\hline

 &  $M_Z$ & $M_{String}$ & & $M_Z$ & $M_{String}$  \\ \hline
$g_1$& 0.41& 0.80&$M_1$& 444&1695 \\
$g_2$& 0.48& 0.80&$M_2$& 619& 1695\\
$g_3$& 1.23& 0.80&$M_3$& 4040& 1695\\
$g_1'$& 0.43& 0.80&$M_1'$& 392& 1695\\
$\Gamma_{Q1}$&0.96 & 0.80&$A_{Q1}$ & 3664&8682\\
$\Gamma_{Q2}$&0.93 & 0.80& $A_{Q2}$&4070&9000\\
$\gamma_{Q3}$&0.27 & 0.08& $A_{Q3}$&5018&1837\\
$\Gamma_{l1}$& 0.30& 0.56&$A_{l1}$ &$-$946&4703\\
$\Gamma_{l2}$& 0.36& 0.56&$A_{l2}$ &$-$707&4532\\
$\Gamma_{l3}$& 0.06& 0.05&$A_{l3}$ &4613&4425\\
$\Gamma_{l4}$& 0.11& 0.13&$A_{l4}$ &4590&4481\\
$\Gamma_{s}$& 0.22& 0.80 &$A$      &1695&12544\\
$m_{Q_c}^2$ & $(2706)^2$ & $(2450)^2$ &
$m_{d_d}^2$ & $(4693)^2$ & $(2125)^2$ \\
$m_{u_c}^2$ & $(2649)^2$ & $(2418)^2$ &
$m_{d_c}^2$ & $(2734)^2$ & $(2486)^2$\\
$m_{\bar{h}_c}^2$ & $(1008)^2$ & $(5622)^2$ &
$m_{h_b'}^2$&$(826)^2$&$(2595)^2$\\
$m_{\varphi_{20'}}^2$ & $-(518)^2$ & $(6890)^2$ &
$m_{\varphi_{22'}}^2$ &$(3031)^2$&$(11540)^2$\\ 
$m_{h_a}^2$ & $(3626)^2$ & $(3982)^2$ &
$m_{h_c}^2$ & $-(224)^2$&$(5633)^2$\\
$m_{h_d}^2$ & $(3666)^2$ & $(4100)^2$ &
$m_{h_e}^2$ & $(4274)^2$ & $(4246)^2$\\ 
$m_{e_a}^2$ & $(2770)^2$ & $(3564)^2$ &
$m_{e_f}^2$ & $(2780)^2$ & $(3958)^2$\\
$m_{e_e}^2$ & $(4195)^2$ & $(4254)^2$ &
$m_{e_h}^2$ & $(4259)^2$ & $(4236)^2$\\
\hline\hline
\end{tabular}
\end{center}
\noindent Table II: $P_1'P_2'P_3'$ flat direction: values of the parameters at 
$M_{String}$ and $M_{Z}$, with $M_{Z'}=735$
GeV and $\alpha_{Z-Z'}=0.005$.  All mass parameters are given in GeV. 





\section{The signature at the Tevatron for the light doubly charged Higgsino
of the supersymmetric left-right model}
\noindent
\centerline{\large\it B. Dutta, R. N. Mohapatra and D. J. Muller}
\medskip
\label{section:Dutta}

\subsection{Introduction}

Supersymmetric left-right models (SUSYLR) where the $SU(2)_R$ gauge
symmetry is broken by triplet Higgs fields $\Delta^c$ with $B-L=2$ have many
attractive features:
1) they imply automatic conservation of baryon and lepton number \cite{moh};
2) they provide a natural  solution to the strong and weak CP problems of
the MSSM \cite{rasin}; 3) they yield a natural embedding of the see-saw
mechanism for small  neutrino masses \cite{gell} where the right-handed
triplet  field ($\Delta^c$) that breaks the $SU(2)_R$ symmetry also gives
heavy mass to the right-handed
Majorana neutrino needed for implementing the see-saw mechanism.

Recently it has been shown that the doubly charged components of the
 triplet Higgs fields are massless
unless there are some  higher dimensional operators (HDO)
\cite{kuchi,aulakh,chacko,goran2}.  This is independent  of how 
supersymmetry breaking is transmitted to the visible sector 
({\it i.e.},  whether 
it is gravity mediated or it is gauge mediated) and also of whether the 
hidden sector supersymmetry breaking scale is above or  below the
$W_R$ scale. In the presence of HDO's, they  acquire masses of order 
$\sim v^2_R/M_{\rm Pl}$.  Since  the measurement of the  Z-width at LEP 
and SLC
implies that such particles  must  have a mass of at least 45\,GeV, this
puts a lower limit on the $W_R$ scale of about
$10^{10}$\,GeV or so. 
For $W_R$ near this lower limit, the masses of the doubly charged
particles are in the 100\,GeV range.
The rest of the particle spectrum
below the $W_R$ scale can be the same as that of the
MSSM with a massive neutrino or it can have an extra pair of Higgs doublets 
in the 10\,TeV  range depending on the structure of the model.

The ordering of the sparticles masses and the existence of the doubly-charged Higgs
fields in the SUSYLR model makes the SUSY
signature distinctive from other SUSY models.
Mass spectra for this type of model have been studied \cite{dm}.
It was found that one of the Higgs fields has a coupling with the third
generation charged leptons which reduces the third generation charged
slepton masses. Because of this, in SUSYLR models with GMSB the lighter
stau is predominantly the NLSP whenever the deltino 
is too massive to play that role. As a result, the decay chains of the SUSY
particles typically lead to the lighter stau. The $\tilde{\tau}_1$ then
decays into a $\tau$ lepton and a gravitino ($\tilde{G}$)
which escapes the detector
undetected (leading to missing energy). Since the gravitino mass is on the
order of eV, the emitted $\tau$ will have high $p_T$ enhancing its
detection possibility. Moreover, pair production of the light doubly charged
Higgsinos always produces four $\tau$ leptons. When the $\tilde{\tau}_1$ is
the NLSP, this occurs through
$\tilde{\Delta}^{c\pm \pm} \rightarrow \tilde{\tau}_1^\pm \tau^\pm$
followed by $\tilde{\tau}_1 \rightarrow \tau \, \tilde{G}$.
When the $\tilde{\Delta}^{c\pm \pm}$ is the NLSP, this occurs through
the stau mediated decay
$\tilde{\Delta}^{c\pm \pm} \rightarrow \tau^\pm \tau^\pm \tilde{G}$.
One can get 
a similar
signal in supergravity motivated LR models with
the gravitinos replaced by the lightest neutralino
(its mass is greater than 33\,GeV),
which will constitute the
missing energy.

Signals involving two or more high $p_T$ $\tau$ leptons are also 
important signals for conventional GMSB models as the lighter stau is 
frequently the NLSP for these models as well. In the SUSYLR model, however,
we find that the production of the deltino can greatly
enhance the signal. In addition, since the deltino decays
into like sign $\tau$ leptons, we find that the distribution in angle
between same sign $\tau$ leptons can be used to distinguish this model
from other GMSB models.

\subsection{Sparticle Masses and Production}
The particle content of this model above the LR scale includes $\phi(2,2,0)$,
$\Delta(3,1,2)$, $\bar \Delta(3,1,-2)$, $\Delta^c(1,3,-2)$, 
$\bar \Delta^c(1,3,-2)$ and a singlet where the numbers in the parentheses 
refer to their transformation properties under
$SU(2)_L\times SU(2)_R\times U(1)_{B-L}$.
After integrating out the fields at the left-right scale, we are left with
the following additional part to the MSSM:
\begin{equation}
W=M_{\Delta}\Delta^{c--}\bar{\Delta}^{c++}+f_il^cl^c\Delta^{c--}
\end{equation}  
where we have assumed that f is diagonal. The PSI experiment \cite{psi} 
has put an upper bound on the product of the first two generation
couplings of
$f_1f_2<1.2\times 10^{-3}$. The magnitude of $f_3$ is unrestrained. The
term $M_{\Delta}$ originates from the nonrenormalizable terms.

The model considered here involves gauge mediated supersymmetry breaking.
In GMSB type models the SUSY breaking is communicated to the observable
sector by the SM gauge interactions. We choose GMSB since
the lighter stau will then have a mass that is almost always below that
of the lightest neutralino due to the presence of the additional coupling $f_3$.
The lighter stau then decays to a $\tau$ lepton
and a gravitino. Since the gravitino  is  very light,
the $\tau$ lepton will typically be very energetic.

There are a number of potential SUSY production mechanisms here. Given the
current lower bounds on the various sparticle masses and the hierarchy
of sparticle masses in GMSB models, the important SUSY production
mechanisms will typically include EW gaugino production. At the Tevatron,
chargino pair ($\chi_1^+ \chi_1^-$) production takes place through
s-channel $Z$ and $\gamma$ exchange and $\chi_2^0 \, \chi_1^\pm$
production is through s-channel $W$ exchange. Squark exchange via the
t-channel also contributes to both processes, but the contributions are
expected to be negligible since the squark masses are large in GMSB
models. The production of $\chi_1^0 \, \chi_1^\pm$ is suppressed due to the
smallness of the coupling involved. 

\begin{figure}[t]
\vspace*{0.1in}
\centerline{\protect\psfig{figure=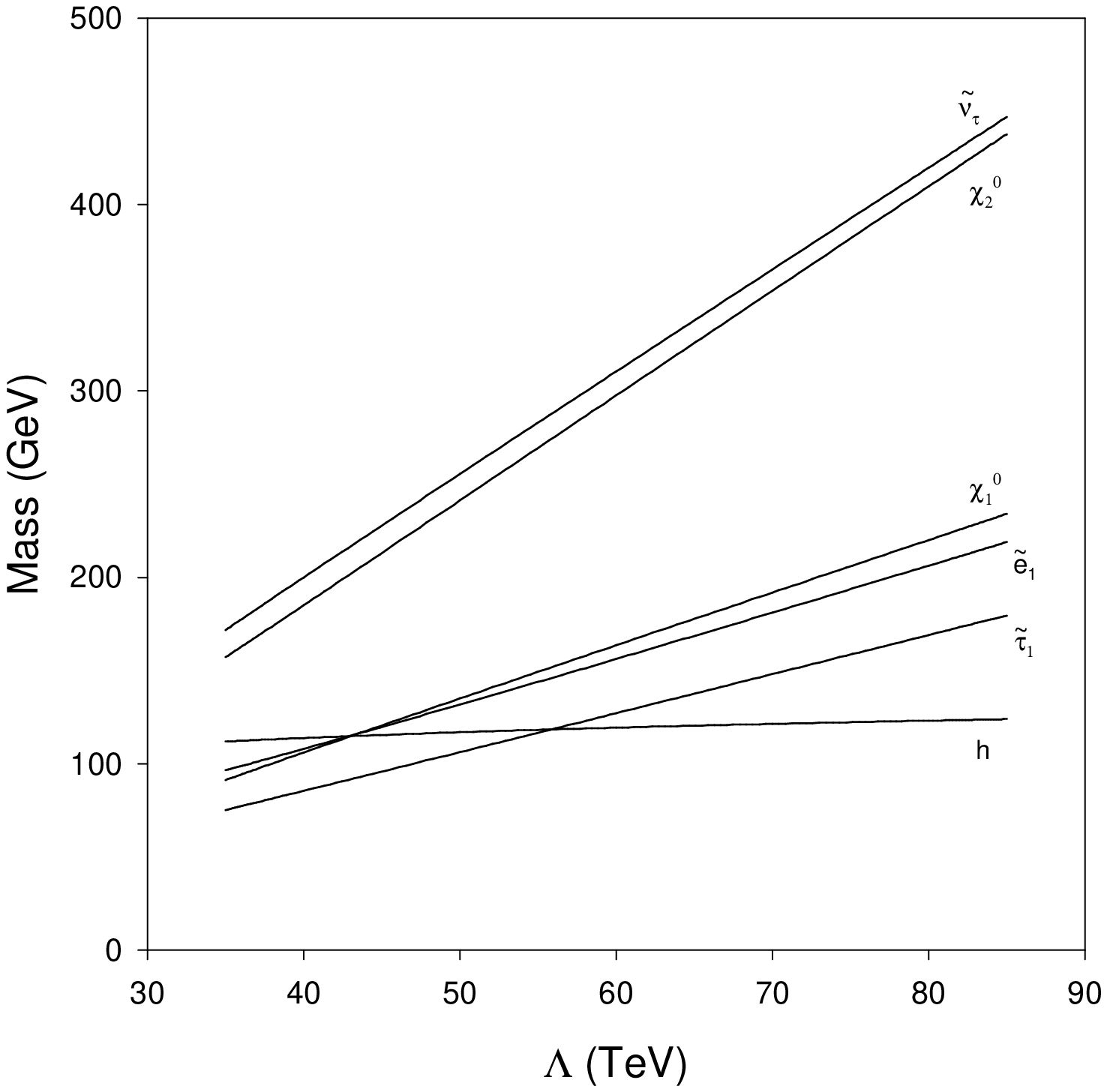,width=3.5in}}
\caption{
Masses of the particles of interest for the input
parameters $tan \beta = 15$, $M/\Lambda$ = 3, $n = 2$,
$f_3 = 0.5$, $f_2 = 0.05$, $f_1 = 0.05$ and 
$M_{\tilde{\Delta}} (M) = 90$\,GeV.}
\label{mass}
\end{figure}

\begin{figure}[t]
\vspace*{0.1in}
\centerline{\protect\psfig{figure=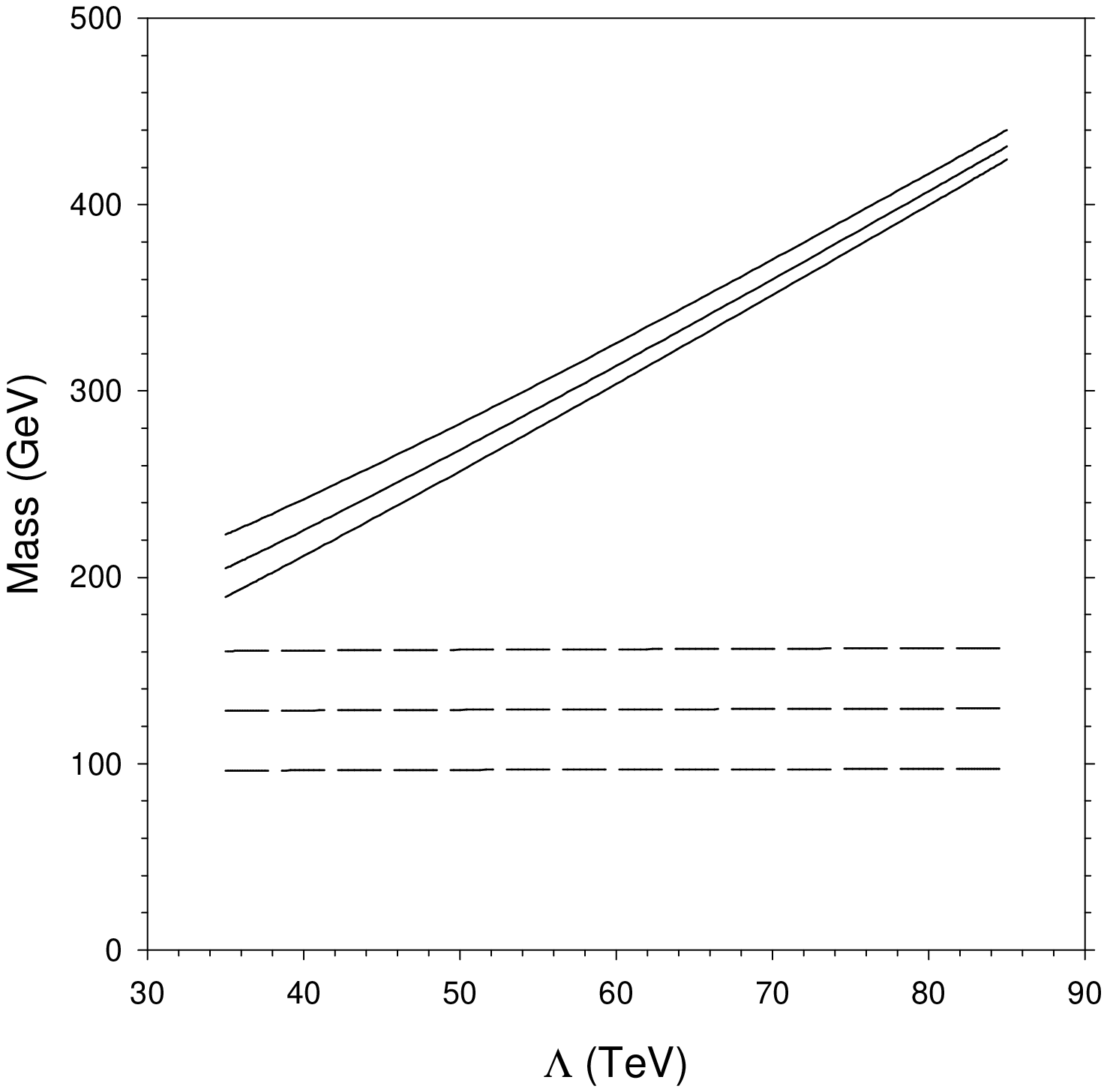,width=4in}}
\caption{
The dashed lines represent the deltino, 
while the solid lines represent the delta boson. From 
bottom to top, the lines in each set are for a messenger scale deltino
mass of 90, 120 and 150\,GeV\@.}
\label{delmass}
\end{figure}

In addition to these usual SUSY production mechanisms of the MSSM, we also 
have deltino
pair ($\tilde{\Delta}^{c++} \tilde{\Delta}^{c--}$) production. This proceeds
through s-channel $Z$ and $\gamma$ exchange. Given that the
$\tilde{\Delta}^{c\pm \pm}$ can be relatively light, it can be a very
important SUSY production mode. In fact, it frequently is the dominant 
mode.
The masses of some of the particles of interest are given
in Fig.~\ref{mass} and Fig.~\ref{delmass}.  In Fig.~\ref{mass} we take
$M_{\tilde{\Delta}} (M) = 90$\,GeV (97 GeV at the weak scale), but the masses of the gauginos and
sleptons (with the exception of the stau) do not vary much with the
messenger scale deltino mass (M is the scale at which the soft
 breaking masses are
introduced in the observable sector) . Fig.~\ref{delmass} gives the masses of the
delta boson and the deltino. The deltino mass is not very sensitive to
the value of $\Lambda$, while the delta boson mass is highly dependent
on $\Lambda$ due to the contributions from the messenger scale loops.

\begin{figure}[t]
\vspace*{0.1in}
\centerline{\protect\psfig{figure=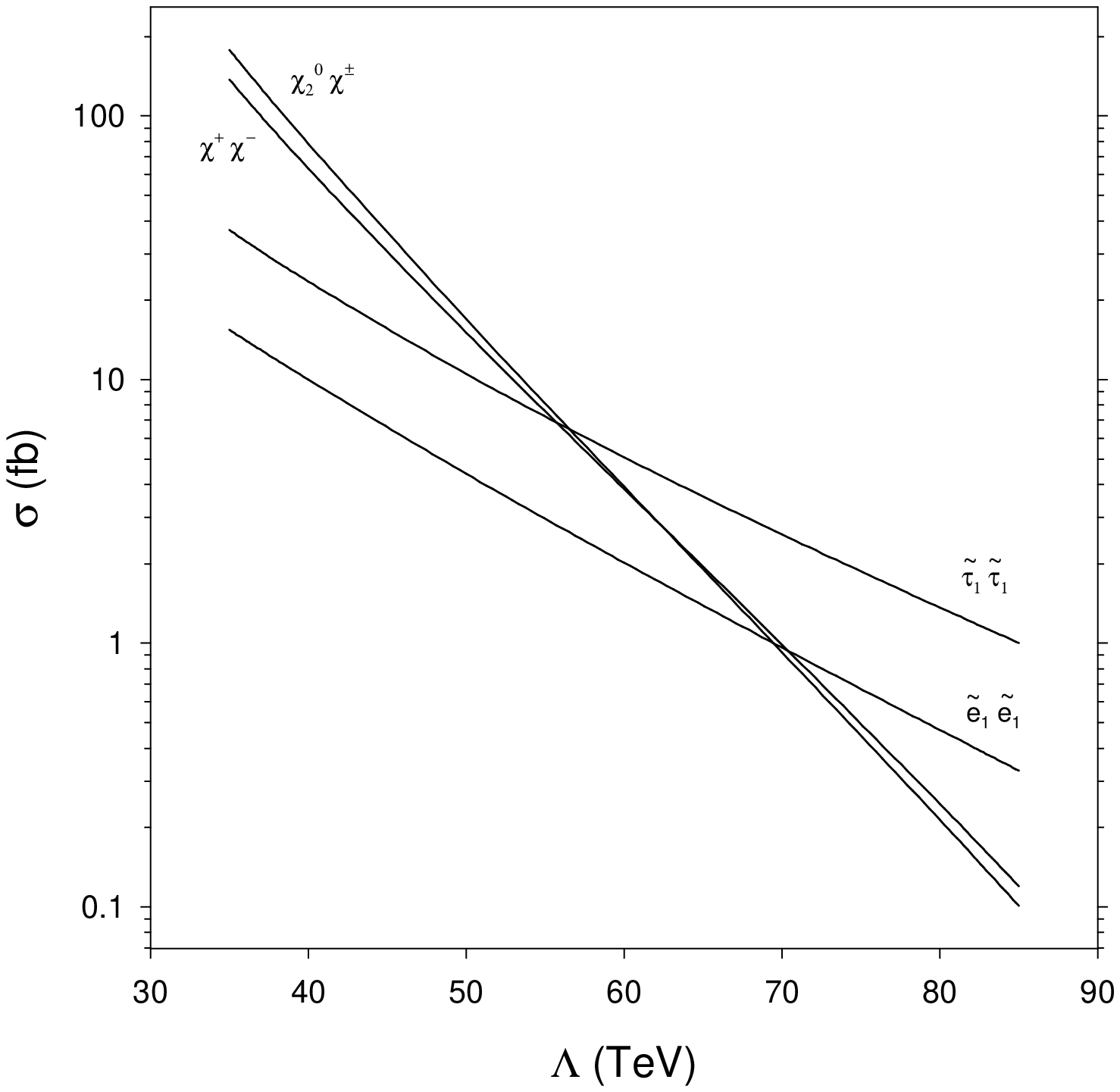,width=4in}}
\caption{
Cross sections for the standard SUSY production modes
for $M_{\tilde{\Delta}} (M) = 90$\,GeV. (97 GeV at the weak scale.)}
\label{cross}
\end{figure}

The cross sections for the more traditional SUSY production modes are given
in Fig.~\ref{cross}. 
We also have deltino pair production; the cross sections for which are
tabulated in
Table~\ref{delcross}. Since the deltino mass does not vary much over the
values of $\Lambda$ ($\Lambda$ is related to the SUSY breaking scale) considered, the cross section for deltino pair
production does not vary much either. This cross section is high enough
for all the deltino masses considered that deltino pair production is
always an important SUSY production mode.

\subsection{Tau Jet Analysis}

We now give an account of the possible $\tau$-jet signatures
for SUSY production at the Tevatron in the context of the left-right
GMSB model.

This analysis is performed in the context of the Main Injector (MI) and
TeV33 upgrades of the Tevatron collider. The center of mass energy is
taken to be $\sqrt{s} = 2$\,TeV and the integrated luminosity is taken to
be 2\,fb$^{-1}$ for the MI upgrade and 30\,fb$^{-1}$ for the TeV33
upgrade.

In performing this analysis, the cuts employed are that final state charged
leptons must have $p_T > 10$\,GeV.
Jets must have $E_T > 10$\,GeV and $|\eta| < 2$. In addition, hadronic
final states within a cone size of
$\Delta R \equiv \sqrt{ (\Delta \phi)^2 + (\Delta \eta)^2 } = 0.4$ are
merged to a single jet. Leptons within this cone radius of a jet are
discounted. For a $\tau$-jet to be counted as such, it must have
$|\eta| < 1$. The most energetic $\tau$-jet is required to have
$E_T > 20$\,GeV\@. In addition, a missing transverse energy cut of
${\rlap/E}_T> 30$\,GeV is imposed.

\runfigd{d90}{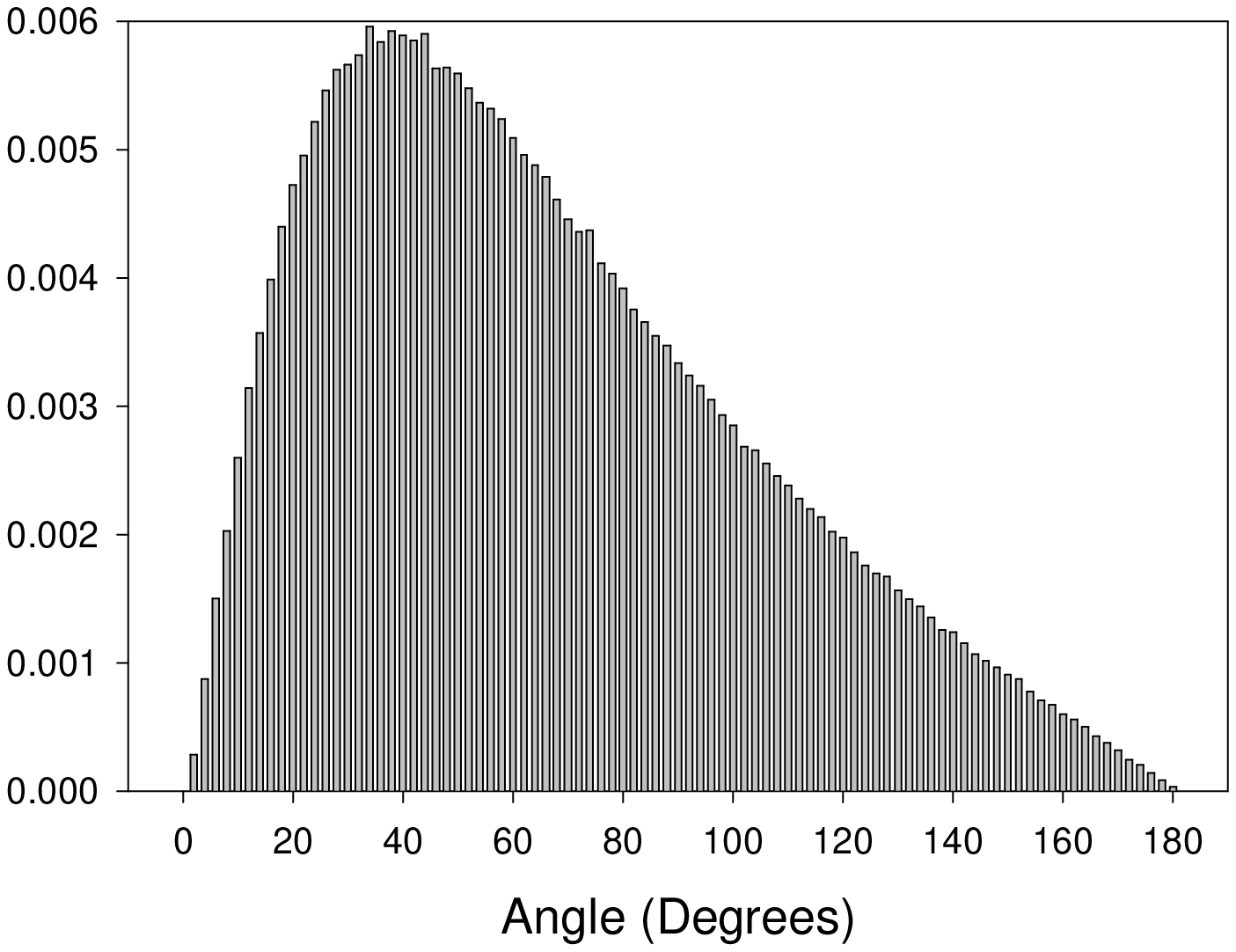}{
Angular distribution between the two most energetic $\tau$-jets
for deltino pair production at the Tevatron. The deltino
mass is about $97\gev$.
This plots gives the distribution
when the $\tau$-jets come from same sign $\tau$ leptons.}

\runfigd{d90b}{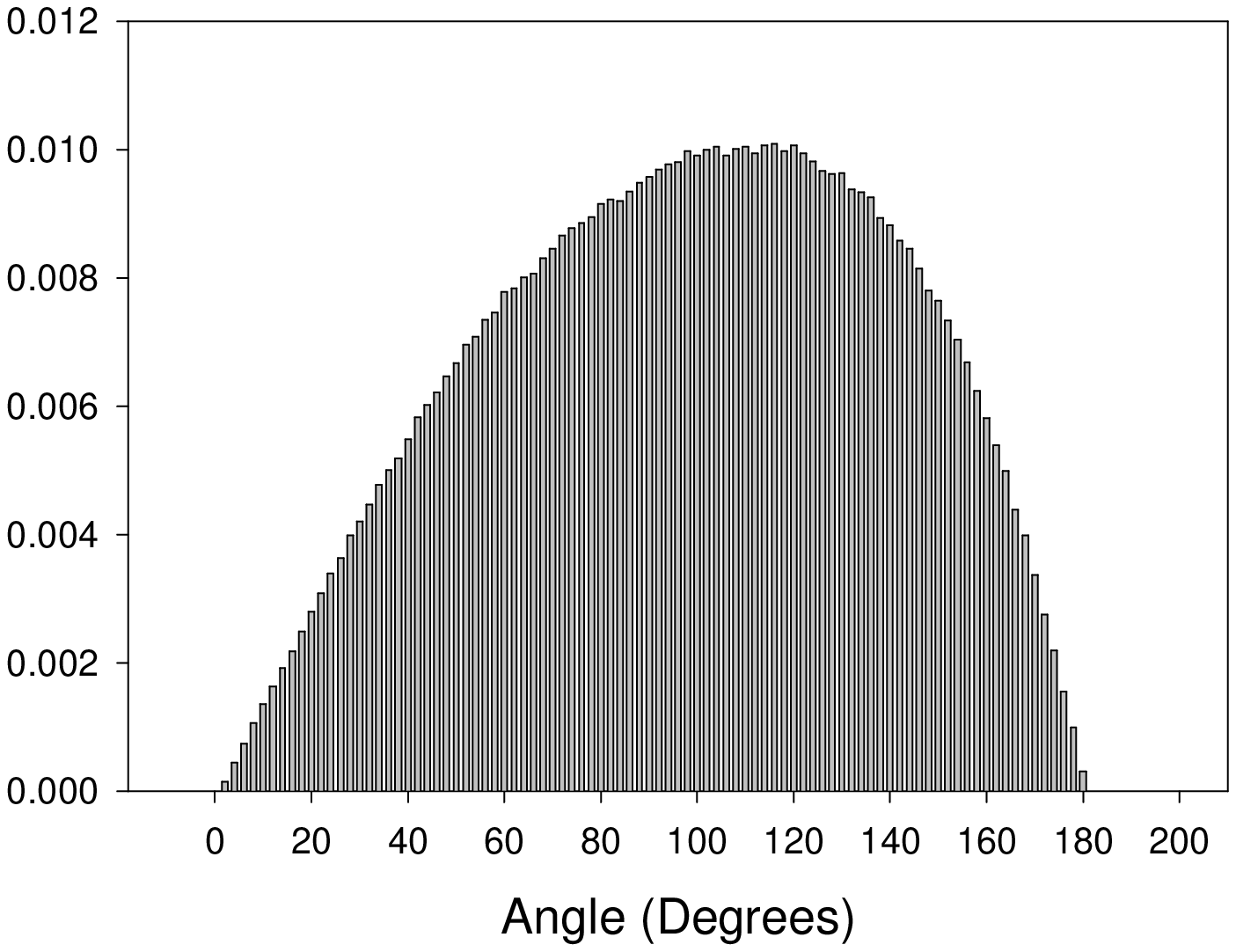}{
Angular distribution between the two most energetic $\tau$-jets
for deltino pair production at the Tevatron. The deltino
mass is about $97\gev$.
This plots gives
the distribution when the $\tau$-jets come from opposite sign $\tau$
leptons.}

\runfigd{g100}{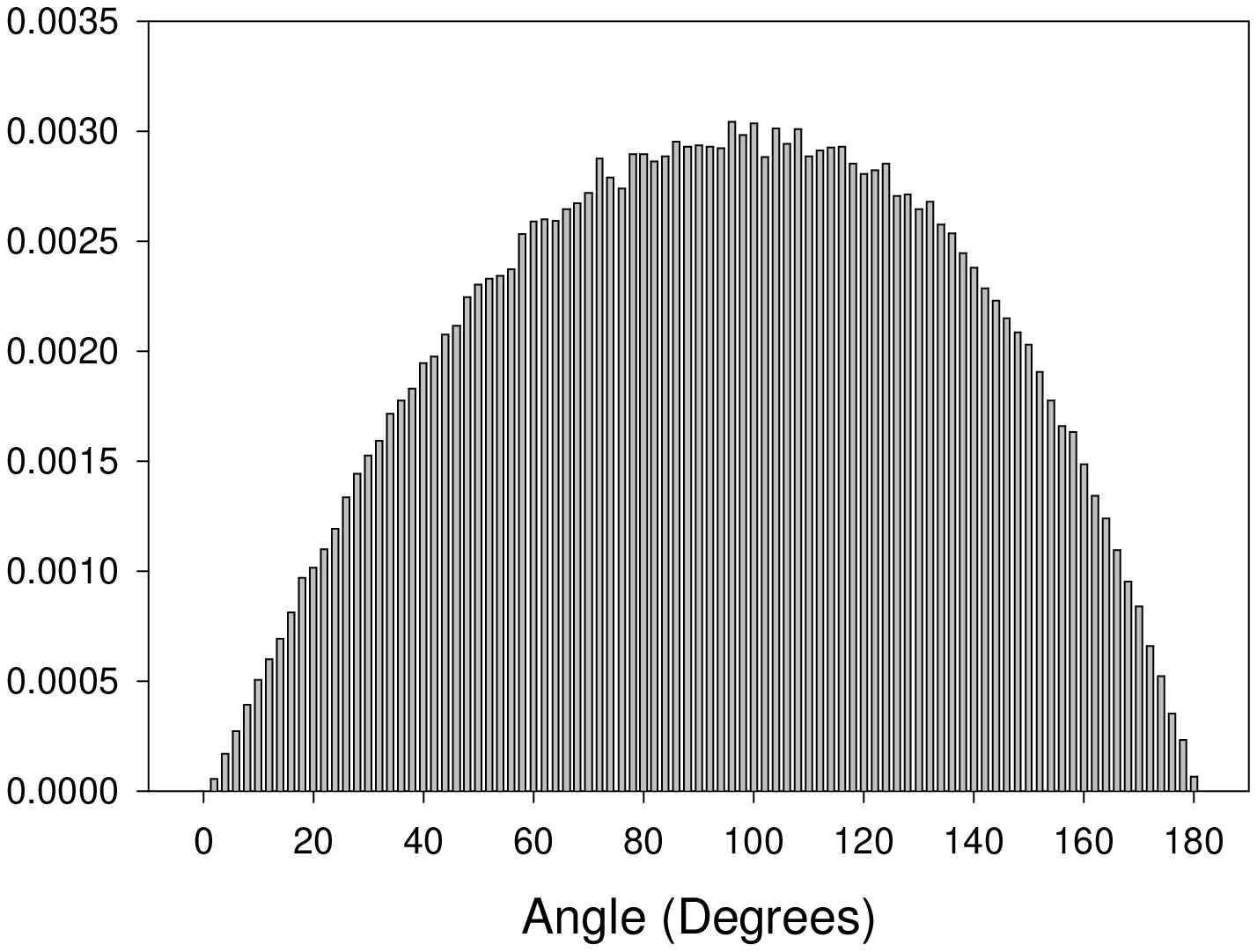}{
Angular distribution between the two most
energetic $\tau$-jets for EW gaugino production at the Tevatron.
The $\chi_2^0$ mass is 100\,GeV\@.
This plot gives the distribution
when the $\tau$-jets come from same sign $\tau$ leptons.}

\runfigd{g100a}{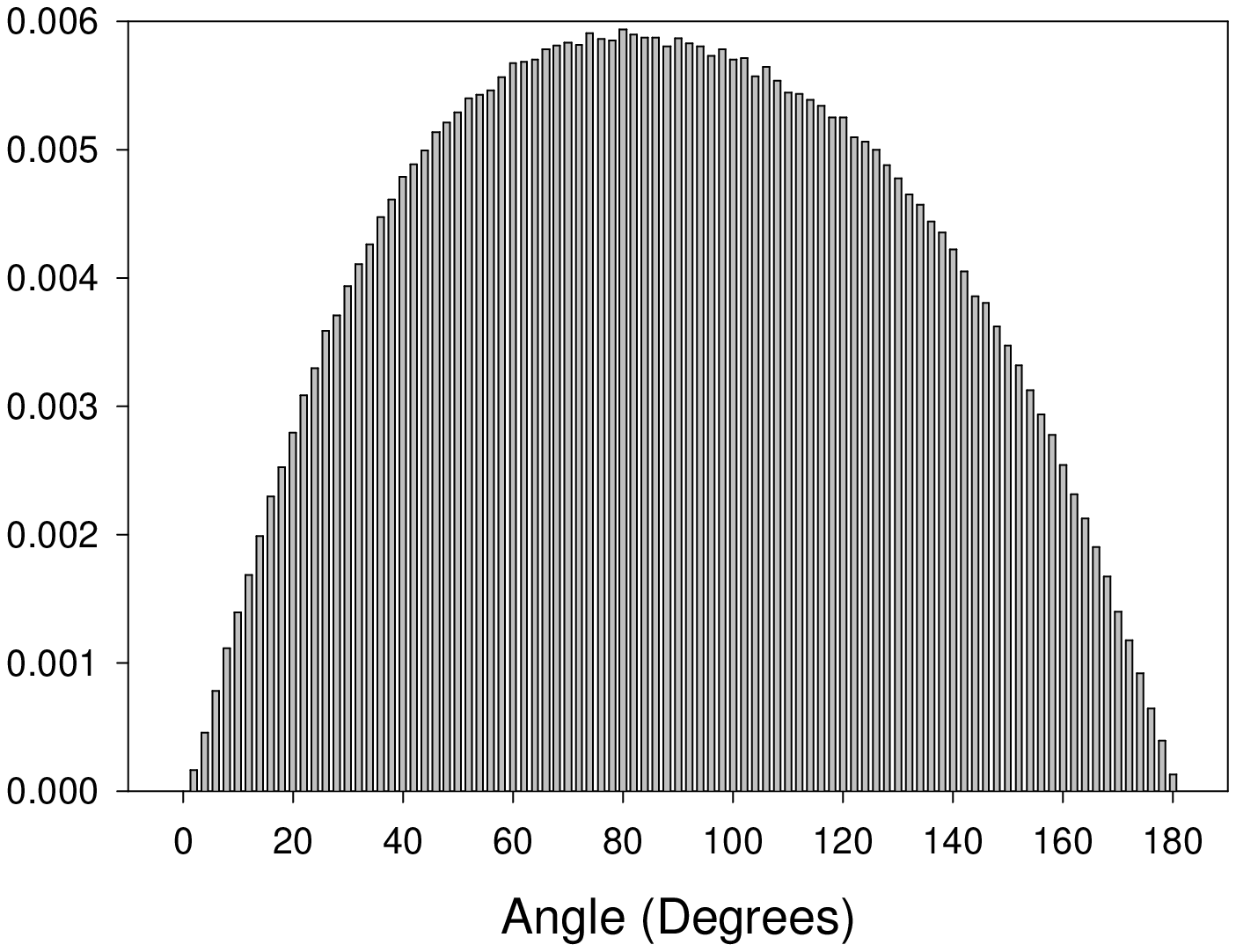}{
Angular distribution between the two most
energetic $\tau$-jets for EW gaugino production at the Tevatron.
The $\chi_2^0$ mass is 100\,GeV\@.
This plot gives the distribution
when the $\tau$-jets come from opposite sign $\tau$
leptons.}

In Table~\ref{sbr90} we give the inclusive
$\tau$-jet production cross sections for a messenger scale deltino mass of
90 GeV, respectively. We include only up
to four $\tau$-jets as the cross sections for more than four $\tau$-jets 
are small. We see that before cuts the
production of
two and three $\tau$-jets are dominant, but the four $\tau$-jet cross section 
is also significant at slightly over 100\,fb.
After the cuts are applied, however, the situation
changes substantially. The one $\tau$-jet mode is now dominant, but the
cross section for two $\tau$-jets is not far below and the three 
$\tau$-jets cross section is not insignificant. For $\Lambda = 35$\,TeV the cross
section for inclusive production of three $\tau$-jets is 32.3\,fb.
For an integrated luminosity of 2\,fb$^{-1}$ (the approximate 
initial value at Run II), this corresponds to about 65 events. For 
30\,fb$^{-1}$, the number of observable events is $\sim 970$.

In comparison to the GMSB model with the MSSM symmetry, 
the two $\tau$-jets and the three $\tau$-jets cross sections
are considerably higher in this model. 
In the GMSB model with MSSM symmetry, the two $\tau$-jets cross section can
be seen at RUN II, but not the three $\tau$ jets \cite{mdn}. 

\subsection{Angular Distributions}

The excess of $\tau$-jets expected in this model does not constitute
an unequivocal signal for this model. $\tau$-jets are part of the
signatures for other models including the minimal GMSB model
when the lighter stau is the NLSP\@. The question then arises as to
whether there is any way to distinguish this model from the minimal
GMSB model. A possible distinguishing characteristic is the
distribution in angle between 
the two highest $E_T$ $\tau$-jets when they come from same sign $\tau$-jets.

Consider deltino pair production.
The deltino tends to decay to like sign $\tau$ leptons. This occurs
directly when the deltino is the NLSP and so decays 
via the three-body decay
$\tilde{\Delta}^{\pm \pm} \rightarrow \tau^\pm \tau^\pm \tilde{G}$.
When the two-body decay of the deltino
$\tilde{\Delta}^{\pm \pm} \rightarrow \tilde{\tau}^\pm_1 \tau^\pm$
occurs, then the second
like sign $\tau$ lepton comes from the subsequent decay of the stau.
In the rest frame of the deltino, the $\tau$ leptons are widely
distributed. In the lab frame, however, the deltinos are quite energetic
and have a large velocity, especially if their masses are small. As a
consequence of this, the decay products of the deltino tend to be collimated
in the direction in which the deltino was moving. Thus when the two most
energetic $\tau$-jets have the same sign in deltino pair production,
the angle between them tends to be smaller than when the two most energetic
$\tau$-jets have opposite sign charges.

Fig.~\ref{d90} gives the distribution in angle between the two most
energetic $\tau$-jets for deltino pair production. This example is for
a weak scale deltino mass of about 97\,GeV\@. We can see that the
distribution in angle for like sign $\tau$-jets, which is given in
Fig.~\ref{d90}(a), peaks at about $40^\circ$. Fig.~\ref{d90}(b) gives the
distribution in angle between the two most energetic $\tau$-jets when they
come from opposite sign $\tau$ leptons. In stark contrast to the previous
case, here the peak occurs at $110^\circ$.

The situation changes as the deltino mass gets larger. This is in part
due to the fact that the deltino pair production cross section gets
smaller and so production of charginos and neutralinos can have a
larger impact on the distributions. In addition, a larger deltino mass
means the deltinos will typically be moving slower. 

Fig.~\ref{g100} shows the angular distributions
for combined $\chi_2^0 \, \chi_1^\pm$ and $\chi_1^+ \chi_1^-$ production
for  the weak scale
$\chi_2^0$ mass is $\sim 100$\,GeV\@.
The distribution for same sign $\tau$-jets is given in Fig~5(a).
We see that the peak occurs at about
$110^\circ$. In $\chi_2^0 \, \chi_1^\pm$ production,
one of the same sign $\tau$-jets generally comes from the chargino and the
other from the neutralino.
We now consider the angular distribution for opposite sign $\tau$-jets
which are given in Fig.~5(b).
In $\chi_2^0 \, \chi_1^\pm$ production, opposite sign $\tau$-jets
frequently come from the neutralino, while in
$\chi_1^+ \chi_1^-$ production one of the $\tau$-jets comes from one of
the charginos and the other $\tau$-jet comes from the other chargino.
Since there is a strong possibility that the opposite sign
$\tau$-jets come from the same particle ($\chi_2^0$), the distribution
should peak at a lower angle than for same sign $\tau$-jets.
We see from the figure that the peak occurs at about $85^\circ$. This feature
does not change with the increase in the gaugino mass.

\subsection{Conclusion} 

In conclusion, we have found that the doubly charged Higgs bosons of LR
models can be potentially observable at Run II of the Tevatron through 
the production of $\tau$-jets. In a
GMSB type theory, SUSYLR models typically produce large numbers of
two and three $\tau$-jet final states. 
This large $\tau$-jet signal is 
also due in large part to pair production of the doubly charged 
Higgsino.
It is also due to the relatively low
mass of the lighter stau (which is frequently the NLSP) in these models,
which is due to the additional coupling $f$. 
We have also shown that the distribution in angle between the two highest
$E_T$ $\tau$-jets is different from other models which do not have this
doubly charged Higgsino.

\begin{table}
\centering
\caption{\label{delcross}  Cross sections (in fb) for deltino pair production
for various values of $\Lambda$. The other parameters used are
$tan \beta = 15$, $n = 2$ and $M/\Lambda$ = 3.}
\vskip 0.25cm
\begin{tabular}{l c c c}
$M_{\tilde{\Delta}} (M)$   &   $\Lambda = 35$\,TeV  &  $\Lambda = 60$\,TeV
    &  $\Lambda = 85$\,TeV  \\
\hline
90\,GeV    &   643.0  &  629.9  &  621.5  \\
120\,GeV   &   228.2  &  222.9  &  219.6  \\
150\,GeV   &   91.6   &  89.2   &  87.7   \\
\end{tabular}
\end{table}
\begin{table}
\caption{\label{sbr90} Inclusive $\tau$-jet for
a messenger scale deltino mass of 90\,GeV\@}
\vskip 0.25cm
\begin{tabular}{l c c c c c c c}
$\Lambda$ (TeV)  &  35  &  40  &  50  &  60  &  70  &  80  &  85  \\
\cline{2-8}
    &  \multicolumn{7}{c}{$\sigma \cdot {\rm BR}$ (fb)}  \\
\hline
1 $\tau$-jet: before cuts  &  198.6  &  132.9  &  73.11  &  71.64  &  70.86
             &  70.34  &  70.14  \\
after cuts   &  168.7  &  156.2  &  97.49  &  91.90  &  90.15  &  89.33
             &  89.05  \\
\hline
2 $\tau$-jets: before cuts &  362.6  &  277.3  &  203.6  &  196.7  &  196.3
             &  194.8  &  194.2  \\
after cuts   &  124.6  &  93.11  &  89.59  &  82.91  &  80.59  &  79.63
             &  79.33  \\
\hline
3 $\tau$-jets: before cuts &  306.0  &  273.5  &  253.6  &  244.7  &  240.9
             &  238.8  &  238.0  \\
after cuts   &  32.31  &  17.75  &  32.45  &  29.02  &  27.53  &  26.88
             &  26.73  \\
\hline
4 $\tau$-jets: before cuts &  119.4  &  116.9  &  126.42  &  116.5  &  113.1
             &  111.6  &  111.1  \\
after cuts   &  3.21  &  1.18  &  5.26  &  4.28  &  3.67  &  3.45  &  3.39  \\
\end{tabular}
\end{table}

\def\nn{\noindent}

\def\Re{{\cal R \mskip-4mu \lower.1ex \hbox{\it e}\,}}
\def\Im{{\cal I \mskip-5mu \lower.1ex \hbox{\it m}\,}}
\def\ie{{\it i.e.}}
\def\eg{{\it e.g.}}
\def\etc{{\it etc}}
\def\etal{{\it et al.}}
\def\ibid{{\it ibid}.}
\def\sub#1{_{\lower.25ex\hbox{$\scriptstyle#1$}}}
\def\tev{\,{\rm TeV}}
\def\gev{\,{\rm GeV}}
\def\mev{\,{\rm MeV}}
\def\to{\rightarrow}
\def\lsim{\mathrel{\mathpalette\atversim<}}
\def\gsim{\mathrel{\mathpalette\atversim>}}
\def\atversim#1#2{\lower0.7ex\vbox{\baselineskip\zatskip\lineskip\zatskip
  \lineskiplimit 0pt\ialign{$\matth#1\hfil##\hfil$\crcr#2\crcr\sim\crcr}}}
\def\undertext#1{$\underline{\smash{\vphantom{y}\hbox{#1}}}$}
\def\be{\begin{equation}}
\def\ee{\end{equation}}
\def\bea{\begin{eqnarray}}
\def\eea{\end{eqnarray}}
\def\IJMP #1 #2 #3 {Int. J. Mod. Phys. A {\bf#1},\ #2 (#3)}
\def\MPL #1 #2 #3 {Mod. Phys. Lett. A {\bf#1},\ #2 (#3)}
\def\NPB #1 #2 #3 {Nucl. Phys. {\bf#1},\ #2 (#3)}
\def\PLBold #1 #2 #3 {Phys. Lett. {\bf#1},\ #2 (#3)}
\def\PLB #1 #2 #3 {Phys. Lett. {\bf#1},\ #2 (#3)}
\def\PR #1 #2 #3 {Phys. Rep. {\bf#1},\ #2 (#3)}
\def\PRD #1 #2 #3 {Phys. Rev. {\bf#1},\ #2 (#3)}
\def\PRL #1 #2 #3 {Phys. Rev. Lett. {\bf#1},\ #2 (#3)}
\def\PTT #1 #2 #3 {Prog. Theor. Phys. {\bf#1},\ #2 (#3)}
\def\RMP #1 #2 #3 {Rev. Mod. Phys. {\bf#1},\ #2 (#3)}
\def\ZPC #1 #2 #3 {Z. Phys. C {\bf#1},\ #2 (#3)}

\section{Indirect Signals for Extra Dimensions}
\noindent
\centerline{\large\it J.L. Hewett}
\medskip
\label{section:Hewett}

One manifestation of theories of low scale quantum gravity
is the existence of a Kaluza Klein (KK) tower of massive gravitons which can
interact with the SM fields on the wall.  In this section we examine the 
indirect effects of these massive gravitons being exchanged 
in Drell-Yan production.  We consider a
novel feature of this theory, namely, the contribution of
gluon-gluon initiated processes to lepton pair production.  

The effective theory below the effective Planck scale in the bulk, $M_{eff}$,
consists of the SM fields on the wall and gravity which
propagates in the full $4+n$ bulk.  
The interactions of these fields are given by
\be
\label{ndints}
\int d^{4+n}x\, T^{\widehat\mu\widehat\nu}{h_{\widehat\mu\widehat\nu}(x^\mu,x^a)
\over M_{eff}^{n/2+1}}\,,
\ee
where $T^{\widehat\mu\widehat\nu}$ is the symmetric, conserved stress-energy tensor in 
the bulk and $h_{\widehat\mu\widehat\nu}$ is the
graviton field-strength tensor, which can be decomposed into
spin-2, 1, and 0 fields.  
Here the indices $\widehat\mu$ extend over the full $4+n$ dimensions, $\mu$ over
the $3+1$ dimensions on the wall, and $a$ over the $n$ bulk dimensions.  
The interactions with the SM matter fields are obtained by decomposing
the above into the 4-dimensional states.  The bulk fields 
$h_{\widehat\mu\widehat\nu}$ appear as Kaluza-Klein towers in the 4-dimensional space 
arising from a Fourier analysis over the cyclic boundary conditions of the
compactified dimensions.  Performing this decomposition, we immediately
see that $T_{\mu a}=0$ and hence the spin-1 KK states don't
interact with the wall fields.  The scalar, or dilaton, states couple 
proportionally to the trace of the stress-energy tensor.  For interactions
with fermions, this trace is linear in the fermion mass, while for gauge
bosons it is quadratic in the boson mass.  Hence, the dilaton does not
contribute to the processes under consideration here.  

We thus only have to consider the interactions of the KK 
spin-2 gravitons with the SM fields.  All the gravitons in the KK tower,
including the massless state, couple in an identical manner.  Hence we
may use the couplings to matter as obtained in the case of linearized general
relativity\cite{gr}.  In this linearized theory, the matrix element for
$q\bar q\to \ell^+\ell^-$ generalized for the case of $n$ massive graviton 
exchanges can be written as
\be
\label{mekk}
{\cal M}  =  {1\over M_{Pl}^2} \sum_n {T^e_{\mu\nu}P^{\mu\nu\lambda\sigma}
T^f_{\lambda\sigma}\over s-m^2_{gr}[n]} \,,
\ee
where the sum extends over the
KK modes.  $P_{\mu\nu\lambda\sigma}$ 
represents the polarization sum of the product of two graviton fields
and is given in \cite{gr}.  The terms in the polarization sum that are 
quadratic and quartic in the transferred momentum do not contribute
to the above matrix element since $T_{\mu\nu}$ is
conserved.  Likewise, the terms which go as $\eta_{\mu\nu}
\eta_{\lambda\sigma}$ lead to terms proportional to $T_\mu^{e\mu}
T_\lambda^{f\lambda}$ which vanish in the limit of zero electron mass.
The remaining terms are $P_{\mu\nu\lambda\sigma}={1\over 2}[\eta_{\mu\lambda}
\eta_{\nu\sigma}+\eta_{\mu\sigma}\eta_{\nu\lambda}-\eta_{\mu\nu}
\eta_{\lambda\sigma}]$ and are exactly
those present in the massless graviton case; they are thus universally 
applicable to all of the states in the KK tower.  Since the spacing of
the KK states is given by $\sim 1/r$, the sum over the states in (\ref{mekk}) 
above can be approximated by an integral which is log divergent for $n=2$
and power divergent for $n>2$.  A cut-off must then be applied to regulate
these ultraviolet divergences, and is generally taken to be the scale of the 
new physics.  For $n>2$ it can be shown\cite{summ} that the dominant
contribution to this integral is
of order $\sim M^2_{Pl}/M_s^4$, where we have taken the cut-off to be the 
string scale, while for $n=2$ this result is multiplied by a factor of order 
$\log(M_s^2/E^2)$, where $E$ is the center-of-mass energy of the process
under consideration.  The exact computation of this integral can only be 
performed with some knowledge of the full underlying theory.  Combining these 
results yields the matrix element
\bea
\label{ffff}
{\cal M} & =  & {\lambda\over M_s^4}  \left\{  \bar q(p_1)\gamma_\mu q(p_2)
\bar \ell(p_3)\gamma^\mu \ell(p_4)(p_2-p_1)\cdot (p_4-p_3)\right.\\
& & \left. \bar q(p_1)\gamma_\mu q(p_2)\bar \ell(p_3)\gamma_\nu \ell(p_4)
(p_2-p_1)^\nu(p_4-p_3)^\mu\right\}\,.\nonumber
\eea
Here, the momentum flow is defined with $p_{1,2}$ into the vertex and
$p_{3,4}$ outgoing.  Note that graviton exchange is $C$ and $P$ conserving,
and is independent of the flavor of the final state.
The coefficient $\lambda$ is of ${\cal O}(1)$ and cannot 
be explicitly calculated without knowledge of the full quantum gravity theory.  
It is dependent on the number of extra dimensions, how they are compactified,
and is in principle a power series in $s/M_s^2$.  However, we neglect this
possible energy dependence in $\lambda$ and note that the limits obtained 
here, which go as $|\lambda|^{1/4}$, are only very weakly dependent on its
precise value and hence on the specific model realization.  In principle the
sign of $\lambda$ is undetermined and we examine the constraints that can be
placed on $M_s$ with either choice of signs.

The angular distribution for $q\bar q\to \ell^+\ell^-$ with massless
leptons is then calculated to be
\bea
\label{dsdz}
{d\sigma\over dz} & = & N_c{\pi\alpha^2\over 2s}\left\{ P_{ij}
\left[A^q_{ij}A^\ell_{ij}(1+z^2)+2 B^q_{ij}B^\ell_{ij}z\right]\right.
\nonumber\\
& & -{\lambda s^2\over 2\pi\alpha M_s^4}P_i\left[2z^3v^q_iv^\ell_i
-(1-3z^2)a^q_ia^\ell_i\right] \\
& & \left. +{\lambda^2s^4\over 16\pi^2\alpha^2M_s^8}\left[
1-3z^2+4z^4\right]\right\} \,,
\nonumber
\eea
where the indices $i,j$ are summed over $\gamma$ and $Z$ exchange, $z=\cos
\theta$, $P_{ij}$ and $P_i$ are the usual propagator factors (defined in \eg,
\cite{leptos}), $A^f_{ij}=(v_i^fv_j^f
+a_i^fa_j^f)$\,, and $B^f_{ij}=(v_i^fa_j^f+v_j^fa_i^f)$.
Note that the total cross section from $q\bar q$ annihilation
is {\it unaltered} by graviton exchanges, independently of fermion flavor 
up to terms of order $s^4/M_s^8$, and hence
the angular distributions will be most sensitive to these
new exchanges.  

In addition to the $q\bar q$ channel,
gravitons can also mediate gluon-gluon contributions to lepton pair production
via $s$-channel exchange, as noted above.  
Such gluon initiated processes are a remarkable
consequence of this theory and have the potential to modify the Drell-Yan
spectrum in a unique manner.
Following an analogous procedure as outlined above for the four-fermion case, 
the matrix element for $gg\to\ell^+\ell^-$ via graviton exchanges is found to be
\bea
\label{ggll}
{\cal M}&=&{-\lambda\over 4M_s^4}  
\bar \ell(p')[(p'-p)_\mu\gamma_\nu+(p'-p)_\nu\gamma_\mu]
\ell(p)\left\{k'_\alpha(k_\mu\eta_{\beta\nu}+k_\nu\eta_{\beta\mu})
+k_\beta(k'_\mu\eta_{\alpha\nu}+k'_\nu\eta_{\alpha\mu})\right.\nonumber\\
& & \left. -\eta_{\alpha\beta}(k'_\mu k_\nu+k_\mu k'_\nu)
+\eta_{\mu\nu}(k'\cdot k
\eta_{\alpha\beta}-k_\beta k'_\alpha)-k\cdot k'(\eta_{\mu\alpha}
\eta_{\nu\beta}+\eta_{\mu\beta}\eta_{\nu\alpha}\right\}\epsilon^\beta_g
(k')\epsilon^\alpha_g(k) \,,\nonumber\\
\eea
where the momentum flow is defined with both $k\,, k'$ flowing into the
vertex and $p\,, p'$ being outgoing and $\epsilon$ represents the gluon
polarization vector.  Because the graviton couplings and the
summation over the KK tower of states for $2\to 2$ processes are universal,
$\lambda$ is the same ${\cal O}(1)$ coefficient as in Eq. (\ref{ffff}).
This matrix element yields the $gg\to\ell^+\ell^-$ 
differential cross section for massless leptons
\be
\label{glue}
{d\sigma\over dz} = {\lambda^2\widehat s^3\over 64\pi
M_s^8} (1-z^2)(1+z^2)\,,
\ee
which has a remarkably simple form.
While the overall numerical coefficient appears to be very small, 
it must be compared to $\sim\alpha^2$ which appears in the usual contributions 
to Drell-Yan production.  In addition, the large parton luminosity
for gluons at higher energy colliders may also compensate for the small
numerical factor.
Since this cross section is also even in $\cos\theta$,
the gluon-gluon contributions will only affect the total cross section and
not the forward-backward asymmetry.  Also note that the ambiguity in the
sign of $\lambda$ does not affect the gluon-gluon contributions
as they do not interfere with the $q\bar q$ initiated process.

The bin integrated lepton pair invariant mass distribution and forward-backward
asymmetry $A_{FB}$ 
is presented in Fig. \ref{haddists}.
In each case the solid histogram represents the SM expectations,
and the `data' points include the graviton exchanges with the error bars
representing the statistics in each bin.  The rapidity cuts, parton density
parameterizations, and assumed integrated luminosity are as labeled, and we
have summed over electron and muon final states.  Here
we present the sample case of $M_s=800$ GeV and the sign ambiguity
in $\lambda$ is visible in the forward-backward asymmetry.  
Since the graviton exchanges only affect 
the invariant mass distribution at order $\lambda^2/M_s^8$, we expect
only minor modifications to this spectrum and we see that this holds true.
For higher energy colliders which have a larger gluon parton luminosity,
such as the LHC,
large string scales do have a sizable effect
on the $M_{\ell\ell}$ spectrum\cite{jlh}.
However, the deviations in $A_{FB}$,
are more pronounced at the Tevatron, where even the two cases 
$\lambda=\pm 1$ are statistically distinguishable from each other 
for this sample case.  The resulting 95\% C.L. search reach
are given in Fig. \ref{hadres}.
We also find that present Tevatron data from Run I with 110
${\rm pb}^{-1}$\, of
integrated luminosity excludes a string scale up to 980 (920) GeV at 95\%
C.L. for $\lambda=-1(+1)$.

\runfig{haddists}{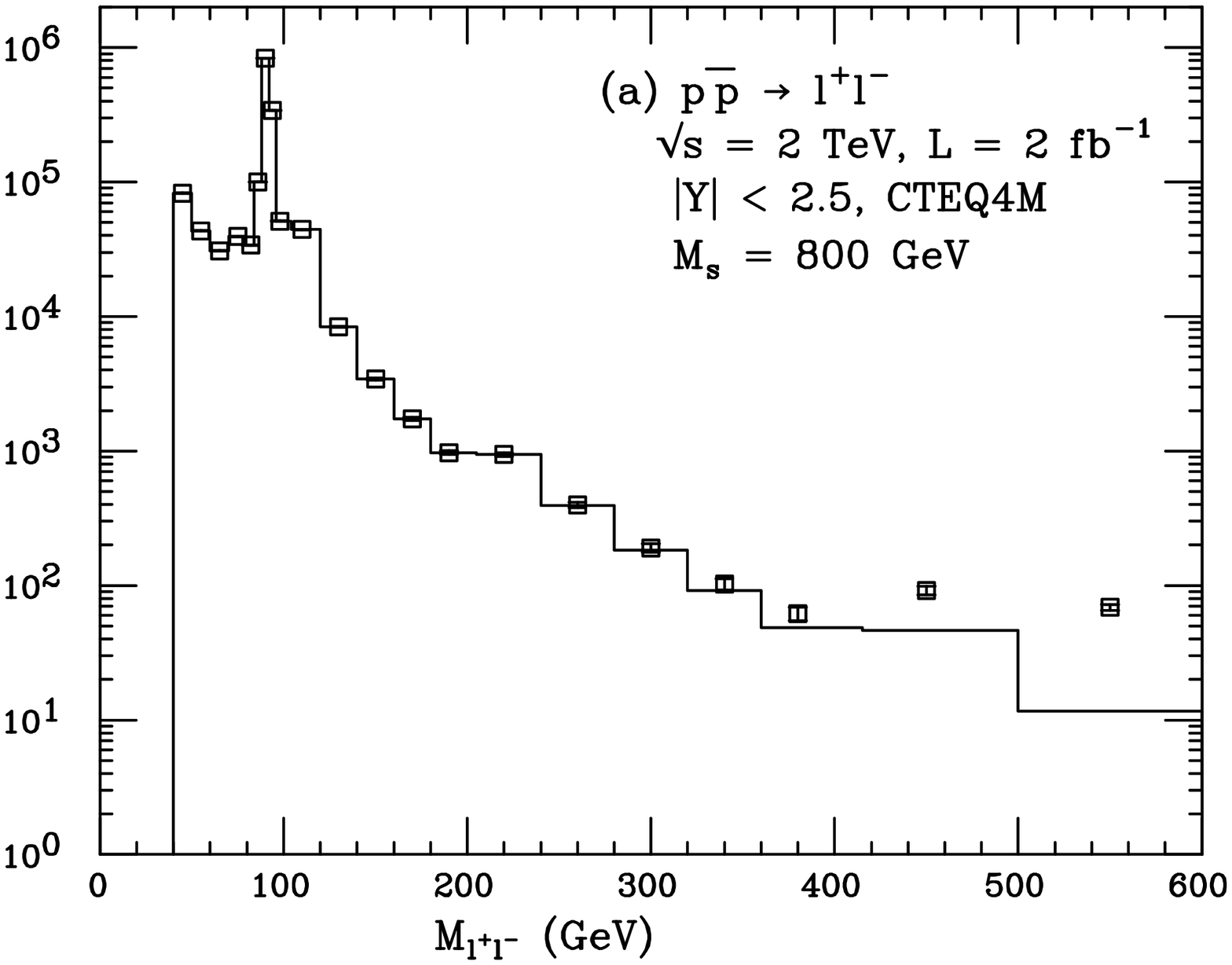}{
Bin integrated lepton pair invariant mass distribution
for Drell-Yan production at the Main Injector.
The SM is represented by the solid histogram.  The data points
represent graviton exchanges with $M_s=800$ GeV and $\lambda=+1$ or $-1$.}

\runfig{haddists2}{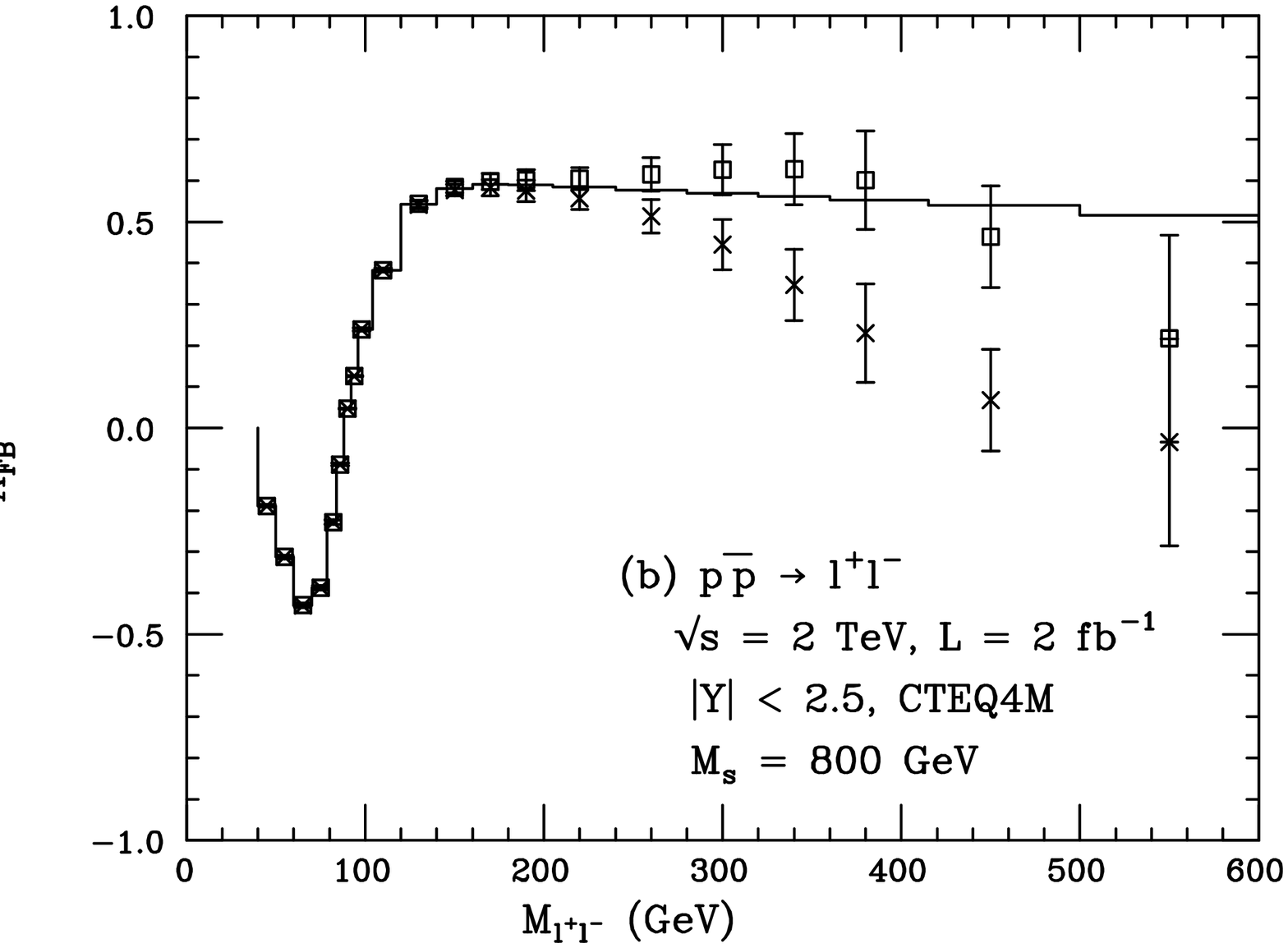}{
Bin integrated lepton pair 
forward-backward asymmetry for Drell-Yan production at the Main Injector.
The SM is represented by the solid histogram.  The data points
represent graviton exchanges with $M_s=800$ GeV and $\lambda=+1$ or $-1$.}

\runfig{hadres}{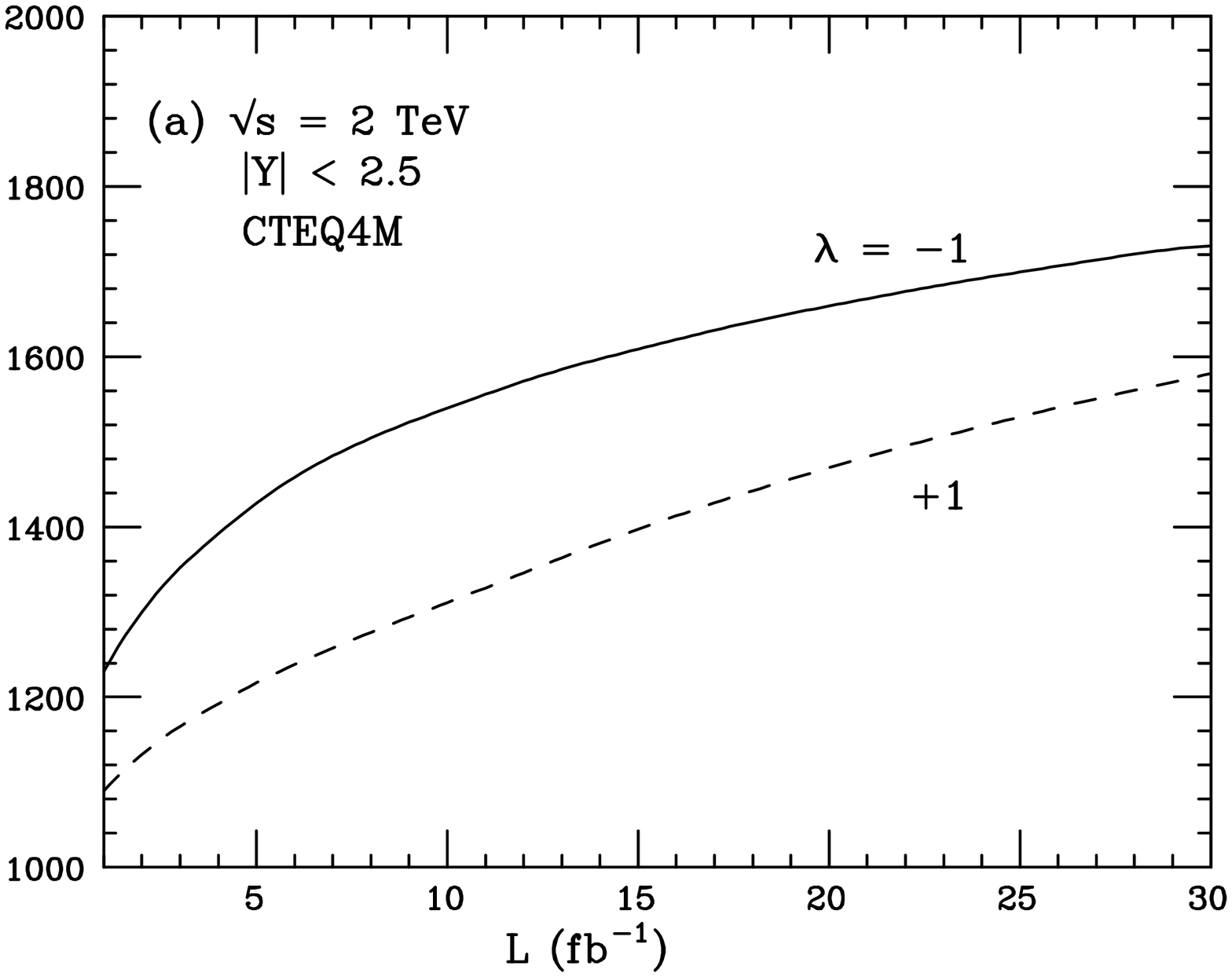}{
95\% C.L. search reach for the string scale as a function of
integrated luminosity at the Tevatron with the sign of $\lambda$ as
labeled.}




\section{Higgs Bounds in Three and Four Generation Scenarios}
\centerline{\large\it D.~Dooling, K.~Kang and S.K.~Kang}
\medskip
\label{section:Dooling}

\subsection{Introduction}

 The search for the Higgs boson being one of the major tasks along with that for supersymmetric sparticle and fourth generation fermions at future accelerators such as LEP200 and LHC makes it a theoretical priority to examine the bounds on the Higgs boson mass in the SM and its supersymmetric extension and to look for any distinctive features.
 The actual measurement of the Higgs boson mass could serve to exclude or at least to distinguish between the SM(3,4) and the MSSM(3,4) models for electroweak symmetry breaking.
Recently, bounds on the lightest Higgs boson mass were calculated in \cite{1,2,3,4,5,6,7,8,9}.
It was found that for a measured $M_{H}$ lying in a certain mass range, both the SM vacuum stability lower bound and the MSSM upper bound are violated, thus shaking our confidence in these theories just as the final member of the mass spectrum is observed.
One method of curing this apparent illness is to take a leap of faith by adding another fermion generation, to fortify these theories with another representation of the gauge group.
This additional matter content, for certain ranges of its mass values, has the desired effect of raising the MSSM3 upper bound above that of the SM lower bound and avoids the necessity of being forced to introduce completely new physics.

In this work, we use the latest LEP Electroweak Working Group data as well as the most recent experimental lower limits on the masses of fourth generation fermions to see how the Higgs boson mass bounds are affected.
With only the standard three generations, a violation of both the SM and the MSSM bounds occurs for a broad range of $M_{H}$ values, signalling the need for an additional generation.
Our presentation is organized as follows.
We first present a summary of our one-loop effective potential (EP) improved by the two-loop renormalization group equations (RGE) analysis.
An expression for the Higgs boson mass is derived up to the next-to-leading logarithm order.
Bounds on $M_{H}$ are obtained by imposing different boundary conditions on the Higgs self-coupling $\lambda$.
Finally, we present our results and also obtain information on the possible mass range of the fourth generation leptons, $M_{L}$.

\subsection{Effective Potential Approach }

As shown in \cite{10}, in order to calculate the Higgs boson mass 
 up to the next-to-leading logarithm approximation,
we must consider the one-loop EP 
improved by two-loop RGE for the $\beta$ and $\gamma$ functions of the
running coupling constants, masses and the $\phi$ field for the Higgs boson \cite{11}.

The two-loop RGE improved one loop EP of the SM4 is given by
\begin{equation}
 V_{1} = V_{(0)} + V_{(1)} 
\end{equation}
where
\begin{eqnarray}
 V_{(0)} &=& -\frac{1}{2}m^{2}(t)\phi_{c}^{2}(t) + \frac{1}{24}\lambda(t)
              \phi_{c}^{4}(t) \\
V_{(1)} &=& \sum_{i=1}^{5} \left(-\frac{\kappa_{i}}{64\pi^{2}}\right)
            h_{i}^4(t)\phi_{c}^{4}(t)
            \left[\ln\frac{h_{i}^{2}(t)\zeta^{2}(t)}{2}-\frac{3}{2}\right] 
\end{eqnarray}
Here $\phi_c$ is the classical field corresponding to the physical
Higgs boson $\phi$, $i = (t,T,B,N,E)$
, $~\kappa_{i}=3$ for $i = (t,T,B)$ and $\kappa_{i}=1$ for $i = (N,E)$ and $h_{i}$ is the Yukawa coupling of the i$^{th}$ fermion to the Higgs field.
In addition,
\begin{eqnarray}
\zeta(t) &=& \left(-\int_{0}^{t} \gamma_{\phi}(t^{\prime})dt^{\prime}\right)
\\ 
\phi_{c}(t) &=& \phi_{c} \zeta(t)
\end{eqnarray}
and
\begin{equation}
\mu(t)=\mu\exp(t)
\end{equation}
where $\mu $ is a fixed scale.

Starting with the above expression for the SM(3,4) EP, one may follow the analysis in \cite{1,4} and obtain the expression for the Higgs boson mass:

\begin{eqnarray}
m_{\phi}^{2} &=& \frac{1}{3}\lambda v^{2}+\frac{\hbar v^{2}}{16 \pi^{2}}\left\{ \frac{\lambda^{2}}{3} +2\lambda (h_{t}^{2}+h_{T}^{2}+h_{B}^{2}) +\frac{2\lambda}{3}(h_{E}^{2}+h_{N}^{2})\right. \nonumber \\
& & -\frac{\lambda}{2}(3 g_{2}^{2}+g_{1}^{2}) + \frac{9}{8}g_{2}^{4} +\frac{3}{4}g_{1}^{2}g_{2}^{2} + \frac{3}{8}g_{1}^{4} \nonumber \\
& &-6h_{t}^{4}\ln\frac{h_{t}^{2}\zeta^{2}}{2} -6h_{T}^{4}\ln\frac{h_{T}^{2}\zeta^{2}}{2}-6h_{B}^{4}\ln\frac{h_{B}^{2}\zeta^{2}}{2} \nonumber \\
& & \left. -2h_{E}^{4}\ln\frac{h_{E}^{2}\zeta^{2}}{2}-2h_{N}^{4}\ln\frac{h_{N}^{2}\zeta^{2}}{2} \right\} + O(\hbar^{2})
\end{eqnarray}
where the three generation case is obtained by simply letting the fourth generation Yukawa couplings go to zero.

Following the method similar of Kodaira et al. \cite{4}, we arrive at the appropriate energy scale at which to evaluate $\frac{\partial^{2} V}{\partial \phi_{c}(t)^{2}}$ by requiring $\frac{\partial V}{\partial \phi_{c}(t_{v})} =$ 0 at the scale $t_{v}$ where $\phi_{c}(t_{v}) = v = (\sqrt{2} G_{F})^{-\frac{1}{2}}$ = 246 GeV.
$t_{v}$ is found to satisfy:
\begin{displaymath}
t_{v} = \ln \frac{v}{\mu} + \int_{0}^{t_{v}} \gamma_{\phi}(t^{\prime})dt^{\prime}.
\end{displaymath}
We then evaluate the first and second derivatives of the EP at the scale $t_{v}$ where $\phi_{c}(t_{v}) = v$.
In the above relation satisfied by $t_{v}$, $\mu$ is a fixed, constant mass scale.
In the MSSM theories, we will take $\mu$ to be $M_{susy}=1$ or 10 TeV, while the SM lower bounds will be derived after choosing $\mu= \Lambda= 10^{19}$ GeV.

 We define the running Higgs mass as:
\begin{displaymath}
m_{h}^{2}(t)= \frac{m_{h}^{2}(t_{v}) \zeta^{2}(t_{v})}{\zeta^{2}(t)}.
\end{displaymath}
The physical, pole masses are related to the running masses via the following equations:
\begin{displaymath}
M_{i}= \left[ 1 +\beta_{i} \frac{4 \alpha_{3}(M_{i})}{3 \pi} \right] m_{i}(M_{i})
\end{displaymath}
\begin{displaymath}
M_{H}^{2} = m_{h}^{2}(t) + Re\Pi(M_{H}^{2}) - Re\Pi(0)
\end{displaymath}
where $\beta_{i}$ = 0 for i = (N,E) and = 1 for i = (t,T,B).
$\Pi(q^{2})$ is the renormalized electroweak self-energy of the Higgs boson.

The method of solving the RGE and the appropriate boundary conditions for the couplings is explained in \cite{1}.
In this update, we use $M_{Z} = 91.1867$ GeV and $\alpha_{3}(M_{Z}) = .119$.

\subsection{Bounds on $M_{H}$}

Let us first discuss the procedure for determining a lower bound on the Higgs boson mass in the SM \cite{5,12}
Working with the two-loop RGE requires the imposition of one-loop boundary conditions on the running parameters.
As pointed out by Casas et al. \cite{5,7}, the necessary condition for vacuum stability is derived from requiring that the effective coupling $\tilde{\lambda}(\mu)>$ 0  rather than $\lambda > 0$ for $\mu(t) < \Lambda$, where $\Lambda$ is the cut-off beyond which the SM is no longer valid. 
The effective coupling $\tilde{\lambda}$ in the SM4 is defined as:
\begin{displaymath}
\tilde{\lambda}=\frac{\lambda}{3} -\frac{1}{16 \pi^{2}}\left\{ \sum_{i=1}^{5} 2 \kappa_{i} h_{i}^{4} \left[ \ln \frac{h_{i}^{2}}{2} - 1 \right] \right\}
\end{displaymath}
where the three generation case is simply the same as the above expression without the fourth generation Yukawa coupling contributions.
Choosing $\Lambda = 10^{19}$ GeV and $M_{top} = 172$ GeV, we arrive at a vacuum stability lower bound on $M_{h}$ of $\sim$ 134 GeV for the SM with three generations.
Allowing $M_{top}$ to be as large as 179 GeV increases the lower bound on $M_{H}$ to $\sim$ 150 GeV.

To compute the MSSM upper bound on $M_{H}$, we assume that all of the sparticles have masses $O(M_{susy})$ or greater and that of the two Higgs isodoublets of the MSSM, one linear combination is massive, also with a mass of $O(M_{susy})$ or greater, while the other linear combination, orthogonal to the first, has a mass of the order of weak-scale symmetry breaking.
With these two assumptions, it is clear that below the supersymmetry breaking scale $M_{susy}$, the effective theory is the SM.
This fact enables us to use the SM effective potential for the Higgs boson when we treat the lightest Higgs boson in the MSSM.

In the MSSM(3,4), the boundary condition for $\lambda$ at $M_{susy}$ is
\begin{displaymath}
\frac{\lambda}{3}(M_{susy})=\frac{1}{4}\left[g_{1}^{2}(M_{susy})+g_{2}^{2}(M_{susy})\right]\cos^{2}(2\beta)+\frac{\kappa_{i}h_{i}^{4}(M_{susy})}{16 \pi^{2}}\left( 2\frac{X_{i}}{M_{susy}^{2}}-\frac{X_{i}^{4}}{6 M_{susy}^{4}} \right)
\end{displaymath}
where $\kappa_{i}$ = 3 for $i = (t,T,B)$ and $\kappa_{i}$ = 1 for $i = (N,E)$ and $X_{i}$ is the supersymmetric mixing parameter for the ith fermion.
Zero threshold corrections correspond to $X_{i}$ = 0.
Maximum threshold corrections occur for $X_{i} = 6 M_{susy}^{2}$.

\runfig{kang1}{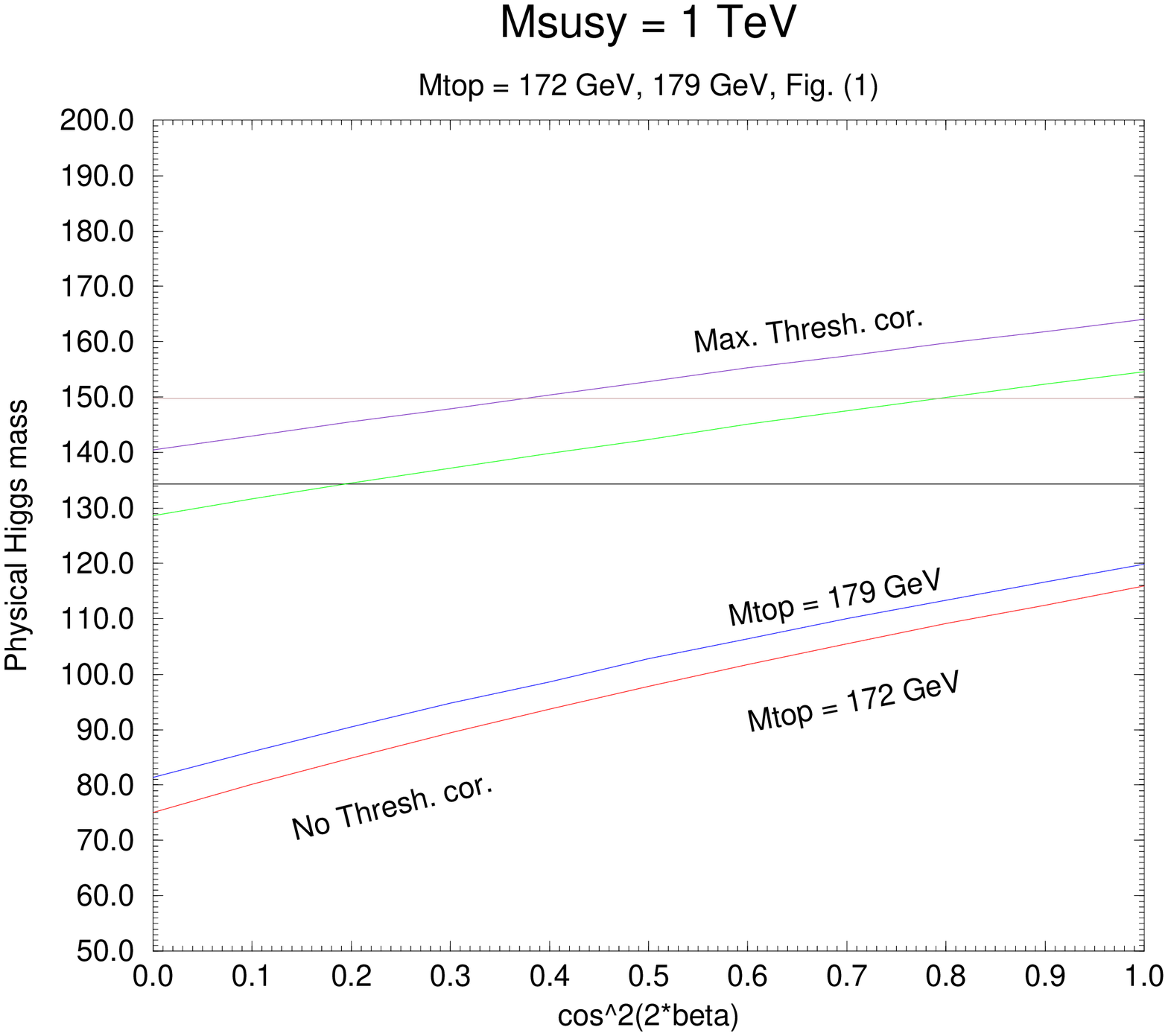}{
The lightest Higgs boson mass $M_{H}$ as a function of $\cos^{2}(2\beta)$.
The bottom two curves correspond to MSSM upper bounds with no threshold corrections, for $M_{top}$ = 172 GeV and 179 GeV, respectively.
 The two upper curves correspond to MSSM upper bounds with maximum threshold corrections, for $M_{top}$ = 172 GeV and 179 GeV, respectively.
The two horizontal lines are the $\cos^{2}(2\beta)$-independent SM3 vacuum stability bounds.
The lower horizontal line corresponds to $M_{top}$ = 172 GeV, while the other horizontal line was computed with $M_{top}$ = 179 GeV.}

\runfig{kang2}{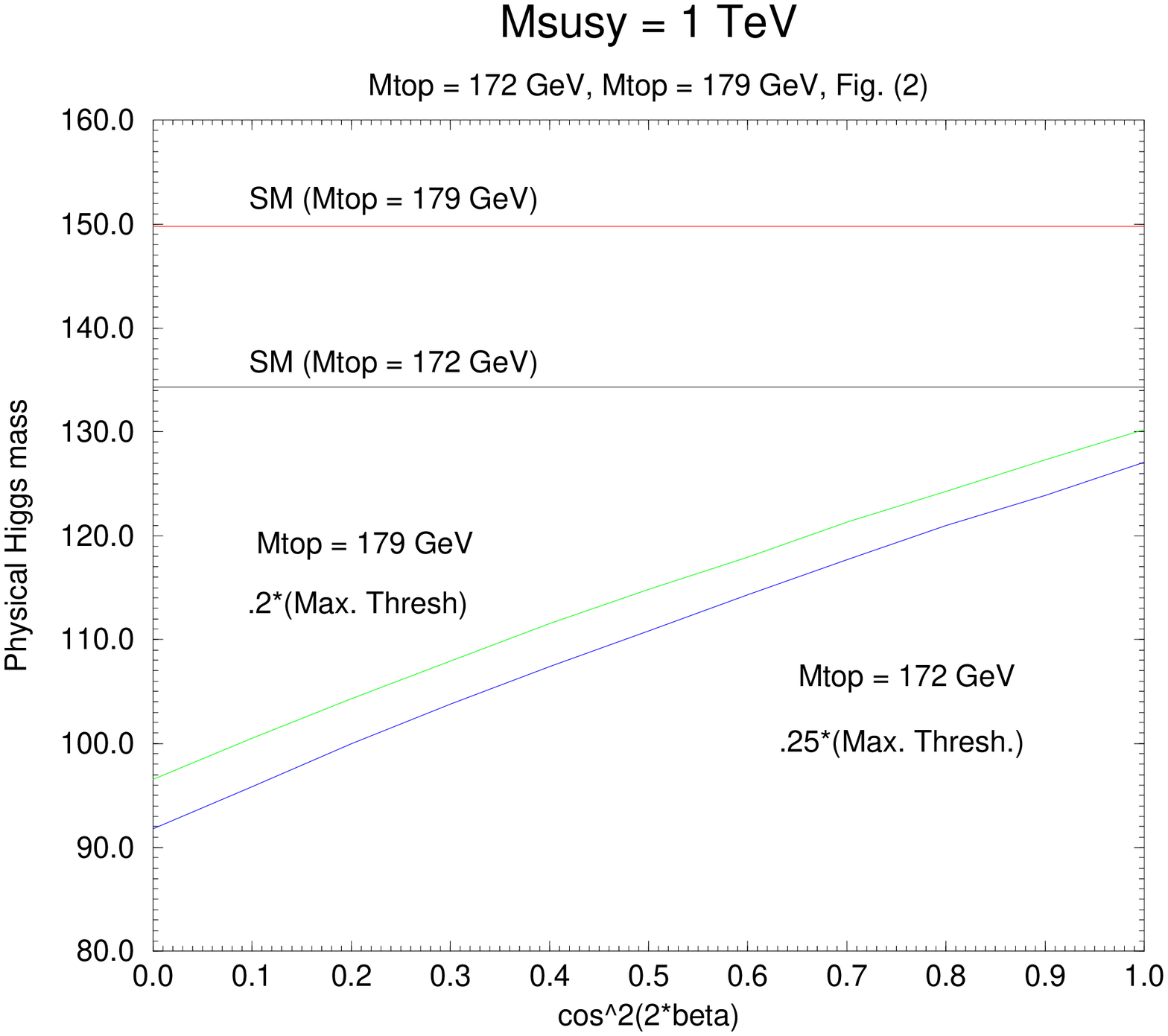}{
Same as Figure 1, but now the MSSM bounds correspond to the minimal threshold corrections consistent with the experimental lower limit on $M_{H}$.}

\runfig{kang3}{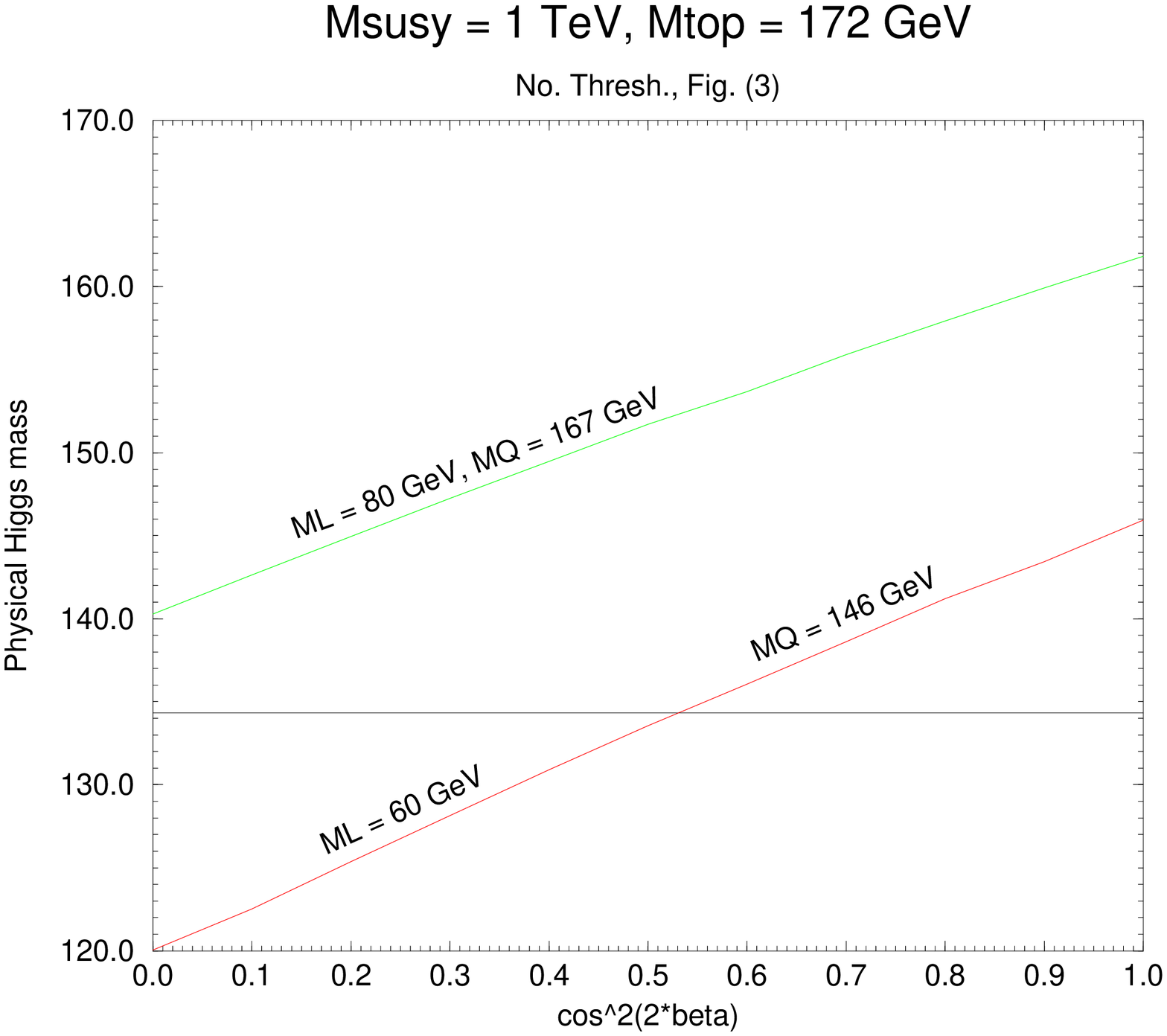}{
Plots of the physical Higgs boson mass as a function of $\cos^{2}(2\beta)$. The $\cos^{2}(2\beta)$-independent flat line is the MSSM3 vacuum stability lower bound for $M_{top}$ = 172 GeV. The lower curve is the MSSM4 upper bound for the same value of $M_{top}$, no threshold corrections and the indicated values for $M_{L}$ and $M_{Q}$. Similarly for the upper curve. Figure 4: Same as Figure 3, but with $M_{top}$ = 179 GeV.}

\runfig{kang4}{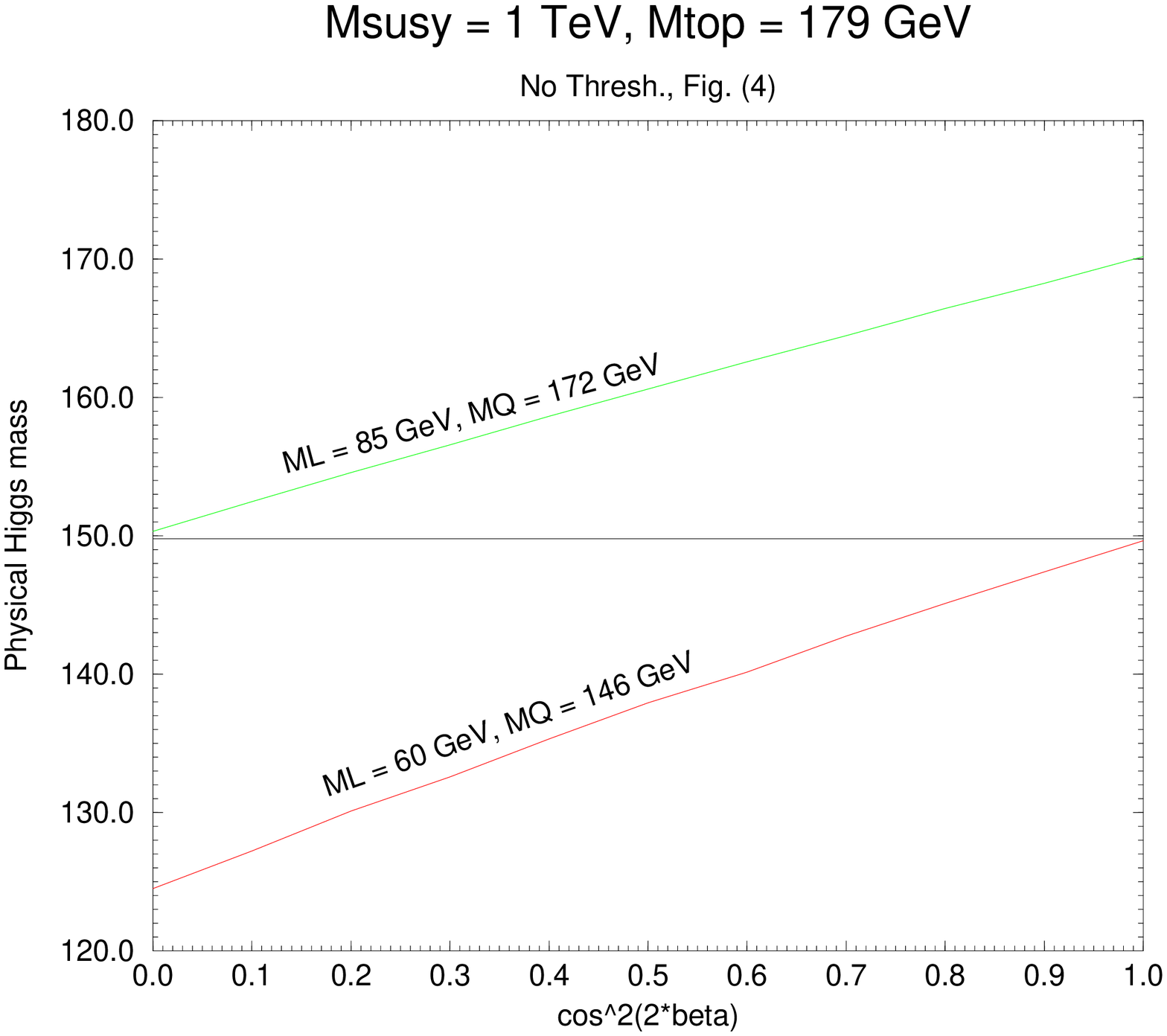}{See caption of Fig.~\ref{kang3}.}

In Fig. (1) we present our numerical two-loop results for the lightest Higgs boson mass bounds in the SM and the MSSM3 as a function of the supersymmetric parameter $\cos^{2}(2\beta)$.
The bottom two curves correspond to the MSSM3 upper bound for the two cases $M_{top} = 172$ GeV and the slightly greater upper bound that results when $M_{top}$ = 179 GeV and with no threshold corrections.
When the case of maximum threshold corrections is considered, these two curves are translated upwards by $\sim$ 55 GeV - 60 GeV, illustrating the strong dependence of the upper bound on the precise value of the threshold corrections.
Yet even with such a dramatic increase in the upper bounds with increasing threshold corrections, we observe that the SM lower bound exceeds the MSSM upper bound for $M_{top} = 172$ GeV and $ 0 < \cos^{2}(2\beta) < .2$ for all values of the threshold correction contribution.
Similarly, for $M_{top} = 179$ GeV, the troublesome situation is only exacerbated, as the SM lower bound exceeds the MSSM upper bound for $0 < \cos^{2}(2\beta) < .38$ independent of the threshold corrections.

In Fig.(2) we present the problem more clearly.
Taking into account the present experimental lower limit on $M_{H}$ of $\sim$ 90 GeV at 95$\%$ CL, we find the value of the threshold correction that gives a smallest upper bound consistent with the experimental lower limit.
Clearly, for this phenomenologically determined lower limit of the threshold contributions, there is a large area in $M_{H} \times \cos^{2}(2\beta)$ space that is inconsistent with both the SM and the MSSM.
For $M_{top} = 172$ GeV, the region 92 GeV $ < M_{H} < $ 134 GeV invalidates both theories independent of $\cos^{2}(2\beta)$, while for $M_{top} = 179$ GeV, the range of mutual invalidation is 92 GeV $ < M_{H} < $ 150 GeV.

\subsection{Fourth Generation}

To resolve the above conundrum, one would like to either raise the MSSM upper bounds, lower the SM lower bounds, or both.
Adding more massive fermions to the theory only increases both bounds, so it is readily apparent that the way out of the area of inconsistency is to consider the MSSM4 and see if the additional matter of the MSSM4 results in MSSM4 upper bounds that exceed the SM3 lower bounds.

We now discuss restrictions on the possible fourth generation fermion masses \cite{2,13,14,15}.
The close agreement between the direct measurements of the top quark at the Tevatron and its indirect determination from the global fits of precision electroweak data including radiative corrections within the framework of the SM imply that there is no significant violation of the isospin symmetry for the extra generation.
Thus the masses of the fourth generation isopartners must be very close to degenerate \cite{14}; i.e.
\begin{displaymath}
\frac{\|M_{T}^{2}-M_{B}^{2}\|}{M_{Z}^{2}} \lesssim 1, \frac{\|M_{E}^{2}-M_{N}^{2}\|}{M_{Z}^{2}} \lesssim 1
\end{displaymath}
Recently, the limit on the masses of the extra neutral and charged lepton masses, $M_{N}$ and $M_{E}$, has been improved by LEP1.5 to $M_{N} > 59$ GeV and $M_{E} > 62$ GeV.
Also, CDF has yielded a lower bound on $M_{B}$ of $\sim$ 140 GeV.

In our previous work, we considered a completely degenerate fourth generation of fermions with mass $m_{4}$.
We derived an upper bound on $m_{4}$ in the MSSM4 by demanding perturbative validity of all the couplings out to the GUT scale \cite{16}.
This constraint led to an upper bound on $m_{4}$ of $\sim$ 110 GeV.
The above experimental lower limit on $M_{B}$ naturally forces us to now a consider a fourth generation where degeneracy only holds among the isodoublets separately.
We therefore consider a fourth generation with masses $M_{L}$ and $M_{Q}$.

In Fig.(3), we present the SM lower bound, the MSSM4 upper bound with the fourth generation masses at their experimental lower limits and with fourth generation masses large enough to remove the problem area for all values of $\cos^{2}(2\beta)$.
The MSSM bounds were calculated with no threshold corrections, and $M_{top}$ is fixed at 172 GeV.
Fig.(4) shows the same information for $M_{top}$ = 179 GeV.
The MSSM4 upper bounds are much more sensitive to $M_{Q}$ than they are to $M_{L}$.
This qualitative behavior is readily understood from inspection of the equation for $m_{\phi}^{2}$.
For this reason, it is necessary to increase $M_{Q}$ appropriately in order to generate a MSSM4 upper bound that is greater than the SM lower bound for all values of $\cos^{2}(2\beta)$.
In fact, keeping $M_{Q}$ at 146 GeV and allowing $M_{L}$ to be 110 GeV does not resolve the problem.
But increasing both $M_{Q}$ and $M_{L}$ as indicated in the figures does remove the problem.
Because all of the bounds increase as $M_{L}$ and $M_{Q}$ increase, and because the upper bounds on $m_{4}$ from the previous work are saturated when the masses of the fourth generation reach some critical values from below, we can conclude that $M_{L}$ must still be $< 110$ GeV.
This conclusion follows because it is $h_{N}$ that violates perturbative validity, so in the non-degenerate case, it is $M_{L}$ that must still respect this upper bound if gauge coupling unification is still to be achieved in the MSSM4.

\subsection{Concluding remarks}

In conclusion, we have studied the upper bounds on the lightest Higgs boson mass $M_{H}$ in the MSSM with four generations by solving the two-loop RGE's and using the one-loop EP.
We have considered a fourth generation of quarks and leptons with degenerate masses $M_{Q}$ and $M_{L}$.
For certain values of $M_{Q}$ and $M_{L}$, the area of mutual inconsistency between the SM and MSSM3 Higgs mass bounds is found to be consistent with the MSSM4 upper bounds.




\section{Fourth Family Status and Prospects}
\noindent
\centerline{\large\it D.~McKay}
\medskip
\label{section:McKay}

\subsection{The Bottom Line}

\medskip

 In light of the general expectation  
from SUSY considerations and from precision electroweak fits that the 
lightest Higgs is light,
    I believe that the inclusion of the essential Higgs FCNC channel in the simulation of the various
    decay channels of $b'$ is the the fourth family project that deserves a strong effort among the BTMSSM
    searches. Analysis of the $b' \rightarrow b + Z$ mode is well advanced at both CDF and D0.  The 
    inclusion of the Higgs mode, with the subsequent $H \rightarrow b + \overline{b}$ decay, will remove
    the last loophole in the $b'$ search up to the top mass.  A good reason for stressing this issue is that
    it ties in directly to the crucial Higgs search mission at Run II, which makes the $b'$ search
    more subtle and more interesting.

    A small but important project is the re-fitting of electroweak parameters to include relatively light
    new physics (50-150 GeV) effects.  A preliminary look is shown in the ``mass bounds'' graph in Section 3.

    More detail on these points is given in Sec. 3 below.
     
\subsection{Words of Motivation}

\medskip

1)  There is no theoretical reason against having a fourth family.  Since
    there is no solid theoretical understanding of the family pattern, there
    is motivation to simply look to see if a new one shows up. Unless
    stated otherwise, fourth family means a family with a lepton doublet
    with mass greater than  $ m_{Z}$/2 but with an otherwise sequential,
    chiral set of leptons $(\nu',\tau')$ and quarks $(t',b')$. (It is assumed
    in all of the discussion that $V_{b't'} \gg V_{b't}\simeq V_{t'b}\gg
    V_{b'c}$
      
\medskip

2)  Though it is not in the vein of the fourth family as just defined,
    the proposal by Frampton and collaborators to solve the 
    strong  CP  problem calls for heavy vector-like fermion states, and 
    the search strategies are largely the same as for chiral lepton and
    quark states. ( see Frampton and Kephart, PRL 66, 166; Frampton and Ng,
    PRD 43, 3034).  
              
\medskip

3)  In a four family extension of the standard model, Hung has recently
    argued that unification of the gauge couplings can be obtained at
    a high enough scale to avoid proton decay. The scale for the fourth
    family can be higher than in the MSSM, but it is still constrained to
    rather low values - lepton  masses below $m_{Z}$ and quark masses at
    around $m_{t}$ -  by the requirement that all couplings stay perturbative
    up to the unification scale. If this motivation applies to SUSY, it applies to
    this four family SM.  The hierarchy problem is left unsolved, of course.
    ( P.Q. Hung, PRL 80, 3000 (1998); see also P. Frampton and P.Q. Hung,
    Phys. Rev. D 58, 057704 (1998) for phenomenology of quasi-stable
    quarks in the $ \sim m_{t}$ range that could have escaped detection
    in the CDF and D0 analysis to date. For a study of four family consequences
    in the MSSM with SUGRA boundary conditions, see J. Gunion, D. McKay and
    H. Pois, PRD 53, 1616,(1996) ).

\medskip

4) For those with a taste for dynamical breaking of electroweak symmetry,
   new heavy quark flavors in a fourth family may have some appeal, since
   the heavy quark condensate mechanism for EW breaking may have a chance if 
   there are extra, heavy quarks in the range 150-230 GeV  to help top do
   the job. (see for example  Hung and Isidori, P.L. B402, 122). 

\medskip

5) Tying in directly with the Higgs boson search and BTMSSM theme of this workshop,
D. Dooling, K. Kang and S.-K. Kang (hep-ph/9710258, to be published, and
contribution to this workshop) have argued that the lower mass bound on the SM
Higgs boson and the upper limit on the lightest Higgs  in the MSSM are relaxed in
four family extensions of these models.  This consideration is especially timely
as the limit on the SM Higgs mass has recently been pushed up to nearly 100 GeV at
LEP II.
 
\bigskip

\subsection{Experimental Facts that tightly constrain a sequential Fourth Family}
 
\medskip 

1) The S parameter in precision electroweak analysis is now $ S = -0.16\pm0.14 (-0.10)$
   as listed by the PDG '98 discussion of "Electroweak model and constraints on 
   new physics"(European Physical Journal C 3,1 (1998)).  A new, heavy (several times $ M_{Z}$)  
   degenerate, chiral family adds $2/3\pi$ to S and this excludes an extra, heavy 
   family of fermions at the 99.2\% C.L. when S is allowed to be arbitrary and fixed by the
   values of the  EW parameters fit at the Z (primarily). Requiring S to be positive with a given
   Higgs mass relaxes this bound somewhat. If some of the masses of the family members are in
   the neighborhood of $m_{Z}$, then the theoretical value of S is less than $2/3\pi$, and a more
   general analysis of the data is called for (I. Maksymyk, C.P. Burgess and D. London, Phys. Rev.
   D 50, 529 (1994)).  In the ``mass bounds'' figure, the S parameter bound  excludes all of
   the  region to the right of the curve in this plot at the 95\% C.L.  $m_{Q}/m_{Z}$ is plotted 
   on the y axis vs $m_{L}/m_{Z}$ on the x axis.  The Q and L designate mass degenerate
   quark and lepton doublets,  respectively. ($ S\leq 0.15$) is the constraint.  Clearly the (chiral)
   lepton doublet must be very light to satisfy this requirement.  The constraint is insensitive 
   to the quark masses because they nearly, but not quite, saturate the bound by themselves for
   large quark mass values, so they have little effect when $m_{Q}/m_{Z} > 1.5$.  

\rotaterunfig{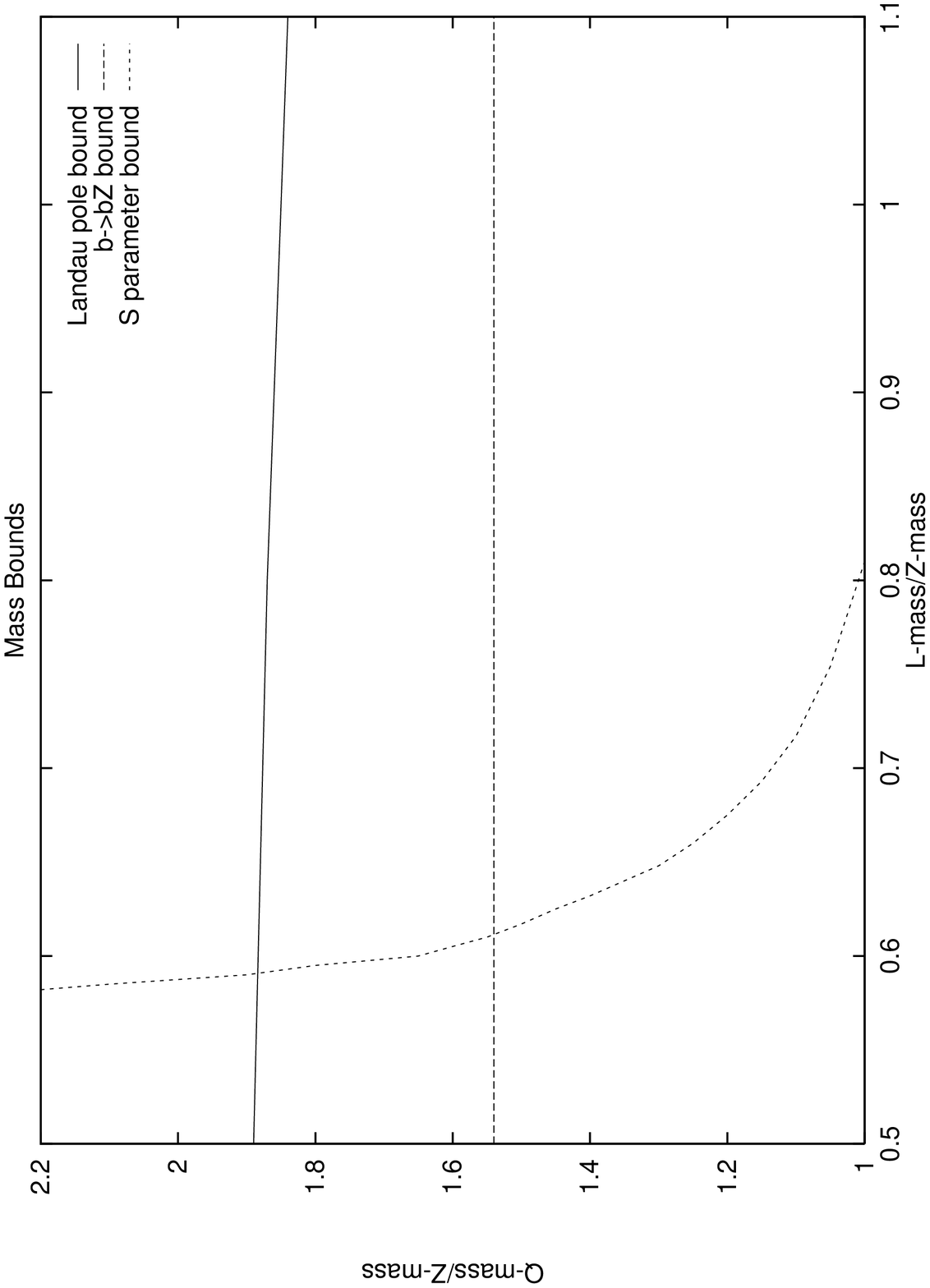}{mckay_massbounds.ps}{
The region below the $b'->b+Z$ curve, to the right of the S-parameter
curve, and above the Landau pole curve are excluded.  This leaves the small
rectangular allowed region at the far left and center of the figure.}

   A completely consistent re-fit to the data with relatively light, fourth family doublets needs
   to be done and is in the works.

\medskip

2)  The direct searches for $ b' \rightarrow b+V(\gamma, Z, g)$ have severely narrowed the options
    below $ M_{b'} = 140 GeV$. The D0 Collaboration ( PRL 78, 3817 (1997)) reported the $ b'\rightarrow
    b+\gamma$ bound of $m_{b'} < 95 GeV$ at 95\% C.L.  95 GeV is the limit for this mode, since the larger
    b+Z mode sets in here. Results of a study of the $b' \rightarrow b+Z$ mode were reported by Joao
    Guimaraes da Costa of CDF at
    the April, '98 AAPT/APS meeting (see also the contribution by Dave Stuart to this workshop).  A limit $m_{b'} > 140 GeV$ at 95\% C.L. is achieved, assuming that
    $\Gamma(b' \rightarrow b+Z) = 100\%$. This bound is shown by the horizontal line in the ``mass bounds''
    figure.  A slip knot in this noose tightening around  b' could be the $b' \rightarrow  b+H$
    mode, which is likely to be as large as or larger than the b+Z mode. This would seriously degrade the $b'$
    mass bounds quoted unless the Higgs final state were included in the tagged sample with favorable efficiencies.
    Failing this, roughly speaking, when the b+Higgs decay mode is accessible to $b'$, the $b+\gamma$
    quoted number of expected  events would be halved and the expected b+Z events would be quartered.
    But in fact, as argued by Guimaraes da Costa (private communication), if the Higgs + b final state
    is present, it will be included in the sample as long as the b + Z branching fraction is not zero
    so that the lepton tagging of the Z is operative.  Moreover the efficiency for tagging the b's from
    Higgs decay is good enough that the factor 4 decrease in events from the competing channel is compensated.
    The bottom line is that if $m_{Z} \sim m_{H}$, the bound $m_{b'} < 140 GeV$ is still likely to obtain.
    The b+Z and b+H decay rates are roughly proportional to $m_{t}^2 - m_{t'}^2$, so their ratio is 
    insensitive to $m_{t'}$, but when $m_{t'} \sim m_{t}$ with $m_{b'} < m_{t'}$, then $ b'\rightarrow c+W$ 
    or $b' \rightarrow c+W^* $ becomes potentially important.  These issues are thoroughly considered in, for
    example, W-S. Hou and R.G. Stuart, Phys. Rev. D 43, 3669 (1991). What is new is the near consensus 
    from SUSY and precision electroweak fits that the lightest Higgs mass is between $m_{Z}$ and $m_{t}$.
    I believe that the simulation
    of competing effects of the various decay channels of $b'$ and the onset of the $t'$ threshold
    and decays, with inclusion of the essential Higgs FCNC channel, is the project that deserves a strong
    effort in Run II.  The reason for stressing this issue is that it ties in directly to the crucial Higgs
    search mission at Run II. A search for the FCNC  decay modes of $b'$ or $t'$ must include the Higgs effects.
    
    CDF has analyzed its data for the presence of long-lived charged 1/3 $(b')$ and 2/3 $(t')$ quarks and has
    reported mass limits of 195 GeV and 220 GeV at 95\% C.L.  This pretty analysis closes the door on the
    possibility that top search signatures missed stable $ (\gamma \tau > 10^{-8} s)$ $b'$ or $t'$ (
    Dave Stuart, contribution to this workshop).

\medskip

3)  Searches for N and L leptons by the LEPII detector groups limit masses
    of stable charged leptons to be above 90 GeV at 95\% C.L.(eg. OPAL Collaboration, Phys.Lett.B,
    433, 195 (1998).  Analysis of  LEPII data also shows that unstable lepton doublets
    can escape detection only if N and L are degenerate at the level of a GeV or so,
    with the neutral one lighter and stable enough to exit the detector.
    (see S. Thomas and J. Wells, hep-ph/9804359 for a technique to detect such states.
     The motivation is that vector-like doublets with radiative splitting on the order of
     several hundreds of MeV are typical in a variety of motivated  extensions to SM,
     but the analysis applies equally well to ``unmotivated'' doublets).

\medskip 

\subsection{Considerations on 
Unification and Perturbation Theory, SUSY Breaking and Other
Theory Prejudices}

As mentioned above, requiring that perturbation theory remain valid as Yukawa and gauge coupling
constants are run to the grand unification scale tightly constrains new family masses.  The
Yukawa couplings blow up (Landau poles) sooner or later as they are run to higher and higher mass
scales. The bigger they start (big masses at the weak scale), the faster they take off.  Because of
the many superpartners, this effect is exaggerated in the MSSM compared to the SM.  Even the latter
suffers rapid blow up when fourth family quark masses exceed 150 GeV or so (a recent study is 
reported in Yu. F. Pirogov and O.V. Zenin, hep-ph/9808396).  In the context of a fourth family 
extension of the MSSM, the constraint is severe indeed.  Setting the top mass in the range 170-
175 GeV and requiring that all couplings run to the GUT scale at a few times $10^{16}$ GeV, one
finds that $m_{b'} < 100 GeV$ and $m_{t'} < 120 GeV$.  Unless the lightest MSSM Higgs
boson is at the Z mass or less, the D0 and CDF searches eliminate a $b'$ with a mass this low. 
In the framework of gravity mediated SUSY breaking and the SUGRA boundary conditions, giving up
perturbative unification means giving up the predictions of the SUSY mass spectrum at low energy,
since the mass parameters are all specified at the GUT scale. This is a punishing penalty to pay!
In gauge mediated SUSY breaking, however, the mass parameters are set at the messenger scale - 
roughly 100 TeV. Within this picture of SUSY breaking, one may require perturbative running
of the coupling constants 
${only
to the messenger scale}$, still retain predictive power over the SUSY mass spectrum, and buy a
wider range of theoretically acceptable fourth family mass values.  The result of a preliminary
one-loop RG study  is shown in the ``Landau Pole Bound'' curve on
the ``mass bounds'' figure for the case $tan\beta = 1.3$, $M_{susy} = M_{Z}$ and $\alpha_{s} = 0.119$.  Values
above the curve are excluded.  The constraints are still surprisingly tight!  Only values of
$m_{q'}$ up to about 1.9 $m_{Z}$, or just below $m_{t}$, are allowed with $m_{L}$ in the range 0.5 - 1.1 $m_{Z}$, as shown in the figure. 
Whether a viable SUSY spectrum and retention of spontaneous breaking of EW symmetry breaking can
be achieved in a four family GMSB model upon running mass relations back down from the messenger
scale remains to be seen.  The ``mass bounds'' figure allows a small rectangular region at the
lowest, degenerate values of the lepton doublet, L, after application of the mixture of theoretical
and experimental constraints.  The lower bound of 140 GeV from the $ b'\rightarrow b+Z$ search 
should be set with light ($m_{h}\sim 100-150 GeV$) Higgs effects included, which brings us
back to our ``bottom line''.  

In connection with the above discussion, it should be remarked that a clever re-identification of the
$t'$ as the top signal and a proposed ``hidden'' top with $m_{t}\sim m_{W}$ and a top decay to stop 
plus LSP in the fourth family extension of the MSSM is probably safely eliminated by the failure to
find $t\overline{t}$ production at LEP II and the necessary light stop to go with it. (M. Carena, H.
Haber and C. Wagner, hep-ph/9512446 and Nucl. Phys. B433, 195; for a stop limit of 70-80 GeV, see
The ALEPH Collaboration, Phys. Lett. B434, 189(1998)).

This summary includes contributions through conversations, email, published papers and
unpublished talks by the following people (but errors and distortions in this summary are strictly
my doing):Jack Gunion, Wei-Shu Hou, Robin Stuart, Dave Gerdes, Joao Guimar$\overline{a}$es da Costa, Bhaskar Dutta,
Dave Stuart, Kara Hoffman, Mel Shocket, Tom Diehl, Taka Yasuda,  Kaori Maeshima, Howie Haber,
P.Q. Hung, Paul Frampton, Kyungsik Kang and Sin Kyu Kang.





\section{Light Gluino Predictions for Jet Cross Sections \\in Tevatron
Run II}
\bigskip
\noindent
\centerline{\large\it L. Clavelli}
\medskip
\label{section:Clavelli}

The CDF collaboration at Fermilab has published \cite{Abe1} a study of
the jet inclusive transverse energy cross section in $p {\overline p}$
cross sections at $1.8 TeV$ which suggest the possibility of anomalous
behavior in both the low and high transverse energy regions. D0 has
not published results in the low transverse energy region but has
presented data at high transverse energy which appear to be consistent
with either the CDF result or the standard model. The apparent anomaly
at high $E_T$ seen by CDF could, therefore, be a statistical
fluctuation. It has also been suggested
\cite{Lai} that these results are compatible with the standard model if
the gluon distribution at high x is appreciably higher than expected on
the basis of previous fits. On the other hand, the anomalous behavior
observed by CDF in both the low and high $E_T$ regions is also
consistent with that expected if the gluino of supersymmetry is light
(below $10 GeV$ in mass is sufficient) \cite{CT}. Although all direct
searches for a light gluino have turned up negative, many indirect
indications of such a light color octet parton have been noted. A
partial list is contained in the references of \cite{CT}.

The measured inclusive cross section at center of mass energy $\sqrt
s$ to produce a jet of transverse energy $E_T$ averaged over a certain
rapidity interval is theoretically expected to have the form
\be
 d\sigma/dE_T  = \alpha_s(\mu)^2 s^{-3/2} F(X_T,\frac{\Lambda}{\sqrt s},
     \frac{m}{\sqrt s})
    + {\cal{O}}(\alpha_s^3)
\ee
Here $\mu$ is the scale parameter, $X_T=2E_T/\sqrt s,\quad \Lambda$ is
the QCD dimensional transmutation parameter, and m represents any of
the masses of the strongly interacting particles in the theory. Taken
to all orders the cross section is independent of $\mu$ but at finite
order the theoretical result depends on $\mu$ which must therefore be
treated as a parameter of the theory. The CDF best fits correspond to
$\mu=E_T/2$. At high energy the scaling function F depends only on
$X_T$ . The CDF data for this cross section compared to the
next-to-leading order (NLO) QCD predictions are below unity at low
$E_T$ and rise dramatically above unity at high $E_T$. In the
Supersymmetry (SUSY) treatment of \cite{CT2} this behavior was
attributed to three phenomena.
\begin{itemize}
\item  With a light gluino the strong coupling constant runs more slowly
being higher than the standard model at high $\mu$ and lower at low
$\mu$.
\item The production of gluino pairs increases F by a roughly
uniform factor of 1.06 for all $E_T$
\item A squark, if present, will cause a bump in the cross section at
about ${m_{\tilde Q}}/2$.
\end{itemize}

The fit of \cite{CT2} used the CDF suggested value of $\mu$ and a
value of $\Lambda$ corresponding to $\alpha_s(M_Z)=0.113$ and a squark
mass of about $106 GeV$. The theoretical ratio of the SUSY prediction
relative to the standard model prediction is relatively insensitive to
higher order corrections since both will have roughly equal higher
order enhancements. In this work the CTEQ3 parton distribution
functions (pdf's) were used. In a later study \cite{Bhatti}, CDF
considered the scaled ratio of the inclusive jet $E_T$ cross sections
at $630 GeV$ and $1.8 TeV$.
\be
  r(X_T) = \frac{s^{3/2} d\sigma/dE_T \quad(\sqrt{s} = 630 GeV)}{
              s^{3/2} d\sigma/dE_T \quad(\sqrt{s} = 1800 GeV)}
\ee

Since at both energies, $\sqrt{s}$ is much greater than the QCD scale
parameter $\Lambda$ and all the quark masses of the standard model
(except the top quark which contributes negligibly at these energies),
the standard model prediction modulo residual corrections from higher
order and from scaling violation in the pdf's is just
\be
        r(X_T) = \frac{\alpha_s^2(\lambda X_T \cdot 0.630GeV/2)}
              {\alpha_s^2(\lambda X_T \cdot 1.8 TeV/2)}
\ee

We have assumed here that the appropriate choice of $\mu$ is $\lambda
E_T$ with $\lambda=1/2$ being the result of the CDF best fit to the
$1.8 TeV$ data. The full standard model prediction with corrections
incorporated seriously overestimates the CDF data. In addition there
is a possible structure in r that, if real, might suggest the
existence of a strongly interacting particle in the $100 GeV$ region
with a production cross section many times larger than that of top. As
always, there is the possibility that the anomaly is due to systematic
errors although it would be surprising if such errors induced
structure in $E_T$. In fact, the D0 experiment does not confirm the
existence \cite{D0} of structure in r suggesting, therefore, an
explanation in terms of systematic errors. Although the systematic
errors could easily affect the normalization of the r parameter, it
would be surprising if they affect the point to point errors. These
systematic errors derive primarily from the lower energy (630 GeV)
data and hence the existence or non-existence of structure should be
definitively resolved by comparing the ratio of the $2 TeV$ data which
will be available beginning in the year 2000 with the $1.8 TeV$ data.
Although the energy step is small, the greatly increased luminosity in
run II coupled with the small systematic errors in the $1.8 TeV$ data
should guarantee sufficient sensitivity to settle the question.

The features observed by CDF in the scaling ratio are those expected
in the light gluino scenario \cite{CT2}. The slower fall-off of
$\alpha_s$ predicts that the r parameter should be generally lower
than the standard model expectations in agreement with the data. In
addition a squark in the $100 GeV$ mass range would provide a bump in
each cross section at roughly fixed $E_T={m_{\tilde Q}}/2$. This would
lead to a dip-bump structure separated by a factor of 1.8/0.63 in
$X_T$ in qualitative agreement with the CDF data. If the bump had
occurred at lower $X_T$ than the dip there would have been no
possibility of a fit in any model where the structure was attributed
to a new particle. Reference \cite{CT2} provided two fits to the CDF
data. The first used the CTEQ3 pdf's and the scale choice $\mu=E_T/2$
with a squark mass of $130 GeV$. In the CTEQ3 pdf's there are, of
course, no initial state gluinos so the cross section bump derives
from the reaction
\be
                   q g \rightarrow q {\tilde g} {\tilde g}
\label{qg}
\ee
with an intermediate squark in the $q {\tilde g}$ channel. The
dynamics are such that the initial state gluon splits into two
dominantly collinear gluinos one of which interacts with the initial
state quark to produce the intermediate squark. Other non-resonant
light gluino contributions to the cross section come from the parton
level processes
\be
      q {\overline q} \rightarrow {\tilde g} {\tilde g}\\
      g g \rightarrow {\tilde g} {\tilde g}
\ee

If the gluino is light it should have a pronounced presence in the
proton dynamically generated from the gluon splitting discussed above.
Two groups \cite{BB,RV} have analyzed deep inelastic scattering
allowing for a light gluino and presented fits to the gluino pdf as
well as modifications of the other pdf's due to the gluino presence.
In ref. \cite{CT2} we compared the scaling violation using the
R\"uckl-Vogt pdf's with that using the CTEQ3 set. With intrinsic
gluinos there are extra contributions to the jet inclusive cross
sections from the processes
\be
       g {\tilde g} \rightarrow g {\tilde g} \\
       q {\tilde g} \rightarrow q {\tilde g} \\
       {\tilde g} {\tilde g} \rightarrow {\tilde g}{\tilde g}\\
       {\tilde g} {\tilde g} \rightarrow g g\\
       {\tilde g} {\tilde g} \rightarrow q {\overline q}
\ee
The second process replaces the higher order reaction of Eq.\
(\ref{qg}) and provides a direct channel pole at the mass of the
squark leading to a peak in the transverse energy cross sections. We
treat the squark as a resonance in the quark-gluino channel. Each of
these reactions of course is subject to higher order corrections but
these tend to cancel in the scaling ratio and in the ratio of the SUSY
transverse energy cross section to that of the standard model. In this
second fit the scale $\mu$ was chosen to be the parton-parton CM
energy.

The purpose of the current work is to return to the inclusive
jet transverse energy cross section and seek a combined fit to this
plus the scaling curve allowing for intrinsic gluinos in the proton.
Fitting both the scaling curve $r(X_T)$ and the $1.8 TeV$ cross
section is equivalent to fitting the transverse energy cross section
at both $1.8 TeV$ and $0.63 TeV$. Using the parameters of this
combined best fit we then present the predictions for the $E_T$ cross
section at $2 TeV$ CM energy of run II and the scaling curve for $1.8
TeV / 2 TeV$. The primary parameters of the combined fit are the scale
$\mu$, the QCD $\Lambda$ parameter or equivalently $\alpha_s(M_Z)$,
and the squark mass $m_{\tilde Q}$. We find the optimal values
\be
     \mu = 0.6 E_T\\
     \alpha_s(M_Z) = 0.116\\
     m_{\tilde Q} = 133 GeV
\ee
In the fit we estimate NLO corrections by the K factor $1 +10
\alpha_s(\mu) /\pi$ and we simulate resolution smearing by increasing the
width of the squark by a factor of 2 from its SUSY QCD prediction $2
\alpha_s m_{\tilde Q}/3$. In addition it is known that the systematic
errors in the $630 GeV$ data form a fairly broad band \cite{Geer}. We
therefore allow the scaling data to float by a uniform factor near
unity. The results are presented in figures 1-4.

Figure 1 shows the
fit to the $1.8 TeV$ jet inclusive $E_T$ cross section averaged over
the CDF rapidity range $0.1 < |\eta| < 0.7$. In order to compare with
the data of ref. \cite{Bhatti}, the light gluino prediction is
plotted relative to the QCD prediction given to us in a private
communication by the author of that reference. At high $E_T$ the fit
goes through the lower range of the CDF errors which suggests it is
also consistent with the D0 data. The fit qualitatively reproduces the
dip at low $E_T$ and shows a peak at low $E_T$ due to the $133 GeV$
squark. Figure 2 shows the scaling function $r(X_T)$ as given in the
light gluino scenario with a $133 GeV$ squark and as given by the
standard model. The data has been moved up by a uniform factor of 1.2
which is consistent with the effect of systematic errors in the $630
GeV$ data. The height and width of the dip-bump structure is in
qualitative agreement with the expectations of the light gluino plus
$133 GeV$ squark model. One might expect that a full simulation
including hadronization and detector acceptance could somewhat shift
this mass.

If a squark exists at $133 GeV$ it should be apparent in
the $e^+e^-$ annihilation cross section through the
quark-squark-gluino final state \cite{CCFHPY}.
The L3 data \cite{L3} shows what is possibly an upward
statistical fluctuation in the hadronic cross section in $e^+ e^-$
annihilation in the $130 GeV$ region. Since the gluino decays are
expected to leave very little missing energy, the quark-squark-gluino
final state might also explain an apparent surplus in the visible
energy cross section at high $E_{vis}$ \cite{L32}. In addition, a SUSY
symmetry breaking scale of $133 GeV$ would, in the light gluino
scenario, predict stop quarks in the region just above the top and
could explain some anomalies in the top quark events
\cite{CG} and lead to an enhancement in the deep inelastic cross section
at high $Q^2$ and high hadronic mass \cite{CC}. If there is indeed a
light gluino and a squark in the $100 \sim 135 GeV$ region, a dip-bump
structure should also be found at LEP II in the scaling ratio of the
inclusive dijet cross section in $e^+ e^-$ annihilation.
\be
        r(M^2/s) = \frac{ s^{2} d\sigma/dM^2 \quad (\sqrt{s}=E_1)}
                    { s^{2} d\sigma/dM^2 \quad (\sqrt{s}=E_2)}
\ee
where both $E_1$ and $E_2$ are above the squark mass. Since the squark
decays in the present model into quark plus gluino, the excess should
be in the four-jet sample but should not appear in the pair production
of two high mass states.

In figure 3, we show the predictions for the
jet inclusive $E_T$ cross section in $p {\overline p}$ collisions at
the energy $2 TeV$ relative to the standard model expectations. The
curve shows a pronounced peak at ${m_{\tilde Q}}/2$ and is generally
$5$ to $10 \%$ below unity due to the slower running of $\alpha_s$ in
the light gluino case and to the scaling violations in the parton
distribution functions. In Figure 4 the scaling ratio
\be
  r(X_T) = \frac{s^{3/2} d\sigma/dE_T \quad(\sqrt{s} = 1.8 TeV)}{
              s^{3/2} d\sigma/dE_T \quad(\sqrt{s} = 2 TeV)}
\ee
is plotted for the case of light gluino plus $133 GeV$ squark and for
the case of light gluino but no squark present (non-resonant solid
line). The dash-dotted curve gives the prediction of the standard
model.

Although we have not attempted to estimate hadronization
corrections nor resolution smearing (apart from doubling the squark
width), we expect that run II will be sensitive to the predicted peaks
if they exist and will therefore either discover or rule out a squark
in the $100 GeV$ mass region in conjunction with a light gluino. With
additional information on dijet mass and angular distributions
\cite{TC,T,DHR}, the Run II measurements are sensitive to a light
gluino with a squark up to $1 TeV$. Since most of the value of SUSY
would be lost with squarks so high in mass, Run II should definitively
settle the question as to whether the light gluino indications
including those referenced in \cite{CT} are the first signs of SUSY or
merely an amazing string of coincidences attributable to systematic
errors.

\runfig{Fig. 1}{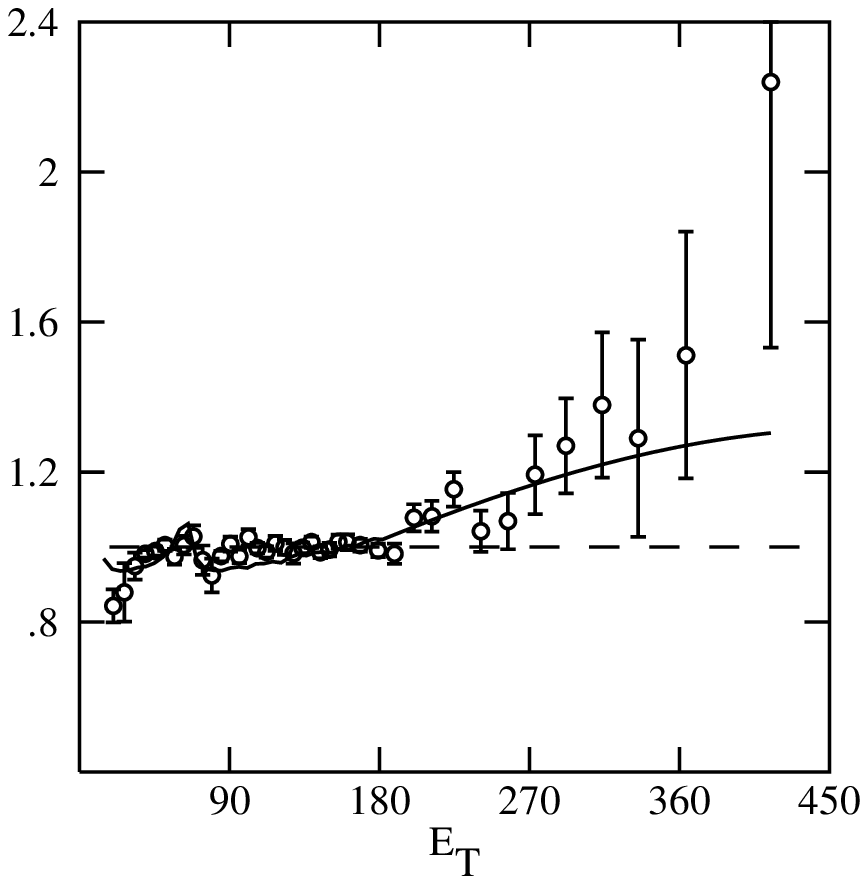}{
Ratio of the inclusive jet transverse energy cross section
with a light gluino and a $133 GeV$ squark to that of the standard
model. CDF data at $1.8 TeV$ is superimposed.}

\runfig{Fig. 2}{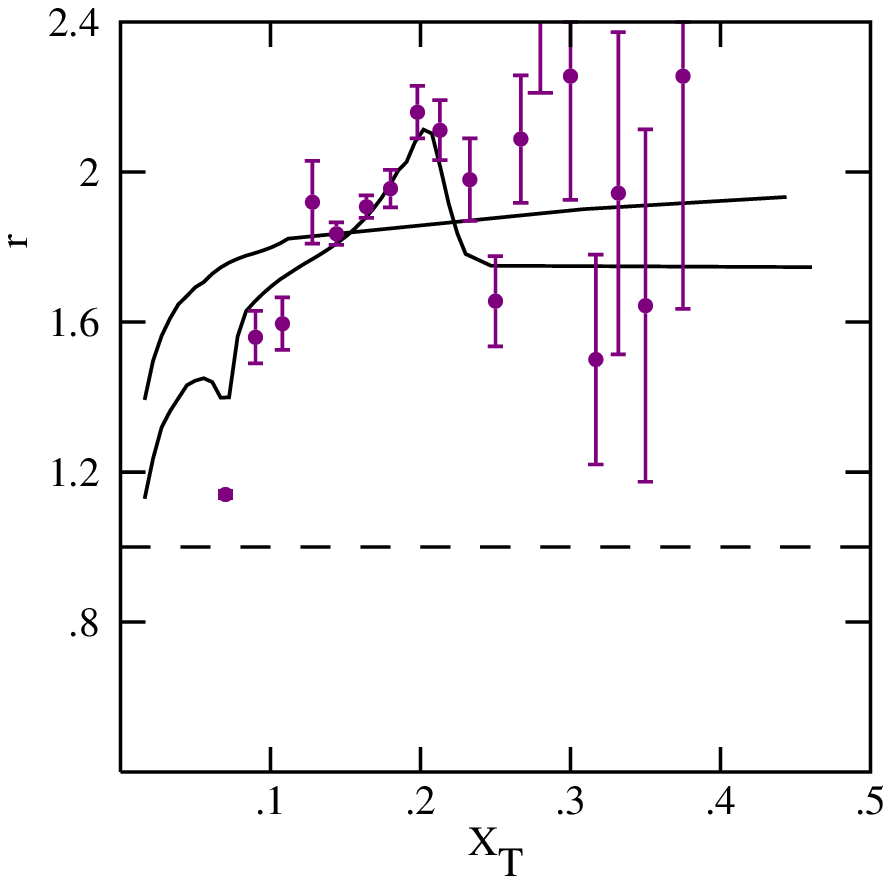}{
CDF data for the scaling ratio of the inclusive jet transverse
energy cross section at $630 GeV$ relative to $1.8 TeV$ compared to
the fit with a light gluino plus $133 GeV$ squark and to the standard
model prediction (structureless curve). The data has been moved up by
$20 \%$ consistent with the systematic errors in the $630 GeV$ data.}

\runfig{Fig. 3}{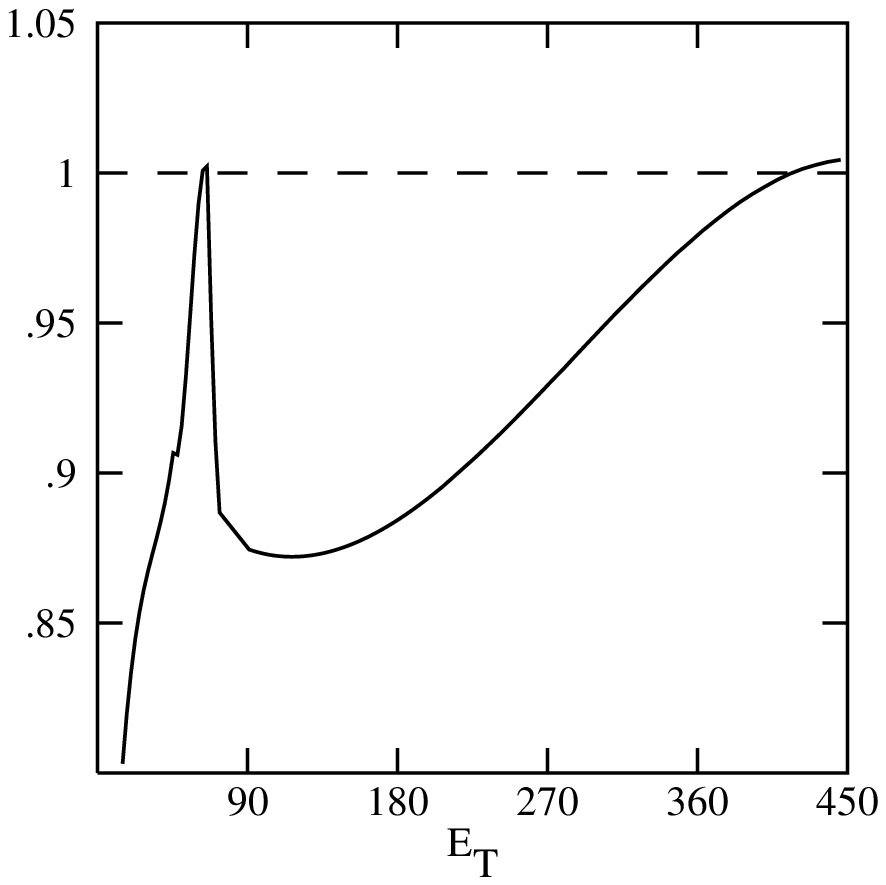}{
Predicted ratio of the inclusive jet transverse energy cross
section with a light gluino and a $133 GeV$ squark to that of the
standard model for $2 TeV p {\overline p}$ collisions.}

\runfig{Fig. 4}{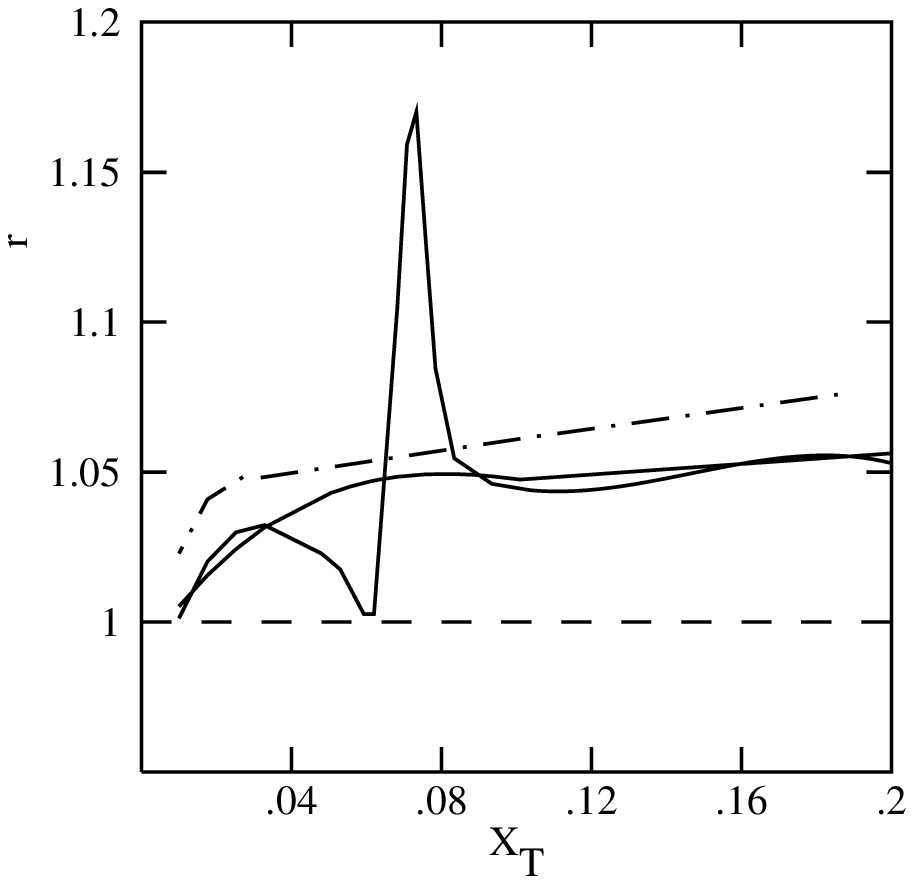}{
Predicted scaling ratio of the inclusive jet transverse
energy cross section at $1.8 TeV$ relative to $2 TeV$ for a) light
gluino only (no squark) and b) light gluino plus $133 GeV$ squark. The
dash-dotted curve shows the standard model prediction.}

\def\topfraction{1.0}
\def\bottomfraction{1.0}
\def\textfraction{0.0}
\def\floatpagefraction{1.0}
\def\met{\mbox{${\hbox{$E$\kern-0.6em\lower-.1ex\hbox{/}}}_T$}} 
\def\MET{\met}
\def\etal{{\it et al.\/}}

\section{Detection of Long-Lived Particles in Run II with D\O}
\noindent
\centerline{\large\it D.~Cutts and G.~Landsberg}
\medskip
\label{section:Cutts}

\subsection{Physics motivation}

Long-lived neutral or charged massive particles appear in many extensions 
of the MSSM. There are two main scenarios which 
result in a long lifetime of some of the SUSY partners. 
In the models that predict a degenerate mass spectrum of SUSY particles, 
light SUSY partners might be stable or long-lived, 
since the phase space would not allow for strong decays modes. 
Some models (see sections

It is particularly interesting if the NLSP is long-lived, 
since the observation of such an NLSP might not be doable 
via standard decay signatures expected for short-lived SUSY particles. 
As an example, many MSSM extensions predict degenerate mass spectrum 
of the neutralinos, in which case the following radiative decay 
could be a dominant decay mode of the second-lightest neutralino:
\begin{equation}
\tilde{\chi^0_2} \to \tilde\chi^0_1 \gamma.
\label{eq:delayed-neutralino}
\end{equation}
Being electromagnetic, this decay mode is suppressed, and for fine splitting 
between the masses of the two lightest neutralinos, 
the decay constant could be small enough to result in a 
long-lived $\tilde\chi^0_2$.

Among other models which allow for a long-lived NLSP, 
are the GMSB scenarios in which a neutral NLSP (usually a neutralino) 
radiatively decays into a gravitino LSP:
\begin{equation}
        \tilde\chi^0_1 \to \gamma\tilde G, \label{eq:delayed-GMSB}
\end{equation}
or a charged NLSP (usually the right-handed stau, $\tilde\tau_R$) decays 
into a $\tau$ and a gravitino NLSP:
\begin{equation}
        \tilde\tau_R \to \tau\tilde G \label{eq:delayed-GMSB-tau}
\end{equation}
In the GMSB scenarios the gravitino mass is given by~(see, e.g., \cite{Feng}):
$$
        M_{\tilde G} = \sqrt{\frac{8\pi}{3}}\frac{F}{M_P},
$$
where $F$ is the vacuum expectation value of the dynamical 
supersymmetry breaking, and $M_P$ is Planck mass. 
Since most of the GMSB models predict $F \sim 10^[14]$~GeV$^2$, 
and $M_P \sim 10^{19}$~GeV, the gravitino is expected to be very light. 
The fact that the gravitino interacts with matter only weakly, 
could make the decays (\ref{eq:delayed-GMSB}) and 
(\ref{eq:delayed-GMSB-tau}) very slow. 
For example, the $c\tau$ for the decay (\ref{eq:delayed-GMSB-tau}) 
is given by~\cite{Feng}:
$$
        c\tau \approx 10~\mbox{km} \times \langle\beta\gamma\rangle \times 
\left[\frac{\sqrt{F}}{10^7~\mbox{GeV}}\right]^4 \times 
\left[\frac{100~\mbox{GeV}}{m_{\tilde\tau_R}}\right]^5,
$$
and could be very large.

Other models (see, e.g. Section~\ref{section:Ellwanger}) 
consider cascade decays of the SUSY particles that can also result 
in displaced vertices. Massive charged long-lived SUSY particles 
are expected in SUSY models with chargino LSP~\cite{Feng1} 
where the lightest neutralino and chargino are nearly degenerate in mass. 
Similar degeneracy automatically occurs in superstring-inspired 
models of SUSY breaking~\cite{Randall}. 
Weakly interacting bound states of gluino LSP also act 
as long-lived massive particles (see, e.g. Section~\ref{section:Baer}).

Generally speaking, in different SUSY models, 
the $c\tau$ of the charged or neutral long-lived 
particles can be from subatomic distances to many kilometers.

\subsection{Detection of the Delayed Decays}
\label{sec:delayed}

Detection and identification of long-lived particles depend on the 
decay path and the charge of this particle. A typical collider detector, e.g. the upgraded D\O\ apparatus~\cite{D0Upgrade}, has 
an inner Silicon Microstrip Tracker (SMT), capable of identifying secondary vertices from long-lived particle decays. The silicon 
detector is surrounded by a less precise Central Fiber Tracker (CFT), which is enclosed in 
the calorimeter. The calorimeter is surrounded by the muon system. Typical outer radii of the vertex detector, 
central tracking detector, calorimeter, and the muon system are $\sim 10$, 100, 200, and 1000~cm, respectively.

In the case of the charged long-lived particles, one can identify 
the secondary decay vertex in the silicon vertex detector 
for $\gamma c\tau$ between $\sim 0.1$~cm and $\sim 10$~cm. Since we expect 
SUSY particles to be heavy, in what follows we will assume $\gamma \approx 
1$, i.e. $\gamma c \tau \approx c\tau$. Both the CDF 
and D\O\ experiments will be capable of identifying such 
secondary vertices and possibly trigger on them. For 10~cm $\lesssim 
c\tau \lesssim 100$~cm, the charged long-lived particle will predominantly decay inside the outer 
tracking volume. The resulting characteristic signature is a kink in the 
outer tracker and large $dE/dx$ in the silicon vertex tracker and the 
inner layers of the outer tracker, typical for a slow-moving charged particle. While triggering 
on the kinks in the outer tracker won't be possible 
in Run II, there is a good chance that such 
kinks can be found offline, if the event is accepted by one 
of the standard triggers. Since a massive slow moving particle 
still has a significant momentum, one would likely trigger on such 
events using a single high-$p_T$ track trigger, 
or a designated $dE/dx$ trigger. For 
100~cm $\lesssim c\tau \lesssim 200$~cm, the charged long-lived particle will decay inside the 
calorimeter, giving a jet-like energy deposition inside it. 
Identification of these particles will therefore rely on 
the fact that such a jet has only one track pointing to 
it (similar to one-prong $\tau$-decays); moreover, this track will have 
high $dE/dx$. In the case of the D\O\ 
detector, an additional $dE/dx$ measurement in the preshower detector can 
be used to aid the identification of such particles both offline 
and at the trigger level. A single high-$p_T$ 
track trigger, a designated $dE/dx$ trigger, or a 
$\tau$-trigger  could be used to trigger on such 
events. Finally, for $c\tau \gtrsim 200$~cm, the long-lived particle will 
look stable from the point of view of the detector, and 
its identification will rely on a high $dE/dx$ track 
in the silicon vertex detector, the outer tracker, the 
calorimeter, and the muon system (if the $c\tau$ 
exceeds it outer radius). An additional time-of-flight (TOF) 
information from a designated TOF system (CDF) or muon system 
scintillators (D\O) could be used to trigger on such events and 
identify them offline. To summarize, both the CDF and 
D\O\ detectors will have very good capabilities for searches for 
charged long-lived particles with lifetimes from a few tens of picoseconds to infinity. 
Identification of the slow-moving particles using $dE/dx$ 
techniques in the case of 
the D\O\ detector is discussed in more detail in 
Section~\ref{sec:HIT-D0}.

The situation is much more complicated for a neutral weakly 
interacting particle. First of all, if such a 
particle decays outside of the calorimeter, it can not be 
detected, and will look like a missing $E_T$ in 
the event, i.e. like the LSP. It's unlikely that one could 
identify the presence of a long-lived NLSP in such events, but 
it still might be possible to identify these events as 
the non-SM ones. If events with such a signature are found in 
Run II, one could conceivably install a wall of scintillating detectors far away from 
the main detector volume and try to 
look for a photon from the radiative decay in this additional scintillator~\cite{Gunion}. 
For the purpose of this report, however, 
we will focus only on the case of $c\tau 
< 2$~m, which roughly coincides with the outer radius of 
the D\O\ calorimeter. The detection technique for such decays relies heavily on the 
fine longitudinal and transverse segmentation of the preshower detector and 
the calorimeter, which is an essential 
and unique feature of the D\O\ detector.

The signature for 
the radiative decay of a long-lived particle (e.g., (\ref{eq:delayed-GMSB}) 
or (\ref{eq:delayed-neutralino})), is a production of 
a photon with a non-zero impact parameter. If 
there was a way to identify the point 
of photon origin, one could single 
out such a delayed radiative decay corresponding to a very distinct 
and low background topology. As is shown in Section~\ref{sec:pointing}, 
the D\O\ calorimeter information, combined with the preshower 
information, can be used to achieve a very precise 
determination of the photon impact parameter. This technique 
would allow to identify long-lived neutral particles with 10~cm 
$\lesssim c\tau \lesssim 100$~cm, in the case of 
the upgraded D\O\ detector. The detectable $c\tau$ range 
can be further extended by a factor of two 
by looking for the photons from the radiative decay in the 
D\O\ hadronic calorimeter. The signature for such photons would look 
like a ``hot cell,'' i.e. an isolated energy deposit in one 
or two hadronic calorimeter cells. The reason that the EM 
energy deposition in the calorimeter is so isolated is 
the fact that each hadronic calorimeter cell contains many radiation lengths 
and completely absorbs an EM shower. 
The main background for this signature is production of high-$p_T$ $K_S^0$-mesons 
which decay into a pair of $\pi^0$'s corresponding 
to the 4$\gamma$ final state. Since the $K_S^0$ 
is significantly boosted, these four photons are highly 
collimated and will be identified as a single 
EM shower in the calorimeter. For a typical $K^0_S$ 
momentum of 25 GeV, which corresponds to $\sim 20$~GeV 
$E_T$ of the resulting $4\gamma$ system, the $\gamma c\tau$ is 
about 130~cm, i.e. most of the $K^0_S \to 4\gamma$ 
decays will occur in the hadronic calorimeter. This 
background, however, can be well predicted and also has a well 
defined $r$-dependence, where $r$ is the radial distance of 
the ``hot cell'' from the detector center. We 
therefore expect this background to be under control in 
Run II. Thus, the upgraded D\O\ detector will have unique 
capabilities for searching for neutral long-lived particles with 
10~cm $\lesssim c\tau \lesssim 2$~m.

Both the CDF and D\O\ could also explore the 0.1~cm $\lesssim 
c\tau \lesssim 10$~cm 
range by looking for a conversion of the photon from 
a radiative decay in the silicon vertex detector. If 
the photon converts, one can determine its direction and the 
impact parameter fairly precisely by looking at the tracks from 
the $e^+ e^-$-pair. One, however, would pay a significant price for being able 
to explore this region of $c\tau$, since the probability of photon 
conversion in the silicon vertex detector is only a 
few per cent. CDF could in principle use the conversion 
technique to explore higher values of the $c\tau$ by looking for 
conversions in the outer tracking chamber. However, due to the 
low conversion probability the sensitivity of the CDF detector in this 
region is by far lower than that of the D\O\ detector.

Apart from being crucial for the study of physics with delayed radiative 
decays, the D\O\ detector ability for photon pointing is very attractive from other 
points of view. In the high luminosity collider environment the average number 
of interactions per crossing exceeds one. (It can be 
as high as five, for the 396~ns bunch spacing expected at the 
beginning of Run II.) Therefore, each event will generally 
have several primary vertices with only one being from the 
high-$p_T$ interaction of physics interest, and the others being 
due to minimum bias $p\bar p$ collisions. The presence of 
multiple vertices creates a problem in choosing the 
right one for determination of the transverse energies of 
the objects. Photon pointing can solve this problem. It is especially 
important at the trigger level when the information about all the 
objects in the events is not generally available, and 
therefore high-$p_T$ objects which produce tracks in the tracking 
chambers can not always be used to pinpoint the 
hard scattering vertex. In some cases, for example the 
$\gamma + \MET$ final state, there are no objects with 
tracks at all, so there is no way to 
determine what vertex the photon originated from without utilizing 
the calorimeter-based pointing. Not knowing which vertex is the one from high-$p_T$ collision 
results in the object $E_T$ mismeasurement, which is especially 
problematic for missing transverse energy calculations. Indeed, \MET\ calculations 
rely heavily on the vertex position and picking the wrong 
one may result in significant missing transverse energy 
calculated in the event which in fact does 
not have any physics sources of real \MET. Not does only 
this affect physics analyses for topologies with \MET, especially the $\gamma + 
\MET$ one, but it also results in a 
worsening of the \MET\ resolution, hence a slower 
turn-on of the \MET-triggers and ergo higher trigger 
rates. It is, therefore, very important to have 
a way of telling the high-$p_T$ primary vertex, and 
photon pointing is the only way of doing this in the 
$\gamma + \MET$ case.

Subsequent sections contain 
technical detail on the high-$dE/dx$ and delayed photon identification.

\subsection{Photon Pointing at D\O} 
\label{sec:pointing}

The importance of photon pointing was appreciated in some Run I analyses, 
particularly the studies of $Z(\nu\nu)\gamma$~\cite{Zg} and $\gamma\gamma$~\cite{monopoles}. We 
have utilized the fine longitudinal segmentation of 
the D\O\ EM calorimeter (which has 
four longitudinal layers) by calculating the c.o.g. 
of the shower in all four layers independently and then fitting the 
four spatial points to a straight line in order to 
determine the impact parameter and the $z$-position of 
the photon point-of-origin~\cite{EMVTX} (see Fig.~\ref{fig:emvtx}). The algorithm used for 
the c.o.g. finding is based on a logarithmic weighting 
of the energy deposition in the EM calorimeter 
cells which belong to the EM shower. The 
spatial resolution of the c.o.g. finding algorithm in 
four calorimeter layers averaged over central (CC) and forward 
(EC) rapidity range, as well as the 
geometrical parameters of the calorimeter layers are given 
in Table~\ref{table:EMVTX}. This study resulted in an algorithm, {\tt 
EMVTX}~\cite{EMVTX}, that has been used in several D\O\ 
analyses involving photons~\cite{Zg,monopoles}.

\begin{figure}[thb]
\vspace*{0.1in}
\centerline{\protect\psfig{figure=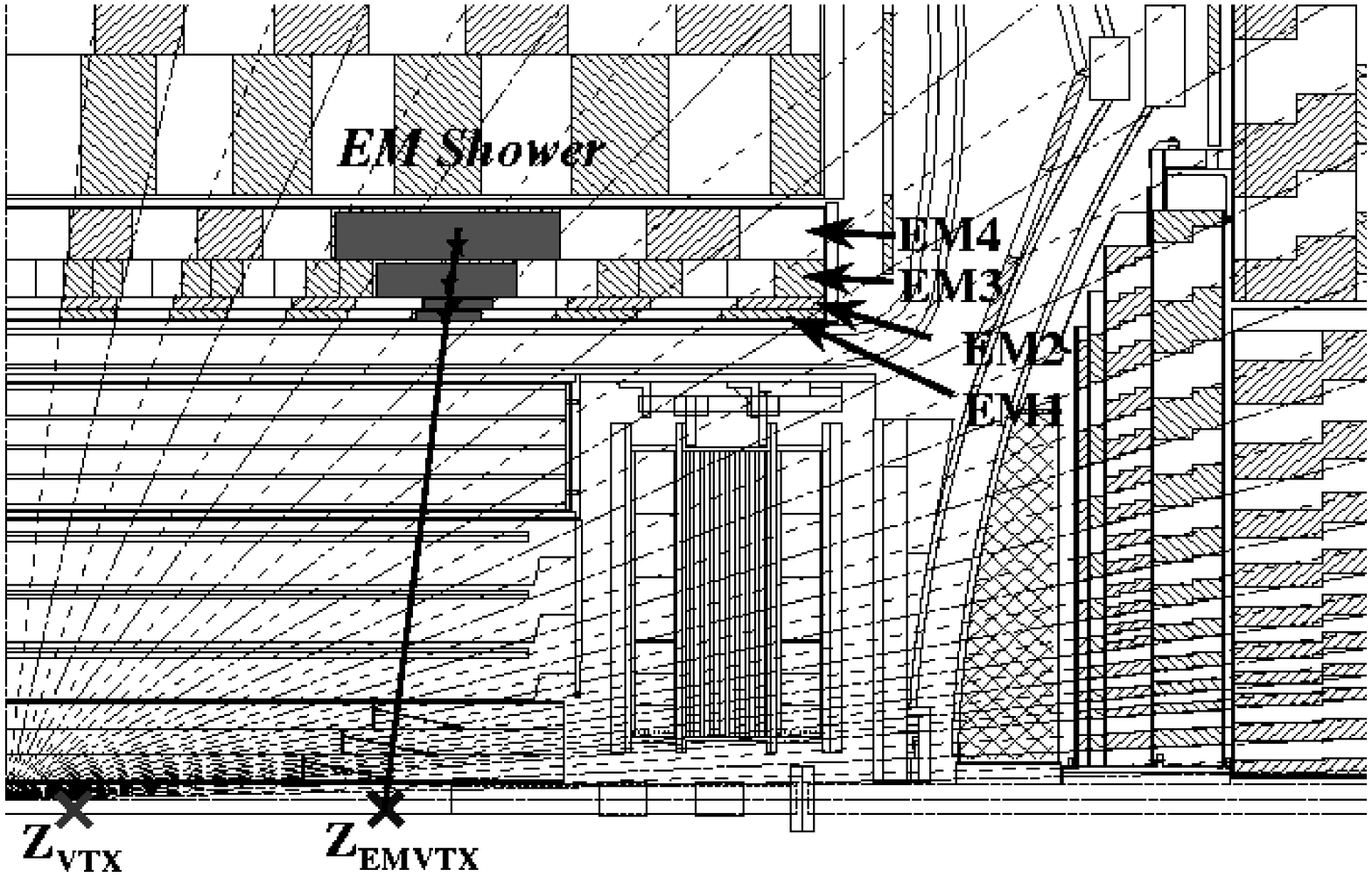,width=\textwidth}}
\caption{Side-view of the D\O\ calorimeter illustrating 
the application of the {\tt EMVTX} algorithm.}
\label{fig:emvtx}
\end{figure}

\begin{table}[htb!]
\begin{center}
\caption{Geometry and average resolutions in the preshower detectors and the 
EM calorimeter layers.}
\begin{tabular}{||l||l|l|l|l|l||}
\hline
Quantity                & Preshower & EM1 & EM2 & EM3 & EM4 \\
\hline\hline
\multicolumn{6}{||c||}{Central Region}\\
\hline
$\langle R \rangle$ 
c.o.g.          & 73.0~cm       & 85.5~cm       & 87.4~cm       & 91.8~cm       
& 99.6~cm       \\
$\sigma_z$              & 2.5~mm        & 20~mm & 20~mm & 7.0~mm        
& 15~mm \\
$\sigma_{r\phi}$        & 1.5~mm        & 17~mm & 17~mm & 3.5~mm        
& 7.5~mm        \\
\hline
Quantity                & CPS & EM1 & EM2 & EM3 & EM4 \\
\hline
\multicolumn{6}{||c||}{Forward Region}\\
\hline
$\langle |Z| \rangle$ 
c.o.g.          & 142.0~cm      & 171.7~cm      & 174.2~cm      
& 179.2~cm      & 189.7~cm  \\
$\sigma_r$              & 1.5~mm        & 8.0~mm        & 8.0~mm        
& 1.5~mm        & 3.5~mm        \\
$\sigma_{r\phi}$        & 2.5~mm        & 7.0~mm        & 7.0~mm        
& 1.0~mm        & 2.8~mm        \\
\end{tabular}
\end{center}
\label{table:EMVTX}
\end{table}

In order to study the improvement of photon pointing made 
possible by the utilization of the fine spatial resolution 
of the preshower detector, we have written a toy Monte Carlo (MC) simulation 
package which takes into account detector geometry, position error in 
calorimeter and preshower, as well as the primary 
vertex distribution. First, we compare the resolution obtained from the 
toy MC with the actual distributions obtained from $W 
\to e\nu$ events collected in Run I. For 
electrons from $W$-events it is possible to determine the point of 
origin by using track information, which is quite precise. 
The difference between the $z$ position of the vertex obtained from 
tracking and from electron pointing as well as the signed 
impact parameter for the electron obtained by pointing are shown in Fig.~\ref{fig:emvtx_w}, 
for central electrons. (The positive sign corresponds to the impact 
parameter to the right of the center 
of the detector observed from the EM cluster location.) The 
distributions are fitted with Gaussian functions with the 
widths of $\sigma_z = 14.0$~cm and $\sigma_r = 9.5$~cm, as determined 
from toy MC. The data agrees well with 
the MC predictions and also appears Gaussian. 
An analogous comparison for forward electrons is shown 
in Fig.~\ref{fig:emvtx_w_ec}. The corresponding resolutions are $\sigma_z 
= 17.0$~cm and $\sigma_r = 4.5$~cm. The resolution 
on impact parameter improves in the forward region because 
the physical size of the calorimeter cells becomes smaller with 
an increase in $|\eta|$. In the $z$-direction, this effect 
is compensated by the small angle of the 
cluster pointing, which dilutes the $z$-resolution.

\begin{figure}[hbt]
\vspace*{-0.5in}
\centerline{\protect\psfig{figure=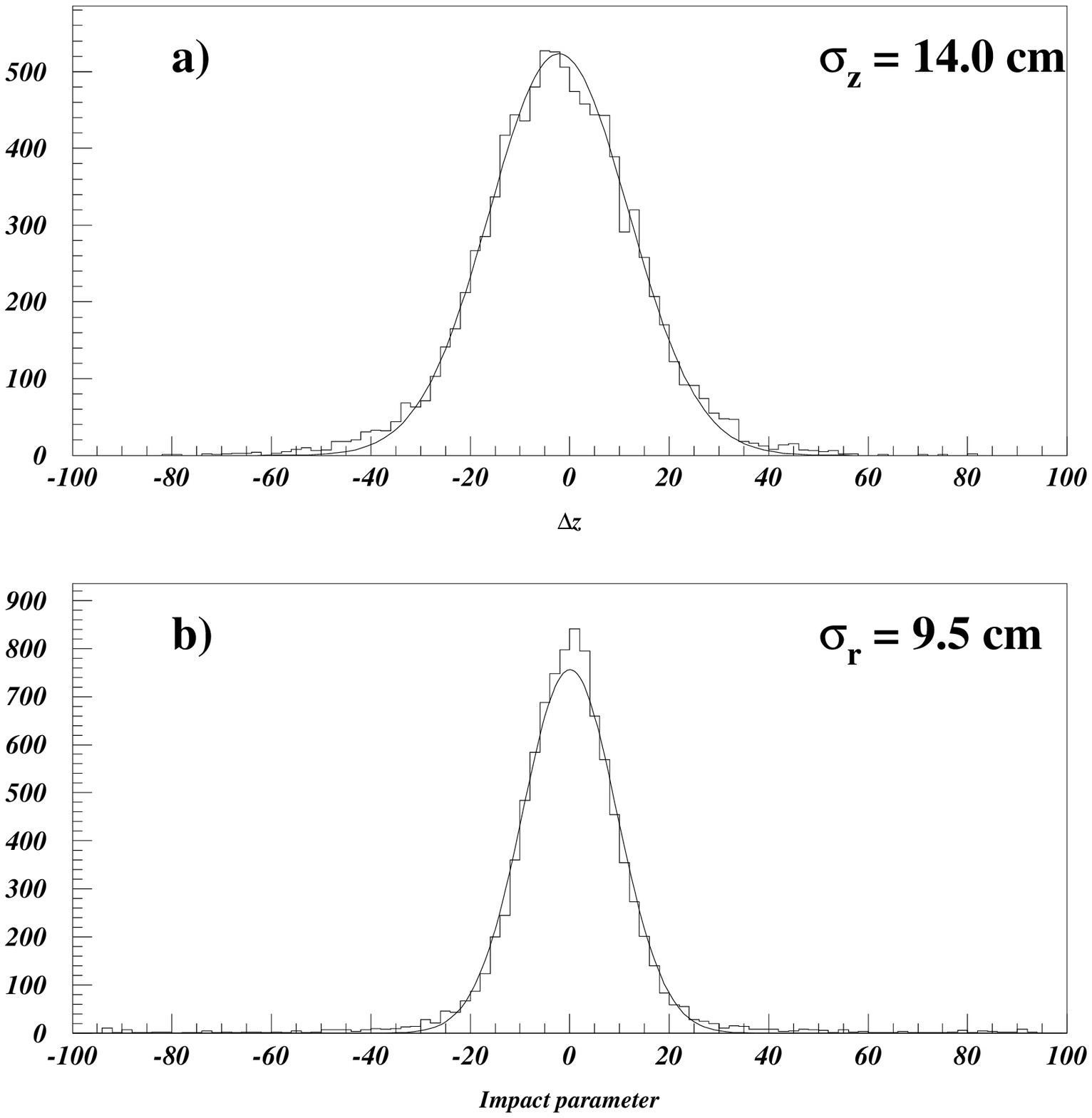,width=4.0in}}
\vspace*{-0.2in}
\caption{Comparison of (a) the error on the $z$-position of the vertex, and 
(b) the signed impact parameter from Run I $W \to e\nu$ data, with 
the results of toy simulations, for central 
electrons.}
\label{fig:emvtx_w}
\end{figure}
\begin{figure}[hbt]
\vspace*{-0.5in}
\centerline{\protect\psfig{figure=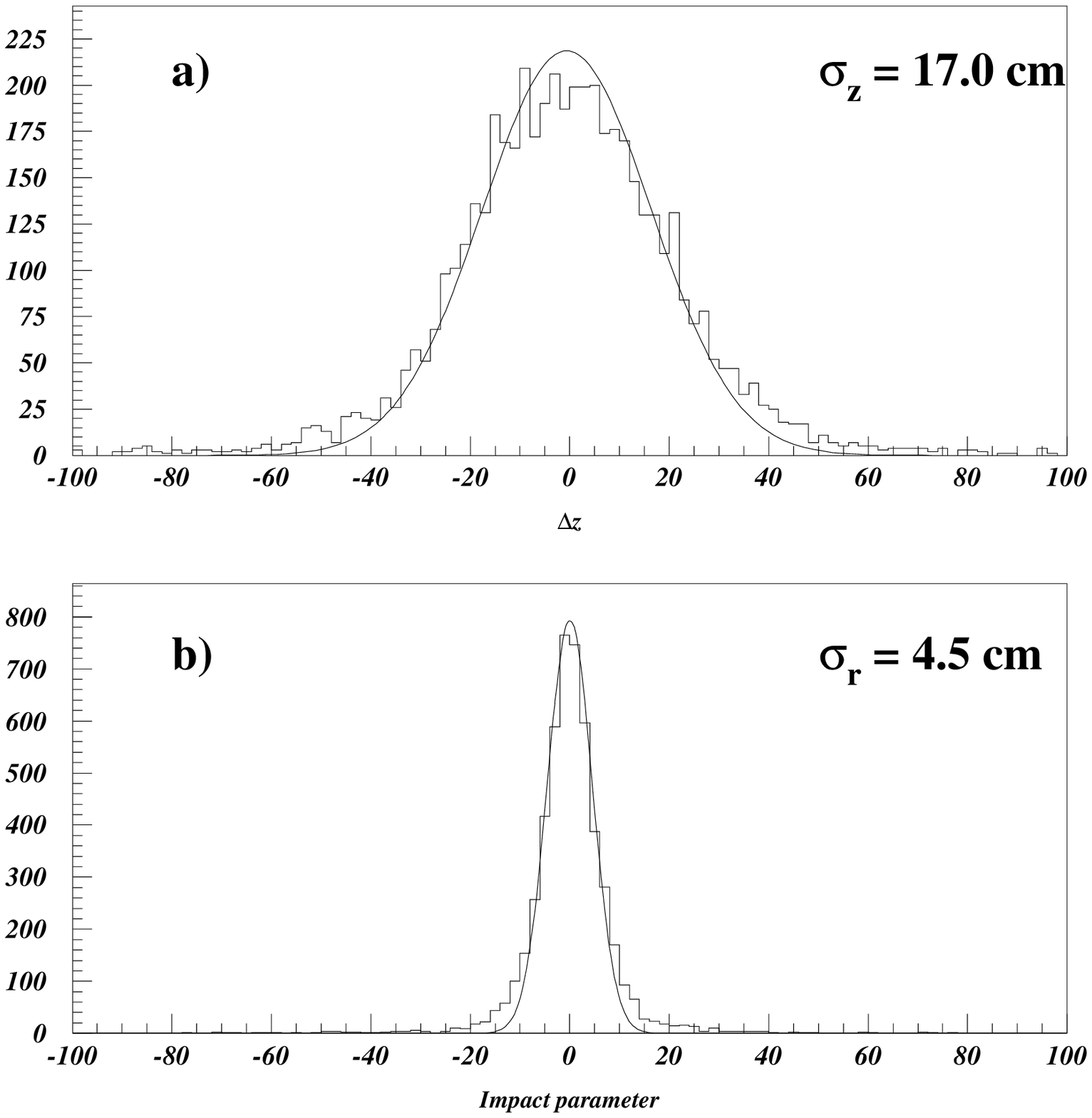,width=4.0in}}
\vspace*{-0.2in}
\caption{Comparison of (a) the error on the $z$-position of 
the vertex, and (b) the signed impact parameter from Run I $W \to e\nu$ data, 
with the results of toy simulations, for forward electrons.}
\label{fig:emvtx_w_ec}
\end{figure}

\begin{figure}[hbt]
\centerline{\protect\psfig{figure=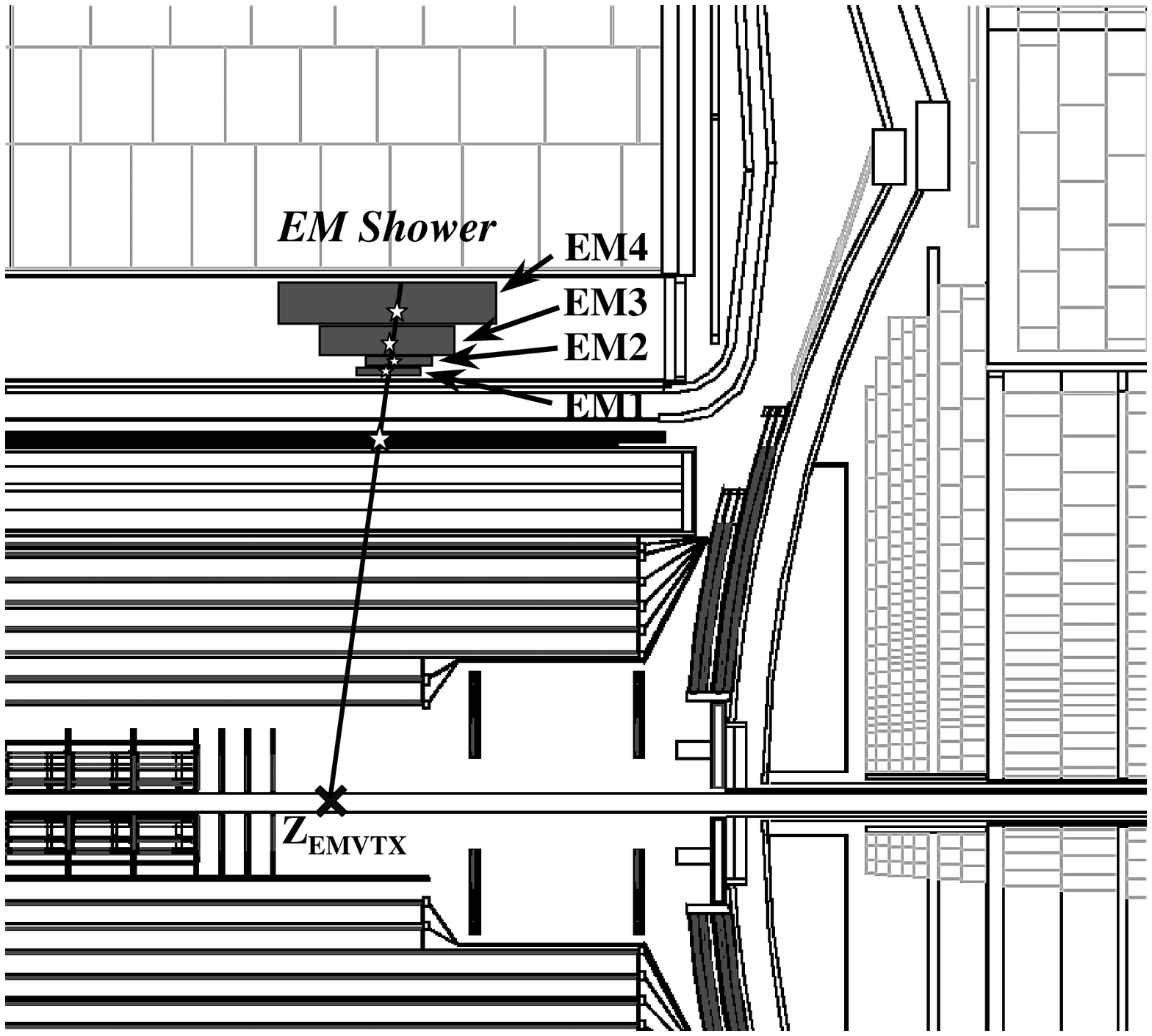,width=\textwidth}}
\caption{Side-view of the D\O\ calorimeter illustrating how the 
{\tt EMVTX} algorithm works.}
\label{fig:emvtx2}
\end{figure}
\begin{figure}[hbt]
\vspace*{-0.5in}
\centerline{\protect\psfig{figure=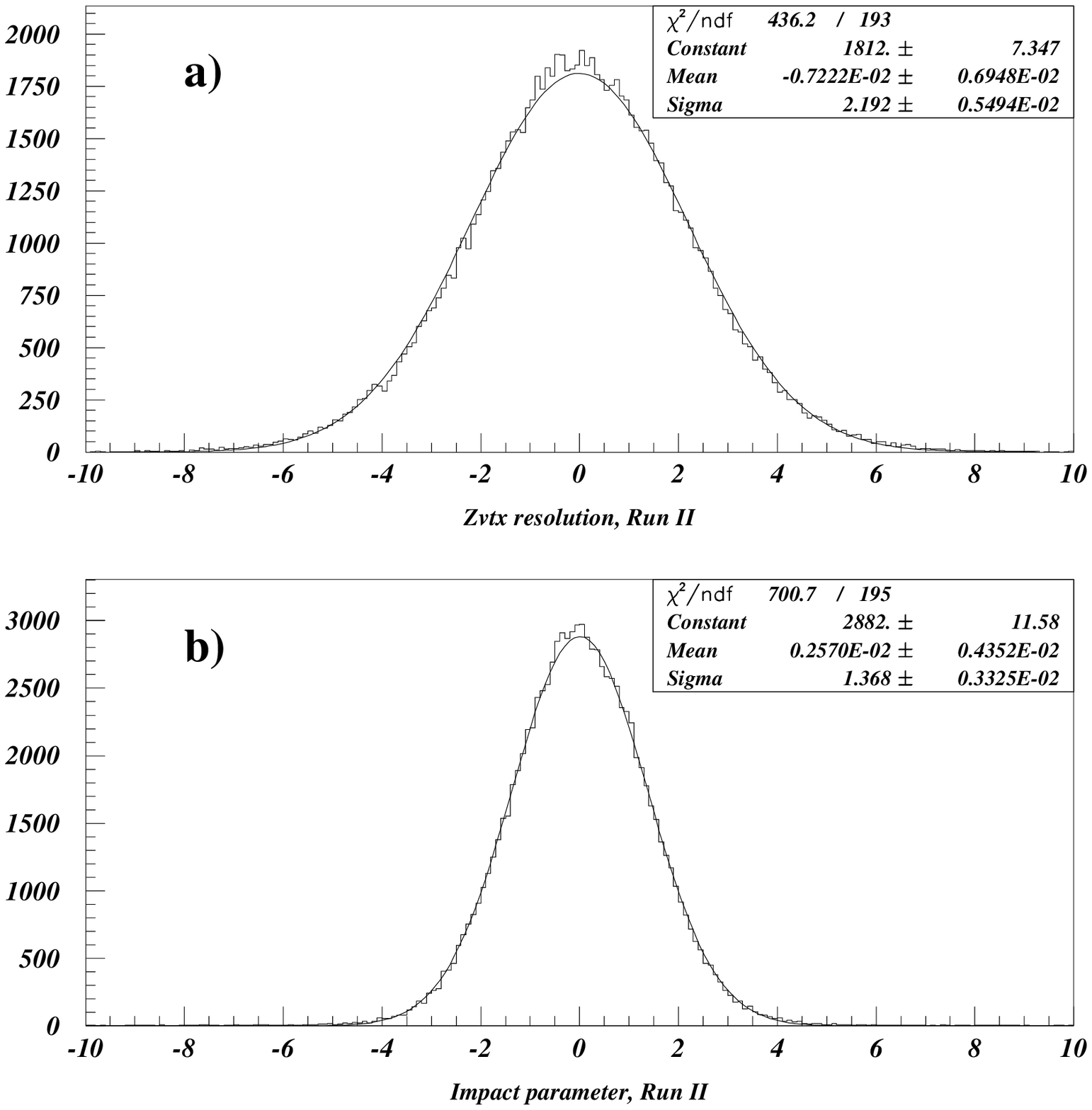,width=4.0in}}
\vspace*{-0.2in}
\caption{Run II simulation of (a) the error on 
the $z$-position of the vertex, and (b) the signed impact 
parameter, for central photons.}
\label{fig:emvtx_run2}
\end{figure}
\begin{figure}[hbt]
\vspace*{-0.5in}
\centerline{\protect\psfig{figure=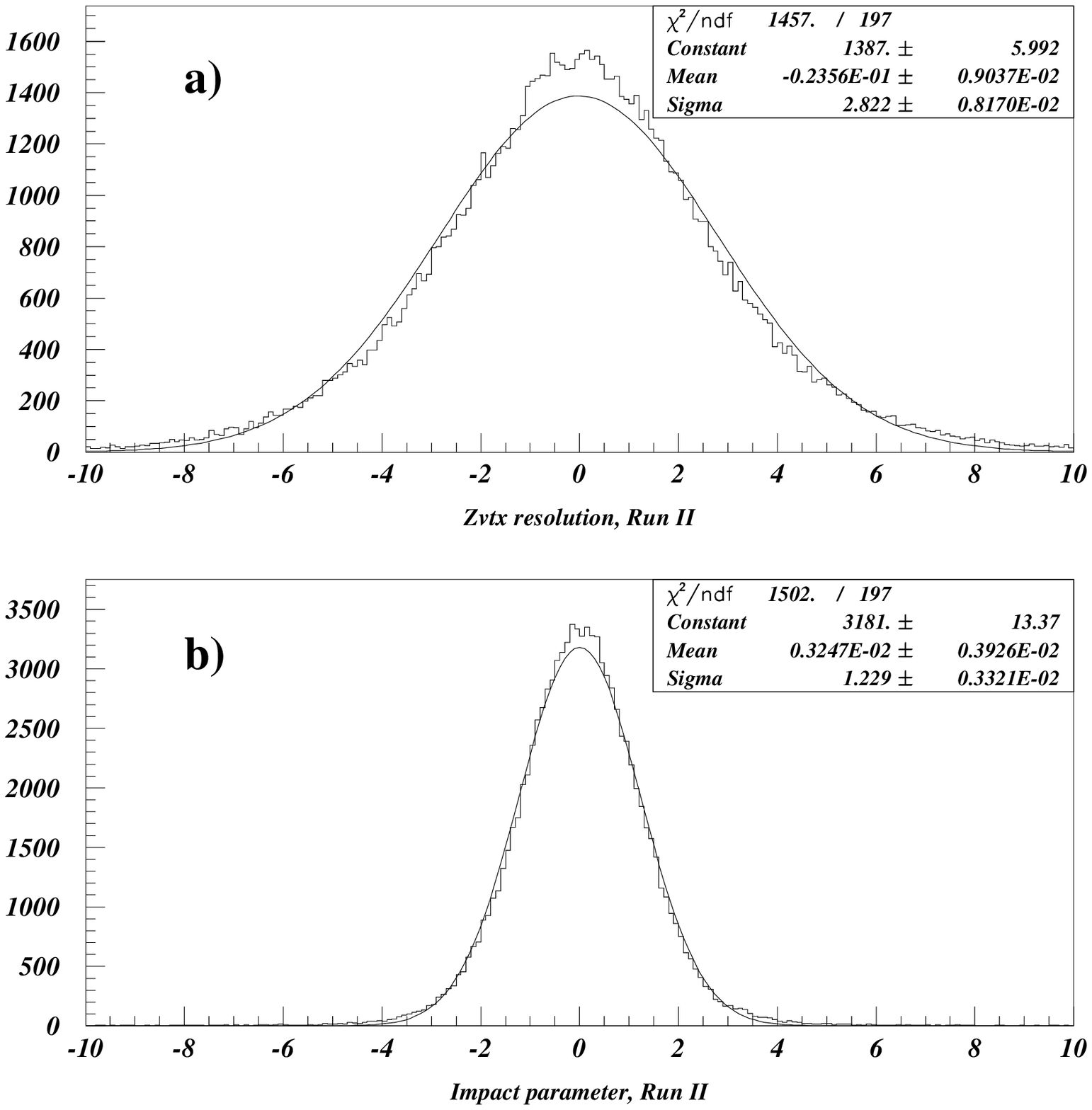,width=4.0in}}
\vspace*{-0.2in}
\caption{Run II simulation of (a) the error on the $z$-position of 
the vertex, and (b) the signed impact parameter, for forward photons. 
The slightly non-Gaussian shape is due to the change in 
the pointing resolution as a function of the photon rapidity.}
\label{fig:emvtx_run2_ec}
\end{figure}

The central and forward preshower detectors of the 
D\O\ Upgrade provide a precision measurement of the photon 
cluster position. The resolution for 
a typical EM shower perpendicular to the preshower 
strip is $\sim 1$~mm. Taking into account the 
crossing angle between the $u$- and $v$-planes in the 
preshower detectors, the following resolutions in $r\phi$ and $z$ ($r$) 
can be obtained: $\sigma_{r\phi} = 1.5$~mm, $\sigma_z = 2.5$~mm 
(CPS) and $\sigma_{r\phi} = 2.5$~mm, $\sigma_r = 2.5$~mm 
(CPS). As one can see, the preshower cluster position 
measurement is superior to that obtained from the EM 
calorimeter; its position is also spatially separated from 
that in the calorimeter, as seen in Fig.~\ref{fig:emvtx2}.

With the additional position measurement coming 
from the preshower detectors, the following pointing resolutions can be 
obtained for central and forward photons in Run II: 
$\sigma_z = 2.2$~cm, $\sigma_r = 1.4$~cm (CC, see Fig.~\ref{fig:emvtx_run2}) 
and $\sigma_z = 2.8$~cm, $\sigma_r = 1.2$~cm 
(EC, see Fig.~\ref{fig:emvtx_run2_ec}). A very significant improvement in photon 
pointing (by a factor of six) is 
achieved by utilizing the additional position measurement provided by 
preshower detectors.

The implementation of the photon-pointing algorithm can 
be done as early as at the trigger Level 3. 
An approximate algorithm that uses only the c.o.g. 
of the EM shower in the preshower and in the 
third layer of the EM calorimeter could be used to decrease 
the amount of calculations and, hence, the decision-making time. Monte 
Carlo simulations show that the impact parameter and $z$-resolutions for 
a simplified algorithm are only 10\% worse than 
those obtained by complete five-point fit, which is 
quite satisfactory for trigger purposes. Having the precise vertex 
information from photon-pointing for photon triggers at Level 3, 
we will also recalculate the \MET based on this vertex, 
and that would significantly improve the turn-on of the 
\MET part of less inclusive triggers which would require an 
EM cluster and \MET.

\subsection{Detection of 
Slow-Moving Massive Charged Particles in D\O}

As described in Section~\ref{sec:delayed}, long-lived charged 
particles have a variety of characteristics which enable 
their identification, particularly in the analysis stage when 
the event's complete data set is available.  Depending on the 
lifetime, a combination of time-of-flight, ionization ($dE/dx$) in several 
different detectors, and the muon-like penetration of a high momentum, isolated 
track all can be 
employed, with the additional presence of a kink where 
there is a decay within the detector volume. Even though 
discovery at the analysis stage seems possible, triggering is 
really crucial, if heavy stable particles produced at the 
Tevatron are to be detected.  In this section we 
discuss the possibilities for triggering, and in particular, the 
use of $dE/dx$ in the hardware trigger as a 
tool for detecting these objects.

The D\O\ trigger system is hierarchical, with 3 
levels, each level passing a small 
subset of the events it examines to the 
next level for further analysis.  Thus Level 1 has 
a high input rate and examines a limited amount 
of information in making its decision, while subsequent levels 
have progressively lower input rates and spend longer analyzing 
more data.  At Level 3, all the data 
digitized for the event is available and the selection 
is made running software algorithms written in high level code 
and derived from the offline analysis.  We expect that the 
massive stable particles will be sufficiently rare, and their characteristics 
clear, such that separating candidates at Level 
3 will be straight forward.  However, we need to understand how 
to identify these events in the hardware triggers so that they 
survive to Level 3.

For Level 1 several tools are available to detect a massive 
stable particle.  The Central Fiber Tracker provides track candidates 
binned in momentum to which it can 
apply an isolation criteria~\cite{Yuri}.  The CFT has 80 
trigger 
segments independently processed by the trigger. For a 
heavy stable particle one can select a high $p_T$ 
track with no other tracks in the home 
and adjacent segments, imposing an isolation of $\pm6.75^\circ$ in azimuthal 
angle~\cite{Yuri}.  Association of a muon with such 
a track would provide an efficient Level 1 trigger for 
massive stable particles, produced centrally and in isolation.

An additional tool~\cite{KJ} at Level 1 is the measurement 
of the time of flight (TOF) from scintillators associated 
with the muon detector and used primarily to 
reduce background in the large area muon detector from 
cosmic ray and beam associated accidentals.  These counters cover much 
of the area just outside the calorimeter, inside the first 
(``A") layer of muon detectors, and completely outside the detector, 
on top of the muon ``C" layer planes.  Electronics 
associated with the scintillation counters provides both a trigger gate 
and a TOF gate, thus giving time windows 
relative to the interaction time.  Scintillator hits received within 
the TOF gate will have the time of flight measurement digitized 
and saved with the data, assuming the event otherwise passes 
Level 1.  Typically the TOF gate will be set sufficiently 
wide to accept slow moving particles, namely at least 100 nsec.  
To contribute to the Level 1 trigger, however, the scintillator hit 
must occur within the trigger gate, which is necessarily much 
shorter (of order 25 nsec) because of the high rate of 
accidentals, particularly in the ``A'' layer counters.  It may be 
possible to run with a considerable wider trigger gate for 
the ``C'' layer counters, given the lower expected accidental rate 
in these counters.  In this case, TOF from the muon 
system would provide a useful tool for Level 1 
triggering on slow particles.  Given a Level 1 trigger, generated 
either through TOF or through an isolated stiff CFT track, 
the TOF data will be a useful tool for the Level 2 and Level 3 triggers.

Beyond Level 1, the new Silicon Microstrip Tracker provides 
interesting possibilities for triggering on slow moving 
particles.  Since energy loss of a slow moving 
particle drops as $1/\beta^2$ as a function 
of its velocity, the excellent $dE/dx$ energy resolution of 
the silicon chip provides a good handle for offline 
identification.  For example, the energy deposited in one silicon 
layer is shown in Fig.~\ref{fig:MSP-ADC}, as measured in 
a test beam~\cite{Maria}.    More importantly for the detection of slow 
moving particles, there are possibilities to exploit these measurements at 
the trigger level, with recent approval of the 
D\O\ Silicon Tracker Trigger (STT)~\cite{STT} as a component of 
the Level 2 trigger system.  The hardware design for the 
STT may allow for the inclusion of several additional backplane 
lines to carry $dE/dx$ information along with the 
other data associated with each cluster of hits~\cite{Uli}.  
We have studied the basic capabilities of such trigger hardware 
to explore its potential for slow particle identification.

\begin{figure}[hbt]
\vspace*{0.1in}
\centerline{\protect\psfig{figure=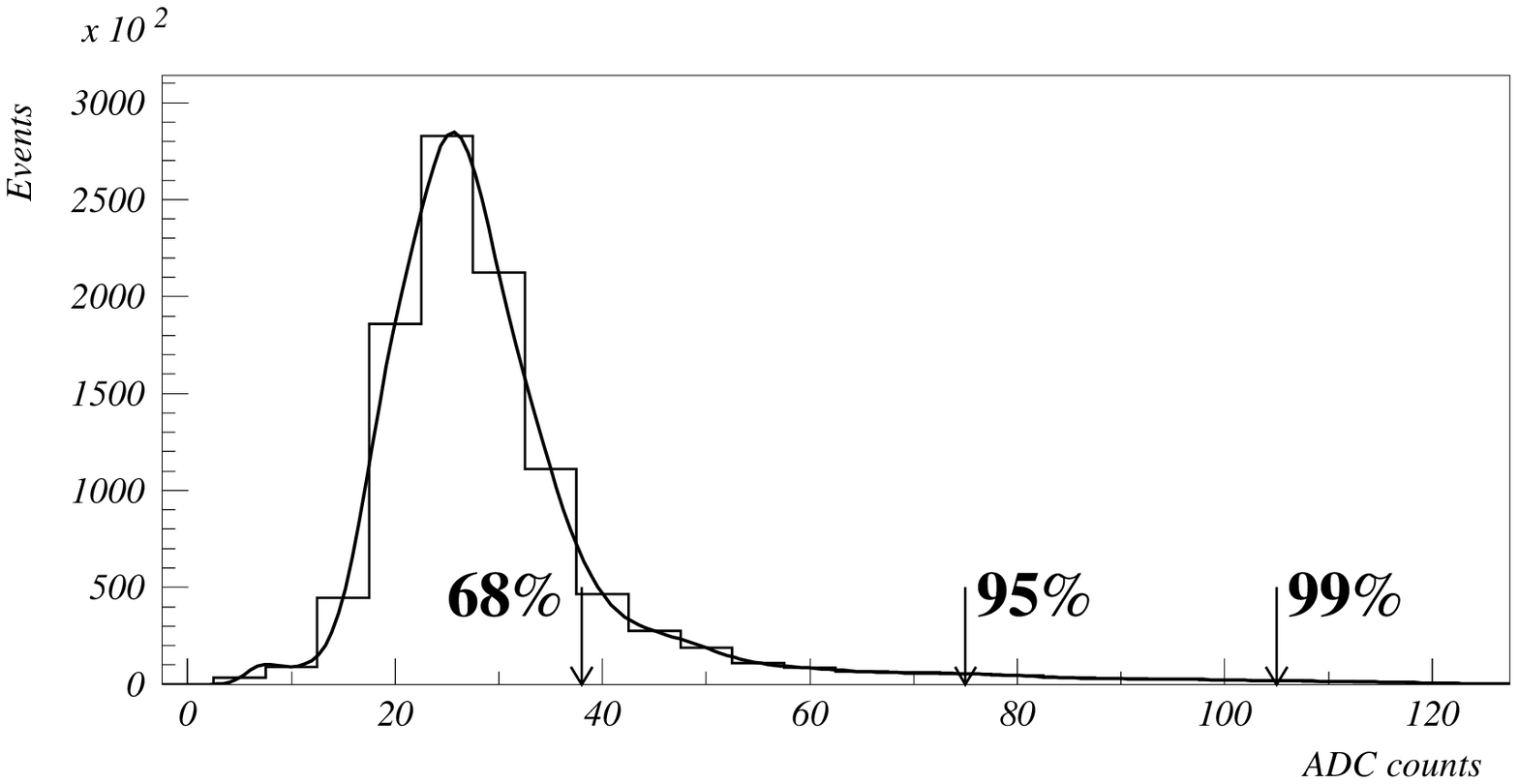,width=\textwidth}}
\caption{Energy deposition in one layer of the Silicon Microstrip 
Tracker in ADC counts, as measured in the test beam.}
\label{fig:MSP-ADC}
\end{figure}

Our simulation of the STT slow particle trigger is based 
on a simple Monte Carlo generator which returns an ADC value appropriate 
for the spectrum shown in Fig.~\ref{fig:MSP-ADC}.  We assume that 
two backplane lines per SMT hit are available, and 
we encode each ADC value into these two bits 
of data, or four bins.  After trying various values 
for the three bin edges, we have chosen to define 
the bins by those ADC values below which lie 67\%, 
95\%, and 99\% of the data. These cutoffs correspond 
to 38, 75, and 105 ADC counts (see Fig.~\ref{fig:MSP-ADC}). 
The STT Level 2 trigger processor finds tracks 
using four layers of silicon; so, in our model, 
the hardware would provide four samples of this two 
bit $dE/dx$ data for each track.  At this point the 
trigger would use some algorithm to combine the four samplings 
most advantageously.  The major concern for the correct identification of 
a slow moving particle is the likelihood of false signals 
from a minimum ionizing particle (MIP), due to an occasional 
response in the very long tail of the Landau 
energy loss distribution.  Based on a few studies, our preliminary 
suggestion is simply to sum the three lowest values, 
rejecting the largest ADC count of the four.  The data 
then would provide a parameter, related to a $dE/dx$ 
of the particle in the silicon tracker, which we call 
``slowness," and which has 10 possible values (0..9).  From our simulation 
we derive the distribution in slowness for 
a $\beta=1$ particle (equivalent to the test beam pion), 
as shown in Fig.~\ref{fig:MSP-slowness}.  

\begin{figure}[hbt]
\vspace*{0.1in}
\centerline{\protect\psfig{figure=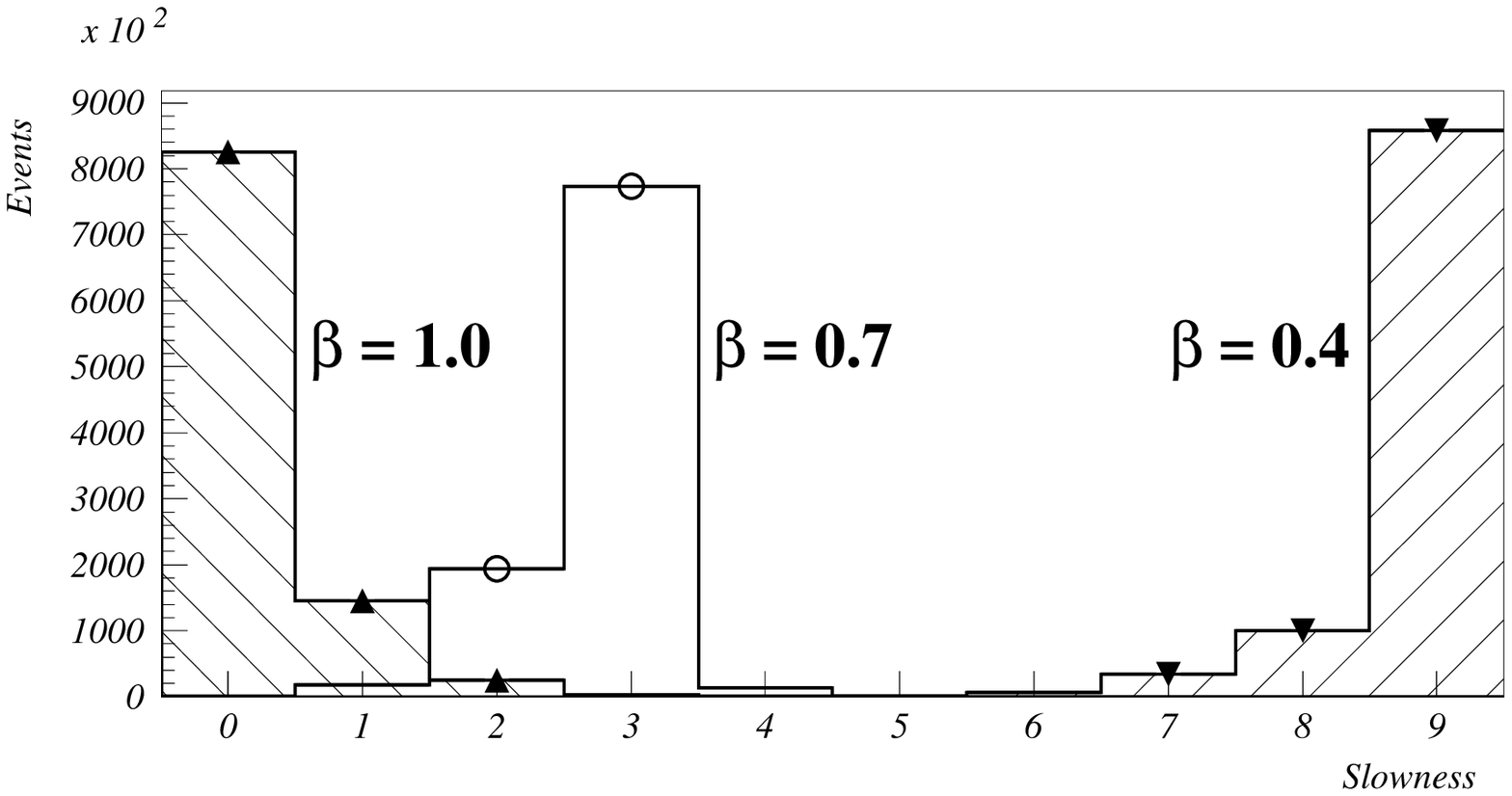,width=\textwidth}}
\caption{Distribution of slowness for massive particles with 
$\beta = 0.4$ and 0.7, and test beam particles with $\beta = 1.0.$}
\label{fig:MSP-slowness}
\end{figure}

The ADC response to the passage of a 
slow moving particle will be similar to the 
ADC distribution of Fig.~\ref{fig:MSP-ADC} scaled by the factor $1/\beta^2$.  
However, because of the very different kinematics 
(the slow moving particle is massive, typically 150 GeV) 
the energy loss distribution will have a less pronounced 
Landau tail.  We make a very conservative assumption and use 
only the Gaussian component in generating ADC values for 
massive slow moving particles.  Several resulting distributions in 
our $dE/dx$ trigger parameter, ``slowness,'' for particles 
with $\beta=0.4$ and $\beta=0.7$, are included in Fig.~\ref{fig:MSP-slowness}. 
There is a clear separation in this parameter 
compared to the $\beta=1$ distribution.

To estimate the effectiveness of the STT $dE/dx$ trigger we consider a selection which tags as a slow particle those whose ``slowness" is greater than or equal to some value.  We vary this selection to study the efficiency for slow particles (at highest possible $\beta$) while maintaining a strong rejection against $\beta=1$ particles.  Figure~\ref{fig:MSP-eff} shows the acceptance as a function of a particle's velocity, for events with slowness $>4$.  The appearance of a few $\beta=1$ particles but not other high $\beta$ events reflects the conservative use, for massive particles, of a Gaussian ADC response, rather than the Landau distribution, which is used to generate the ADC values from a MIP.

\begin{figure}[hbt]
\vspace*{0.1in}
\centerline{\protect\psfig{figure=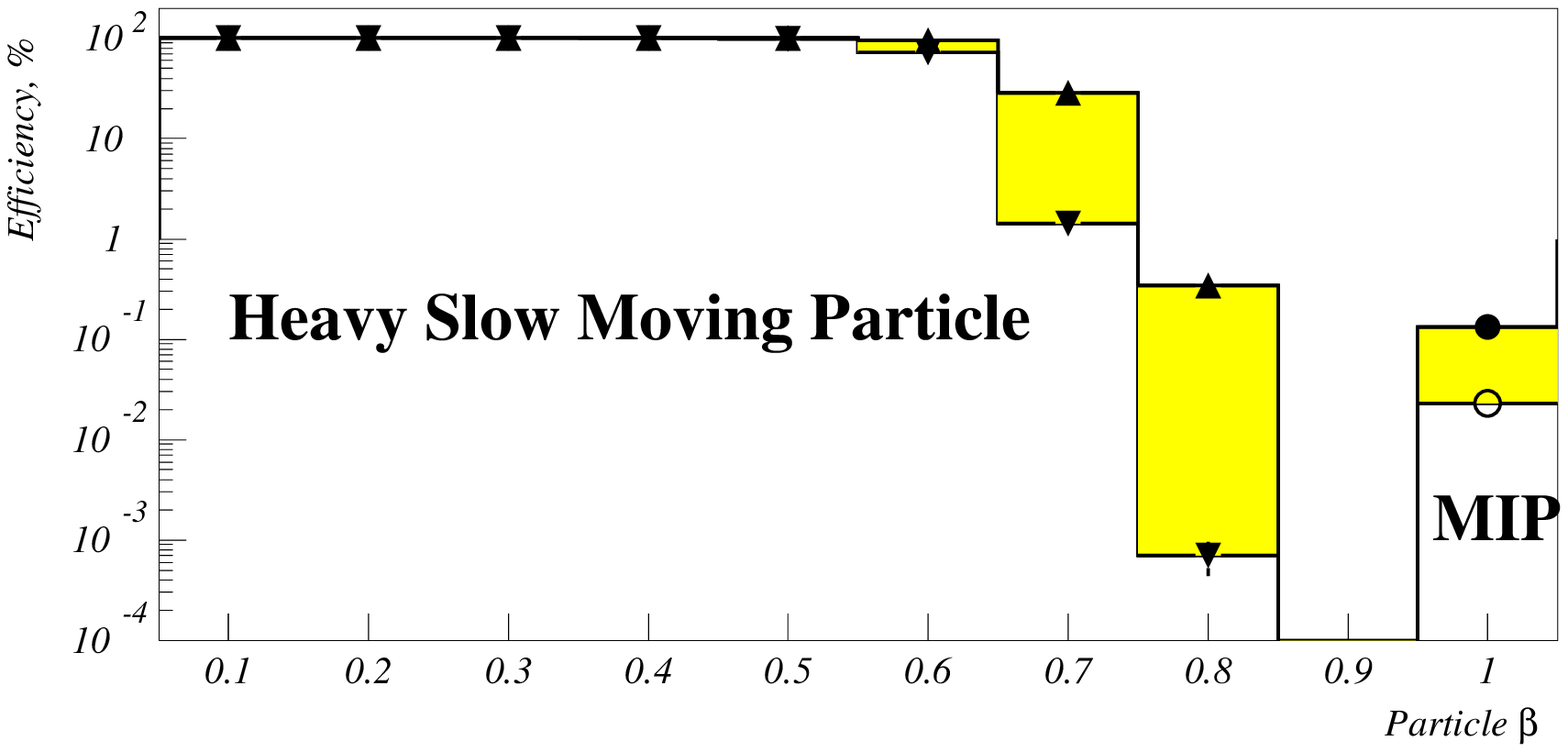,width=\textwidth}}
\caption{Efficiency as a function of particle $\beta$ for events with slowness 
$>3$. The open histogram corresponds to the 
case without angular smearing; the filled 
area shows the change when 
angular smearing of $\pm 45^\circ$ is taken into account.}
\label{fig:MSP-eff}
\end{figure}

\begin{figure}[hbt]
\vspace*{0.1in}
\centerline{\protect\psfig{figure=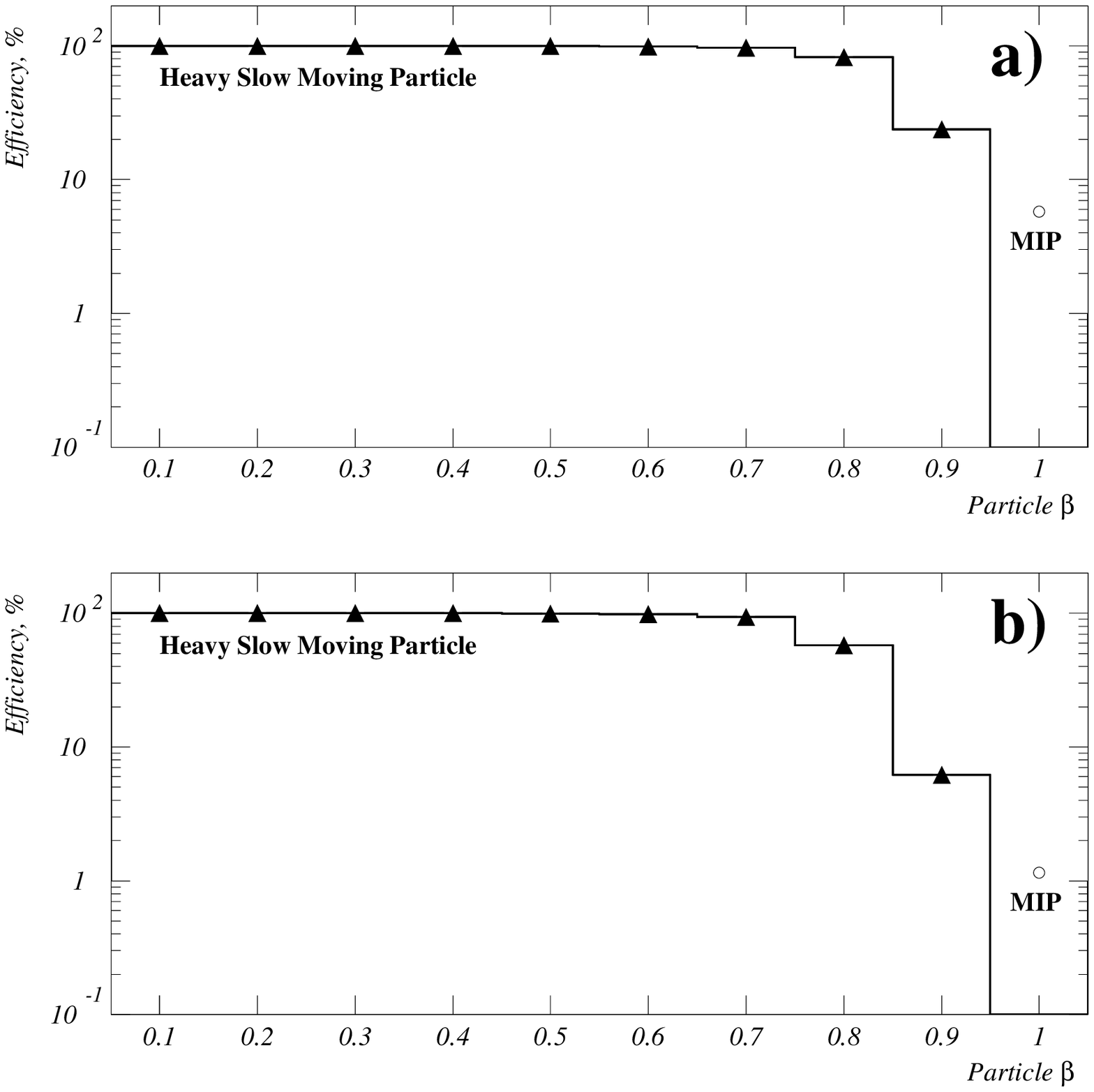,width=\textwidth}}
\caption{Efficiency as a function of particle $\beta$ 
for the events with redefined slowness 
a) $>7$; b) $>8$. Angular smearing of $\pm 
45^\circ$ is taken into account.}
\label{fig:MSP-eff1}
\end{figure}

Despite the intent to trigger only on centrally 
produced objects, tracks will tend to be inclined 
to the silicon, increasing and broadening the $dE/dx$ response.  In 
fact, the STT hardware will search for track candidates 
from adjacent barrels; so that tracks may be inclined to 
the normal by as much as $45^\circ$ although the 
STT will have no information about this angle.  We 
have modeled the effect of inclined tracks by scaling the 
appropriate ADC response by $1/\cos\theta$, where $\cos\theta$ is 
generated uniformly between .707 and 1.0, or by $\eta$, 
where $\eta$ is generated uniformly between 0 
and 0.88.  The differences between smearing based on $\cos\theta$ 
and that based 
on $\eta$ are small.  We present here results using $\cos\theta$ smearing 
for the centrally produced massive stable particles and 
$\eta$ smearing for the $\beta=1$ background.  As seen in 
Fig.~\ref{fig:MSP-eff}, the effect of the variation in ADC 
response due to track inclination is only a small 
reduction in the rejection for $\beta=1$ particles; moreover, 
this effect helpfully raises the cutoff in $\beta$ for 
massive particles.  Overall, this smearing does not appear 
to affect the STT ``slowness" trigger significantly. Including 
track inclination, the study suggests that 
the STT Level 2 $dE/dx$ trigger would provide 
a fully efficient tag for slow particles up to 
$\beta=0.7$ with an acceptance of $\beta=1$ particles 
less than $2\times 10^{-3}$.  

Because of its excellent rejection for $\beta=1$ particles, 
the Level 2 $dE/dx$ trigger will be a good means 
to select slow moving particles, independent of other criteria 
such as TOF or track isolation.  However, 
if the particle is sufficiently long-lived to traverse 
the entire detector, it may be possible to relax 
the $dE/dx$ requirement.  We have studied the effect 
of modifying the hardware ADC sampling, to explore 
widening the acceptance of heavy particles in $\beta$ at the expense of
$\beta=1$ rejection.  There is good physics motivation in doing so, as some GMSB models predict the production of heavy stable particles with $\beta$ in the range 0.8-0.9~\cite{Jianming}. Using ADC bins with edges corresponding to 50\%, 68\%, and 90\% of the distribution, we find good acceptance for massive particles at high $\beta$, as shown in Fig.~\ref{fig:MSP-eff1}, with rejection factors between 20 and 100, for $\beta=1$ particles.

The above study illustrates the potential of a STT $dE/dx$ trigger.  It 
does assume that the ADC distribution for a 
MIP is as seen in the test beam, 
and that differences in the silicon chip response 
over the detector won't significantly affect this distribution.  
The study suggests that track inclination may 
not be a serious problem.  Further, the ``slowness" 
flag derived with the STT from $dE/dx$ can 
in Level 2 be combined with other information (as of 
a straight, non-oblique and isolated muon) to provide a 
global Level 2 trigger.  In summary, it seems promising 
that the STT could provide very useful $dE/dx$ 
information early in a slow particle's lifetime, which 
can be combined with other data to create an efficient 
trigger for these interesting objects.

\label{sec:HIT-D0}

\def\ppbar{\mbox{$p\overline{p}$}}
\def\bbbar{\mbox{$b\overline{b}$}}
\def\ccbar{\mbox{$c\overline{c}$}}
\def\ttbar{\mbox{$t\overline{t}$}}
\def\ee{\mbox{$ee$}}
\def\zzero{\mbox{$Z^0$}}
\def\pelp{\mbox{$e^+$}}
\def\pelm{\mbox{$e^-$}}
\def\pelpm{\mbox{$e^{\pm}$}}
\def\epem{\mbox{$e^+e^-$}}
\def\lplm{\mbox{$\ell^+\ell^-$}}
\def\qbar{\mbox{$\overline{q}$}}
\def\gluino{\mbox{$\tilde{g}$}}
\def\squark{\mbox{$\tilde{q}$}}
\def\sqbar{\mbox{$\bar{\tilde{q}}$}}
\def\mgluino{\mbox{$M(\gluino)$}}
\def\msquark{\mbox{$M(\squark)$}}
\def\csquarkl{\mbox{$\tilde{c}_L$}}
\def\mcsl{\mbox{$M(\csquarkl)$}}
\def\ssb{\mbox{$\squark\overline{\squark}$}}
\def\csquark{\mbox{$\tilde{c}$}}
\def\tsquark{\mbox{$\tilde{t}$}}
\def\stopo{\mbox{$\tilde{t}_1$}}
\def\ttb{\mbox{$\tsquark\overline{\tsquark}$}}
\def\ttbone{\mbox{$\tsquark_1\overline{\tsquark}_1$}}
\def\chione{\mbox{$\tilde{\chi}_{1}^{\pm}$}}
\def\mchione{\mbox{$M(\tilde{\chi}_{1}^{\pm})$}}
\def\mstopo{\mbox{$M(\tilde{t}_1)$}}
\def\mcone{\mbox{$M(\tilde{\chi}_{1}^{\pm})$}}
\def\none{\mbox{$\tilde{\chi}_{1}^0$}}
\def\mchio{\mbox{$M(\none)$}}
\def\lsp{\mbox{$\tilde{\chi}_{1}^0$}}
\def\mz{\mbox{$M_0$}}
\def\mo{\mbox{$M_{1/2}$}}
\def\rpv{\mbox{${R\!\!\!\!\!\:/_p}$}}
\def\rp{\mbox{${R_p}$}}
\def\lamp{\mbox{$\lambda_{121}'$}}
\def\gev{\mbox{$\;{\rm GeV}$}}
\def\tev{\mbox{$\;{\rm TeV}$}}
\def\gevc{\mbox{$\;{\rm GeV}/c$}}
\def\gevcc{\mbox{$\;{\rm GeV}/c^2$}}
\def\tevcc{\mbox{$\;{\rm TeV}/c^2$}}
\def\chis{\mbox{$\chi^{2}$}}
\def\ifb{\mbox{${\rm fb}^{-1}$}}
\def\ipb{\mbox{${\rm pb}^{-1}$}}
\def\met{\mbox{${E\!\!\!\!/_T}$}}
\def\intlum{\mbox{${ \int {\cal L} \; dt}$}}
\def\pt{\mbox{$p_T$}}
\def\et{\mbox{$E_T$}}
\def\modulus#1{\left| #1 \right|}
\def\bp{\mbox{$b'$}}
\def\lxy{\mbox{$L_{xy}$}}
\def\dedx{\mbox{${\rm d}E/{\rm d}x$}}
\def\etal{{\it et al.}}
\def\Journal#1#2#3#4{{#1} {\bf #2}, #3 (#4)}
\def\PRDR#1#2#3{{Phys. Rev. D} {\bf #1}, Rapid Communications, #2 (#3)}

\section{Searches for Beyond the MSSM Phenomena at CDF}
\noindent
\centerline{\it M.~Chertok}
\medskip
\label{section:Chertok}

Two remarks motivate the discussion to follow.
First, if the LSP is charged and has an appreciable lifetime (or is stable),
it can be detected in a search for charged massive particles
(CHAMPs).  Second, while the MSSM makes no predictions regarding the
possibility of extra quark families, there is no theoretical reason
against them \cite{mckay}.  

CDF has performed searches for these
phenomena using data taken with the Run I detector \cite{det}. 
Upgrades of both
the detector and Tevatron are currently underway \cite{tdr}.  These will
provide substantial enhancements for these and other searches in Run
II, scheduled to begin in 2000.  During this run, CDF II will collect roughly
2 \ifb\ of data at $\sqrt{s}=2 \tev$,
corresponding to twenty times the present statistics.  The 10\%
increase in energy corresponds to a 40\% increase in the \ttbar\
yield, and similarly will aid new phenomena searches.


Although the LEP data at the \zzero\ pole exclude extra fermion
generations with light neutrinos \cite{LEP}, models including fourth family
quarks have received recent theoretical attention \cite{mckay,four_theory}.
CDF performs three complementary searches for such quarks, described below.
\subsection{Search for long-lived parent of the \zzero}
CDF performs a general search for long-lived particles decaying to
\zzero\ bosons \cite{lxy}.  
One example is a \bp\ quark decaying via a FCNC 
to a $b$ quark and a \zzero.  This decay can
dominate (depending on the \bp\ mass) but may lead to a long
lifetime.  Another example is low-energy symmetry breaking models with SUSY
\cite{ambrosanio} that predict the decay $\none \to \zzero \tilde{G}$ with 
a long lifetime due to the small coupling constant of the gravitino.

The technique for this search is to select \epem\ pairs
from \zzero\ decay and search the transverse decay length distribution 
for evidence of \zzero\ bosons originating at a displaced vertex.
Electrons ($e^+$ or $e^-$) are required to satisfy $\et>20\gev$,
$\pt>15\gevc$, and $|\eta|<1$, with the pair reconstructing to the
\zzero\ mass: $|M_{ee}-M_{\zzero}|<15\gevcc$.  
Quality cuts are imposed to reduce the effects of
misreconstructed tracks.  The $e^+$ and $e^-$ are required to originate from
a common vertex, and nearly collinear tracks are removed by requiring
$|\Delta\phi-\pi|>0.02$. A high-purity sample of
$J/\psi\to\mu^+\mu^-$ is
used to check that these cuts do not introduce lifetime biases.
In 90~\ipb, 703 events pass the full selection.

The reconstructed transverse decay length \lxy\ distribution from
these events is clustered around the origin, as expected, and can be
modeled with a central gaussian and tails due to tracking errors.  
The negative \lxy\ region is used to estimate the background from 
tracking errors and select an \lxy\ cut which is then used to search 
for a signal in the positive \lxy\ region.  
A cut of $\lxy>1mm$ is chosen, which corresponds to an expectation of
$\leq$ one event based on the central gaussian.
There is one candidate event with $\lxy>+1mm$.
As the data are consistent with background levels, limits are derived
as shown in Figure~\ref{fig:lxy}.  The left figure shows the model
independent cross section times branching ratio limit as a function of 
the lifetime, as well as a limit (insert) for the case of $\bp\to
b\zzero$.  In this case two jets with $\et>10\gev$ and $|\eta|<2$ are
additionally required and the \lxy\ cut is made at $0.1 mm$.  Also
shown (right) is the resulting limit on the \bp\ mass as a function of 
the lifetime.

Plans for the continuation of this analysis in Run II include the
addition of the muon channel and an attempt to perform the search
using only the outer tracking chamber without the SVX.  
The former would increase the cross section sensitivity by approximately a 
factor of two while the latter would flatten out the limit to
$\sim30$ cm.  Folding in the other factors for Run II improvements and
integrated luminosity, the present cross
section limit minimum of 0.5 pb for the model independent case
should be reduced to below 5 fb.  For the \bp\ model, masses up to the 
top quark mass should be excluded in Run II.

\begin{figure}[t]
\begin{minipage}[t]{.50\linewidth}
\centerline{\protect\psfig{figure=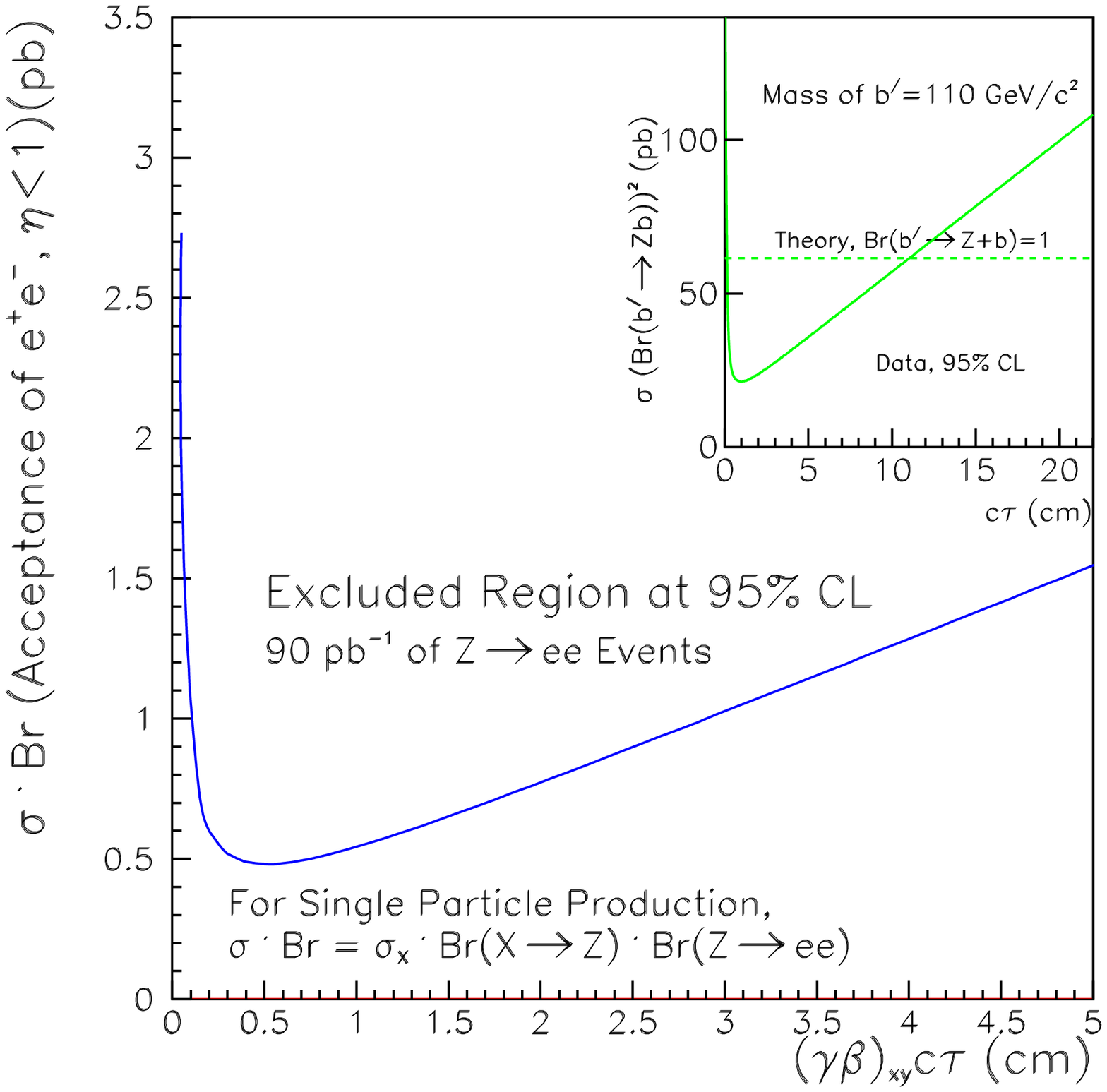,width=\linewidth}}
\end{minipage}\hfill
\begin{minipage}[t]{.50\linewidth}
\centerline{\protect\psfig{figure=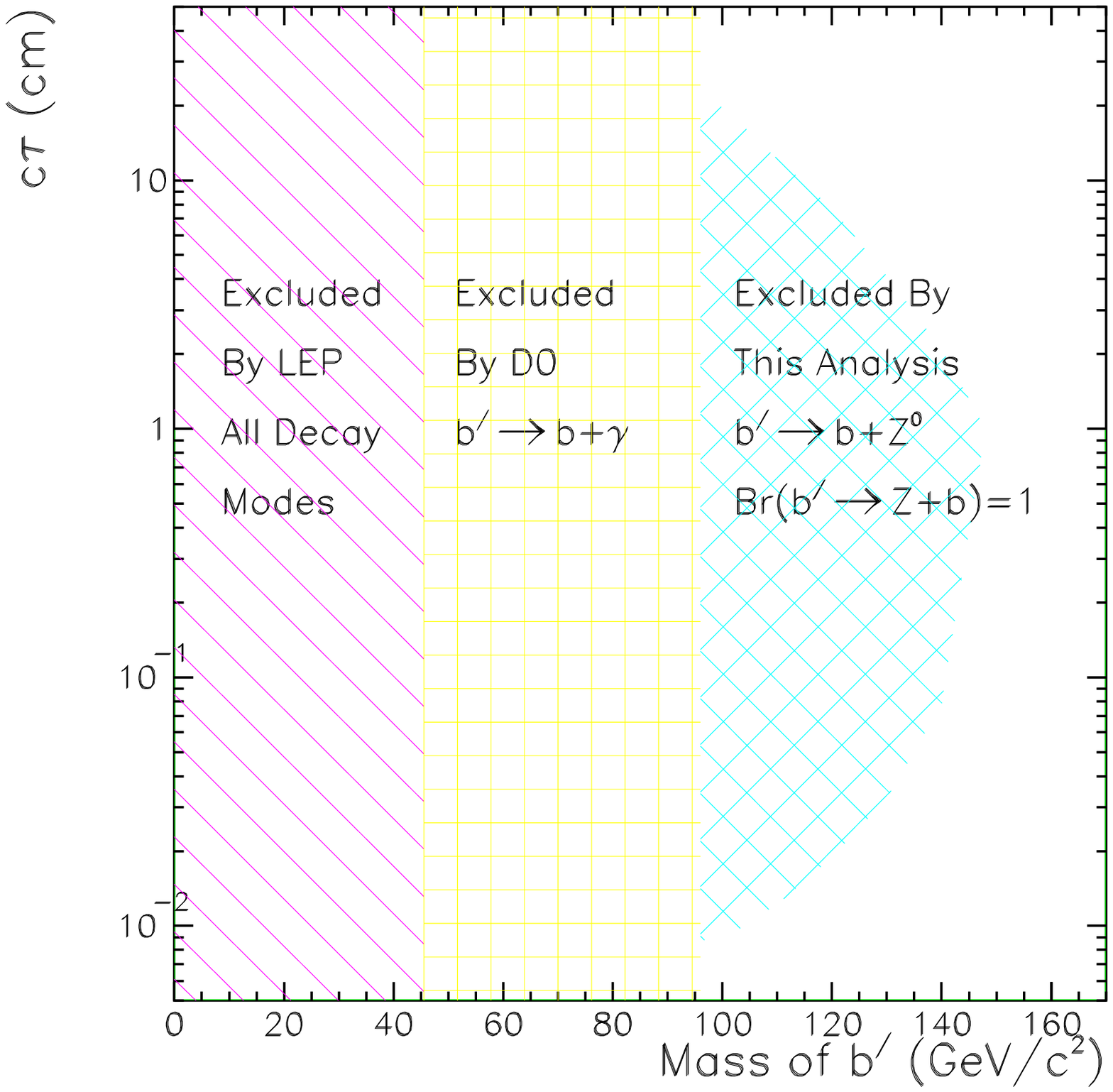,width=\linewidth}}
\end{minipage}
  \caption{Search for a long-lived parent of the \zzero.  
        Cross section limit (left) and mass limit (right).}
  \label{fig:lxy}
\end{figure}

\subsection{Search for 
$b'\bar{b}' \to \zzero\zzero b\bar{b} \to (\ell^+\ell^-)(q\bar{q})(b\bar{b})$}
If a fourth generation \bp\ quark is lighter than both $t$ and $t'$,
then the allowed CC decay $\bp \to cW^-$ is doubly Cabbibo suppressed and the
FCNC decay $\bp \to b\zzero$ dominates if $m_{\bp}>m_{\zzero}+m_b$.
At the Tevatron, \bp\ would be pair produced with a cross section like
that for top with $b'\bar{b}' \to \zzero\zzero b\bar{b} \to
(\ell^+\ell^-)(q\bar{q})(b\bar{b})$ as one decay pattern.
CDF performs a search for such a \bp\ by requiring two electrons
consistent with $\zzero \to \epem$, with $\et(e_1)>20\gev$,
$\et(e_2)>10\gev$, and $75 \gevcc \leq M(\epem) \leq 105 \gevcc$.  In
87 \ipb, 6548 events satisfy this electron selection.  Three or
more jets with $|\eta|<2$ are then required.  For 
the search region $M(\bp)<130\gevcc$,
two jets must satisfy $\et>15\gev$ while the third can have
$\et>7\gev$.  For $M(\bp)>130\gevcc$, all jets must satisfy
$\et>15\gev$.  Also, the sum of the jet \et\ for jets with $\et>15\gev$ must 
scale with the \bp\ mass in the following way: $\sum
\et(\et(j)>15\gev) > M(\bp) - 60 \gev$, motivated by a study of the
backgrounds.  Finally, one $b$ tag (displaced vertex) 
using the SVX detector information is required for the event to pass.
This final requirement removes the 31 events remaining after the sum
\et\ requirement.

The \bp\ signal is generated using HERWIG, with masses in the
range $100\gevcc \leq M(\bp) \leq 170\gevcc$ and passed through the
CDF detector simulation program.  The efficiency for $\zzero \to\epem$
is about 60\% and the efficiency for the 3 jet cut ranges from 25\% to
66\% with $M(\bp)$.  The main
background process for $\bp\to b\zzero$ comes from $\zzero +n\;j$ events, where
$n\geq3$, and are modeled using VECBOS and HERWIG.  These events are
passed through the CDF detector simulation program and filtered with
the same event selection as for the signal.  

Presently, this analysis is being finalized and
the muon channel is being added.  Including this additional acceptance
along with the improvements expected in Run II described above, the ultimate 
sensitivity of the search will reach $M(\bp)\simeq M(t)$. 
It should be noted that the decay $\bp \to bH$ must be considered in
this regime.  However, if $H\to\bbbar$ is appreciable, much of the
signal can be recovered as long as $\bp \to b\zzero$ remains large
enough to provide triggerable leptons from $\zzero\to\lplm$. 

\subsection{Search for long-lived charged massive particles}
Several models outside the MSSM predict new charged massive particles
(CHAMPs) with appreciable lifetimes.  These include fourth generation
leptons and quarks, weak \rp\ violation, and gauge mediated SUSY breaking
models.  A strongly interacting CHAMP would have a large cross section 
(like that for top), which allows large masses to be probed
with little background.  CDF performs such a search using as a
reference model a fourth generation quark with fragmentation to an integer 
charged meson within a jet.

The technique for this search is to use the large ionization energy
loss, \dedx, in the tracking chambers to tag massive (and therefore
slow-moving) particles.  Starting with 90 \ipb\ of data from Run I,
various tracking cuts for the SVX and central 
tracker (CTC) are applied to select high quality tracks. 
Candidate tracks are required to have $|p| > 35\gevc$
and $|\eta|<1$ and to pass ionization cuts in both the SVX and CTC which
correspond to requiring $\beta\gamma < 0.85$. Finally, a mass is calculated
from the momentum and \dedx\ and a sliding cut, $M_{\dedx} > 0.6 M_{CHAMP}$, is
applied for each assumed $M_{CHAMP}$.

The background to this search is due to particles which fake a large
ionization signal, mostly from particles whose tracks overlap.  
The probability of a track faking a signal in
both the SVX and CTC is quite low: of the 20K events passing the
selection cuts, $12.1\pm1.8$ tracks from fakes are expected with
$\beta\gamma<0.85$.  
Since the background is higher at low mass, a tighter cut of 
$\beta\gamma<0.7$ is used for the search region $M_{CHAMP} < 100 \gevcc$.
With this cut, $2.5\pm0.8$ tracks from fakes are expected.  These
background levels agree well with the data as shown in 
Figure~\ref{fig:champ}.  The overall efficiency for this search ranges 
from 0.75\% to 3\% as a function of the mass of the CHAMP and is about
twice as high for assumed charge $+2/3$ as for charge $-1/3$ due to
fragmentation effects.
Systematic uncertainties for this analysis are dominated by
the uncertainty on interactions of a massive quark in the calorimeter which
contribute between 13 and 20\% depending upon the assumed quark charge.
Figure~\ref{fig:champ} also shows the resulting 
cross section limits for this analysis.  Comparing these curves to the 
cross section prediction from PYTHIA, mass limits are obtained at
$190\gevcc$ for \bp\ and $220\gevcc$ for $t'$.
\begin{figure}[t]
\begin{minipage}[t]{.50\linewidth}
\centerline{\protect\psfig{figure=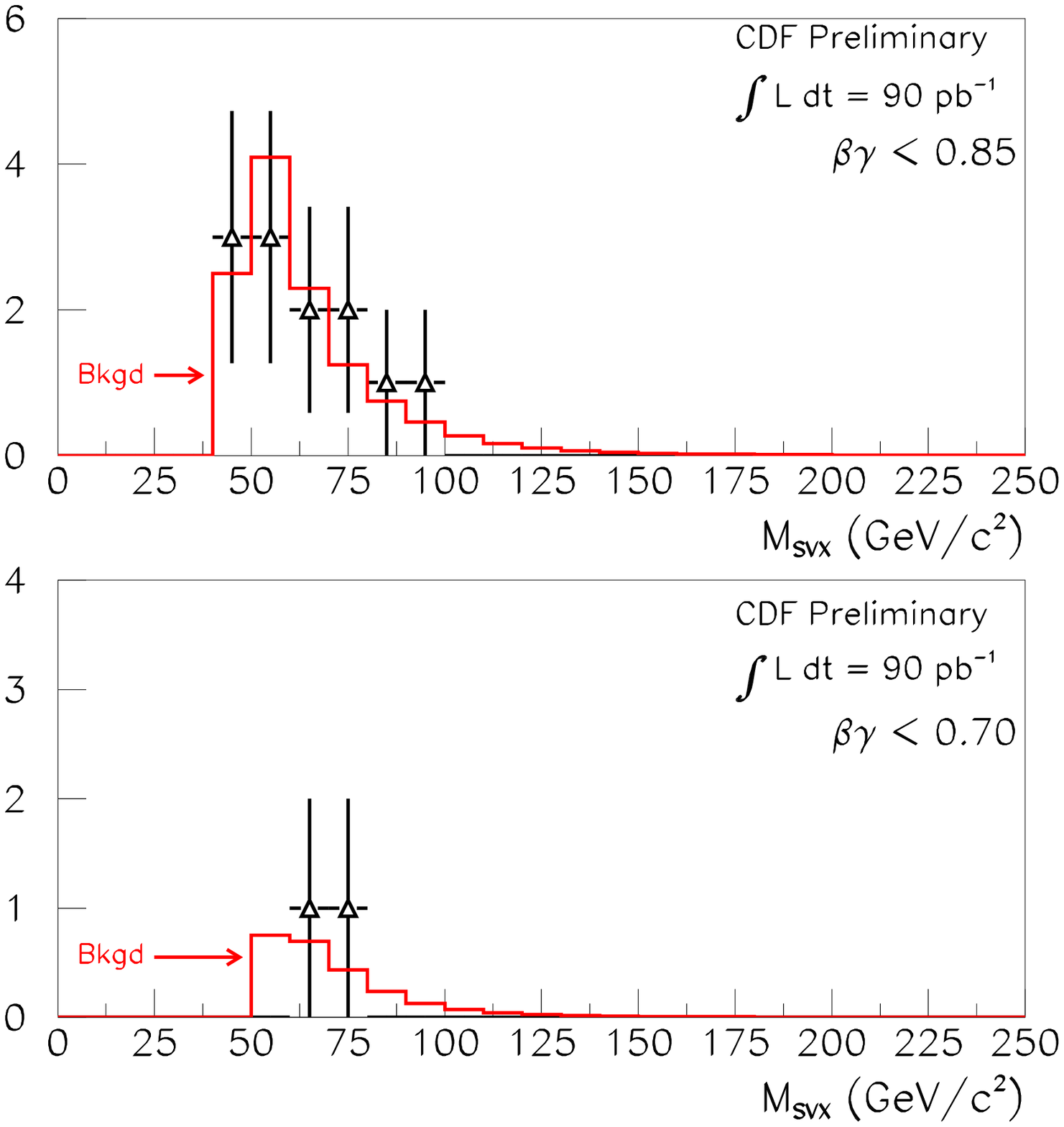,width=\linewidth}}
\end{minipage}\hfill
\begin{minipage}[t]{.50\linewidth}
\centerline{\protect\psfig{figure=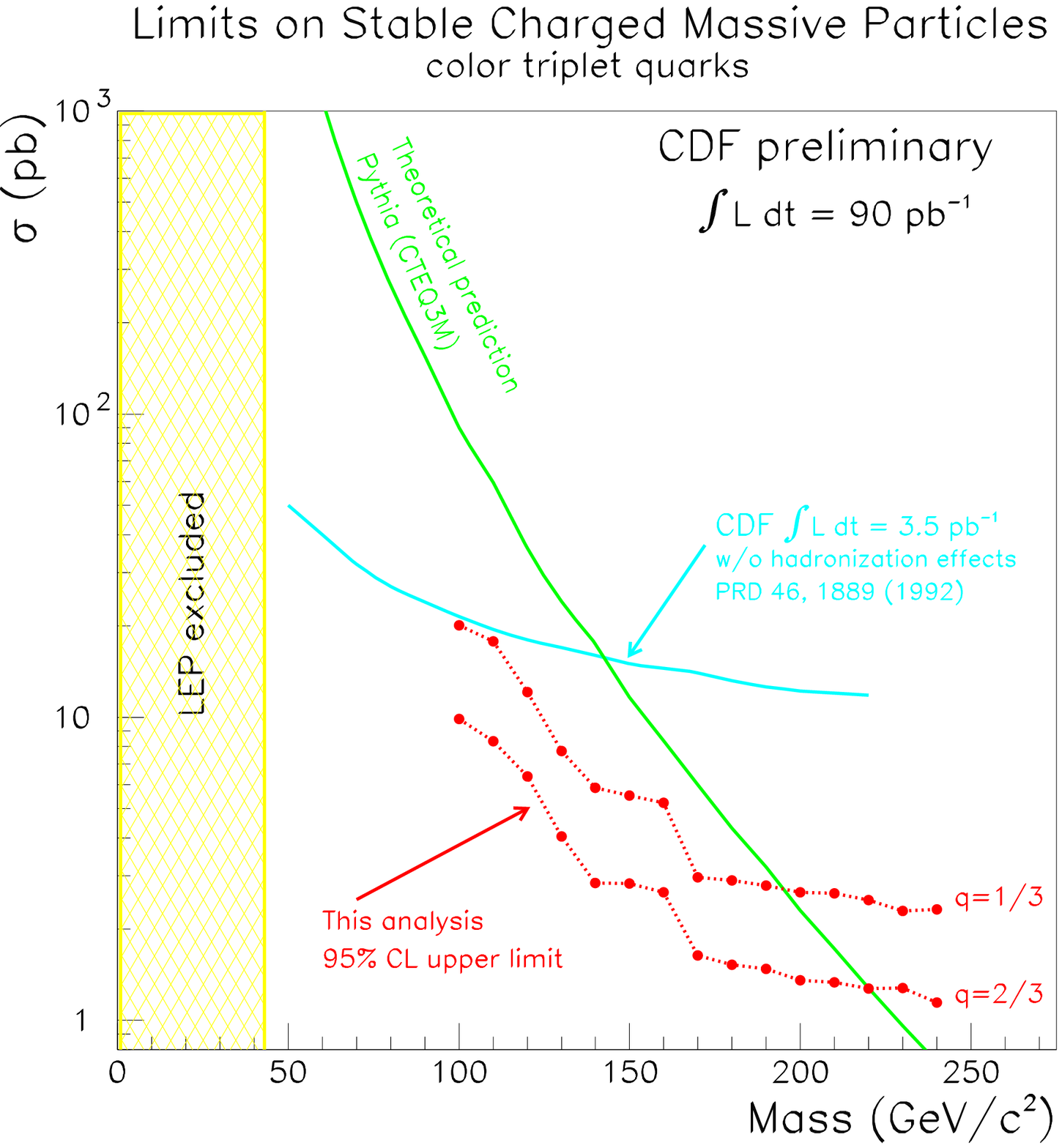,width=\linewidth}}
\end{minipage}
  \caption{Search for strongly interacting 
           CHAMPs.  Masses reconstructed using the SVX for two values
        of the cut on $\beta\gamma$ (left) and
        mass limit (right).}
  \label{fig:champ}
\end{figure}

In Run II, the CHAMP search will be enhanced by the larger integrated
luminosity, higher cross section ($\sim40\%$), and improved detector
acceptance ($\sim80\%$) due to the new tracking chambers.  Moreover, 
there is a proposal to include a time of flight system to the CDF
detector.  This will greatly help searches for weakly interacting
CHAMPs, and will improve the acceptance for 
analysis described here by $\sim50\%$. 
Combining these factors, the Run II cross section limit for strongly
interacting CHAMPs should reduce to 20 fb. 

\section{New Gauge Bosons at the Tevatron Run II}
\noindent
\centerline{\large\it T. Rizzo}
\medskip
\label{section:Rizzo}

\def\Re{{\cal R \mskip-4mu \lower.1ex \hbox{\it e}\,}}
\def\Im{{\cal I \mskip-5mu \lower.1ex \hbox{\it m}\,}}
\def\ie{{\it i.e.}}
\def\eg{{\it e.g.}}
\def\etc{{\it etc}}
\def\etal{{\it et al.}}
\def\ibid{{\it ibid}.}
\def\tev{\,{\ifmmode\mathrm {TeV}\else TeV\fi}}
\def\gev{\,{\ifmmode\mathrm {GeV}\else GeV\fi}}
\def\mev{\,{\ifmmode\mathrm {MeV}\else MeV\fi}}
\def\to{\rightarrow}
\def\slash{\not\!}
\def\mh{\ifmmode m\sbl H \else $m\sbl H$\fi}
\def\mch{\ifmmode m_{H^\pm} \else $m_{H^\pm}$\fi}
\def\mt{\ifmmode m_t\else $m_t$\fi}
\def\mc{\ifmmode m_c\else $m_c$\fi}
\def\mz{\ifmmode M_Z\else $M_Z$\fi}
\def\mw{\ifmmode M_W\else $M_W$\fi}
\def\mws{\ifmmode M_W^2 \else $M_W^2$\fi}
\def\mhs{\ifmmode m_H^2 \else $m_H^2$\fi}   
\def\mzs{\ifmmode M_Z^2 \else $M_Z^2$\fi}
\def\mts{\ifmmode m_t^2 \else $m_t^2$\fi}
\def\mcs{\ifmmode m_c^2 \else $m_c^2$\fi}
\def\mchs{\ifmmode m_{H^\pm}^2 \else $m_{H^\pm}^2$\fi}
\def\ztwo{\ifmmode Z_2\else $Z_2$\fi}
\def\zone{\ifmmode Z_1\else $Z_1$\fi}
\def\mtwo{\ifmmode M_2\else $M_2$\fi}
\def\mone{\ifmmode M_1\else $M_1$\fi}
\def\tb{\ifmmode \tan\beta \else $\tan\beta$\fi}
\def\xw{\ifmmode x\subw\else $x\subw$\fi}
\def\ch{\ifmmode H^\pm \else $H^\pm$\fi}
\def\lum{\ifmmode {\cal L}\else ${\cal L}$\fi}
\def\inpb{\,{\ifmmode {\mathrm {pb}}^{-1}\else ${\mathrm {pb}}^{-1}$\fi}}
\def\infb{\,{\ifmmode {\mathrm {fb}}^{-1}\else ${\mathrm {fb}}^{-1}$\fi}}
\def\epem{\ifmmode e^+e^-\else $e^+e^-$\fi}
\def\ppb{\ifmmode \bar pp\else $\bar pp$\fi}
\def\bsg{\ifmmode B\to X_s\gamma\else $B\to X_s\gamma$\fi}
\def\bsll{\ifmmode B\to X_s\ell^+\ell^-\else $B\to X_s\ell^+\ell^-$\fi}
\def\bstt{\ifmmode B\to X_s\tau^+\tau^-\else $B\to X_s\tau^+\tau^-$\fi}
\def\lamt{\ifmmode \tilde\lambda\else $\tilde\lambda$\fi}
\def\shat{\ifmmode \hat s\else $\hat s$\fi}
\def\that{\ifmmode \hat t\else $\hat t$\fi}
\def\uhat{\ifmmode \hat u\else $\hat u$\fi}
\def\half{\textstyle{{1\over 2}}}
\def\elli{\ell^{i}}
\def\ellj{\ell^{j}}
\def\ellk{\ell^{k}} 
\def\matth{\mathsurround=0pt}
\def\lsim{\mathrel{\mathpalette\atversim<}}
\def\gsim{\mathrel{\mathpalette\atversim>}}
\def\undertext#1{$\underline{\smash{\vphantom{y}\hbox{#1}}}$}

In this section, we 
provide an overview of the capabilities of the Tevatron to both discover 
and explore the couplings of new gauge bosons during Run II. The problems 
associated with identifying a $Z'$ once discovered at the Tevatron and issues 
related to gauge kinetic mixing are also discussed.

\subsection{Conventional $Z'$ and $W'$ Search Reaches}

From the string point of view, new gauge bosons are perhaps one of the most 
natural extensions of the MSSM{\cite {lang}}. The conventional approach 
in searching for new gauge bosons at hadron colliders is via the Drell-Yan 
channel where a resonant, on-shell particle is produced which subsequently 
decays to lepton pairs and is easily observable over any continuum 
background{\cite {over}}. The relevant parameter is thus the product of the 
production cross section times the leptonic branching fraction, $\sigma B$. 
Given a particular extended gauge model with a fixed set of couplings and a 
set of PDF's there is very little theoretical uncertainty associated with the 
calculation of this quantity. This implies that search or exclusion reaches 
are relatively straightforward to establish {\it under the assumption} that 
the new particle can only decay to SM fermion pairs. (It has been shown that 
if new decay modes are responsible for decreasing the leptonic branching 
fraction of a $Z'/W'$ by a factor of 2 it results in a search reach 
degradation of only $\simeq 50-60$ GeV{\cite {over}} at Run II.)

Using the canonical $Z'$ from $E_6$ as an example and remembering that in this 
case the fermionic 
couplings depend on a parameter $\theta$, we show in Fig.~\ref{fig1r} both the 
current exclusion reach from Run I as well as the search reach anticipated 
from Run II. As the $Z'$ couplings vary the search reach also varies over a 
respectable range of $\simeq 150$ GeV. We also see from this figure that this 
spread of values is quite typical given the fairly wide set of $Z'$ models. 
Fig.~\ref{fig1r} also shows that for a 
rather wide range of models, including the $E_6$ and several Left-Right cases, 
the search reach scales almost linearly with the log of the integrated 
luminosity over the range of values relevant for future Tevatron running. In 
fact a fit to the curves in Fig.~\ref{fig1r} 
reveals that the mass reach scales 
with luminosity in a quite model-independent manner as 
$M\simeq M_0+91.3 \log L-2.68 \log^2 L$ GeV with ${\cal L}$ in 
$fb^{-1}$ with $M_0$ being the only model-dependence.

\begin{figure}[htbp]
\centerline{
\protect\psfig{figure=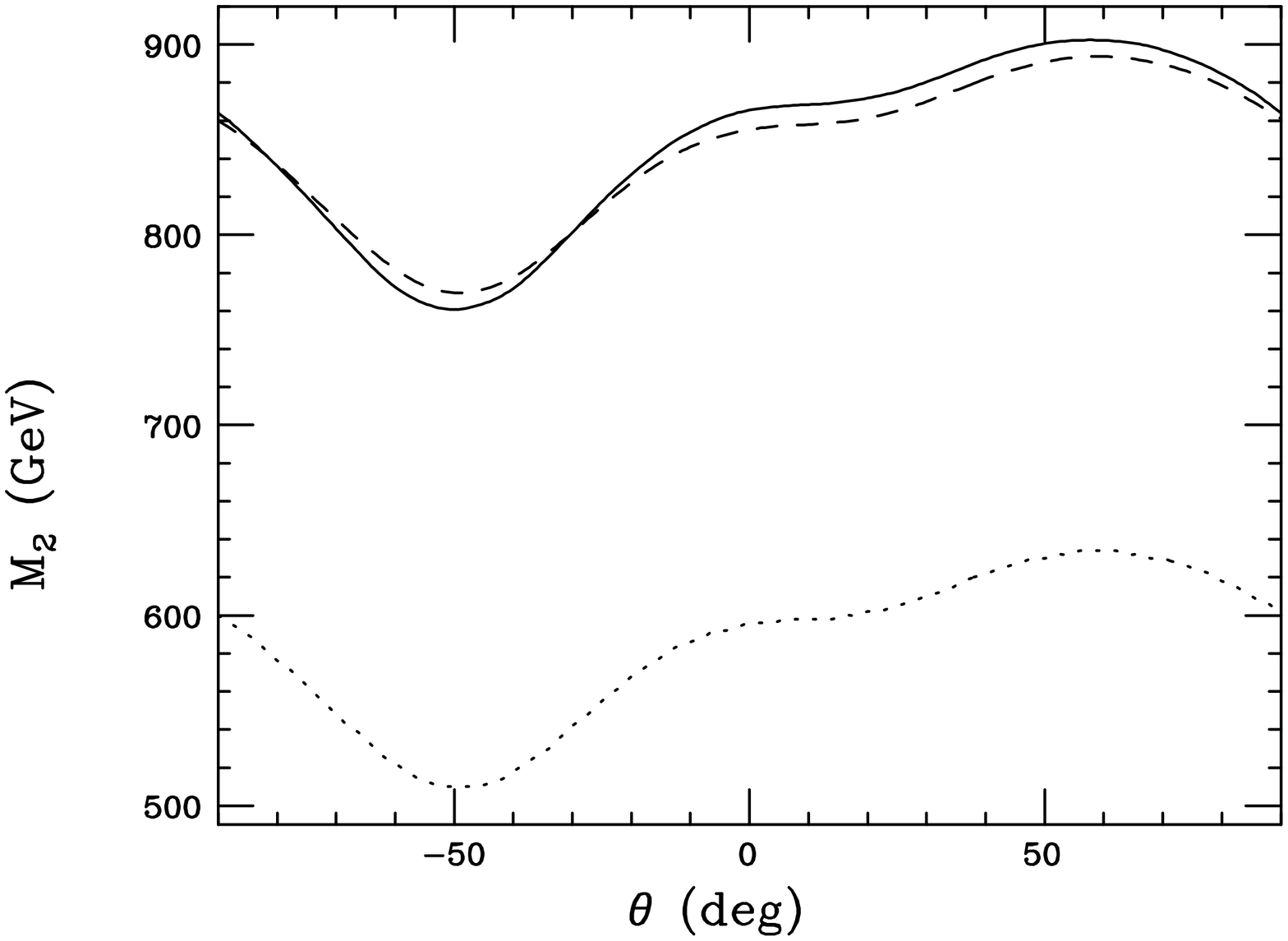,height=8.0cm,width=9.1cm,angle=-90}
\hspace*{-5mm}
\protect\psfig{figure=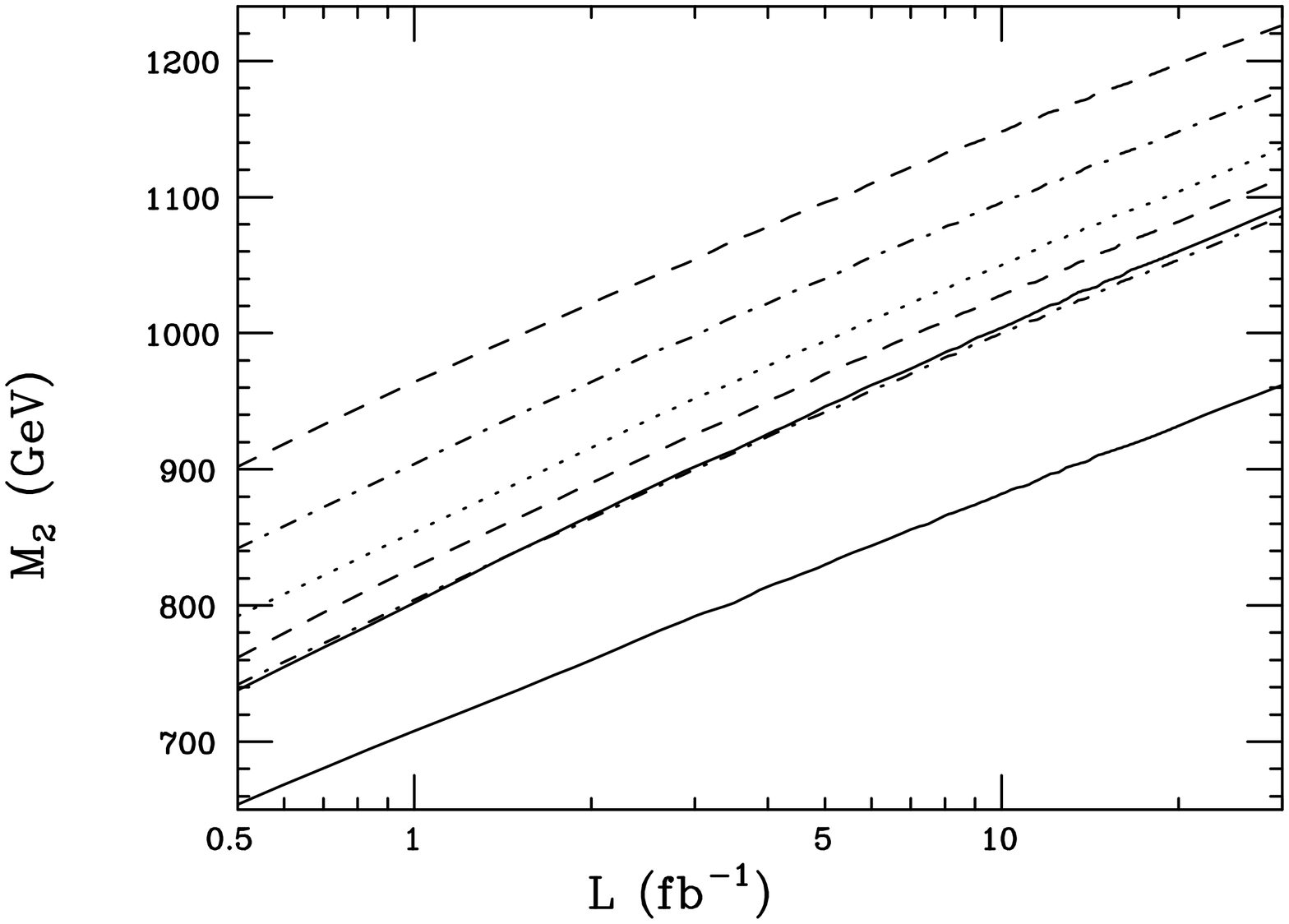,height=8.0cm,width=9.1cm,angle=-90}}
\caption{(Left) Approximate $95\%$CL exclusion limit (dots) from Run I and the 
anticipated Run II (2 $fb^{-1}$) discovery reach (solid and dash) for a $Z'$ 
arising from $E_6$ as a function of the parameter $\theta$ assuming decays to 
only SM fermions. The two curves represent the results obtained by employing 
CTEQ4M and MRST98~\protect\cite{pdf} PDF's. (Right) Scaling behavior of $Z'$ 
search reaches with integrated luminosity with Run II conditions for seven 
different extended models. Note the spread of $\simeq 300$ GeV at any given 
luminosity.}
\label{fig1r}
\end{figure}
\vspace*{0.5mm}

The situation with $W'$ search reaches is somewhat different as the canonical 
example is the Left-Right Model{\cite {over}}. Here not only can the overall 
$W'$ coupling strength, $g_R$, differ from that of the $W$ in the SM, $g_L$, 
but the $W'$ production cross section may be modulated by a distinct, 
right-handed CKM matrix, $V_R$, and its leptonic decay may involve a massive 
neutrino, $N$, which can decay in the detector (and thus not appear as missing 
$p_t$, as has been studied by D0{\cite {d0}}). The variation of the search 
reach with $V_R$, whose structure is {\it a priori} unknown, assuming 
$g_L=g_R$ and massless neutrinos in the final state is shown in 
Fig.~\ref{fig2r}. While the reach in the case $V_R=V_L$ is $\simeq 1040$ GeV, 
a serious degradation is experiences as one scans over other possible forms 
of $V_R$ and is easily reduced by more than $20-30\%$ and perhaps as much as 
$50\%$. Fig.~\ref{fig2r} also 
shows the almost linear scaling of the $W'$ search reach with the log of the 
integrated luminosity in the most naive case; a fit gives to the curve yields 
$M\simeq M_0'+76.3 \log L-1.86 \log^2 L$ GeV for the reach. 

A last point to remember 
regarding the $W'$ reach is that if $N$ is heavier than the $W'$ then 
the Drell-Yan search becomes useless and other modes such as $W'\to WZ$ or 
$jj$ must be employed{\cite {other}}. By scaling existing searches for dijet 
mass bumps to the energy and luminosity of Run II we have estimated that a $W'$ 
can be discovered in this mode up to masses of 880(970) GeV for a luminosity
of 2(30) $fb^{-1}$.

\begin{figure}[htbp]
\centerline{
\psfig{figure=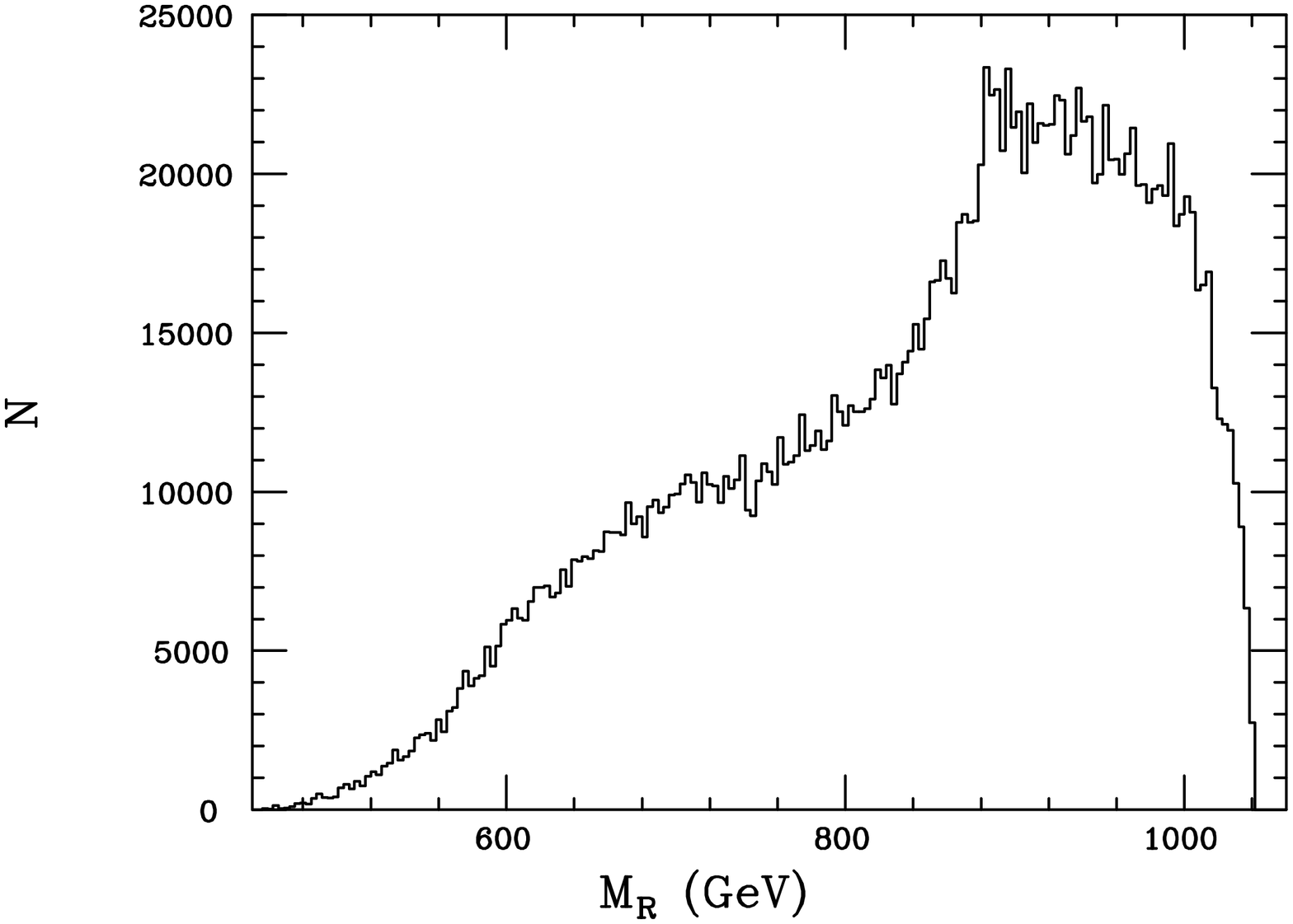,height=8.0cm,width=9.1cm,angle=-90}
\hspace*{-5mm}
\psfig{figure=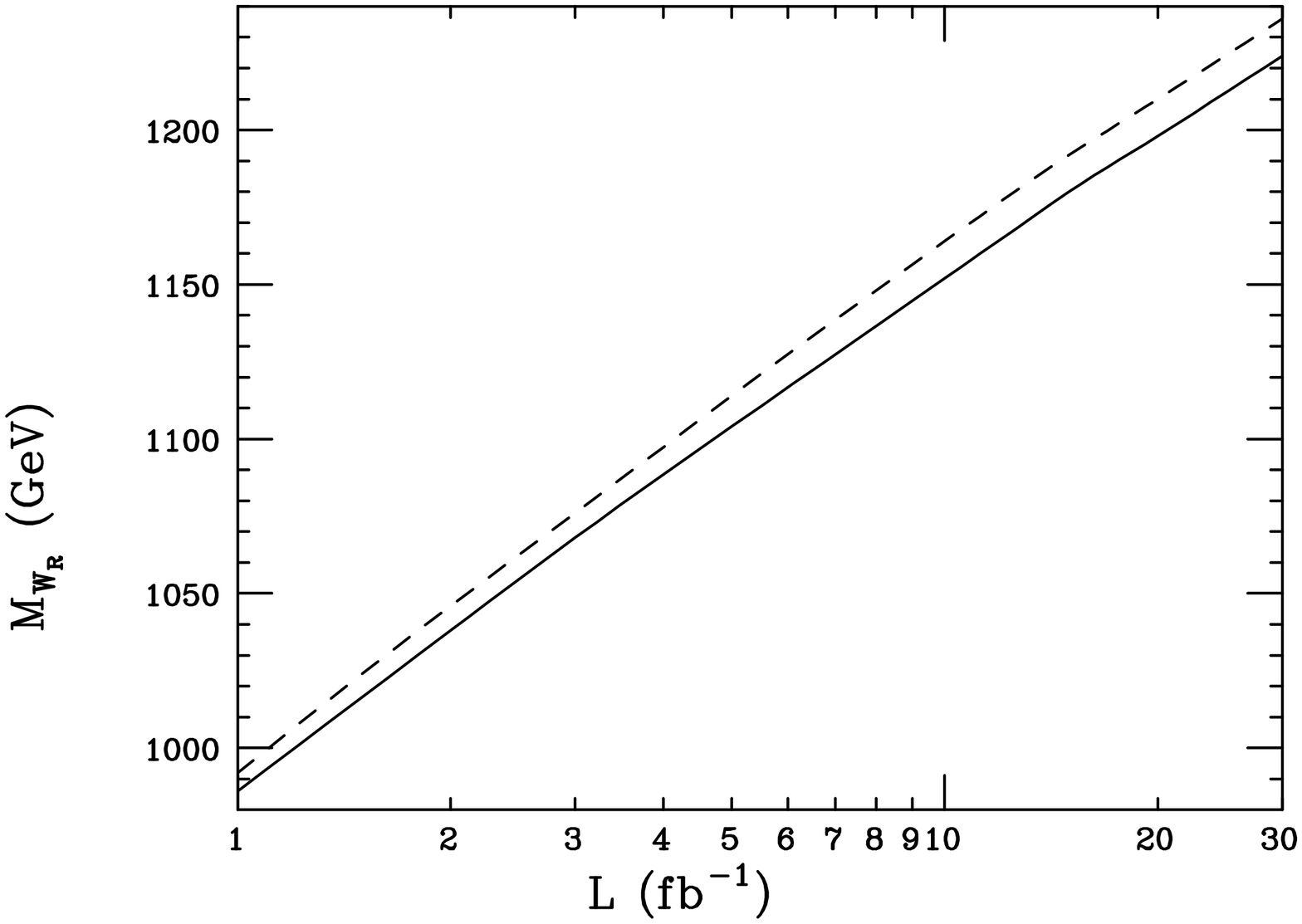,height=8.0cm,width=9.1cm,angle=-90}}
\caption{(Left) Distribution of $W_R$ search reaches at Run II for 
$2\times 10^6$ different sets of mixing angles and phases in the matrix $V_R$. 
(Right) Search reach as a function of integrated luminosity for $W'$ assuming 
the simplest scenario: $\kappa=g_R/g_L=1$, $V_L=V_R$ and essentially massless 
and stable neutrinos in the final state, for the same PDF's as in 
Fig.~\ref{fig1r}.}
\label{fig2r}
\end{figure}
\vspace*{0.5mm}

\subsection{$Z'$ Coupling Determinations and Gauge Kinetic Mixing}

If new gauge bosons do indeed exist not far beyond the present limits then 
they will be easily be discovered during Run II. It then becomes mandatory to 
address the next question--what $Z'$ is it? To do this one needs to measure 
as many of the various $Z'$ properties as possible. While such discussions 
have taken place for the LHC{\cite {over}}, little has been done to address 
these issues at the Tevatron. We would suppose that if a $\sim 700-800$ GeV 
$Z'$ were discovered accumulating additional luminosity would be easily 
justified and that at least a few hundreds of events would eventually be 
collected in both the $e^+e^-$ and $\mu^+\mu^-$ channels. Even with this 
number of events only a few `on-peak' observables will have statistical errors 
less than $\sim 10\%$ and are thus useful for coupling determinations: ($i$) 
the familiar forward-backward asymmetry of the final leptons, $A_{fb}$, 
($ii$) the longitudinal polarization of one of the $\tau$'s in 
$Z\to \tau^+\tau^-$, 
$P_\tau$, and ($iii$) the relative rate for $Z'\to b\bar b$ compared to 
$\ell^+\ell^-$, $R_{bl}${\cite {x2bb}}. To see that indeed these three 
variables, if reasonably well measured, will be able to separate many of the 
more popular $Z'$ models, we compare their correlated values in 
Fig.~\ref{fig3r}. 
Note that the couplings in some cases, such as the Left-Right 
Model, depend upon the values of a single continuous parameter and so their 
predictions lie along specific curves and not at unique points. It is clear 
that a determination of only one of these variables will not be sufficient and 
generally all three are necessary for good model separation. 

\begin{figure}[htbp]
\centerline{
\psfig{figure=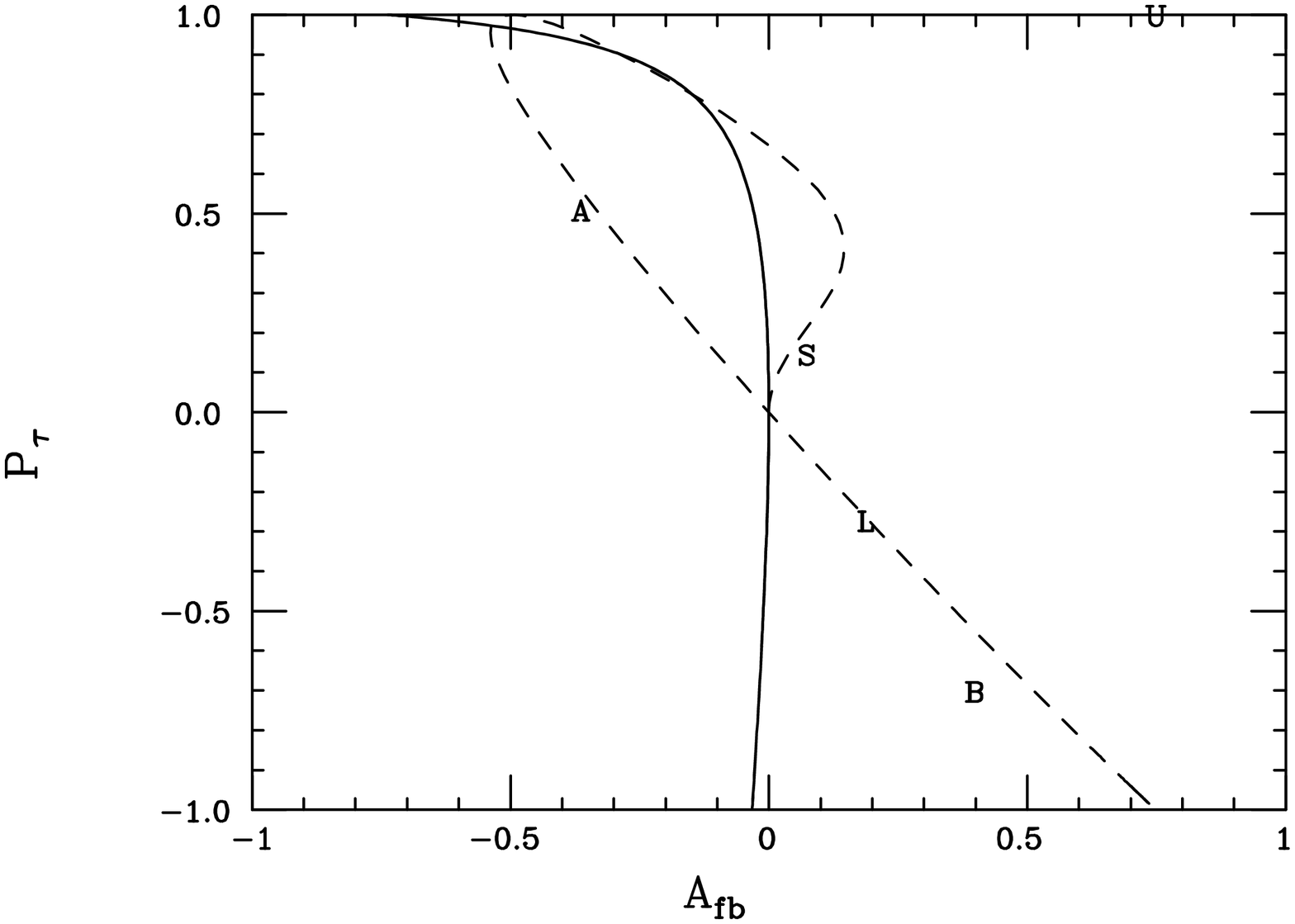,height=8.0cm,width=9.1cm,angle=-90}
\hspace*{-5mm}
\psfig{figure=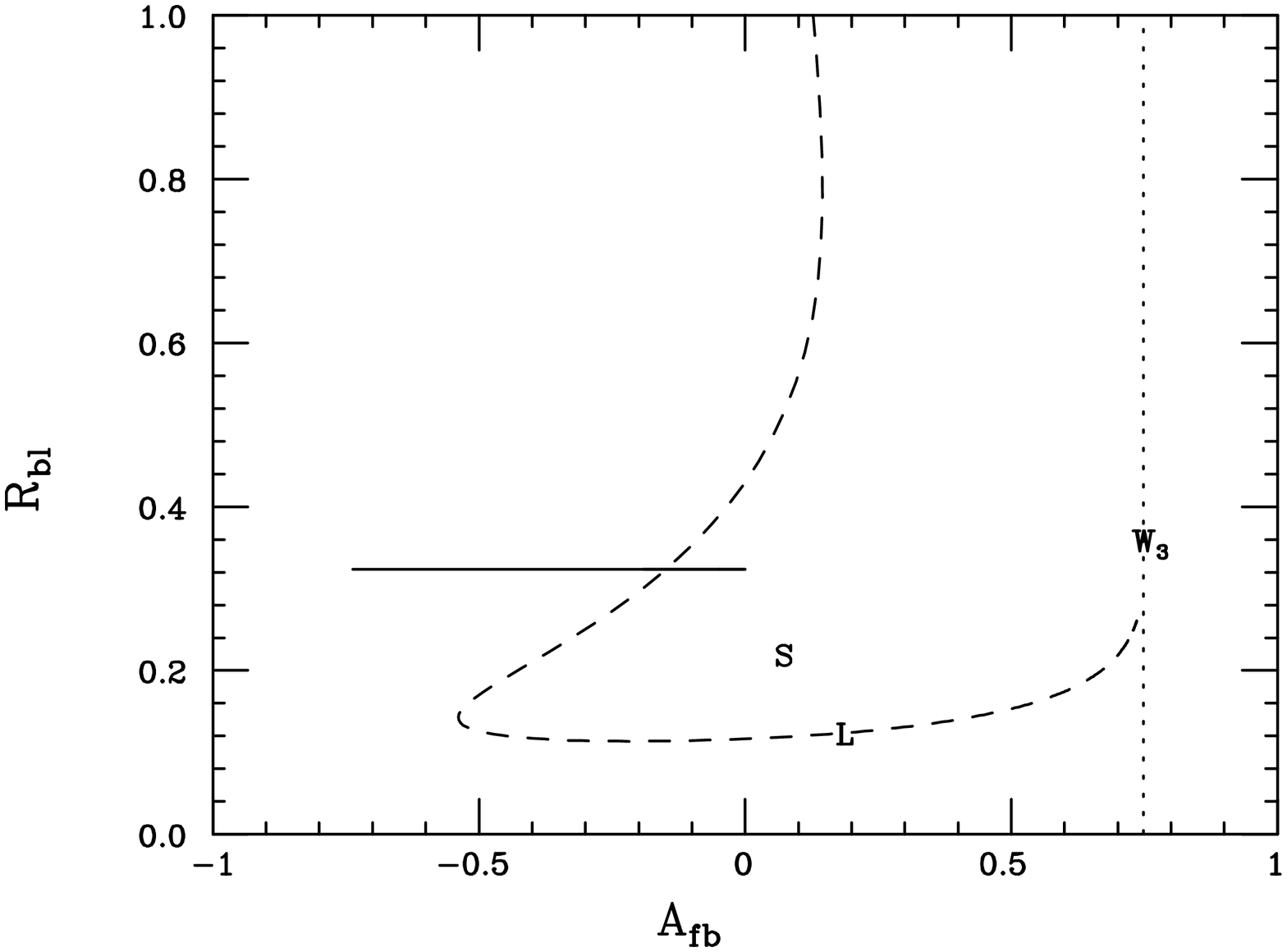,height=8.0cm,width=9.1cm,angle=-90}}
\caption{Values of $P_\tau$, $R_{bl}$ and $A_{fb}$ for a 
number~\protect\cite{over} 
of $Z'$ models at the Tevatron: conventional $E_6$(solid), Left-Right(dash), 
Un-unified(dots or U), Alternative Left-Right(A), L-R with $\kappa=1$(L), 
SM couplings(S), Kaluza-Klein excitations($\rm W_3$ and B). $M_{Z'}=700$ GeV 
has been assumed. Some models are only shown on one panel as their 
predictions lie outside the displayed range.}
\label{fig3r}
\end{figure}
\vspace*{0.5mm}

For extended gauge models based on GUTS with only an additional $U(1)'$ 
factor, such as those arising in the $E_6$ case, the results shown in 
Fig.~\ref{fig3r} can be too optimistic due to the presence of gauge kinetic 
mixing(GKM){\cite {gkm}}. GKM is an induced mixing between the $Z'$ field 
strength and that associated with the SM hypercharge that arises due to 
vacuum polarization-like graphs. At the GUT scale, such graphs cancel since 
complete GUT matter representations exist. Once the GUT symmetry is broken the 
matter content of the theory below that GUT scale no longer lies in complete 
representations. In, \eg, the MSSM only the Higgs doublet superfields remain 
light whereas their associated color triplet partners remain at the GUT scale. 
The existence of incomplete representations then leads to GKM via the RGE's 
and results in a modification of the naive expectations for the $Z'$ couplings 
at the TeV scale. In the case of conventional $E_6$, the $Z'$ couples to a 
charge $Q'_\theta$, which is dictated by group theory and the value of the 
$\theta$ mixing parameter. GKM modifies this coupling as 
$Q'_\theta \to \lambda (Q'_\theta+\delta \sqrt{3\over 5}{Y\over 2})$ where 
$Q_{em}=T_3+{Y\over 2}$. Given a specific matter content of the low energy 
theory below the GUT scale the parameters $\lambda$ and $\delta$ become 
calculable via the RGE's. As far as $Z'$ physics at the Tevatron is concerned 
it is quite fortunate that they cannot take on arbitrary values; for the case 
of the $\eta$-type model($\theta \simeq 37.76^o$) made popular by string 
theory, $\delta={-1\over 3}$ leads to leptophobia. In this case the $Z'$ does 
not couple to leptons and the standard Drell-Yan search technique fails. This 
is evident by the `hole' in the search reach shown in Fig.~\ref{fig4r}. Given 
a reasonable set of assumptions the matter content of the extended $E_6$ 
model below the GUT scale is not arbitrary, there being only(!) 68 possible 
sets of superfields to consider{\cite {gkm}}. For each set the allowed ranges 
of $\lambda$ and $\delta$ are calculable, as shown in Figs.~4 and 5 of 
Ref.{\cite {gkm}}, from which $Z'$ search reaches can be calculated. From this 
analysis it can be shown that at one loop $-0.286 \leq \delta \leq 0.250$ so 
that exact 
leptophobia does not happen. From a complete scan of the parameter space the 
worse case scenario is found to occur for an $\eta$-type model with 
$\lambda=0.862$ and $\delta=-0.286$ which yields a 
search reach of 482(736) GeV assuming a luminosity of 2(30) $fb^{-1}$. 

\nn
\begin{figure}[htbp]
\centerline{
\psfig{figure=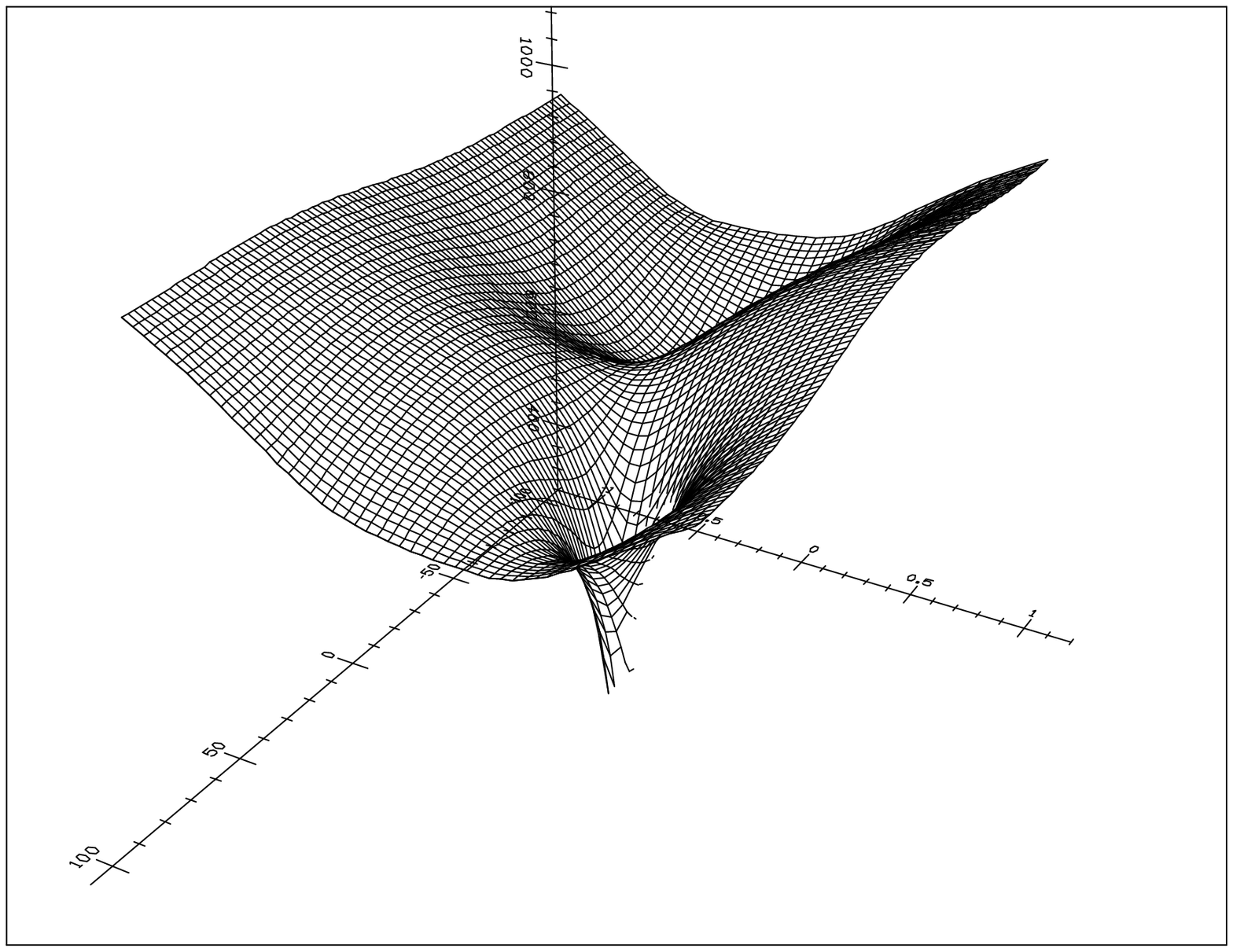,height=11.0cm,width=17cm,angle=-90}}
\vspace*{-1.25cm}
\caption{Search reaches for the $E_6$ $Z'$ at the Tevatron($2~fb^{-1}$) in 
GeV as functions of $\theta$(in degrees on the left axis) and 
$\delta$(right axis) assuming no exotic decay modes; the leptophobic hole is 
evident. The sign of $\delta$ has been reversed in this plot for ease of 
viewing and $\lambda=1$ has been assumed. The dimple in the search reach 
occurs when the $Z'$ coupling to $u$-quarks vanishes.}
\label{fig4r}
\end{figure}
\vspace*{0.5mm}

GKM also affects the $E_6$ predictions for $P_\tau$, $R_{bl}$ and $A_{fb}$ 
since these now depend on the additional parameters $\delta,\lambda$, making 
model separation more difficult. 
Fig.~\ref{fig5r} 
shows the result of a Monte Carlo scan of the allowed space of 
$\theta-\delta-\lambda$ values; we see that the locus of $E_6$ points no longer 
lie along a pair of lines but are now in fairly broad (statically) shaded 
regions. These plots explicitly show the extreme importance of having multiple 
observables available for model separation. While the models labelled by 
A, B, L, U, and $\rm W_3$ remain easily distinguishable from one another and 
$E_6$ if all three observables are well measured, there is now a reasonably 
large overlap between the Left-Right case and $E_6$. Similarly, a $Z'$ with 
SM couplings(S) is now no longer separable from $E_6$ even when all three 
observables are employed. Having more observables available would clearly 
prove useful.

\begin{figure}[htbp]
\centerline{
\psfig{figure=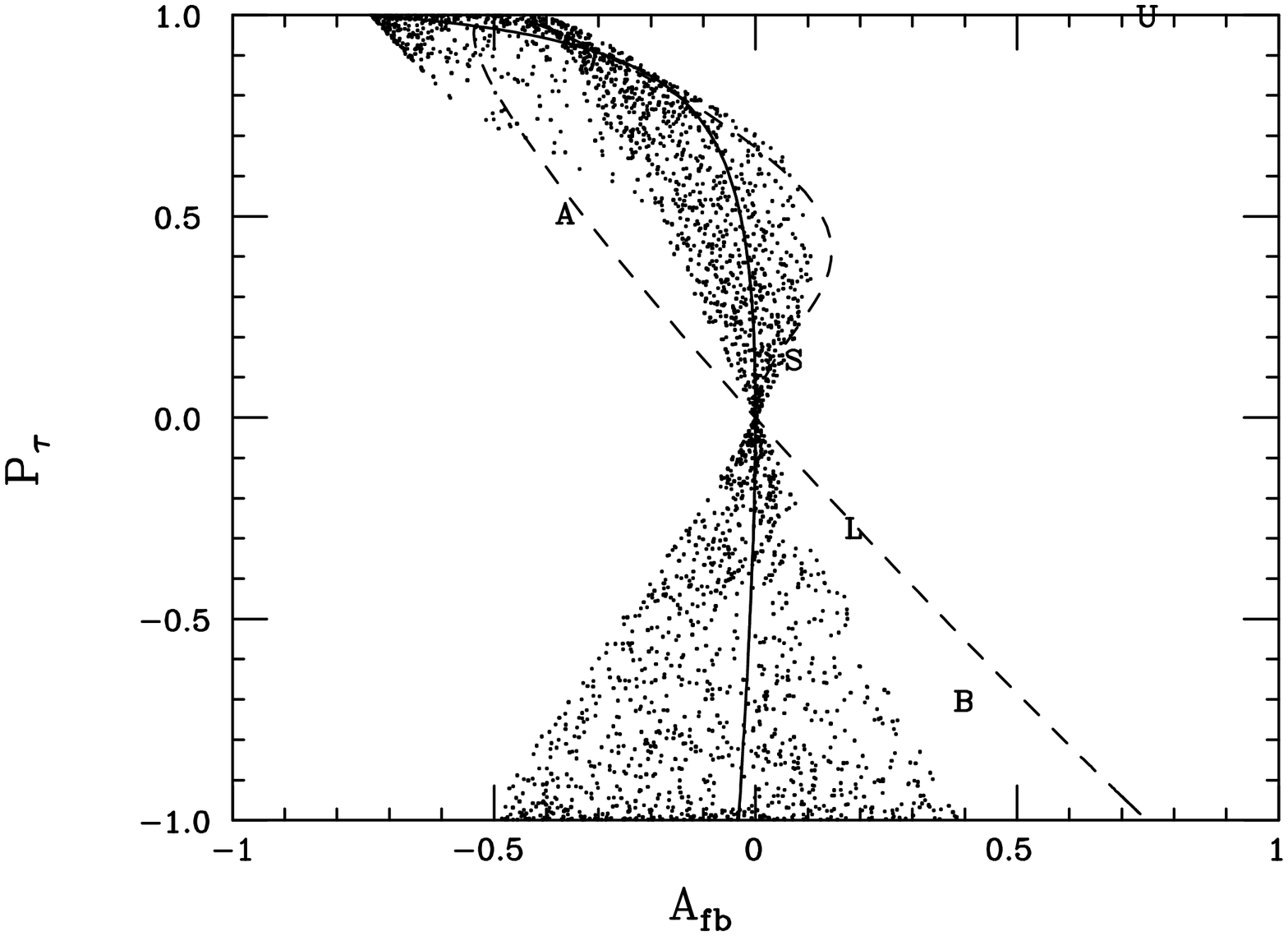,height=8.0cm,width=9.1cm,angle=-90}
\hspace*{-5mm}
\psfig{figure=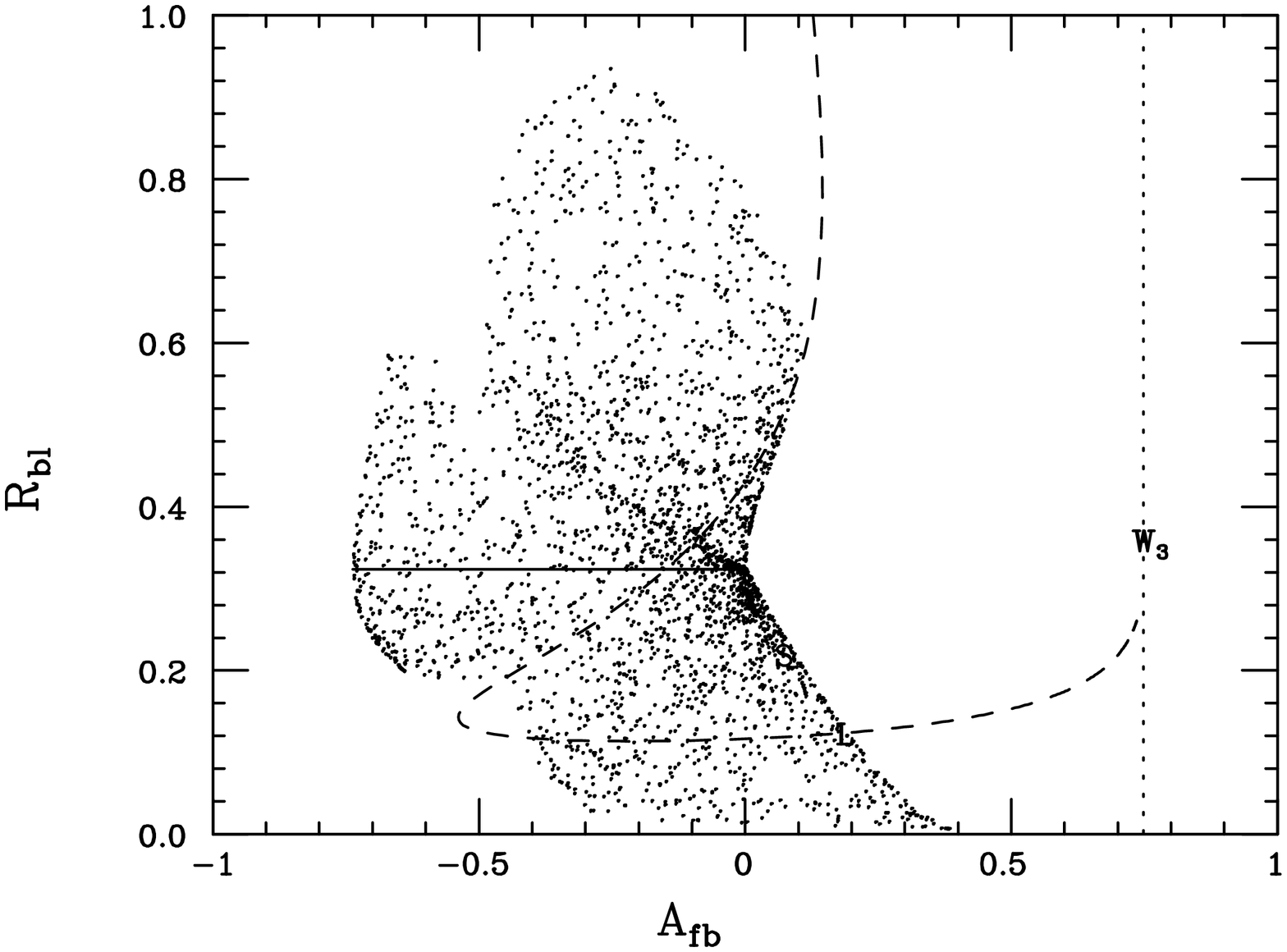,height=8.0cm,width=9.1cm,angle=-90}}
\caption{Same as Fig.~\ref{fig3r} but now showing the effects of gauge 
kinetic mixing on the couplings of the $Z'$ in $E_6$.}
\label{fig5r}
\end{figure}
\vspace*{0.5mm}

\section{Conclusions}
\medskip
\label{section:Conclusions}

We hope that the reader has been impressed with the variety of
different possible phenomenological manifestations for supersymmetry.
We have seen that some versions of supersymmetry could present significant
challenges, while other versions will be either discovered or eliminated
very soon during Run II. It seems very likely that at least a few of
the supersymmetric particles will be observed during Run II.


\end{document}